\documentclass[5p, times, a4paper]{elsarticle}
\usepackage{xcolor}
\usepackage{setspace}
\usepackage[breaklinks, colorlinks]{hyperref}
\usepackage{url}

\usepackage{breakurl}
\usepackage{graphicx}
\usepackage{pdfpages}
\usepackage{caption}
\usepackage{subcaption}
\usepackage{amsfonts}
\usepackage{amsmath}
\usepackage{amsthm}
\usepackage{tabularx}
\usepackage{enumitem}
\usepackage{array}
\usepackage{booktabs}
\usepackage{float}
\usepackage{multicol}
\usepackage{multirow}
\usepackage{ragged2e}
\usepackage{bbding}
\usepackage{pifont}
\usepackage{wasysym}
\usepackage{multicol}
\usepackage{authorbiography}
\usepackage{wrapfig}
\usepackage{array}
\usepackage{amssymb}
\usepackage{diagbox}
\usepackage{pdfrender}
\usepackage{nccmath}
\usepackage{xpatch}
\xpatchcmd{\NCC@ignorepar}{%
\abovedisplayskip\abovedisplayshortskip}
{%
\abovedisplayskip\abovedisplayshortskip%
\belowdisplayskip\belowdisplayshortskip}
{}{}

\newcommand{\saizhuoDrafting}{}
\ifdefined\saizhuoDrafting
\newcommand{\todo}[1]{{\color{blue}TODO: {#1}}}
\newcommand{\saizhuo}[1]{{\color{red}Saizhuo: {#1}}}
\else
\newcommand{\todo}[1]{}
\newcommand{\saizhuo}[1]{}
\fi

\ifdefined\qiComments
\newcommand{\qi}[1]{{\color{green}Qi: {#1}}}
\else
\newcommand{\qi}[1]{}
\fi

\newcommand{\saizhuoFeedback}{}
\ifdefined\saizhuoFeedback
\newcommand{\feedback}[1]{{\color{orange}Reply: #1}}
\else
\newcommand{\feedback}[1]{}
\fi

\newcommand*{\boldcheckmark}{%
  \textpdfrender{
    TextRenderingMode=FillStroke,
    LineWidth=.5pt, 
  }{\checkmark}%
}

\newcommand{\ItemizeGlobal}[1]{
\begin{itemize}[leftmargin=*]
    #1
\end{itemize}
}

\newcommand{\invis}[1]{}

\newcommand{\authorBib}[3]{
\vspace{2.5em}
\setlength\intextsep{0pt} 
\begin{wrapfigure}{l}{0.129\textwidth}
    \centering
    \includegraphics[width=1.2\linewidth]{#1}
\end{wrapfigure}
\noindent \textbf{#2} #3
}

\newcolumntype{P}[1]{>{\centering \arraybackslash}p{#1}}
\newcolumntype{L}[1]{>{\raggedright \arraybackslash}p{#1}}
\newcolumntype{C}{>{\centering \arraybackslash}X}
\newcolumntype{R}{>{\raggedright \arraybackslash}X}
\newcolumntype{M}[1]{>{\centering\arraybackslash}m{#1}}

\AtBeginDocument{\hypersetup{pdfborder={0 0 1}}}

\newenvironment{tightenumerate}{
\begin{enumerate}[leftmargin=*,topsep=2pt]
  \setlength{\itemsep}{2pt}
  \setlength{\parskip}{0pt}
}{\end{enumerate}}

\newenvironment{tightitemize}{
\begin{itemize}[leftmargin=*,topsep=2pt]
  \setlength{\itemsep}{2pt}
  \setlength{\parskip}{0pt}
}{\end{itemize}}

\begin{document}
\null
\includepdf[pages={1}]{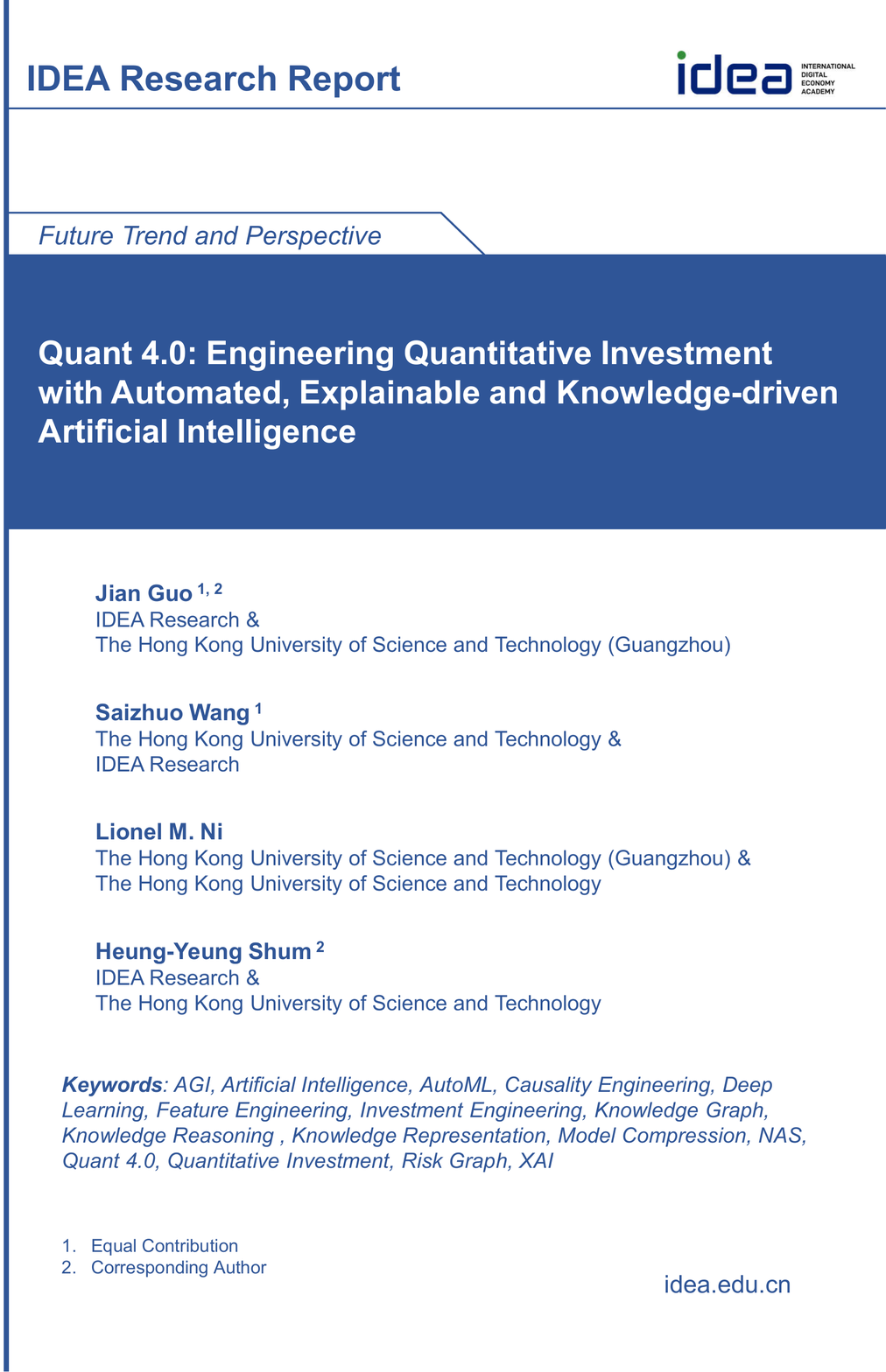}
\newpage
\renewcommand*\listfigurename{\Large \centering List of Figures}
\renewcommand*\contentsname{\Large Table of Contents}
\tableofcontents
\newpage
\setcounter{page}{1}
\title{Quant 4.0: Engineering Quantitative Investment with Automated, Explainable and Knowledge-driven Artificial Intelligence}


\author[1,3]{Jian Guo\fnref{cofirst}\corref{cor1}}\ead{guojian@idea.edu.cn}
\author[1,2]{Saizhuo Wang\fnref{cofirst,intern}}\ead{swangeh@connect.ust.hk}
\author[2,3]{Lionel M. Ni}\ead{ni@ust.hk}
\author[1,2]{Heung-Yeung Shum\corref{cor1}}\ead{hshum@idea.edu.cn}
\cortext[cor1]{Corresponding author.}
\fntext[cofirst]{Equal contribution.}
\fntext[intern]{This work was done during the internship at IDEA Research.}

\affiliation[1]{organization={IDEA Research},
            addressline={International Digital Economy Academy, 5 Shihua Road, Futian District}, 
            city={Shenzhen},
            postcode={518045}, 
            state={Guangdong},
            country={China}}
\affiliation[2]{organization={The Hong Kong University of Science and Technology},
            addressline={Clear Water Bay, Kowloon}, 
            city={Hong Kong},
            postcode={999077}, 
            country={China}}
\affiliation[3]{organization={The Hong Kong University of Science and Technology (Guangzhou)},
            addressline={1st Duxue Road, Nansha District}, 
            city={Guangzhou},
            postcode={518055}, 
            state={Guangdong},
            country={China}}

\begin{abstract}
Quantitative investment (``quant'') is an interdisciplinary field combining financial engineering, computer science, mathematics, statistics, etc. Quant has become one of the mainstream investment methodologies over the past decades, and has experienced three generations: Quant 1.0, trading by mathematical modeling to discover mis-priced assets in markets; Quant 2.0, shifting quant research pipeline from small ``strategy workshops'' to large ``alpha factories''; Quant 3.0, applying deep learning techniques to discover complex nonlinear pricing rules. Despite its advantage in prediction, deep learning relies on extremely large data volume and labor-intensive tuning of ``black-box'' neural network models. To address these limitations, in this paper, we introduce Quant 4.0 and provide an engineering perspective for next-generation quant. Quant 4.0 has three key differentiating components. First, \emph{Automated AI} changes quant pipeline from traditional hand-craft modeling to the state-of-the-art automated modeling, practicing the philosophy of ``algorithm produces algorithm, model builds model, and eventually AI creates AI''. Second, \emph{Explainable AI} develops new techniques to better understand and interpret investment decisions made by machine learning black-boxes, and explains complicated and hidden risk exposures. Third, \emph{Knowledge-driven AI} is a supplement to data-driven AI such as deep learning and it incorporates prior knowledge into modeling to improve investment decision, in particular for quantitative value investing. Moreover, we discuss how to build a system that practices the Quant 4.0 concept. Finally, we propose ten challenging research problems for quant technology, and discuss potential solutions, research directions, and future trends.


\end{abstract}

\maketitle

\thispagestyle{plain}
\pagestyle{plain}

\newcommand{\figureEfficientFrontier}{
    \begin{figure}[!ht] 
        \centering
        \begin{subfigure}{\linewidth}
            \centering
            \includegraphics[width=\linewidth]{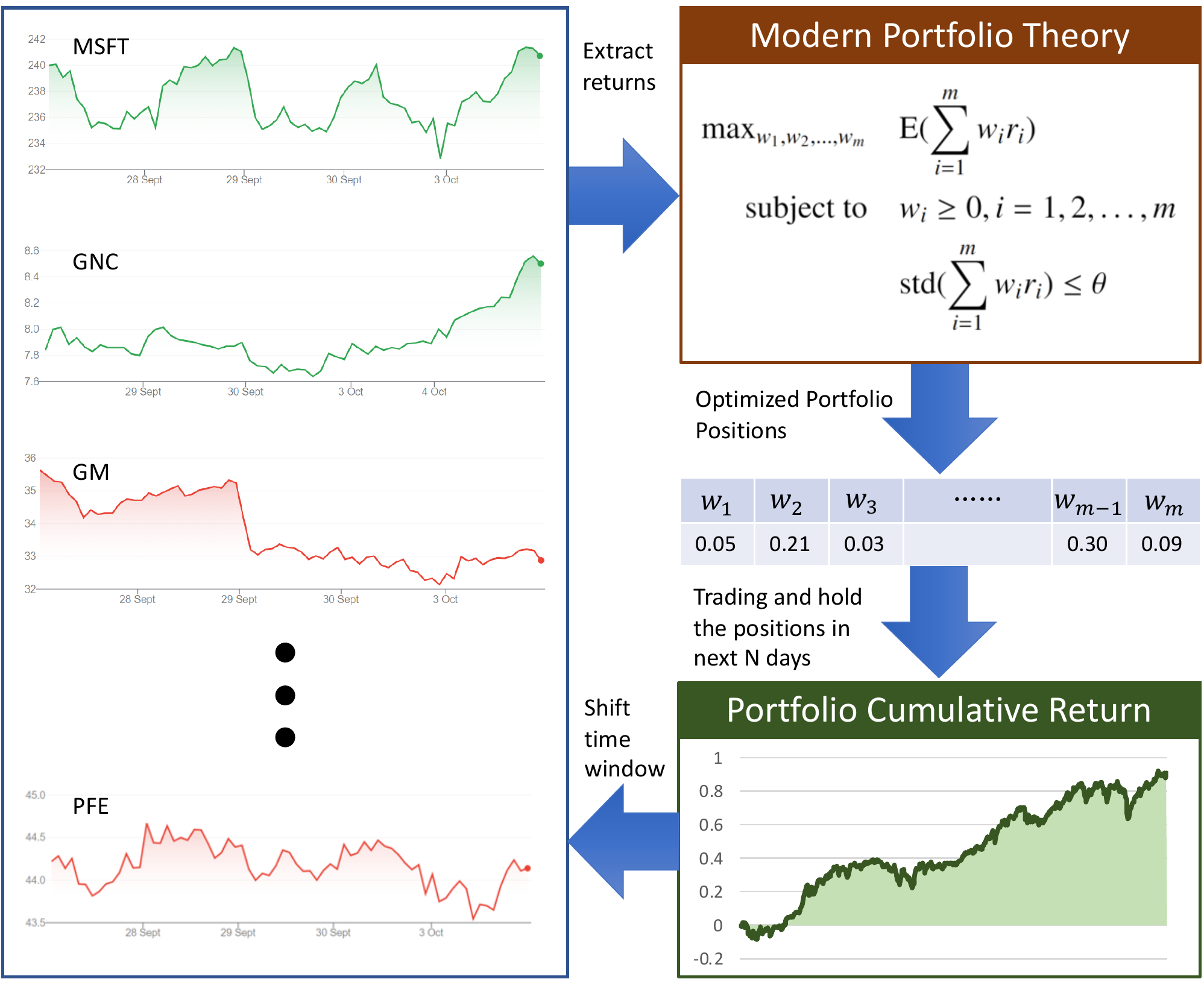}
            \caption{Illustration of Markowitz' portfolio optimization theory.}
        \end{subfigure}
        \hfill
        \begin{subfigure}{\linewidth}
            \includegraphics[width=\linewidth]{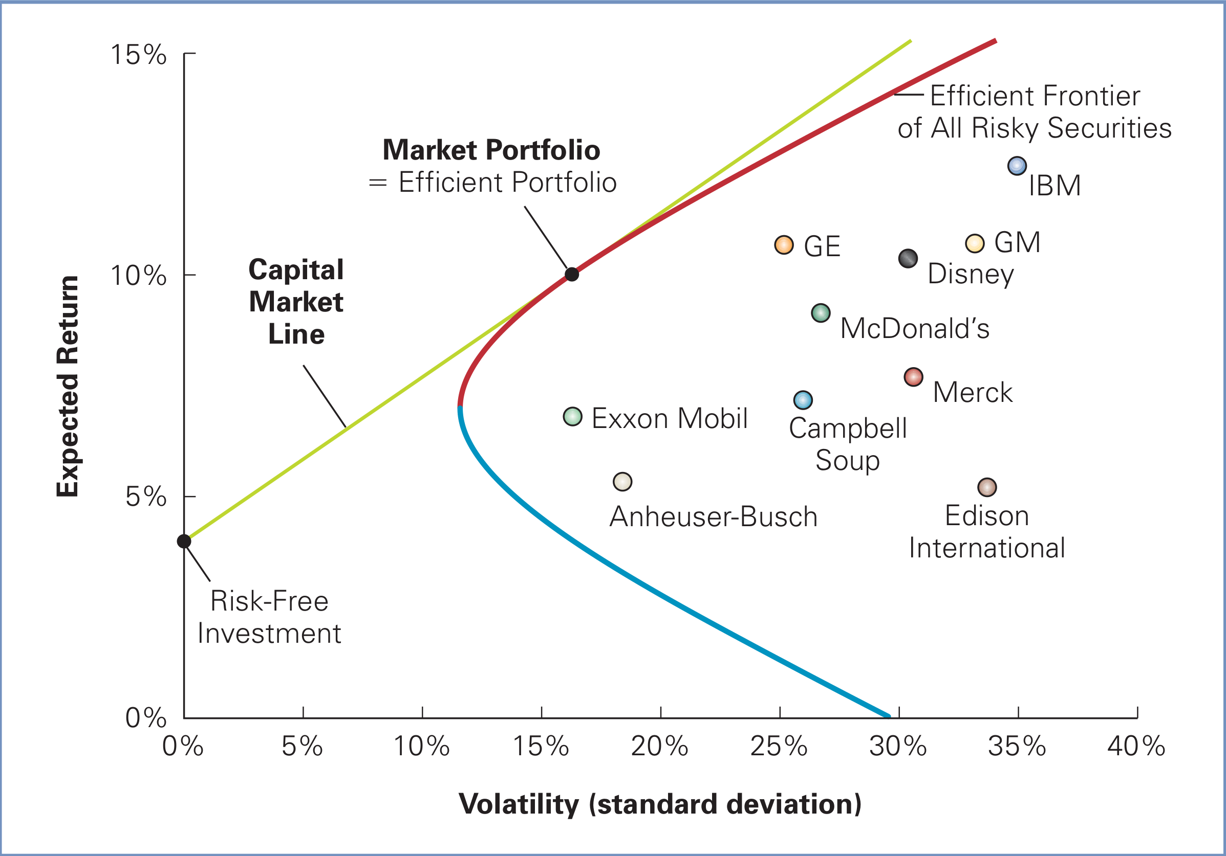}
            \caption{Illustration of efficient frontier first formulated by Harry Markowitz. Figure cited from \cite{berk_corporate_2013}.}
            \label{fig:efficient_frontier}
        \end{subfigure}
        \caption{Portfolio optimization}
        \label{fig:markowitz_theory}
    \end{figure}
}

\newcommand{\figureSystemArch}{
\begin{figure*}[htp]
    \centering
    \includegraphics[height=0.95\textheight]{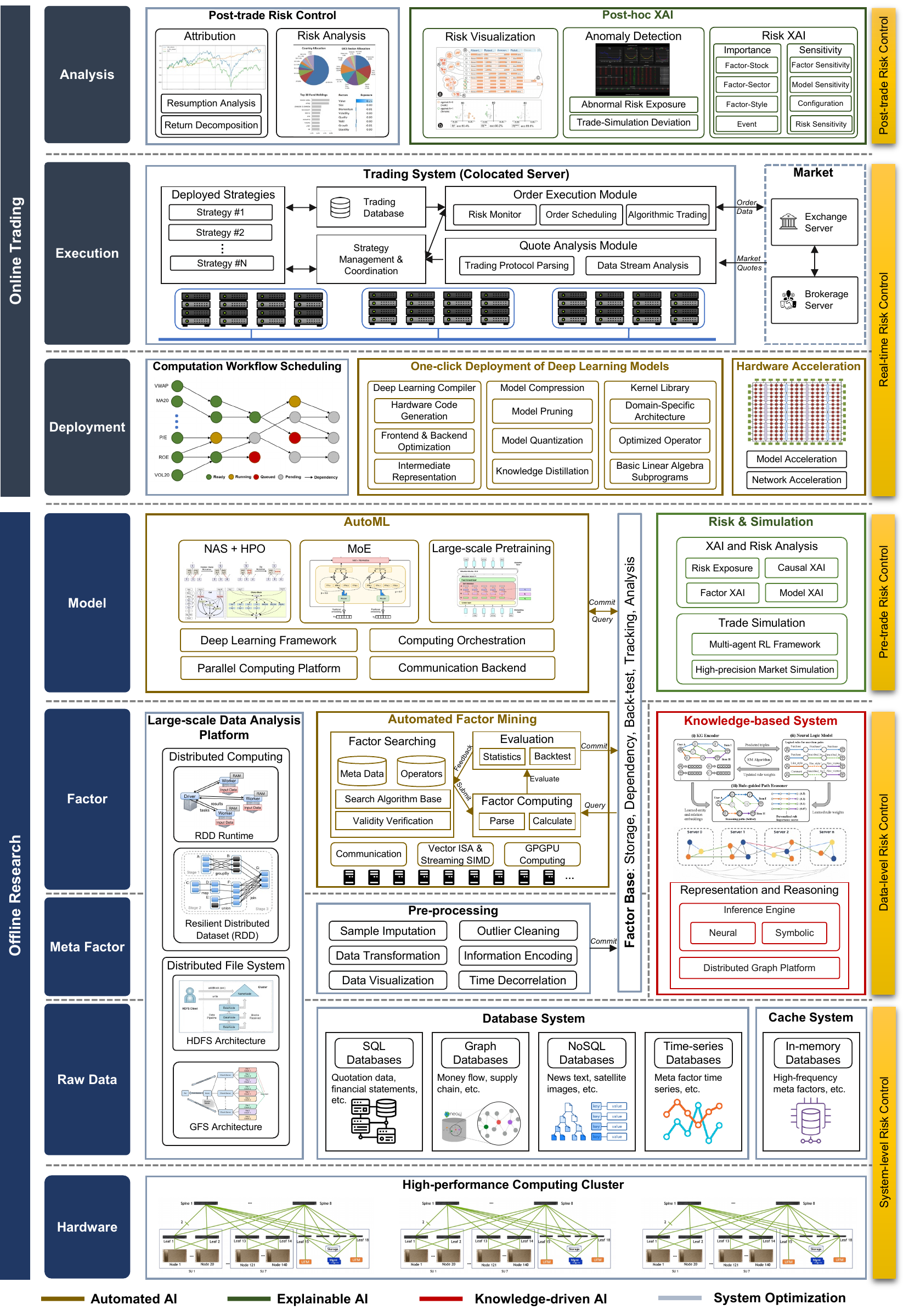}
    \caption{Architecture of an example Quant 4.0 engineering platform for investment research and trading. Part of this figure is cited from \cite{zaharia_resilient_2012, zaharia_spark_2010, wang_visual_2020, ghemawat_google_2003, shvachko_hadoop_2010, fedus_switch_2021, real_regularized_2019, bender_understanding_2018}.}
    \label{fig:system_arch}
\end{figure*}
}

\newcommand{\figureAcademicTimeline}{
    \begin{figure*}[!ht]
        \includegraphics[width=\textwidth]{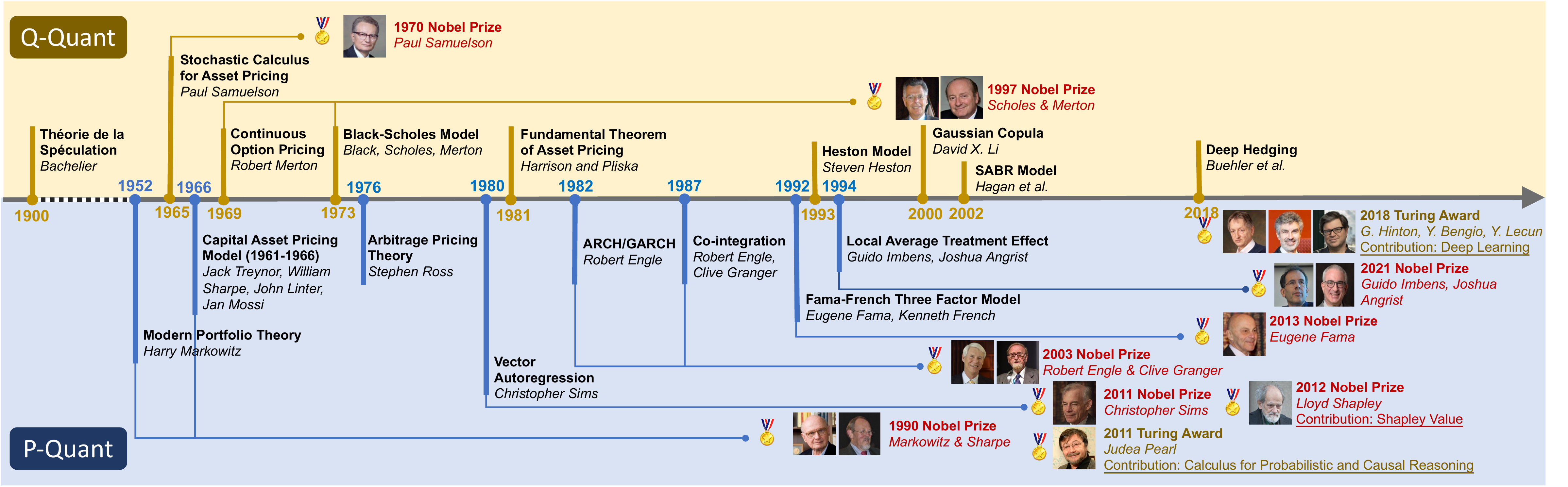}
        \caption{Main academic contributors and their works that deeply influence the development of quantitative investment. Photo credit: Wikipedia.}
        \label{fig:academic_timeline}
    \end{figure*}
}

\newcommand{\figureStrategyClassification}{
    \begin{figure*}[!ht]
        \includegraphics[width=\linewidth]{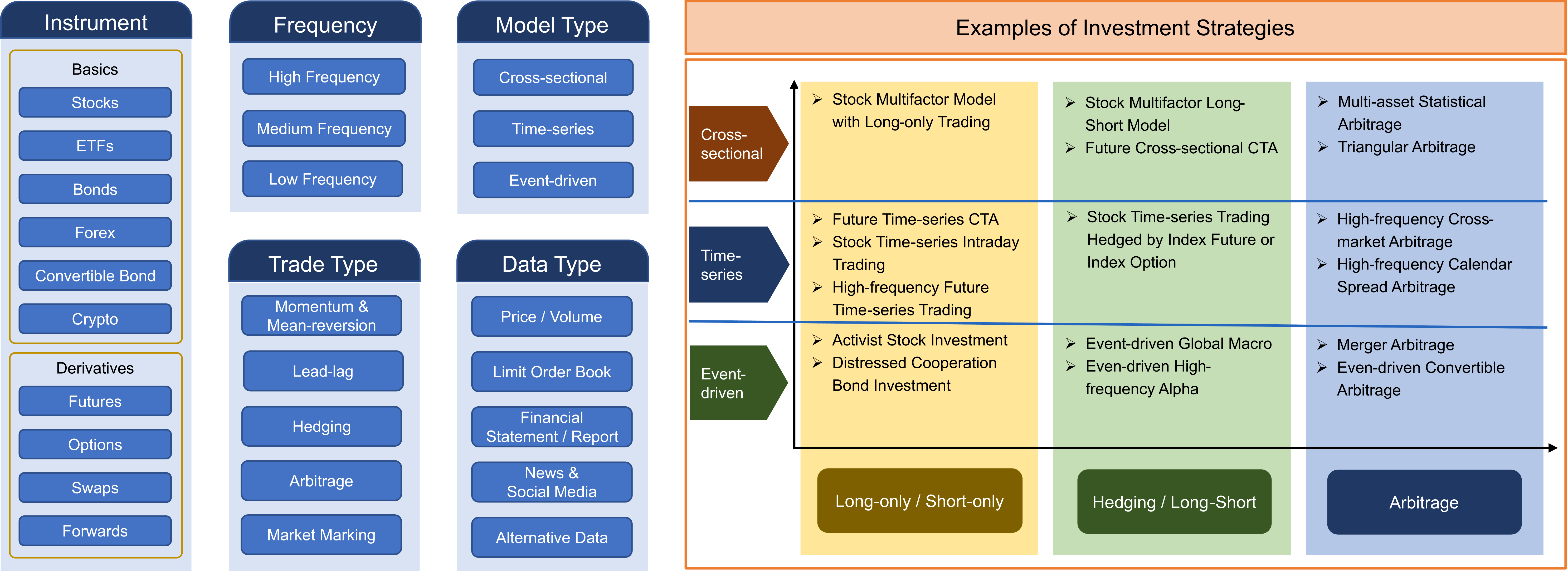}
        \caption{Classification of common strategies and investment instruments. }
        \label{fig:strategy_classification}
    \end{figure*}
}

\newcommand{\figureVersatilityAccuracyExplainability}{
    \begin{figure}[!h]
        \includegraphics[width=\linewidth]{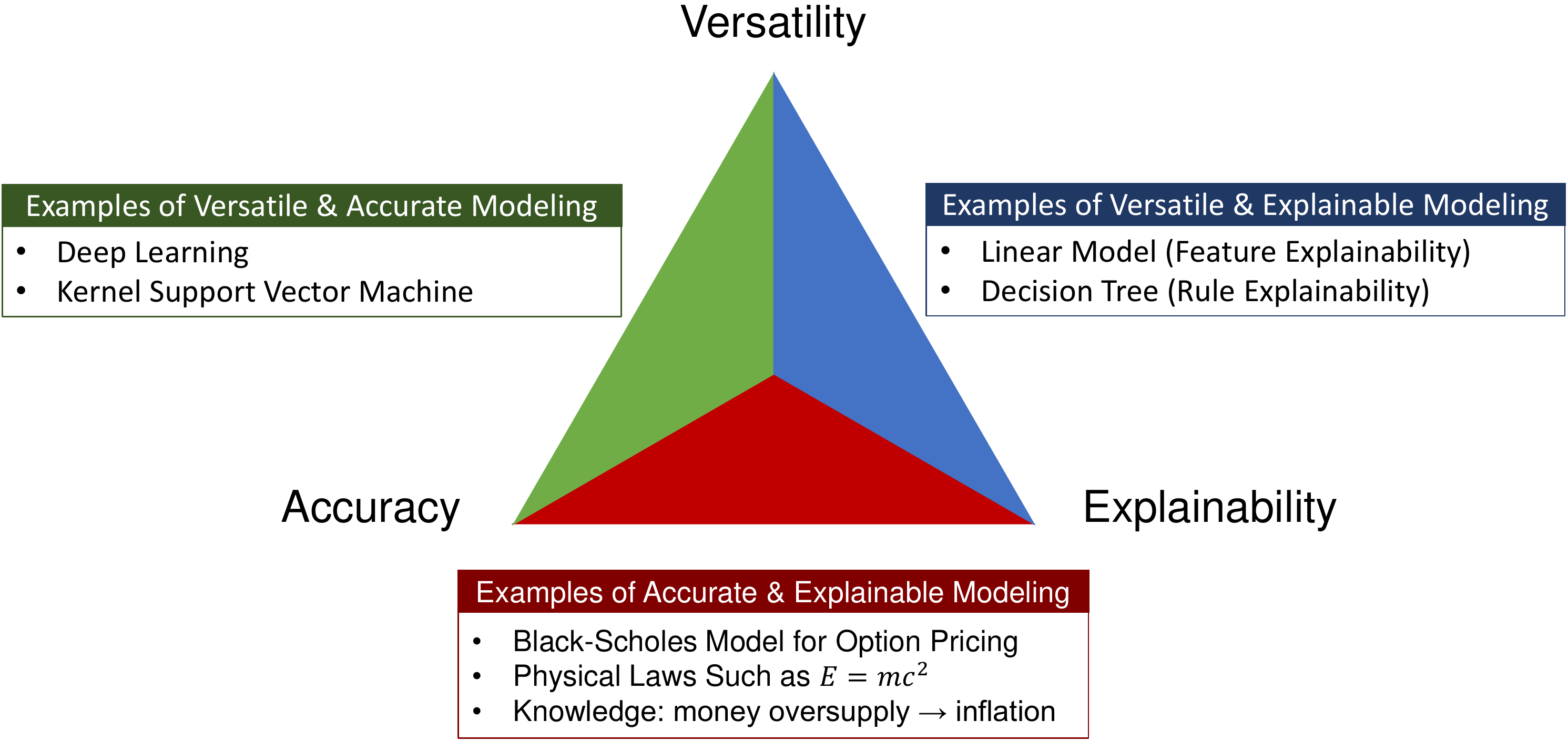}
        \caption{Impossible trinity of versatility, accuracy and explainability in modeling.}
        \label{fig:VersatilityAccuracyExplainability}
    \end{figure}
}

\newcommand{\figureQuantFourKeys}{
    \begin{figure}[!ht]
        \includegraphics[width=\linewidth]{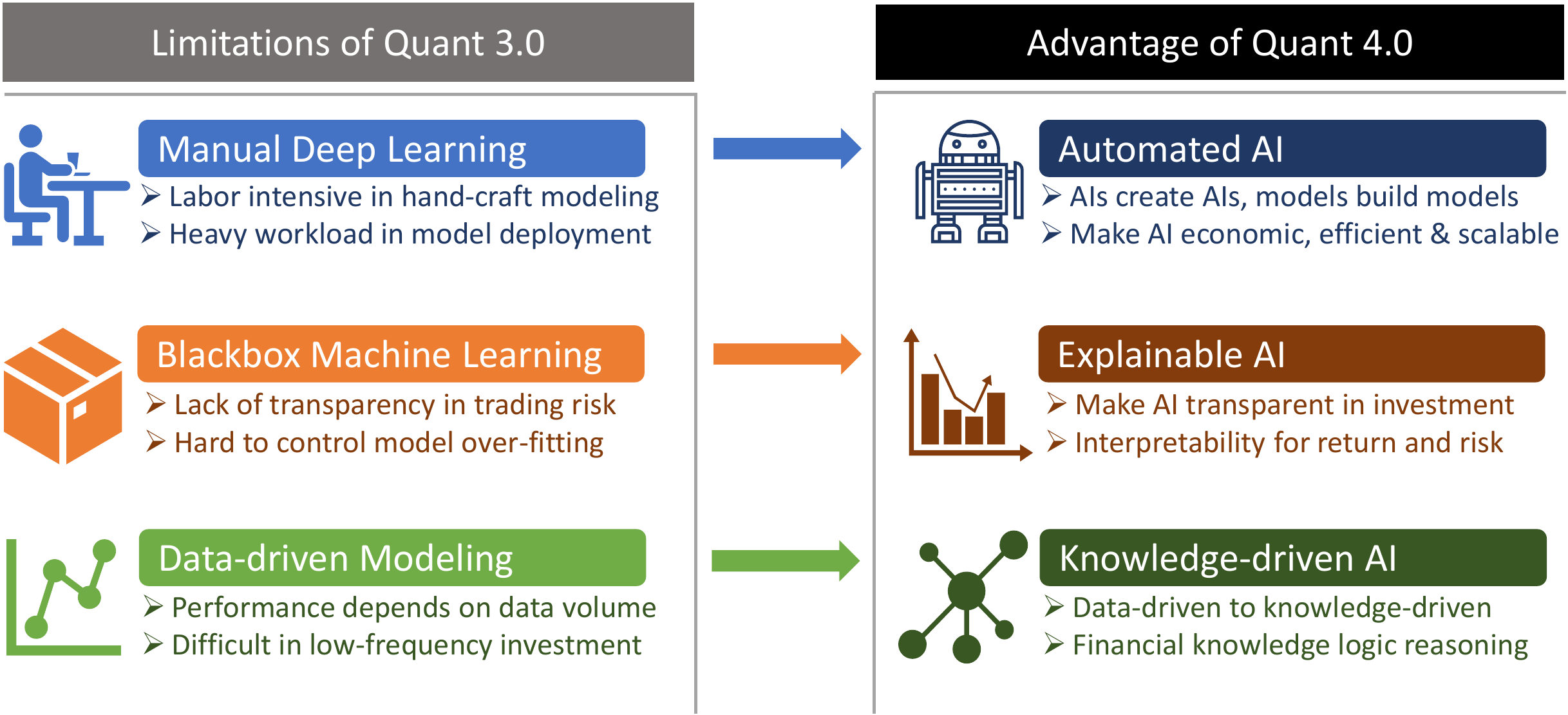}
        \caption{The three key components of Quant 4.0. }
        \label{fig:Quant4Keys}
    \end{figure}
}

\newcommand{\figureSpeedVsCapacity}{
    \begin{figure*}[!t]
        \includegraphics[width=\textwidth]{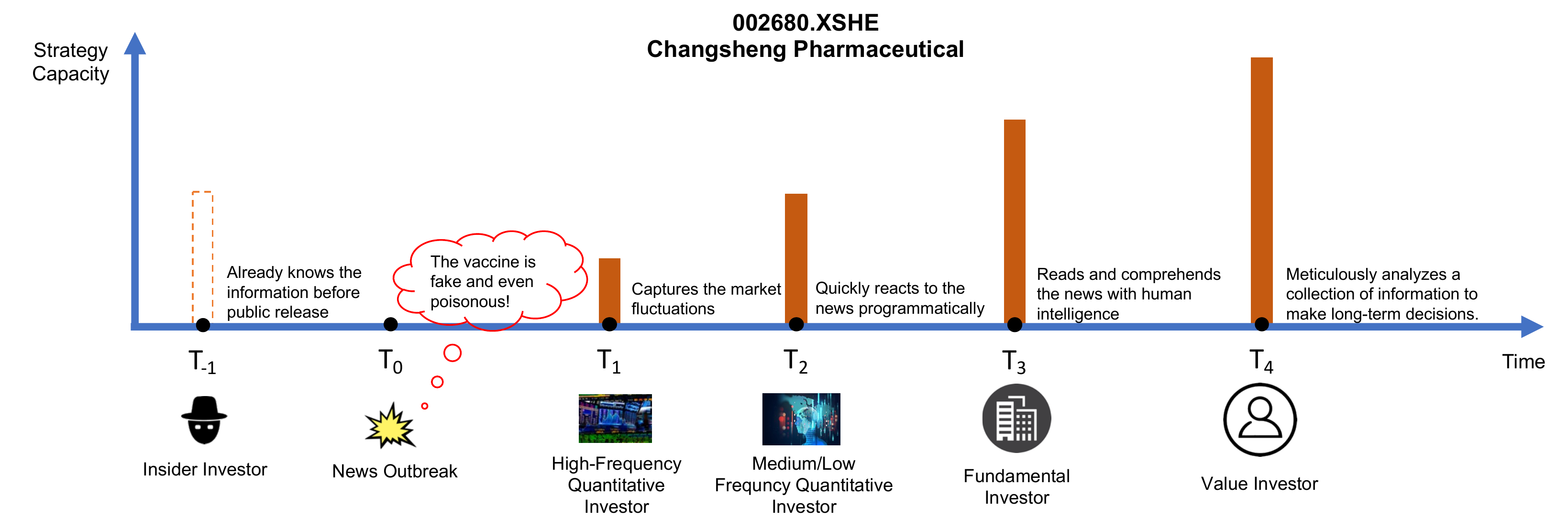}
        \caption{An example of how different participants react to the event: the scandal of Changsheng Pharmaceutical Co. Ltd in 2018 \cite{STchangsheng_unfolding_2018}. Quantitative investment usually reacts faster compared to fundamental investors and value investors since they need more time to analyze and make decisions. However, the market capacity of quantitative investment is usually smaller compared to fundamental investment and value investment. Note that although we include insider traders, the behaviors are illegal and are not discussed in this paper.}
        \label{fig:reaction_time}
    \end{figure*}
}

\newcommand{\figureQuantHistory}{
    \begin{figure*}[ht]
        \centering
        \includegraphics[width=\linewidth]{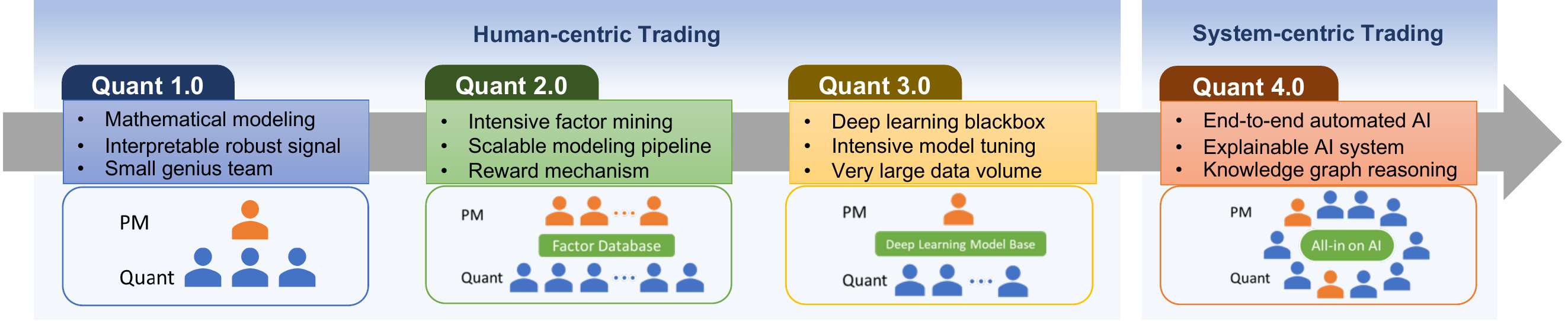}
        \caption{The development history of of quantitative investment in industry, from Quant 1.0 to Quant 4.0.
            \label{fig:quant_history}}
    \end{figure*}
}
\newcommand{\figurePosterior}{
    \begin{figure}[!h]
        \centering
        \includegraphics[width=0.7\linewidth]{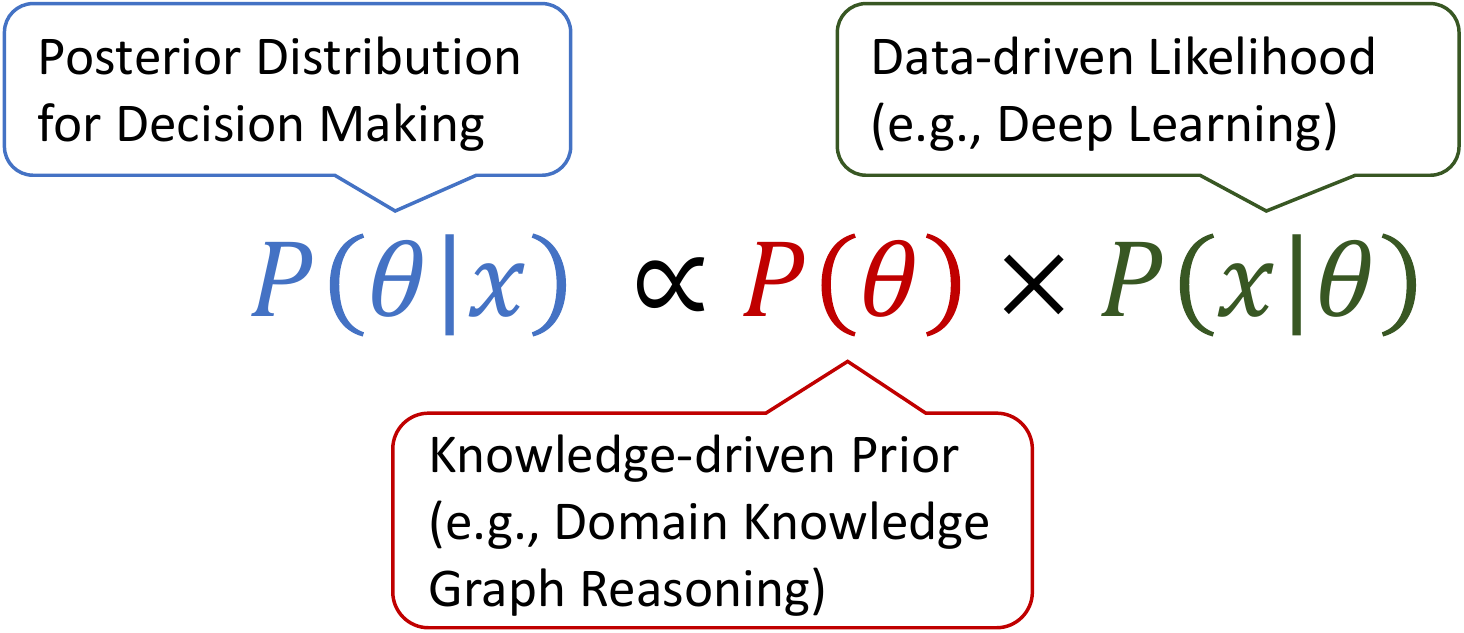}
        \caption{Complementary function of data-driven AI and knowledge-driven AI in decision making.
            \label{fig:posterior}}
    \end{figure}
}

\newcommand{\figureFactorMiningPipeline}{
\begin{figure*}
    \centering
    \includegraphics[width=\textwidth]{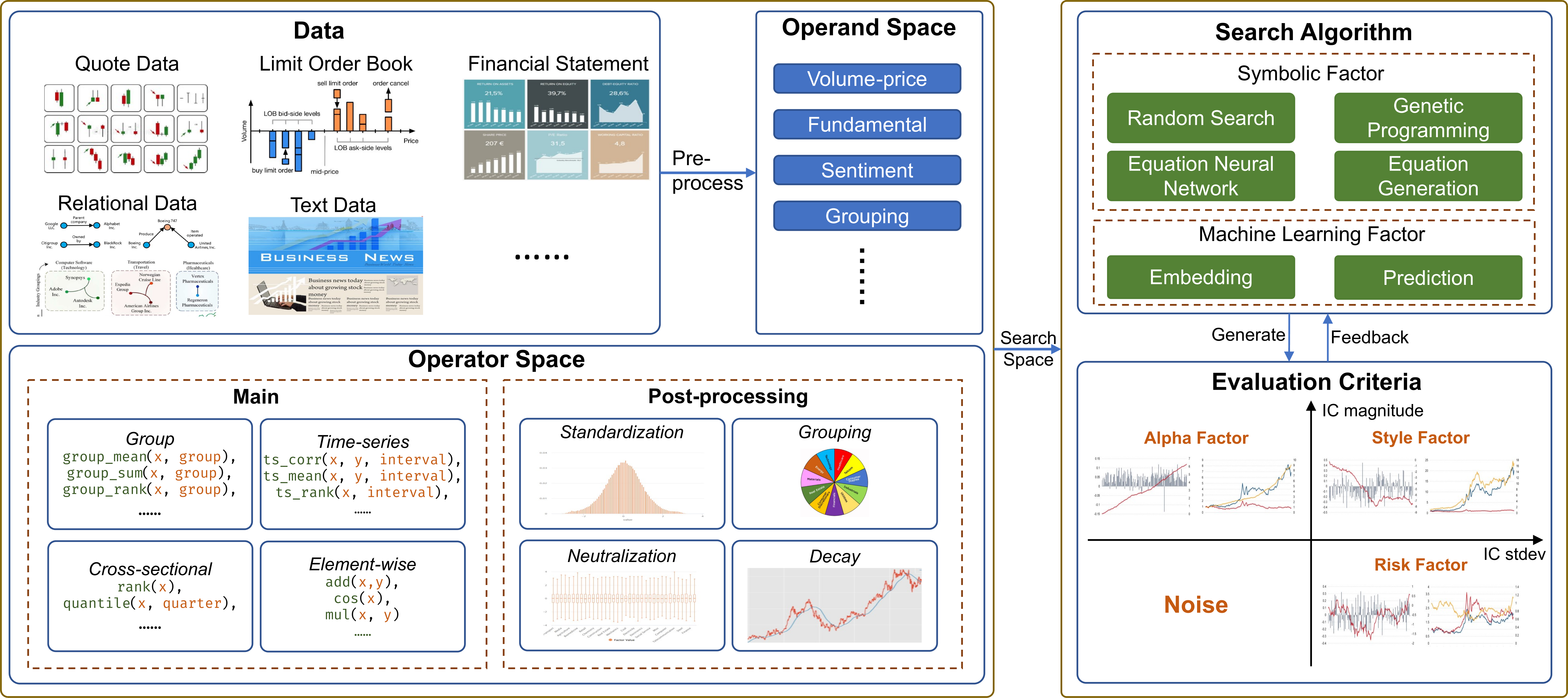}
    \caption{An example factor mining pipeline. The search space is defined by operators and meta-factors, where meta-factors are extracted from raw data in various forms. The search space is explored by search algorithms that are discrete or continuous. The evaluation module provides feedback to search algorithms based on certain criteria that serve as guidance for the next search iteration. Part of this figure is cited from \cite{feng_temporal_2019, sawhney_spatiotemporal_2020, wu_how_2021}.}
    \label{fig:factor_mining}
\end{figure*}
}

\newcommand{\figureQuantWorkflow}{
    \begin{figure*}[!t]
        \centering
        \includegraphics[width=\textwidth]{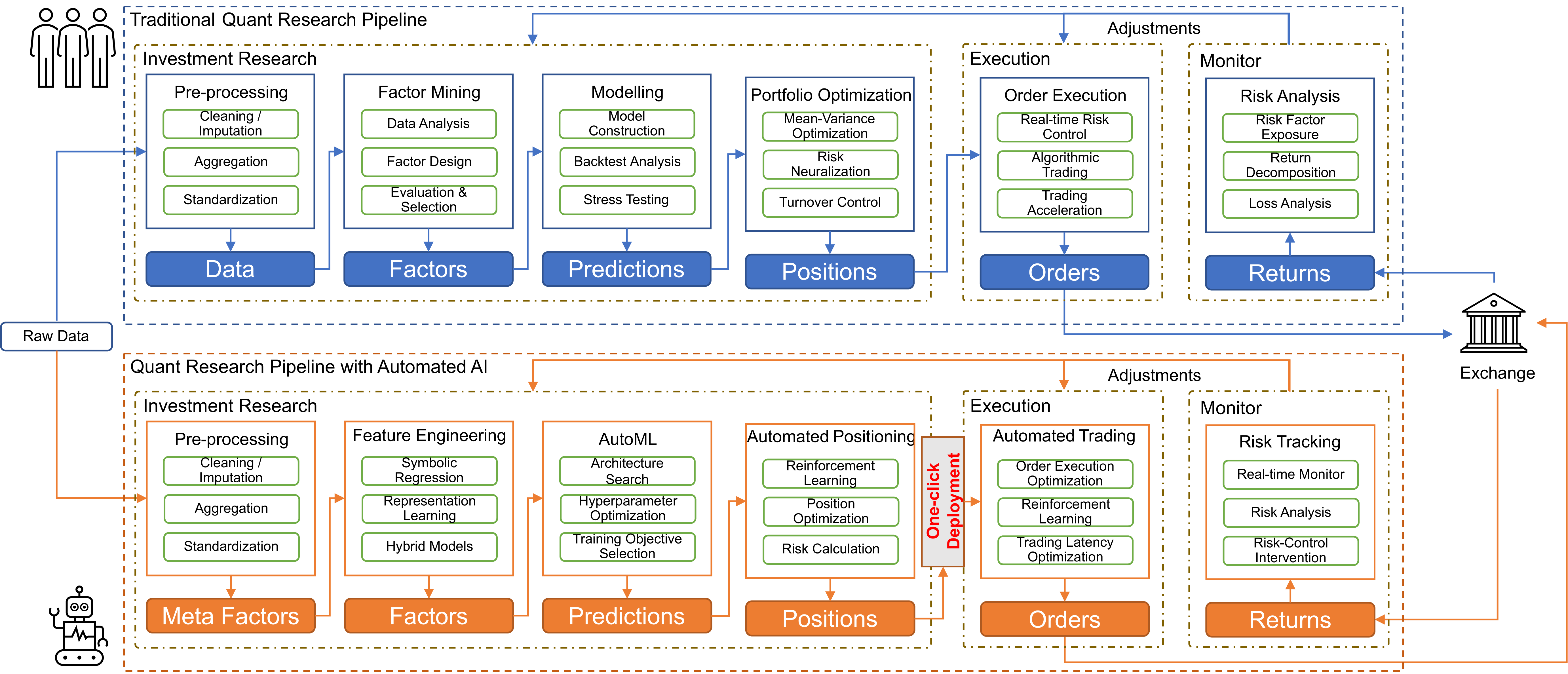}
        \caption{A prototypical workflow of quantitative investment with comparisons between the current quantitative investment system (manual, upper blue part) and AI investment engineering (automated, lower orange part).}
        \label{fig:quant_workflow}
    \end{figure*}
}

\newcommand{\figureTradingTasks}{
    \begin{figure*}[!ht]
        \centering
        \includegraphics[width=\textwidth]{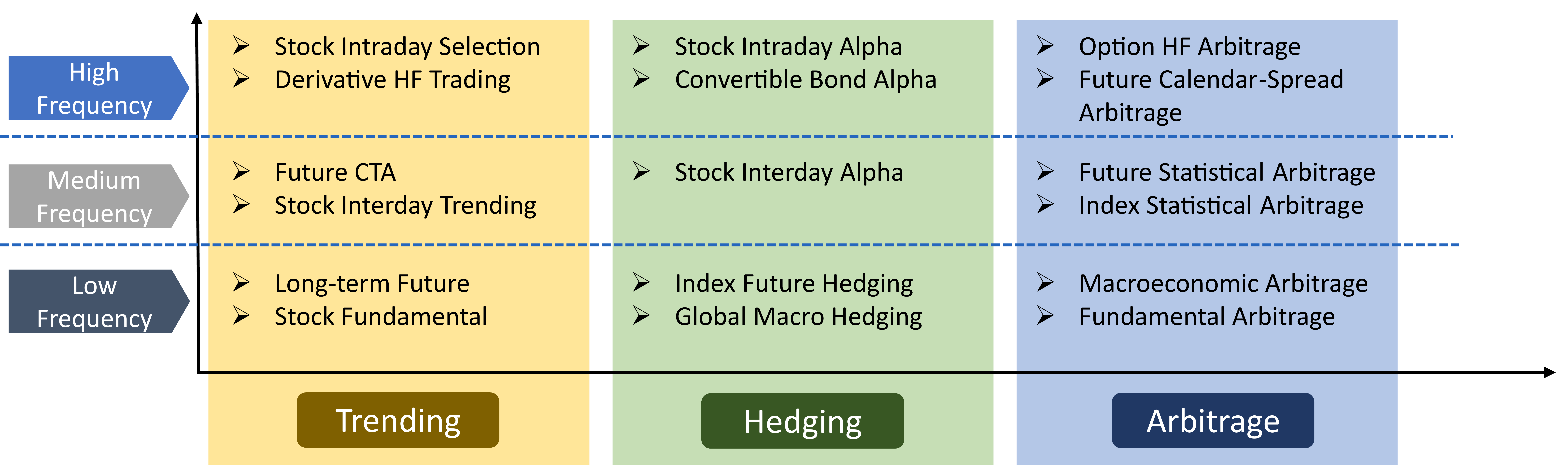}
        \caption{The listing of actual trading tasks for quantitative investment.}
        \label{fig:trading_tasks}
    \end{figure*}
}

\newcommand{\figureMLQuantWorkflow}{
    \begin{figure*}[!t]
        \centering
        \includegraphics[width=\textwidth]{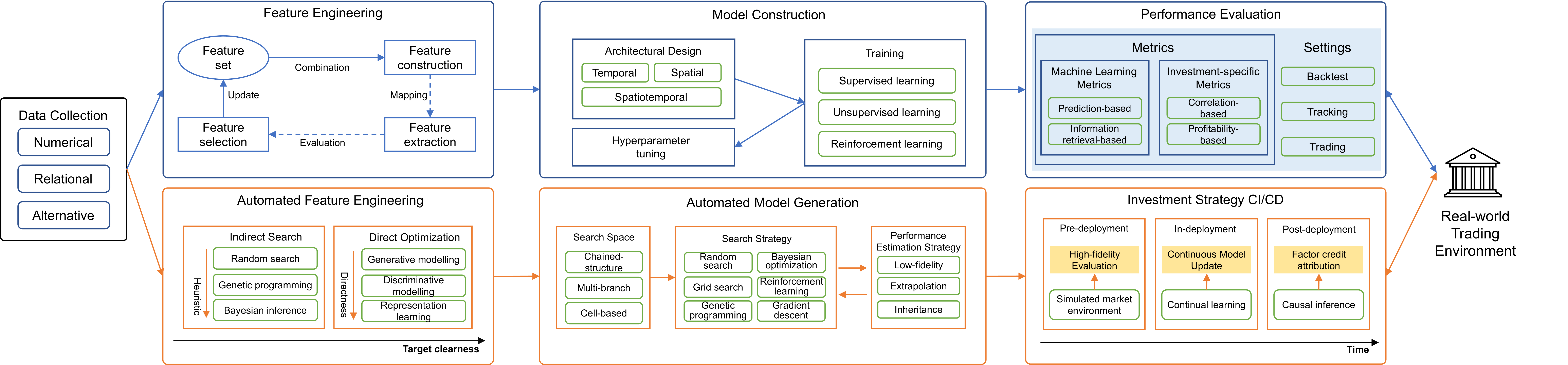}
        \caption{ The quantitative investment workflow with machine learning. The upper part illustrates current practice, while the lower part illustrates the algorithmic design for AI investment engineering.}
        \label{fig:automl_quant_workflow}
    \end{figure*}
}

\newcommand{\figureImpossibleTrinity}{
\begin{figure}
    \centering
    \includegraphics[width=\linewidth]{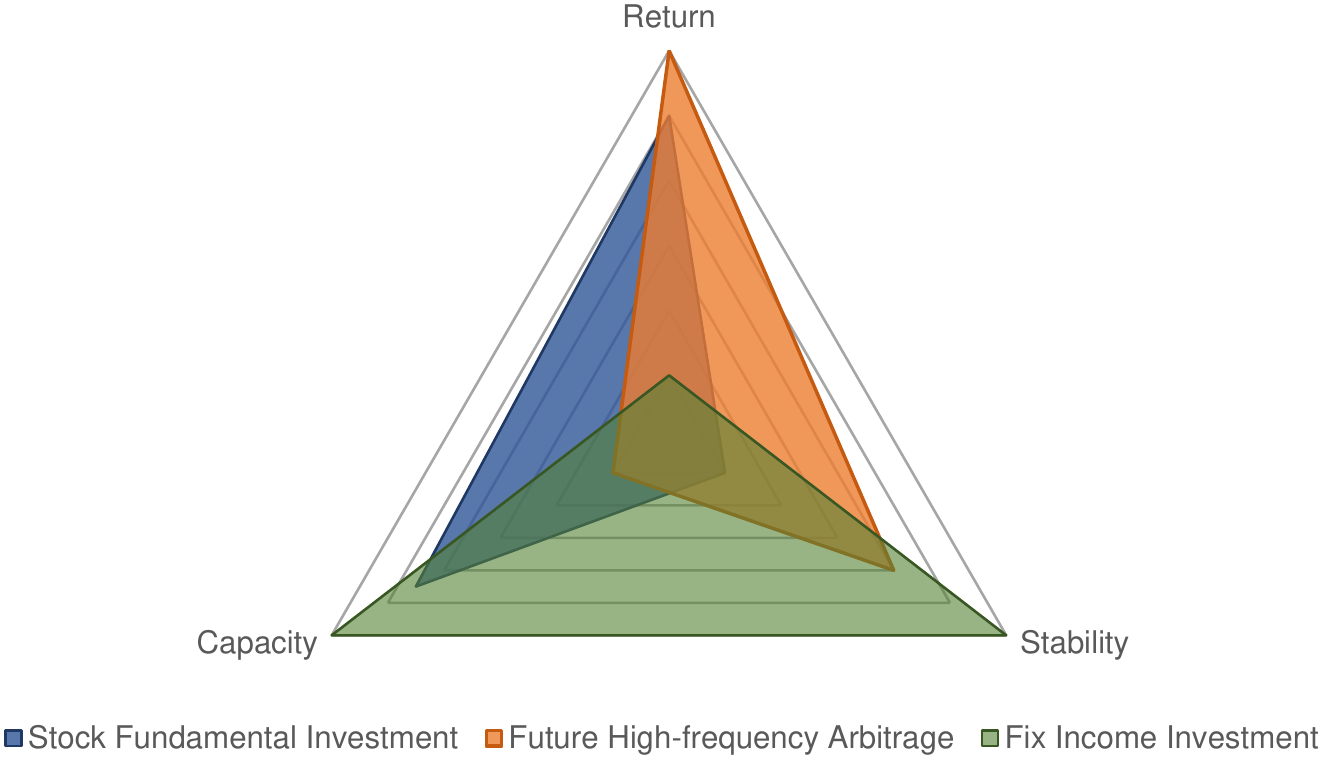}
    \caption{The impossible trinity in quantitative investment}
    \label{fig:impossible_trinity}
\end{figure}
}

\newcommand{\figureFoundamentalLawActiveManagement}{
\begin{figure}[!h]
    \centering
    \includegraphics[width=\linewidth]{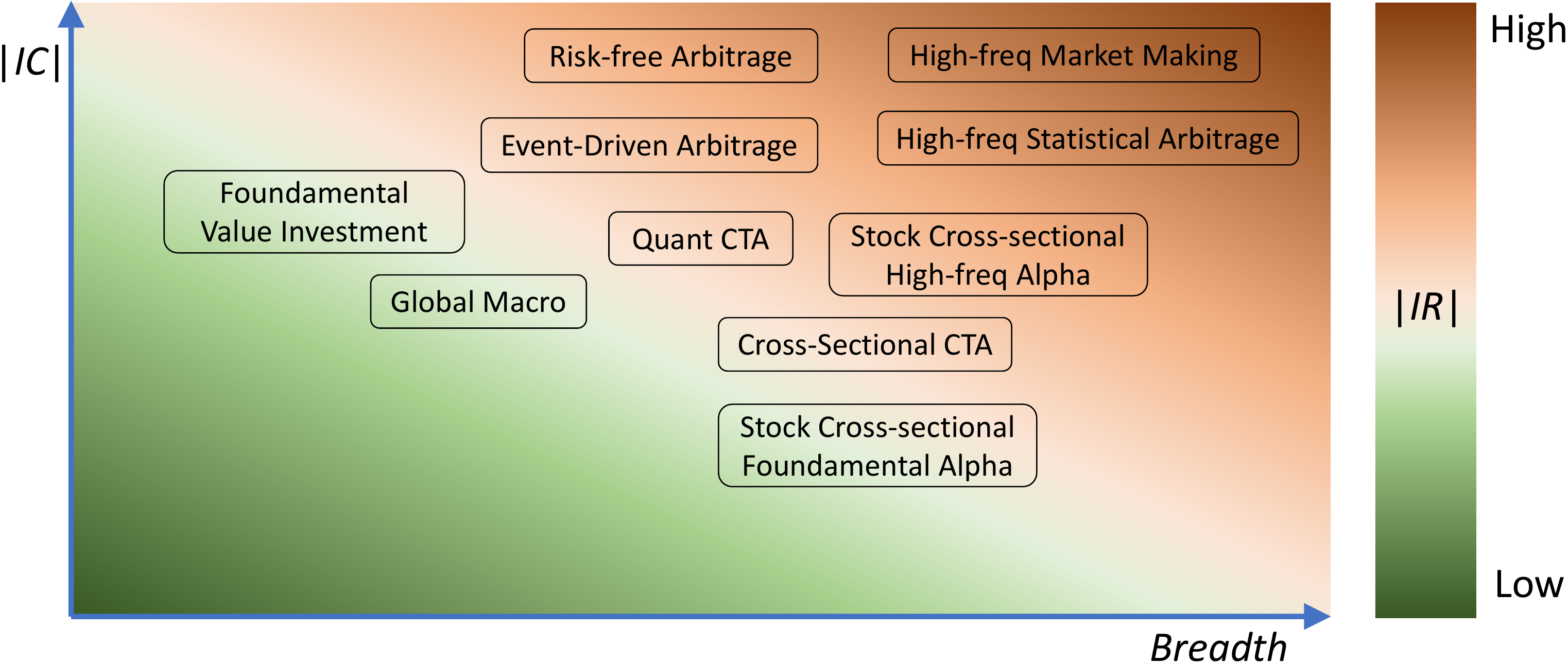}
    \caption{Illustration of different trading strategies under the Fundamental Law of Active Management.}
    \label{fig:FoundamentalLawActiveManagement}
\end{figure}
}

\newcommand{\FigureSystemOneTwo}{
\begin{figure}[!ht]
    \centering
    \includegraphics[width=\linewidth]{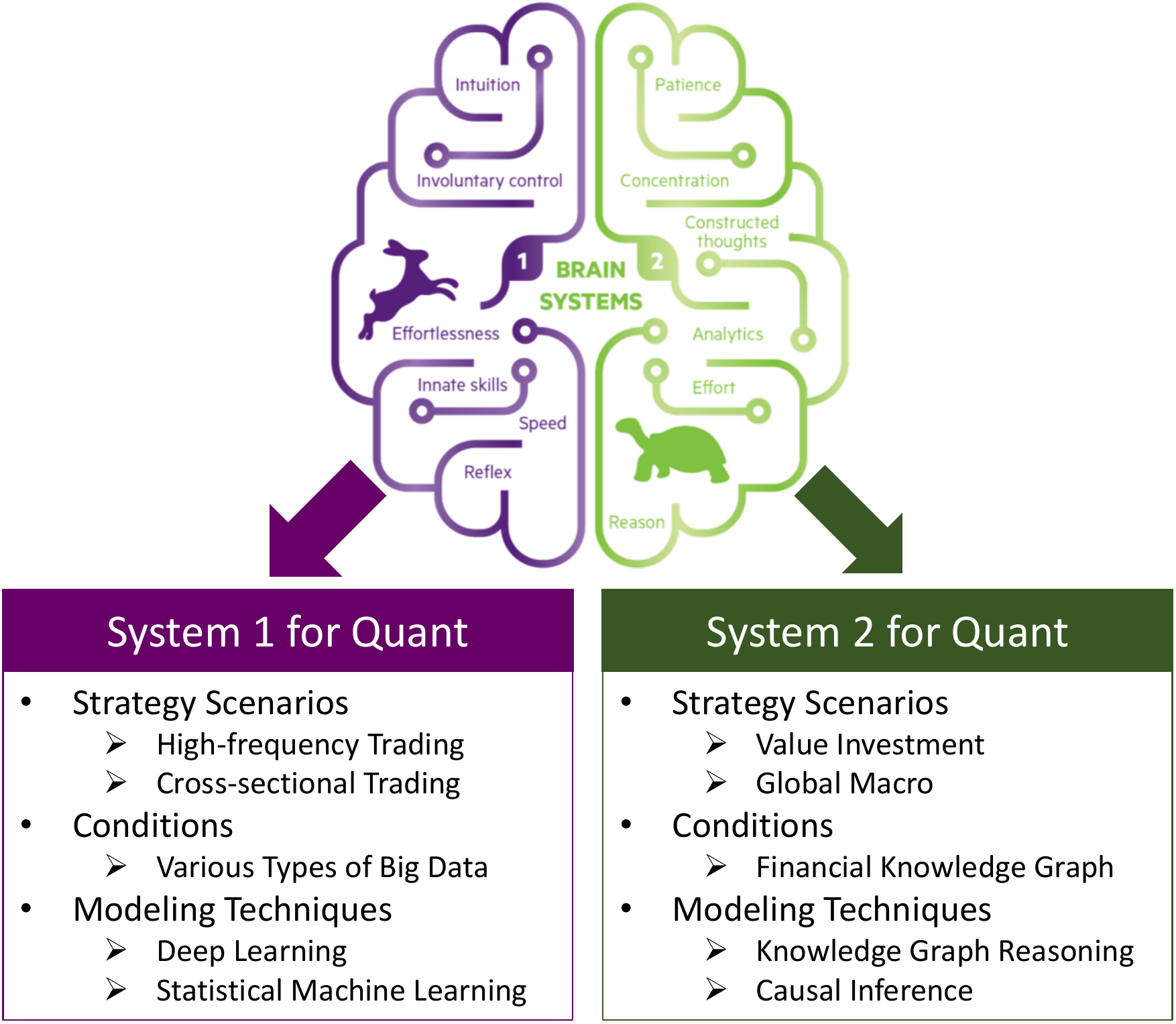}
    \caption{Comparison of System 1 and System 2 in AI technology for quant research.}
    \label{FigureSystem12}
\end{figure}
}

\newcommand{\FigureEndtoEndModeling}{
\begin{figure}[!h]
    \centering
    \includegraphics[width=\linewidth]{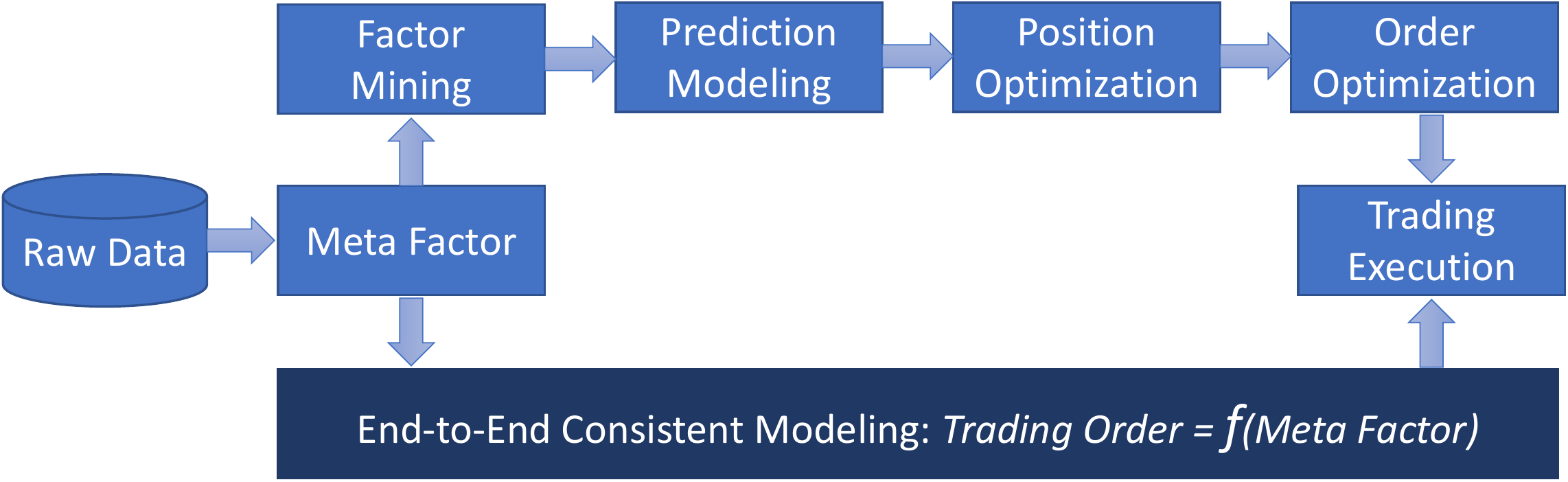}
    \caption{Comparison of traditional quant research pipeline and end-to-end consistent modeling. }
    \label{fig:EndtoEndModeling}
\end{figure}
}

\newcommand{\figureQuantSystem}{
    \begin{figure*}[!htbp]
        \centering
        \includegraphics[width=\textwidth]{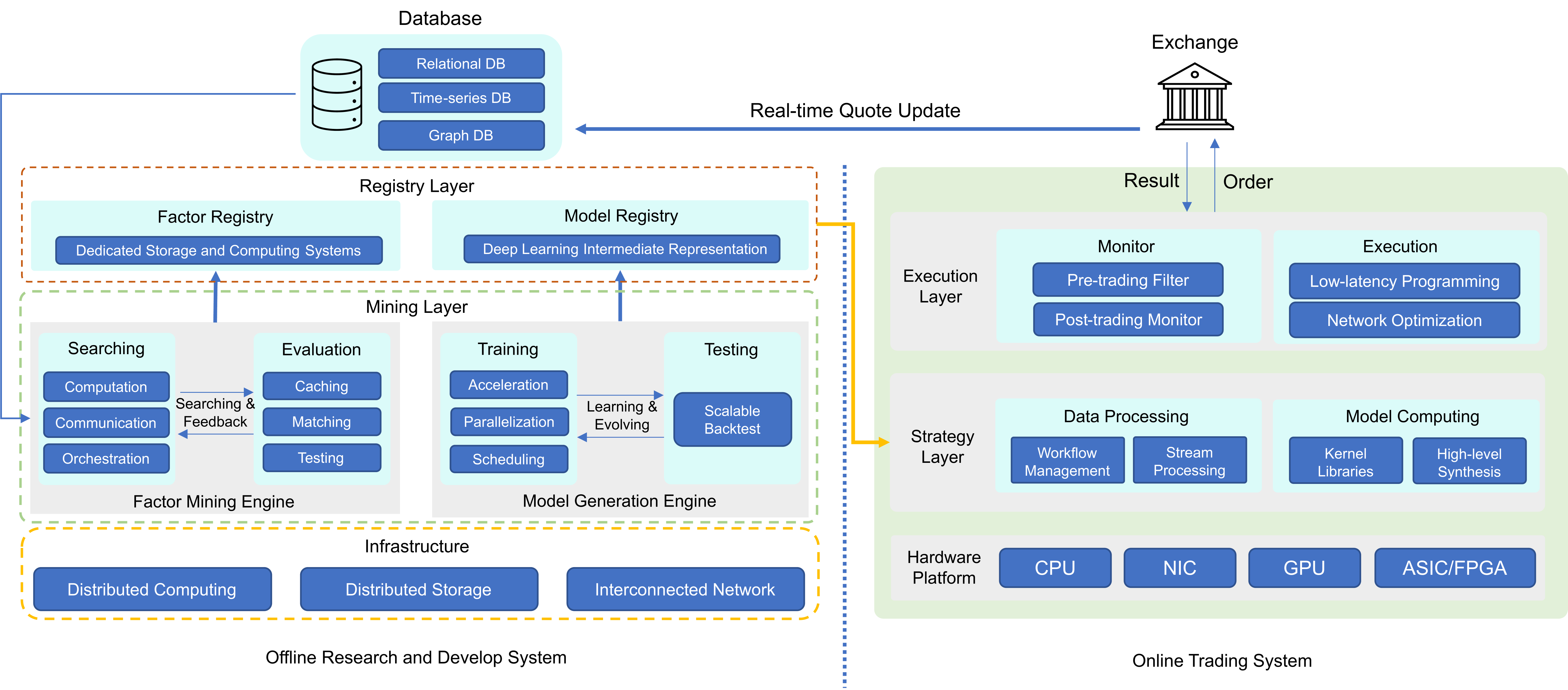}
        \caption{The architecture of a dedicated quantitative investment system.}
        \label{fig:quant_system}
    \end{figure*}
}

\newcommand{\figureExplainRoadmap}{
    \begin{figure*}[!ht]
        \centering
        \includegraphics[width=\textwidth]{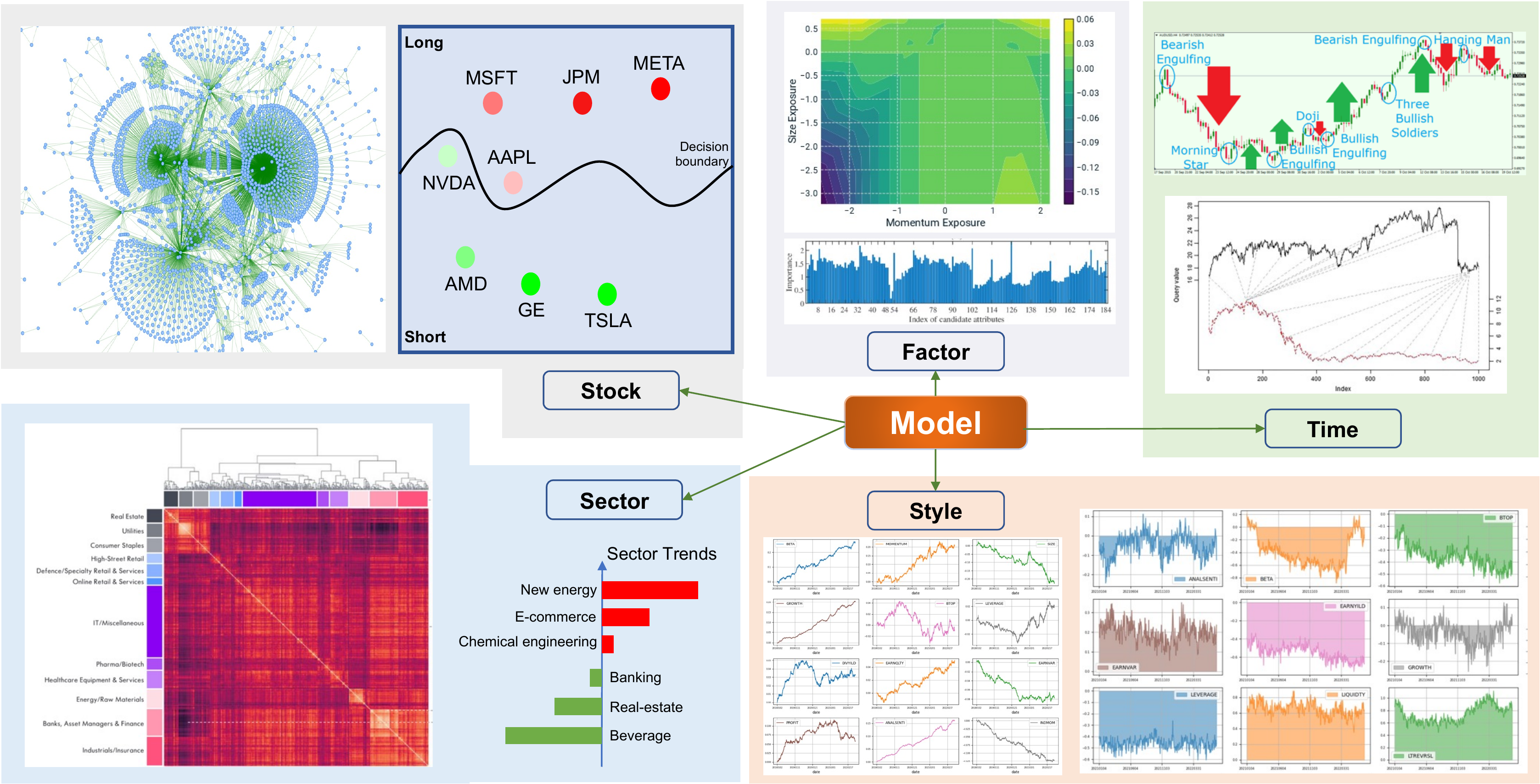}
        \caption{The five dimensions of explainable AI for Quant 4.0.}
        \label{fig:explain_roadmap}
    \end{figure*}
}

\newcommand{\figureXAIMethods}{
\begin{figure}[!htbp]
    \centering
    \includegraphics[width=0.7\linewidth]{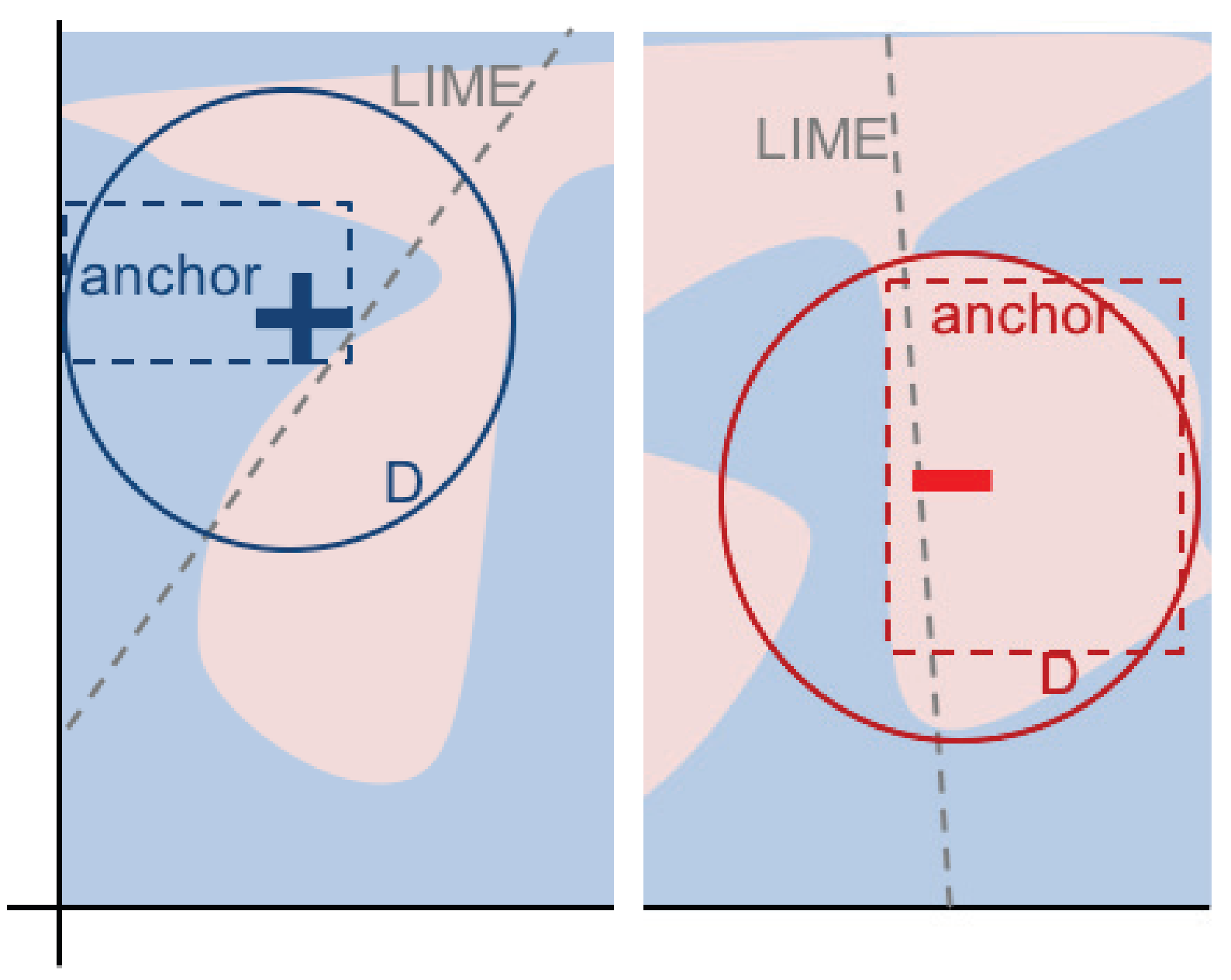}
    \caption{Illustration of LIME \cite{ribeiro_why_2016} and Anchors \cite{ribeiro_anchors_2018}. Figure cited from \cite{ribeiro_anchors_2018}.}
    \label{fig:xai_methods}
\end{figure}
}

\newcommand{\figurePrincipleActiveManagement}{
    \begin{figure*}[ht]
        \centering
        \begin{subfigure}{0.53\textwidth}
            \centering
            \includegraphics[width=\linewidth]{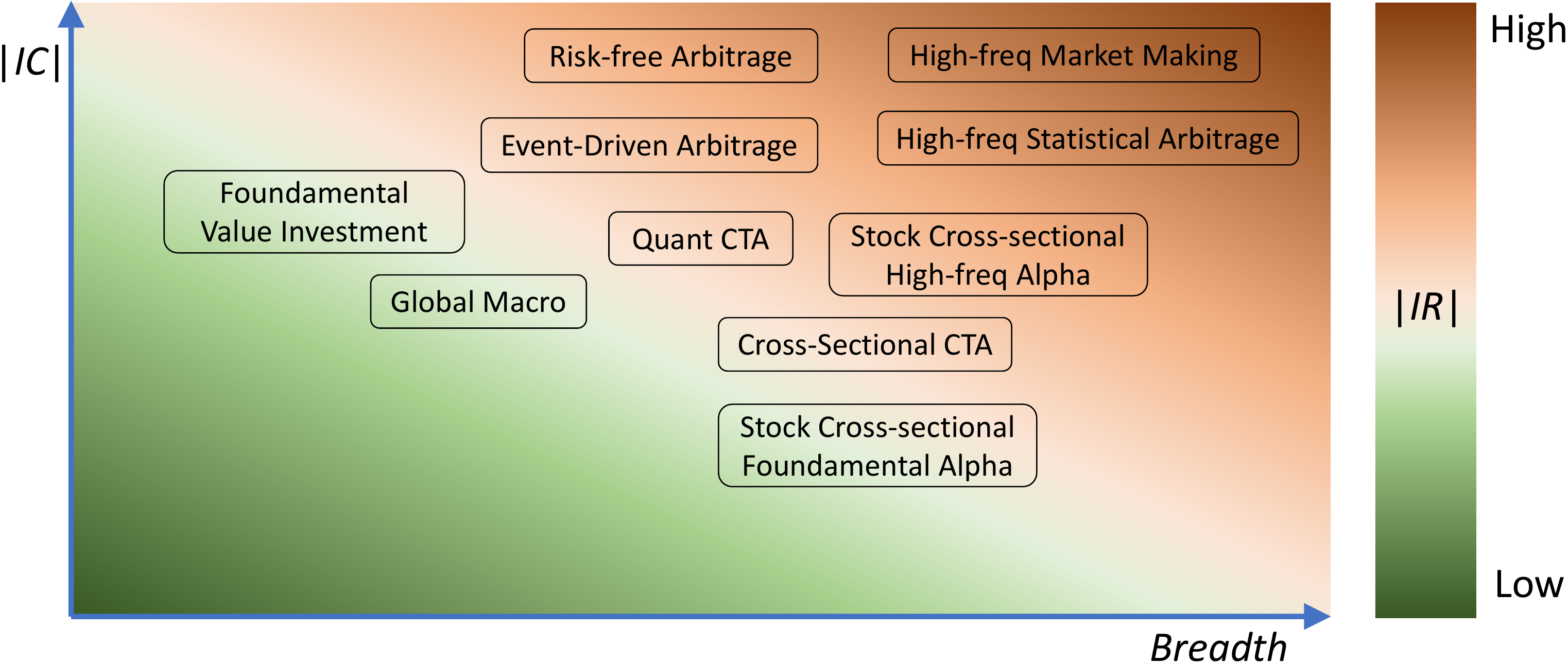}
            \caption{Common investment strategies under the fundamental law of active management. The figure illustrates the relationship between the magnitude of \emph{IR} with \emph{breadth} and the magnitude of \emph{IC} for different strategies.}
            \label{fig:law_of_active_mgmt}
        \end{subfigure}\hfill
        \begin{subfigure}{0.42\textwidth}
            \centering
            \includegraphics[width=\linewidth]{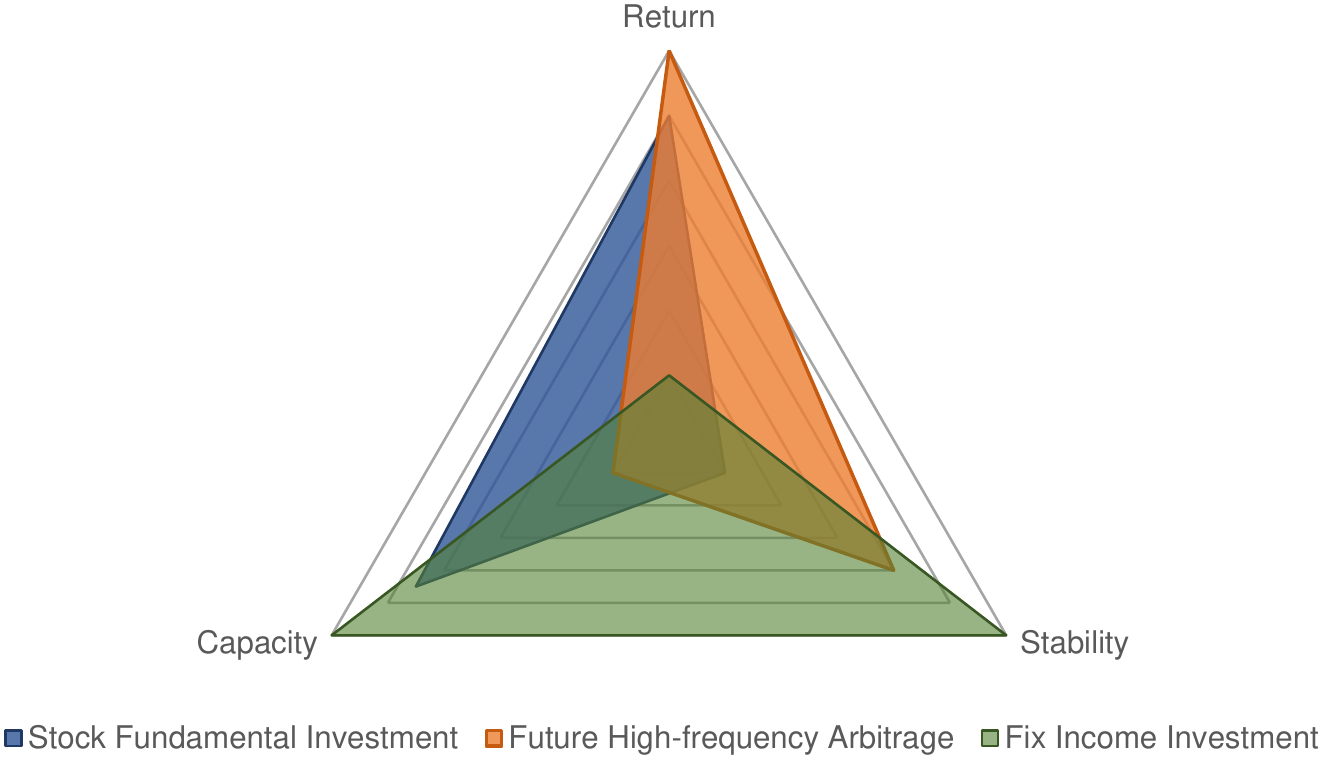}
            \caption{Illustration of the impossible trinity of return, capacity, and stability for active management using three typical strategies: stock fundamental investment, future high-frequency arbitrage and fixed income investment.}
            \label{fig:impossible_trinity}
        \end{subfigure}
        \caption{Illustration of the principles for active investment management with specific strategies.}
        \label{fig:PrincipleActiveManagement}
    \end{figure*}
}

\newcommand{\figureNumericalData}{
    \begin{figure}
        \centering
        \begin{subfigure}{0.25\linewidth}
            \centering
            \includegraphics[width=\linewidth]{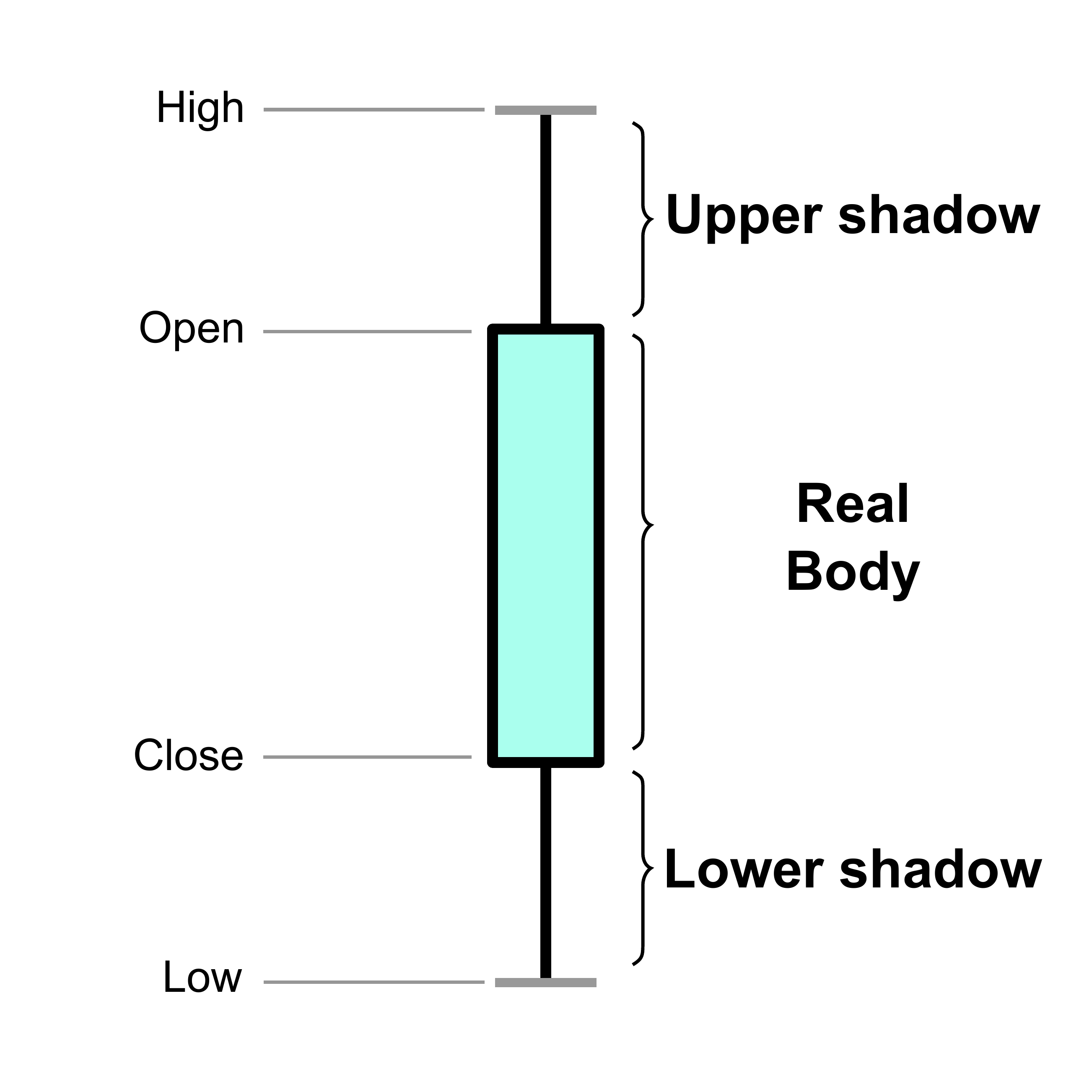}
            \caption{Candlestick chart\cite{noauthor_candlestick_2022} }
        \end{subfigure}
        \begin{subfigure}{0.3\linewidth}
            \centering
            \includegraphics[width=\linewidth]{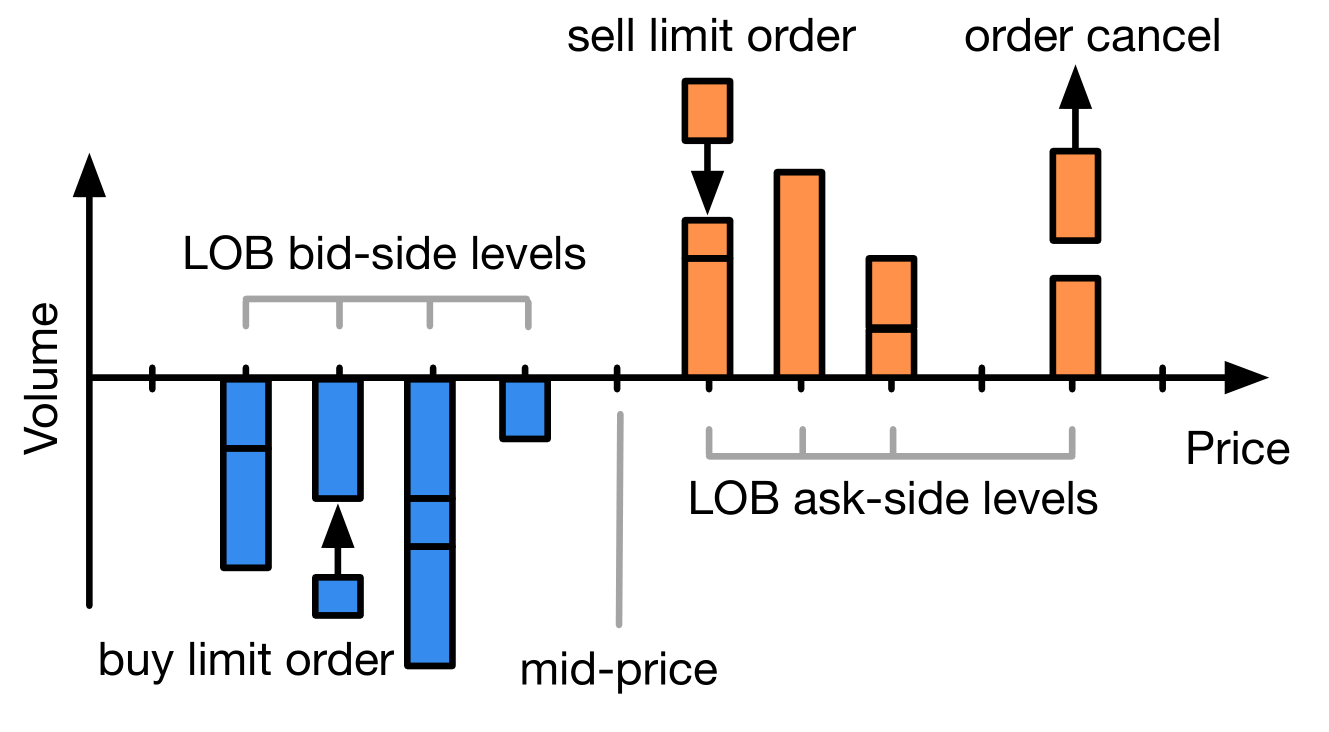}
            \caption{Limit Order Book\cite{wu_how_2021}}
        \end{subfigure}
        \begin{subfigure}{0.4\textwidth}
            \centering
            \includegraphics[width=\linewidth]{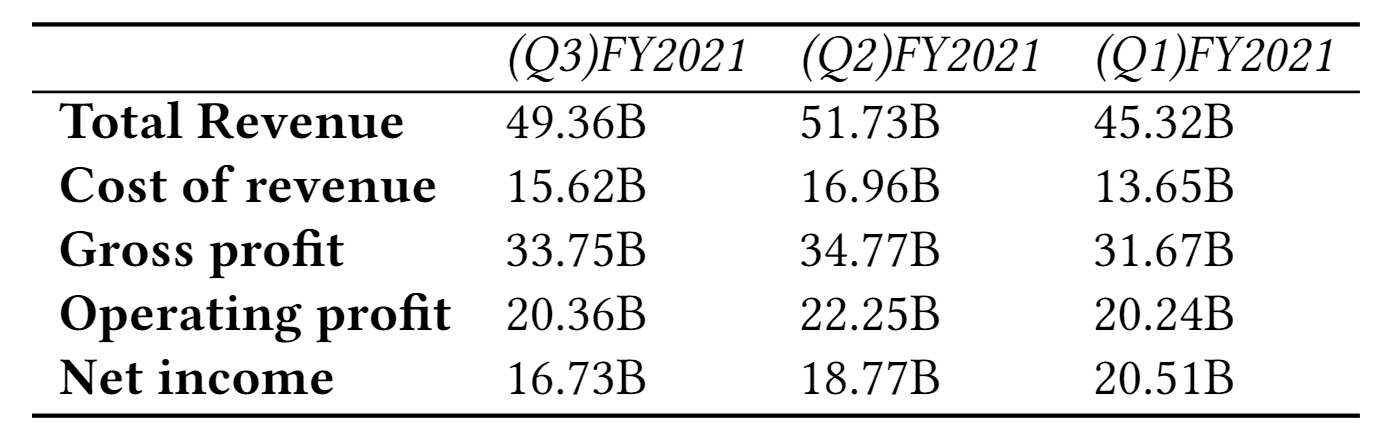}
            \caption{Fundamental data of Microsoft.}
        \end{subfigure}
        \caption{Examples of numerical data.}
        \label{fig:numerical_data}
    \end{figure}
}

\newcommand{\figureRelationalData}{
    \begin{figure}
        \centering
        \begin{subfigure}{0.46\textwidth}
            \centering
            \includegraphics[width=\linewidth]{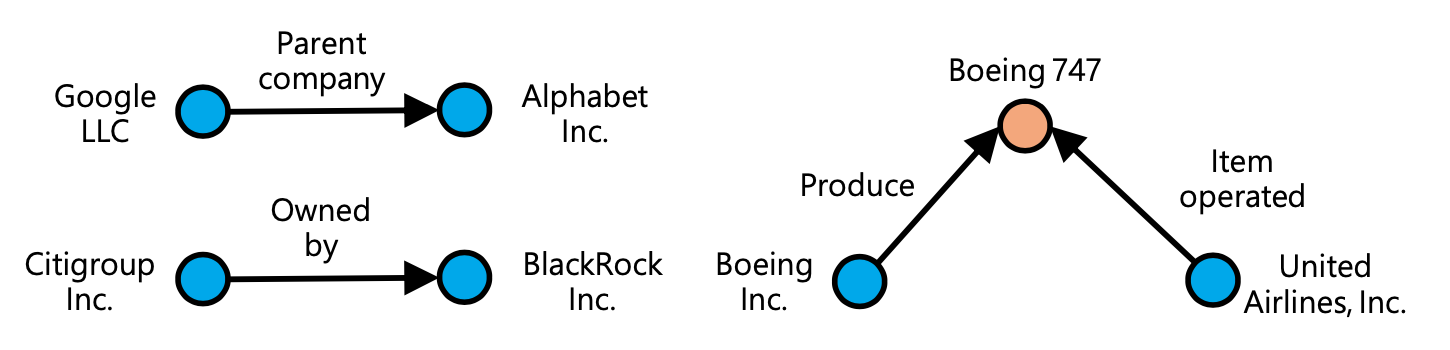}
            \caption{Pairwise edges \cite{feng_temporal_2019}}
        \end{subfigure}
        \begin{subfigure}{0.46\textwidth}
            \centering
            \includegraphics[width=\linewidth]{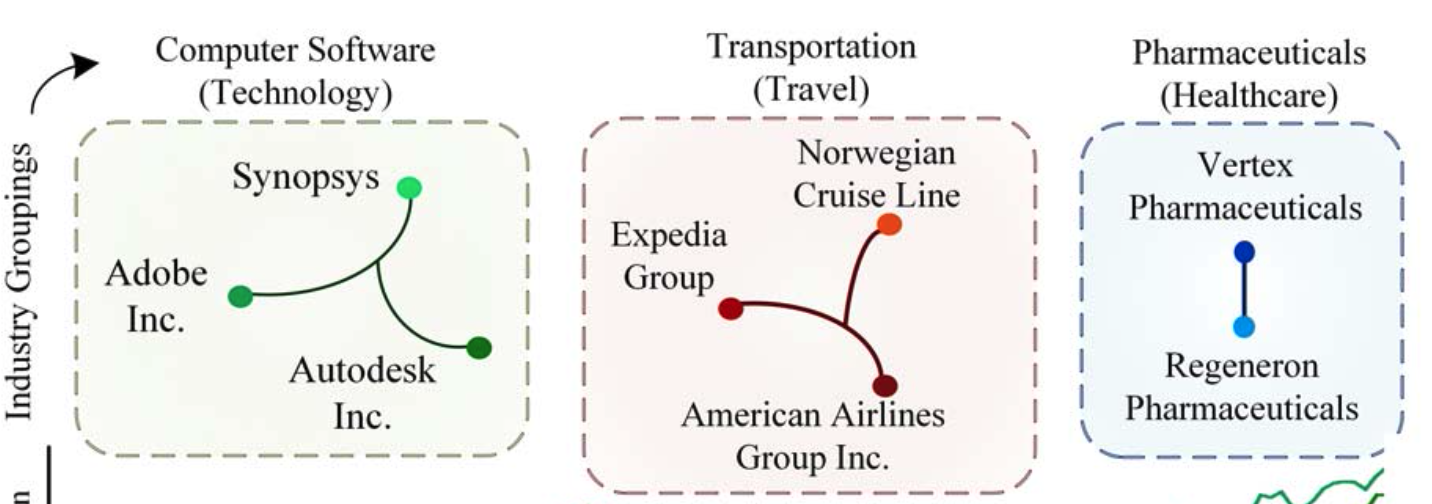}
            \caption{Hyperedges \cite{sawhney_spatiotemporal_2020}}
        \end{subfigure}
        \caption{Relational data examples.}
        \label{fig:relational_data}
    \end{figure}
}

\newcommand{\figureSpatiotemporal}{
    \begin{figure}
        \centering
        \begin{subfigure}{0.46\linewidth}
            \centering
            \includegraphics[width=\linewidth]{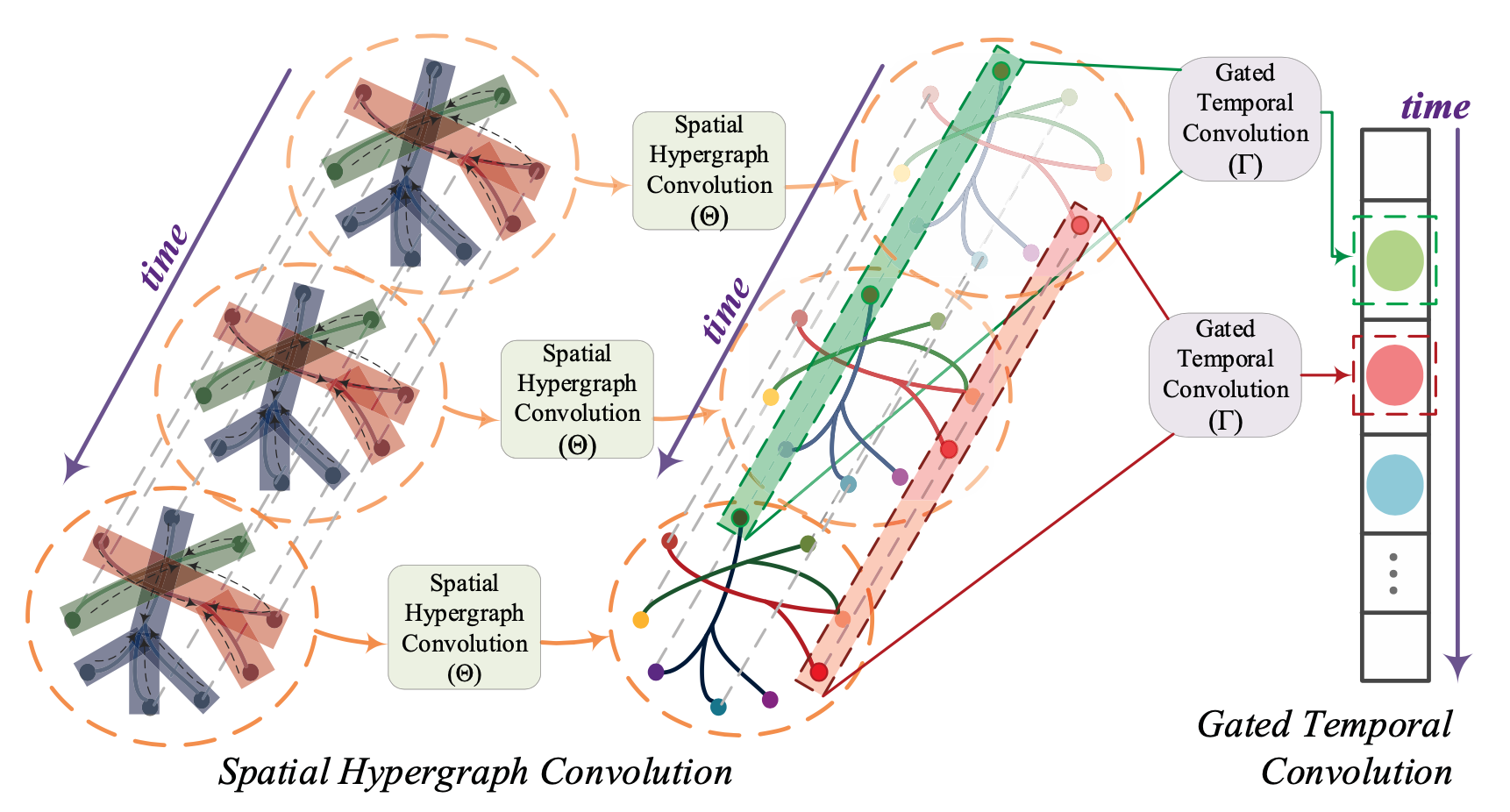}
            \caption{Decoupled spatiotemporal model \cite{sawhney_spatiotemporal_2020}}
            \label{fig:spatiotemporal_decoupled}
        \end{subfigure}
        \begin{subfigure}{0.5\linewidth}
            \centering
            \includegraphics[width=\linewidth]{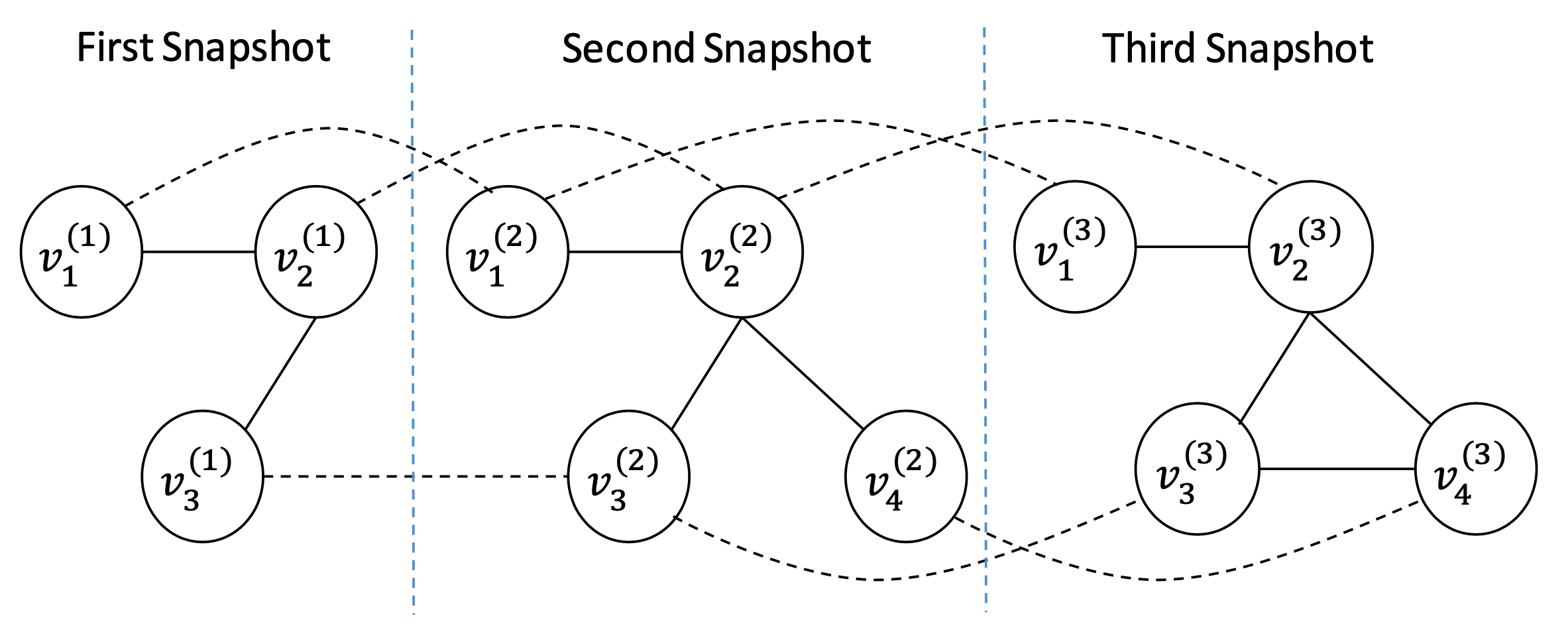}
            \caption{An example of coupled spatiotemporal model using temporal unrolling \cite{kazemi_dynamic_2022}}
            \label{fig:spatiotemporal_coupled}
        \end{subfigure}
        \caption{Examples of spatiotemporal models.}
        \label{fig:spatiotemporal}
    \end{figure}
}

\newcommand{\figureRollingWindow}{
    \begin{figure*}[!htbp]
        \centering
        \includegraphics[width=\textwidth]{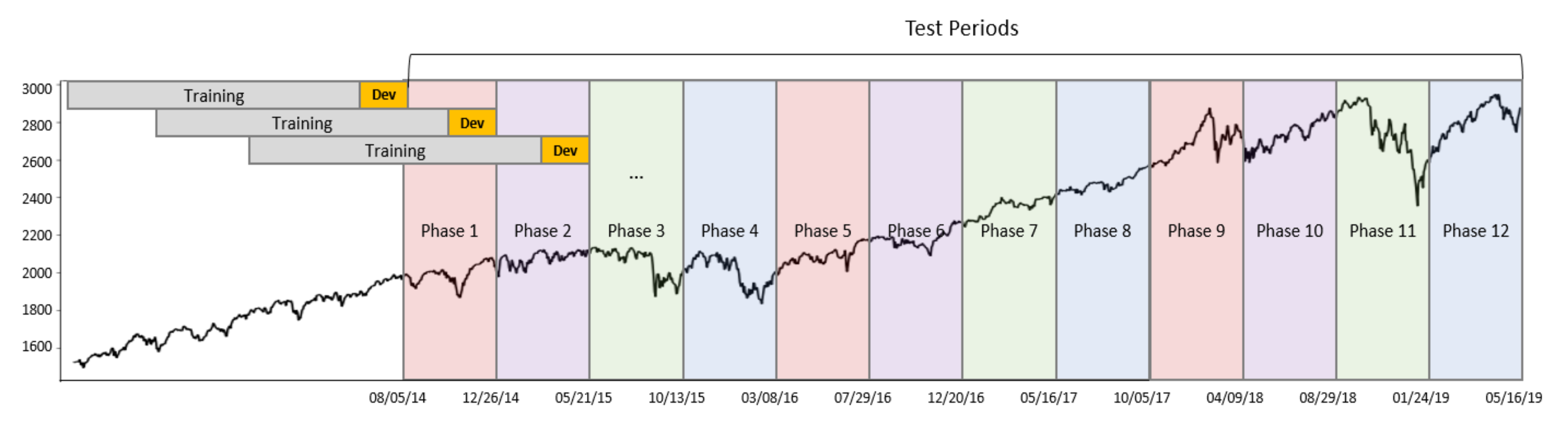}
        \caption{Illustration of the rolling window experiments \cite{kim_hats_2019}.}
        \label{fig:rolling_window}
    \end{figure*}
}

\newcommand{\figureXAILoops}{
\begin{figure}[!t]
    \centering
    \includegraphics[width=0.7\linewidth]{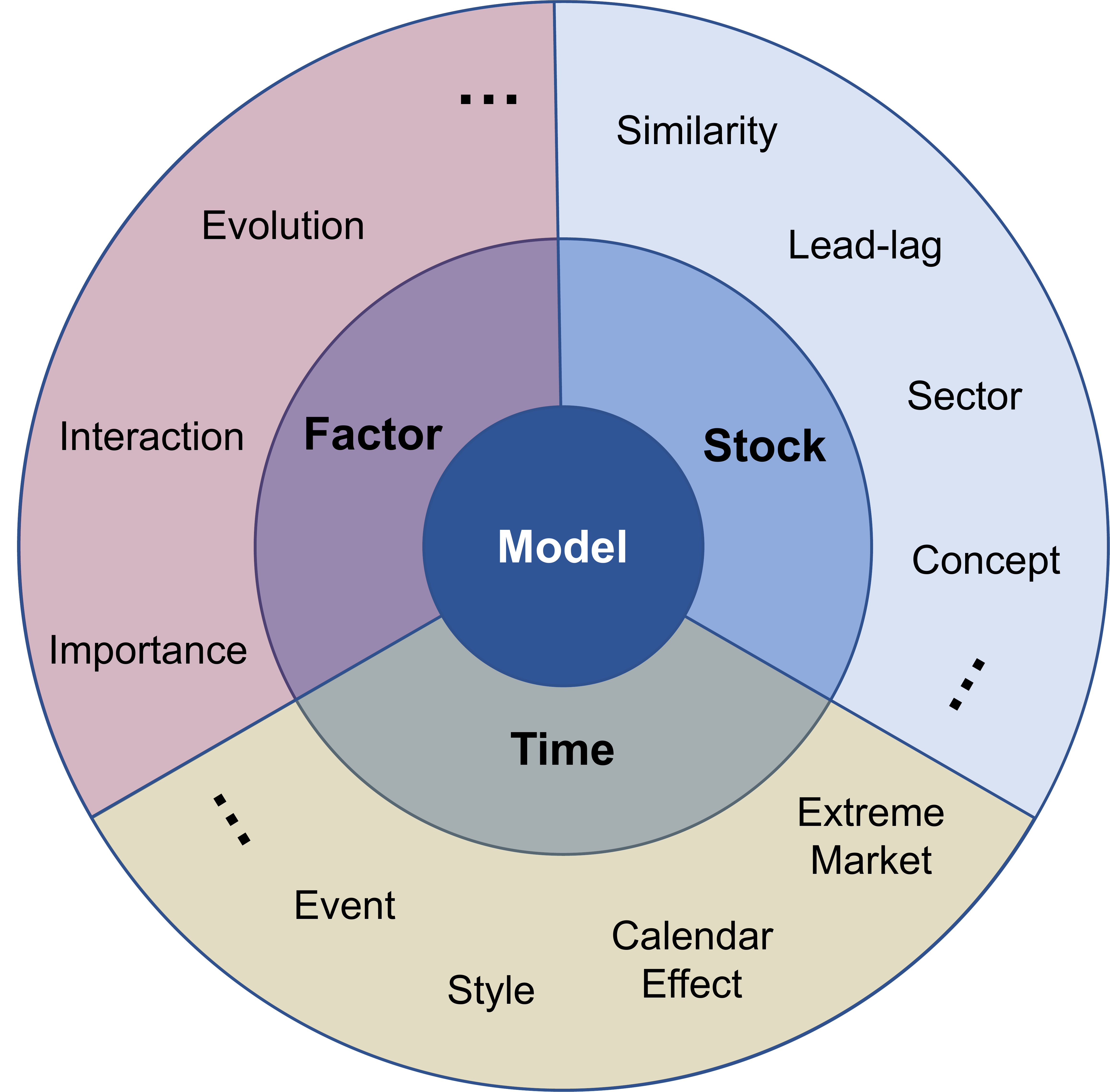}
    \caption{XAI from data dimensions.}
    \label{fig:XAILoops}
\end{figure}
}

\newcommand{\figureAutoFeatureEngineering}{
\begin{figure*}
\centering
\begin{subfigure}{0.48\textwidth}
\centering
\includegraphics[width=\linewidth]{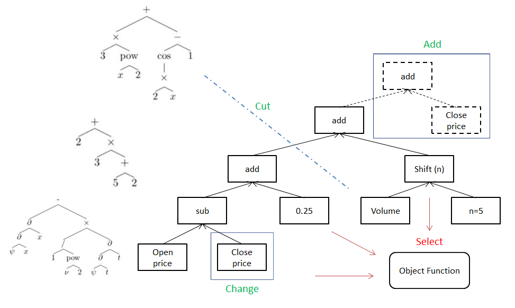}
\caption{Symbolic factors generated via genetic programming}
\end{subfigure}
\begin{subfigure}{0.48\textwidth}
\centering
\includegraphics[width=\linewidth]{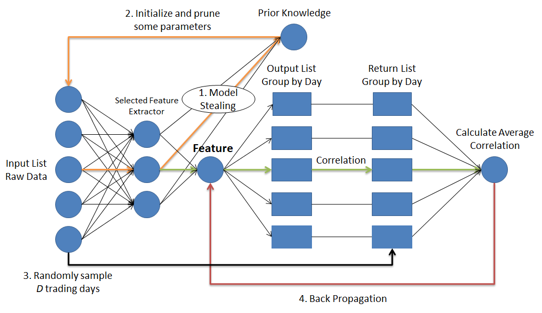}
\caption{Neural factors optimized via minimizing the correlation loss}
\end{subfigure}
\caption{Symbolic and neural factors. Figures from \cite{fang_alpha_2020}}
\label{fig:symbolic_nn_factors}
\end{figure*}
}

\newcommand{\figureModelSearchSpace}{
\begin{figure}
    \centering
    \includegraphics[width=0.8\linewidth]{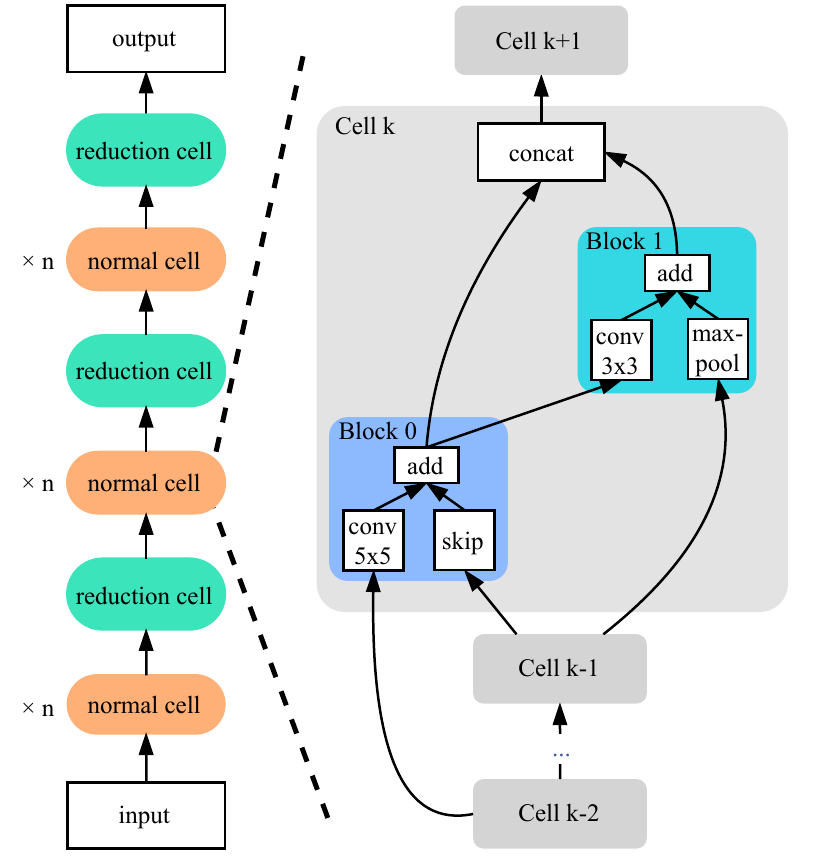}
    \caption{Cell-based search space for neural architecture search. Figure from \cite{he_automl_2021}}
    \label{fig:cell_structure}
\end{figure}
}

\newcommand{\figureGridvsRandom}{
\begin{figure}[!ht]
    \centering
    \includegraphics[width=\linewidth]{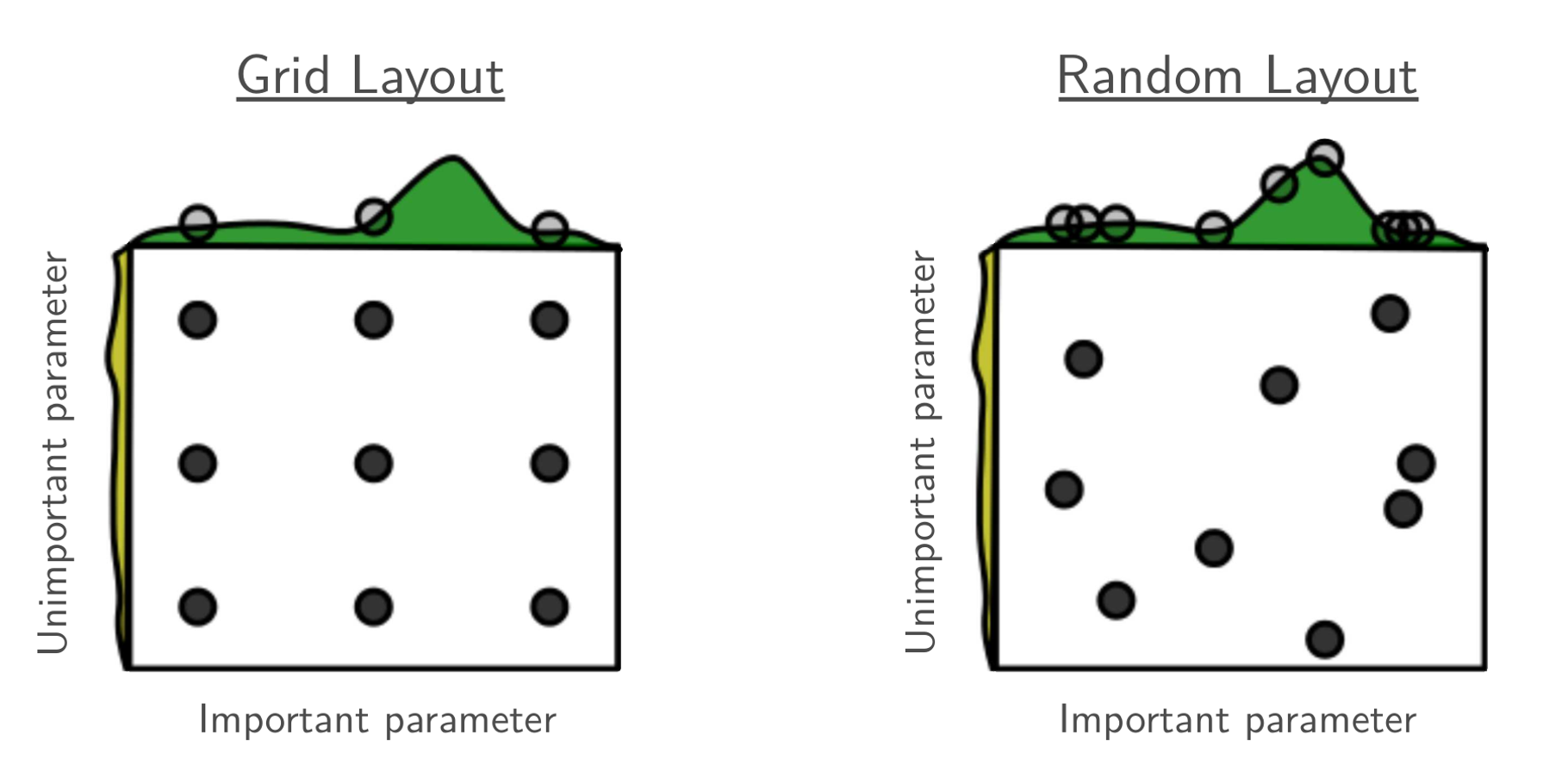}
    \caption{Comparison between random search and grid search in hyperparameter optimization. Figure from \cite{bergstra_random_2012}.}
    \label{fig:grid_random}
\end{figure}
}

\newcommand{\figureNAS}{
\begin{figure*}
    \centering
    \includegraphics[width=\textwidth]{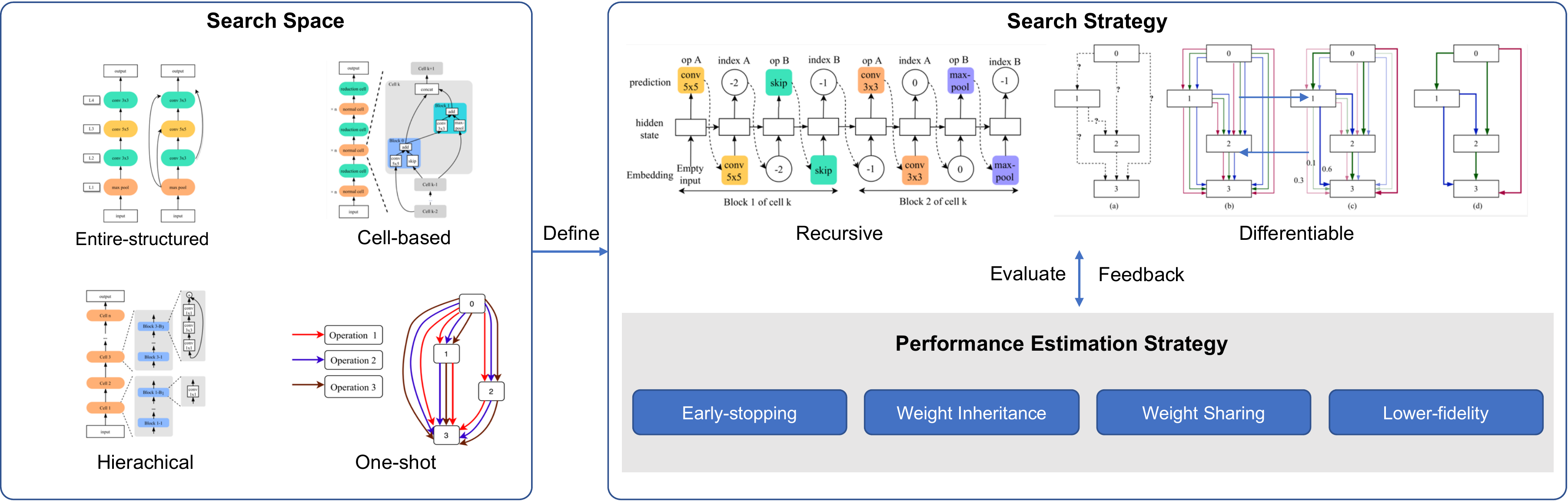}
    \caption{The automated modeling pipeline for architecture search. The structure of this figure is adapted from \cite{elsken_efficient_2019}. Illustrations of search spaces and search strategies are cited from \cite{he_automl_2021, zhang_one-shot_2021}.}
    \label{fig:nas_pipeline}
\end{figure*}
}

\newcommand{\figurePretrain}{
    \begin{figure*}
        \centering
        \includegraphics[width=\textwidth]{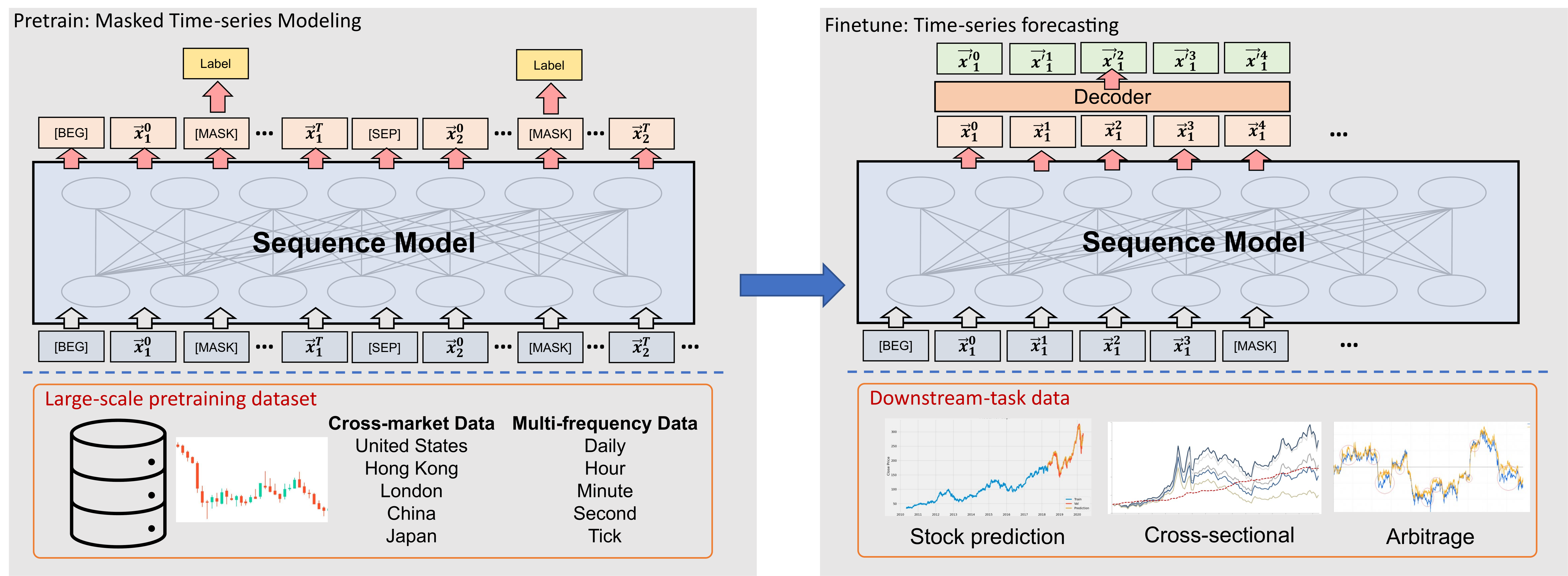}
        \caption{Pretrain-finetune paradigm. Figure is adapted from \cite{devlin_bert_2019, zhang_mg-bert_2021}.}
        \label{fig:pretrain}
    \end{figure*}
}

\newcommand{\figureSearchStrategy}{
\begin{figure*}
    \centering
    \begin{subfigure}{0.48\textwidth}
    \centering
    \includegraphics[width=\linewidth]{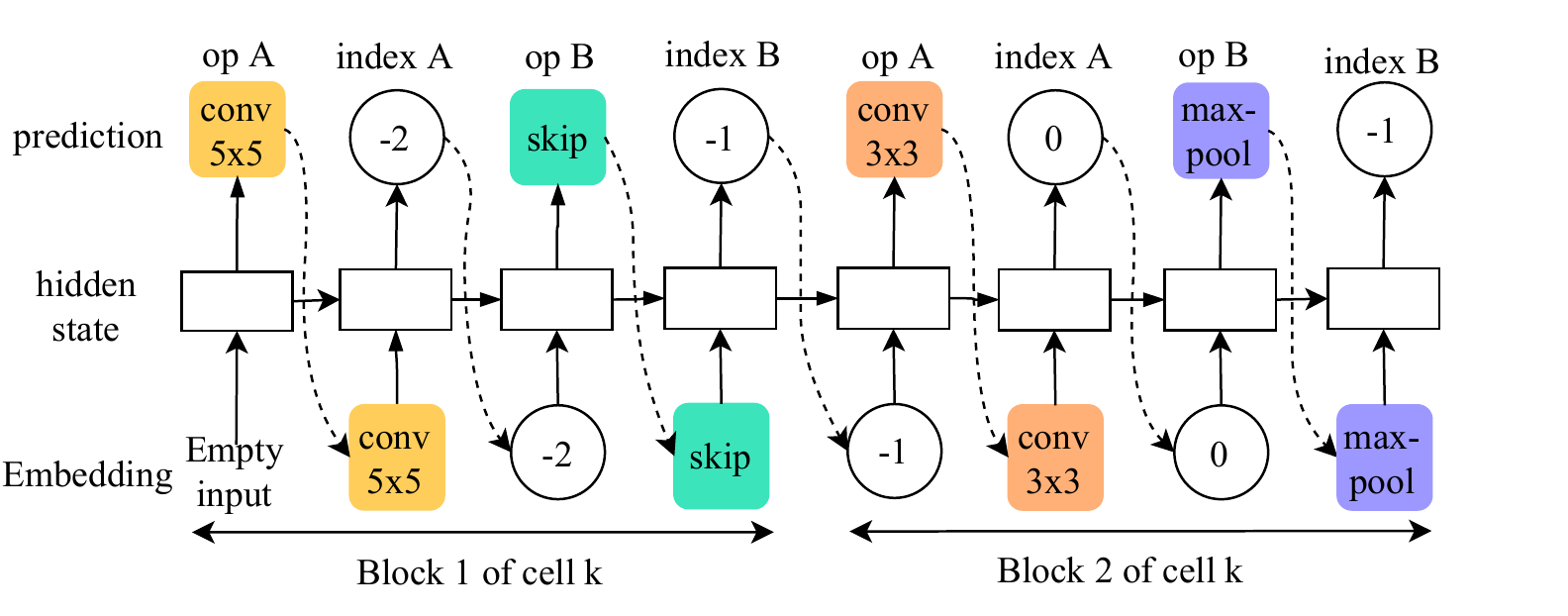}
    \caption{RNN controller generating blocks sequentially. Figure from \cite{he_automl_2021}}
    \label{fig:rnn_controller}
    \end{subfigure}
    \begin{subfigure}{0.48\textwidth}
    \centering
    \includegraphics[width=\linewidth]{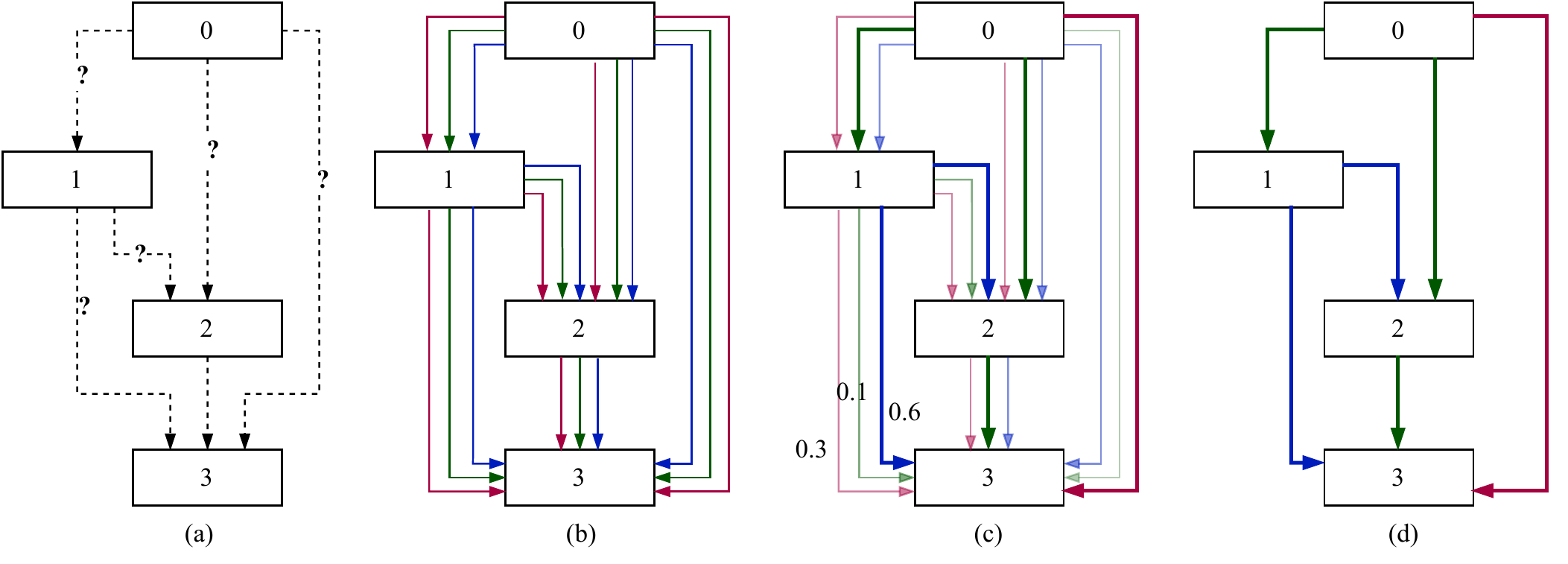}
    \caption{Differentiable architecture search \cite{he_automl_2021}}
    \label{fig:darts}
    \end{subfigure}
    \caption{Architectural search strategies}
    \label{fig:search_strategy}
\end{figure*}
}

\newcommand{\figureBayesianOpt}{
\begin{figure}[!ht]
    \centering
    \includegraphics[width=\linewidth]{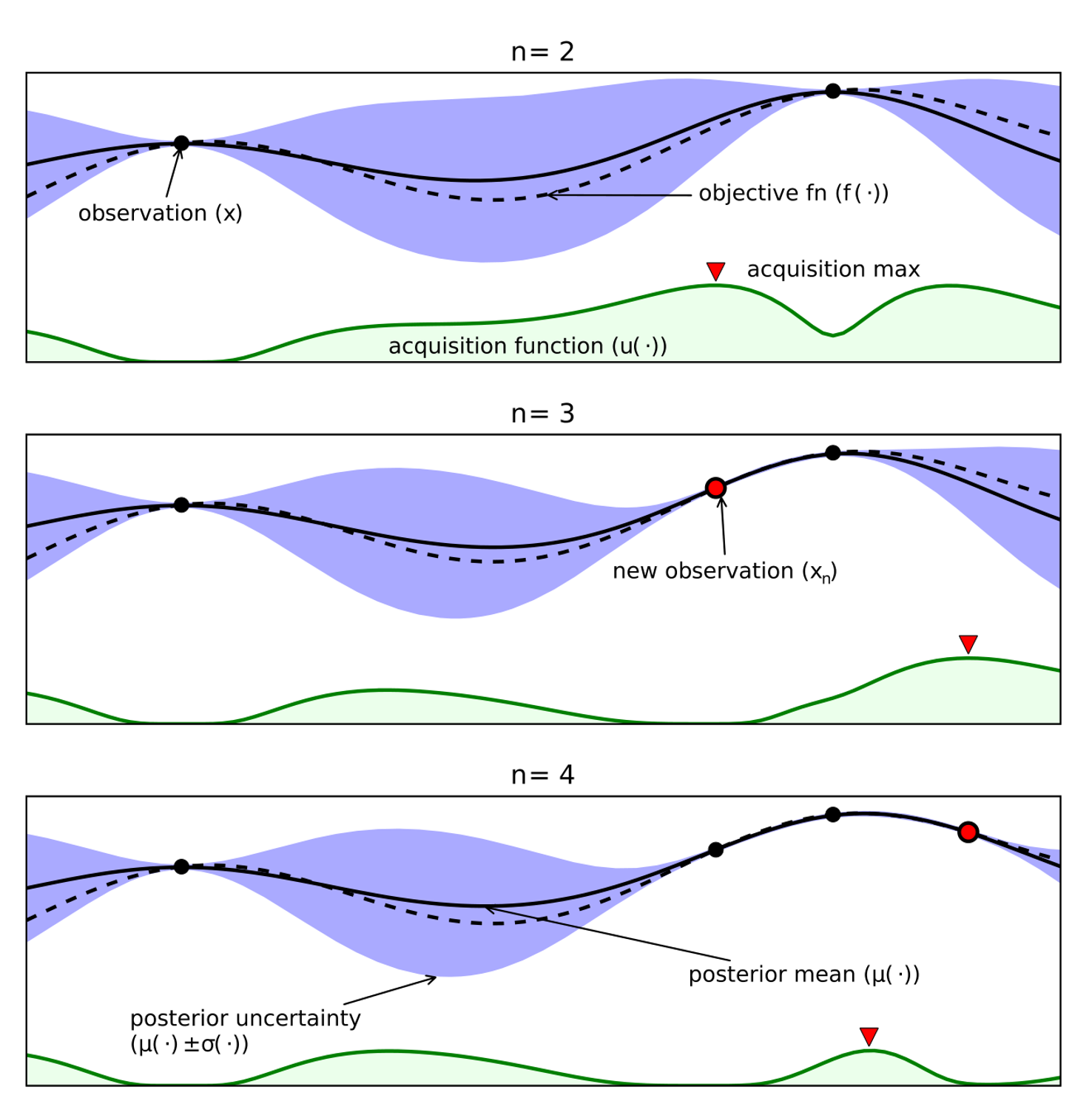}
    \caption{Illustration of Bayesian optimization. Figure from \cite{shahriari_taking_2016}}
    \label{fig:bayesian_opt}
\end{figure}
}

\newcommand{\figureFactorDependency}{
\begin{figure}
    \centering
    \includegraphics[width=\linewidth]{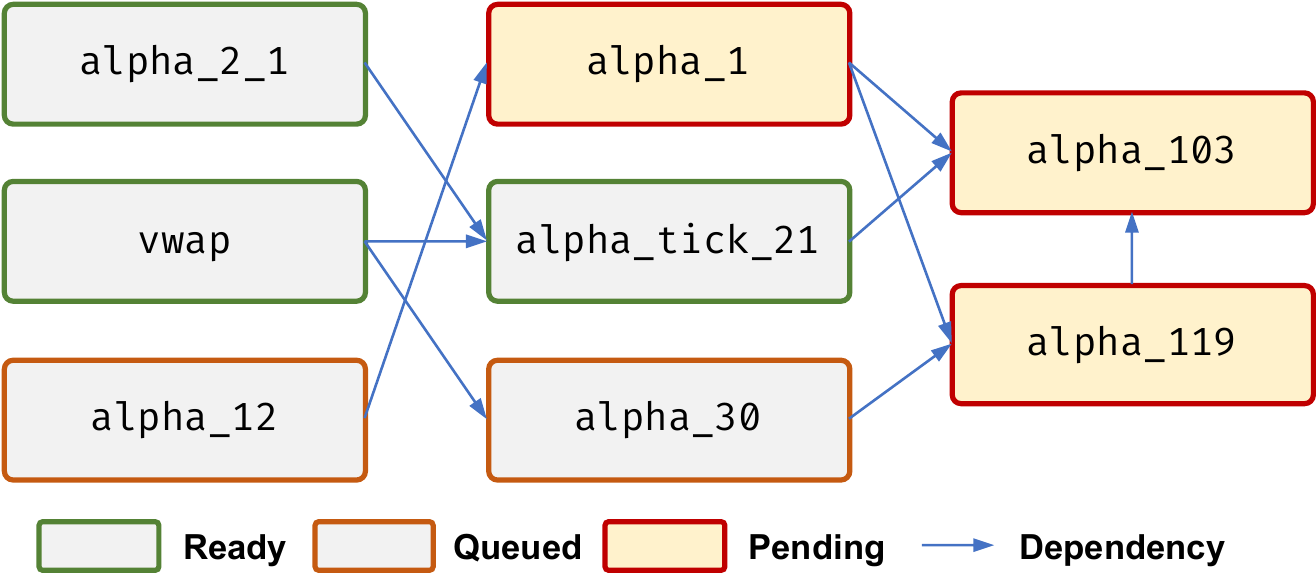}
    \caption{Factor dependency plot}
    \label{fig:factor_dag}
\end{figure}
}

\newcommand{\figureInferenceOpt}{
\begin{figure*}
    \centering
    \begin{subfigure}{0.53\textwidth}
    \centering
    \includegraphics[width=\linewidth]{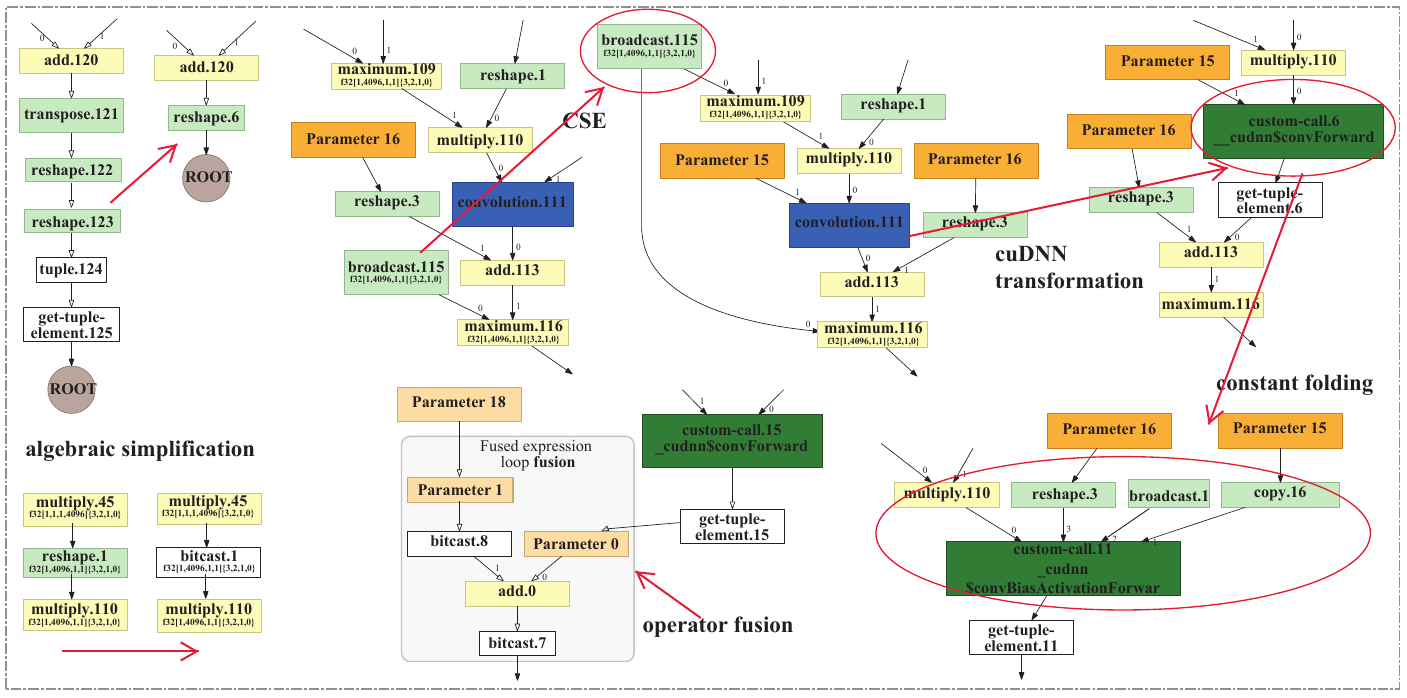}
    \caption{Front-end optimization}
    \label{fig:front_end_opt}
    \end{subfigure}
    \begin{subfigure}{0.46\textwidth}
    \centering
    \includegraphics[width=\linewidth]{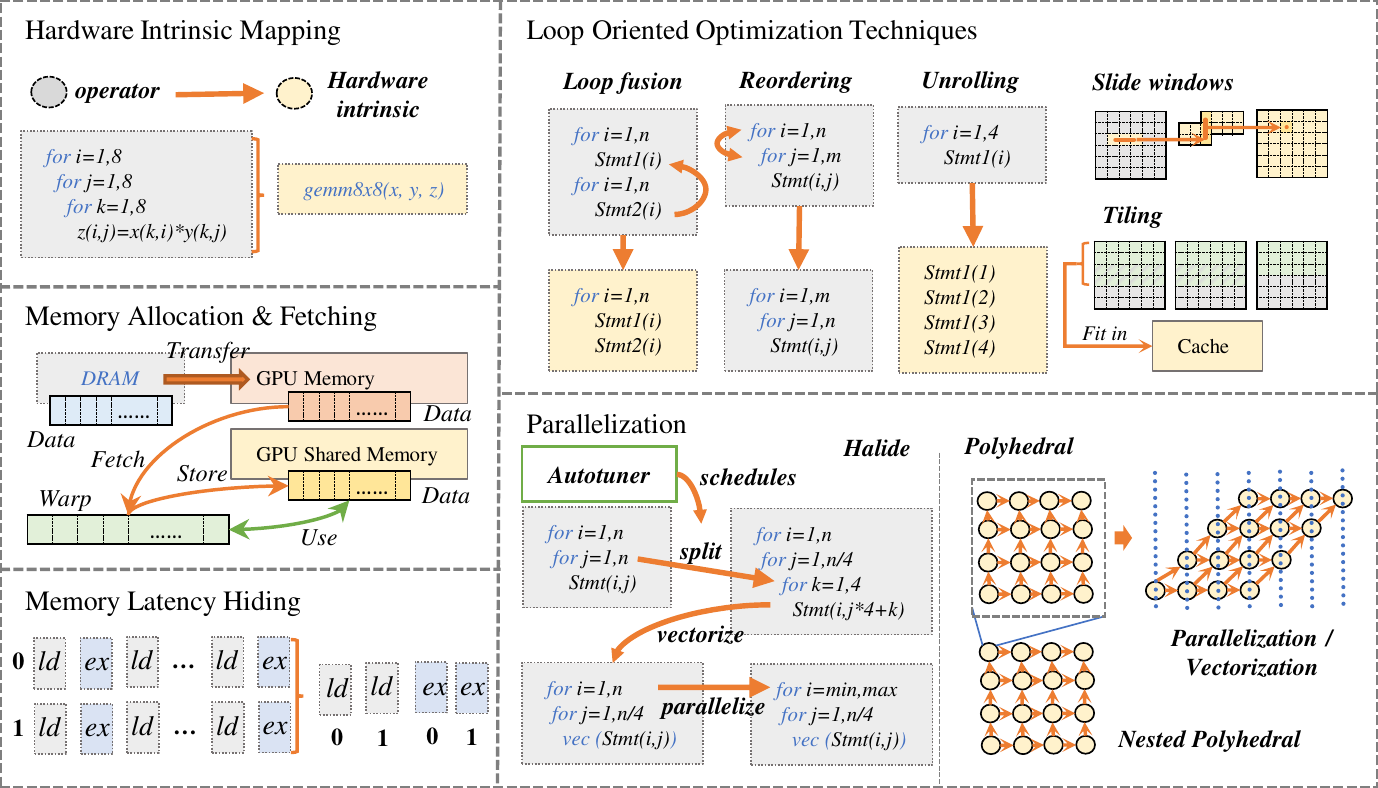}
    \caption{Back-end optimization}
    \label{fig:back_end_opt}
    \end{subfigure}
    \caption{Deep learning compiler optimization techniques. Figure from \cite{li_deep_2021}}
    \label{fig:dl_compiler}
\end{figure*}
}

\newcommand{\figurePruneQuantize}{
\begin{figure*}[!ht]
    \centering
    \begin{subfigure}{0.25\textwidth}
        \centering
        \includegraphics[width=\linewidth]{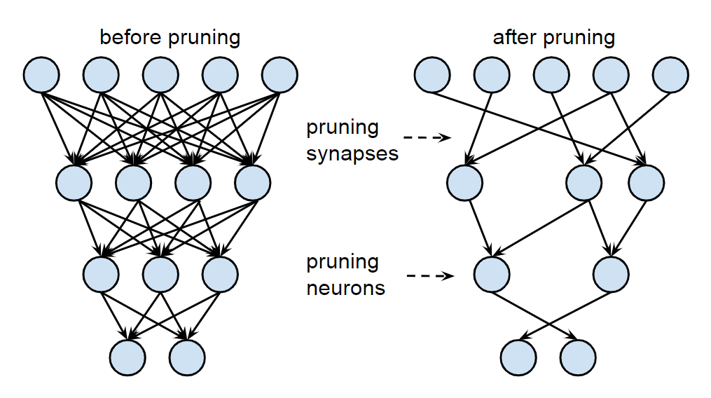}
        \caption{Model pruning. Figure from \cite{han_learning_2015}}
        \label{fig:model_pruning}
    \end{subfigure}
    \begin{subfigure}{0.20\textwidth}
        \centering
        \includegraphics[width=\linewidth]{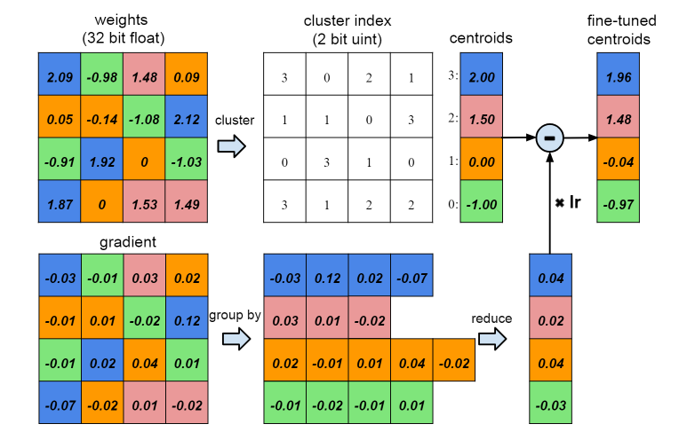}
        \caption{Model quantization. Figure from \cite{han_deep_2016}.}
        \label{fig:model_quantize}
    \end{subfigure}
    \begin{subfigure}{0.3\textwidth}
        \centering
        \includegraphics[width=\linewidth]{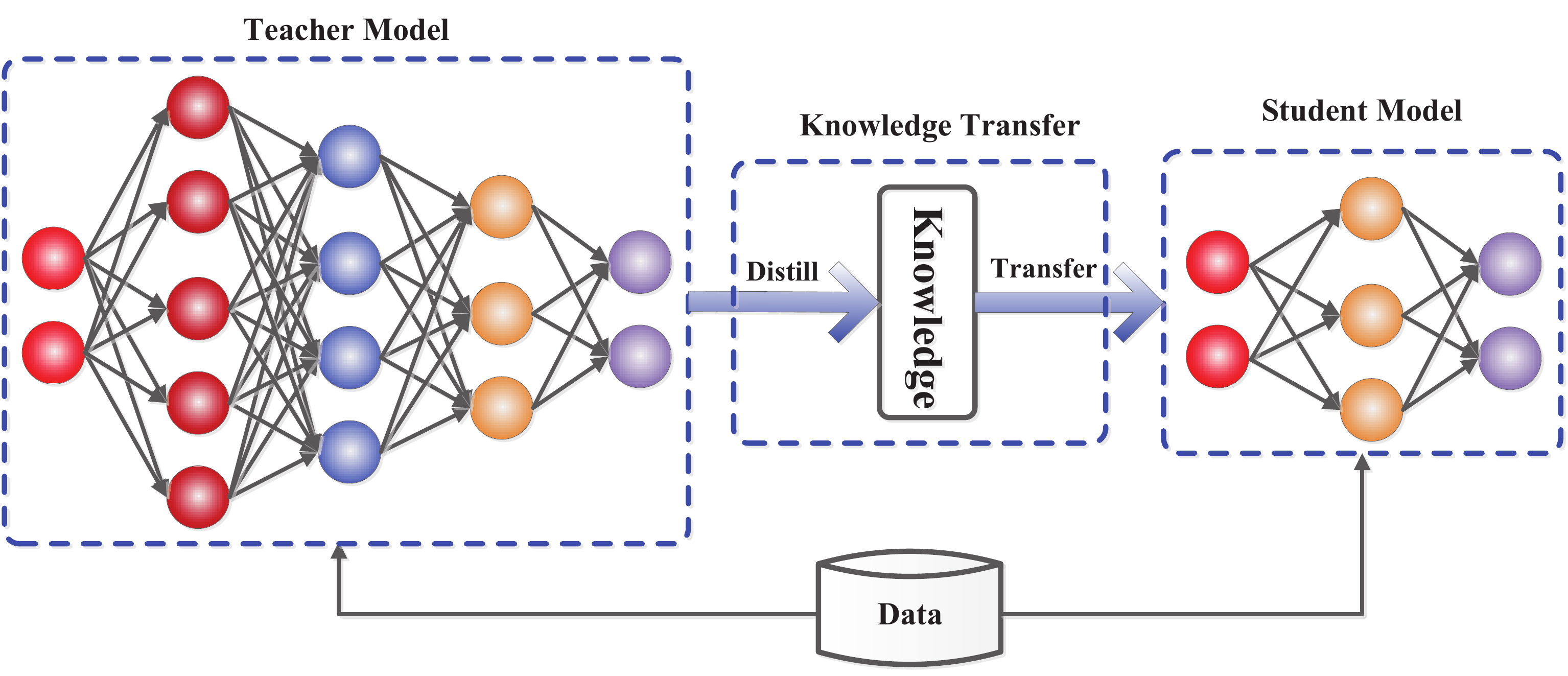}
        \caption{Knowledge distillation. Figure from \cite{gou_knowledge_2021}.}
        \label{fig:knowledge_distill}
    \end{subfigure}
    \begin{subfigure}{0.20\textwidth}
        \centering
        \includegraphics[width=\linewidth]{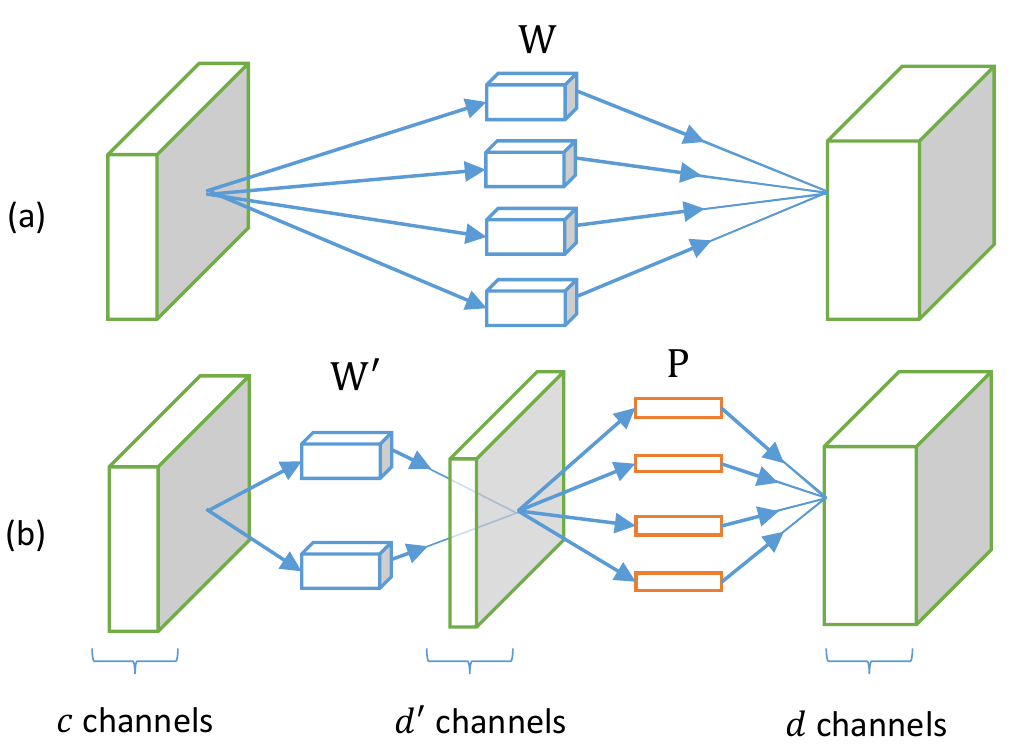}
        \caption{Low-rank approximation of CNN models. Figure from \cite{zhang_efficient_2015}.}
        \label{fig:low_rank}
    \end{subfigure}
    \caption{Model compression techniques}
    \label{fig:prune_quantize}
\end{figure*}
}

\newcommand{\figureKnowledgeDistill}{
\begin{figure}
    \centering
    \includegraphics{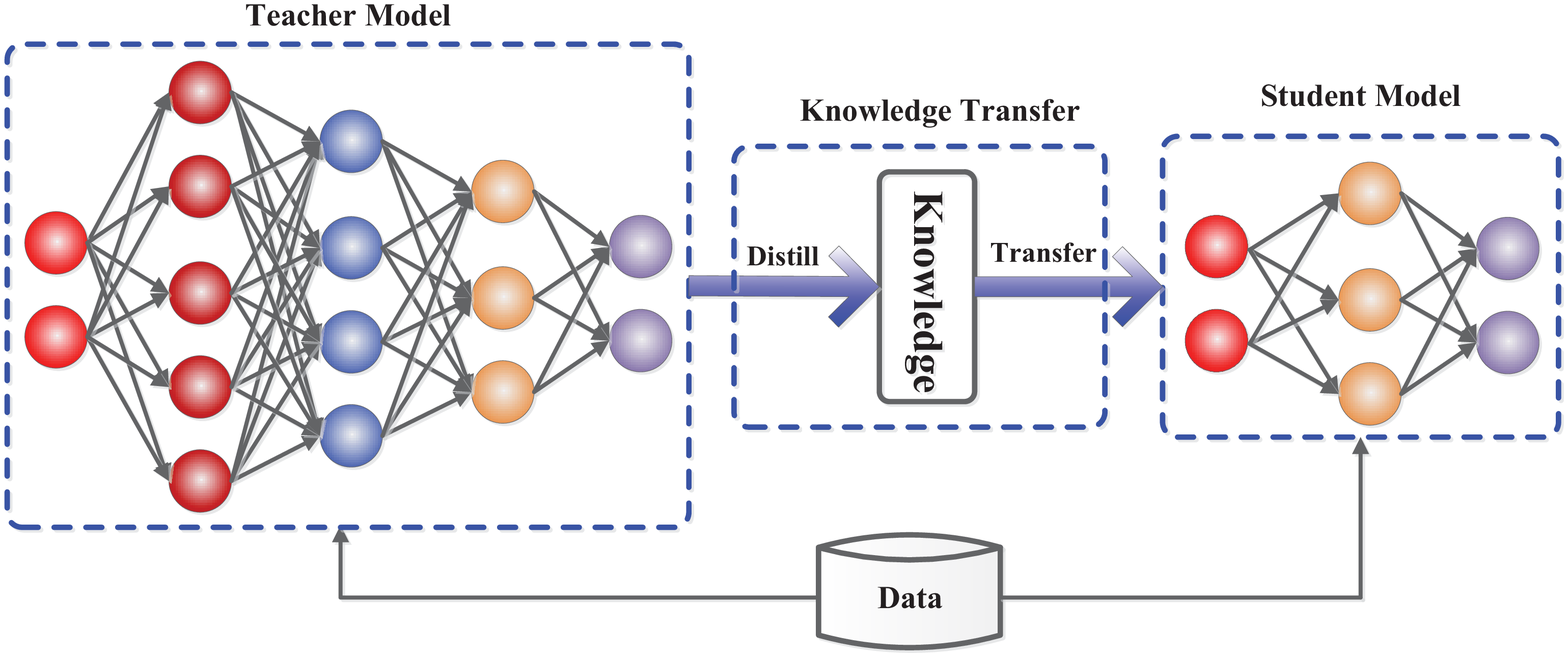}
    \caption{Knowledge distillation}
    \label{fig:knowledge_distill}
\end{figure}
}

\newcommand{\figureFactorMiningSystem}{
\begin{figure}[!ht]
    \centering
    \includegraphics[width=\linewidth]{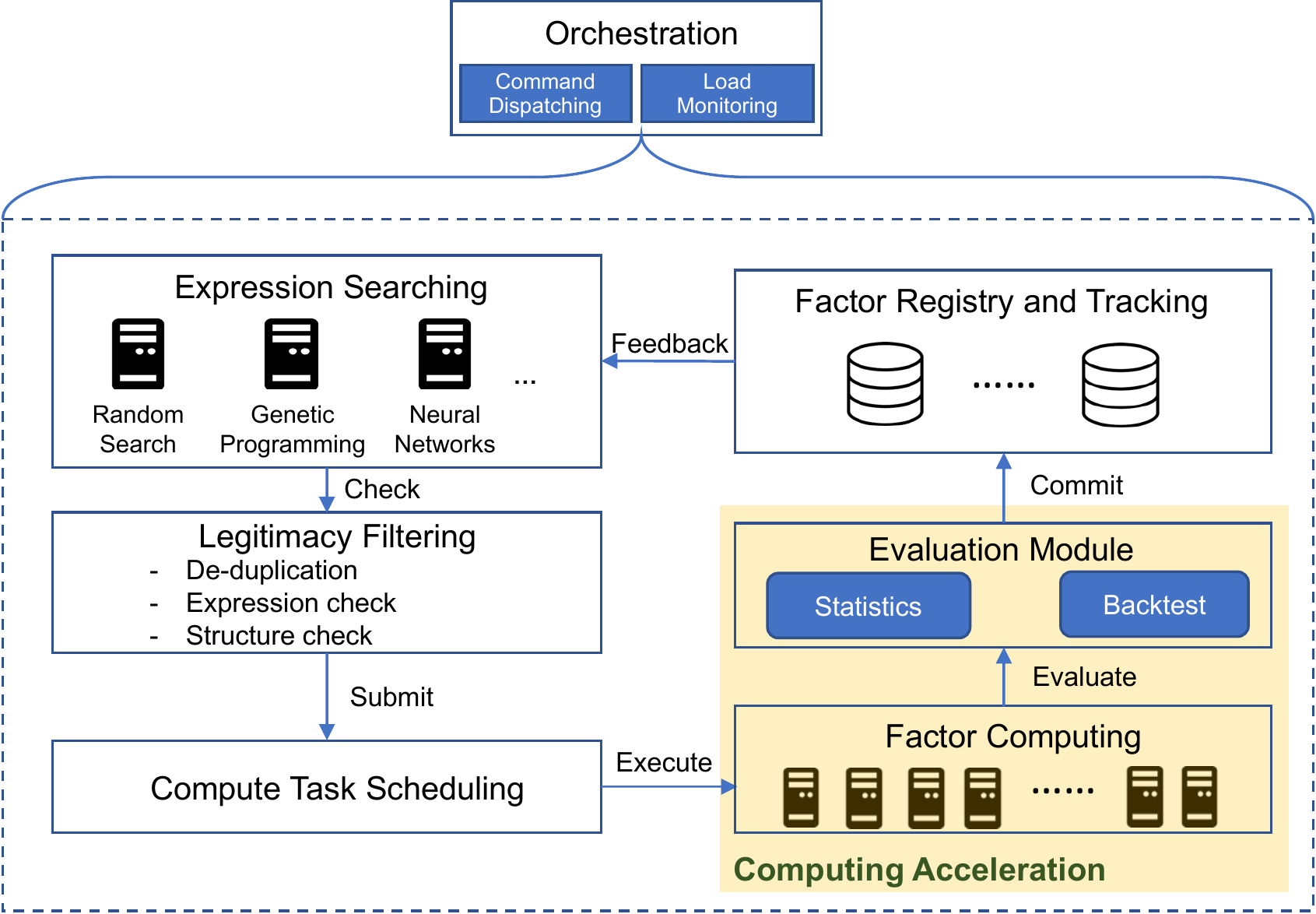}
    \caption{Architecture of the factor mining system}
    \label{fig:arch_factor_mining}
\end{figure}
}

\newcommand{\figureModelTrainingSystem}{
\begin{figure}[!ht]
    \centering
    \includegraphics[width=\linewidth]{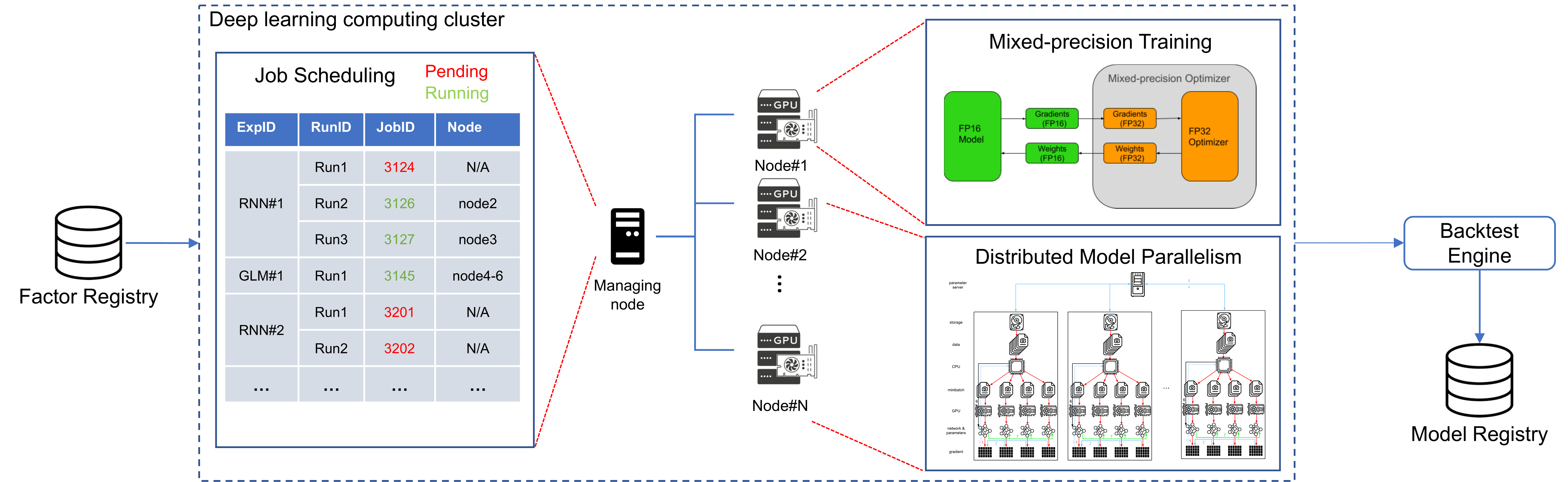}
    \caption{Architecture of the model training system. Part of this figure is cited from \cite{zhang2021dive, mixed_precision_blog}}
    \label{fig:arch_model_training}
\end{figure}
}

\newcommand{\figureStreamProcessing}{
\begin{figure}[!ht]
    \centering
    \includegraphics[width=\linewidth]{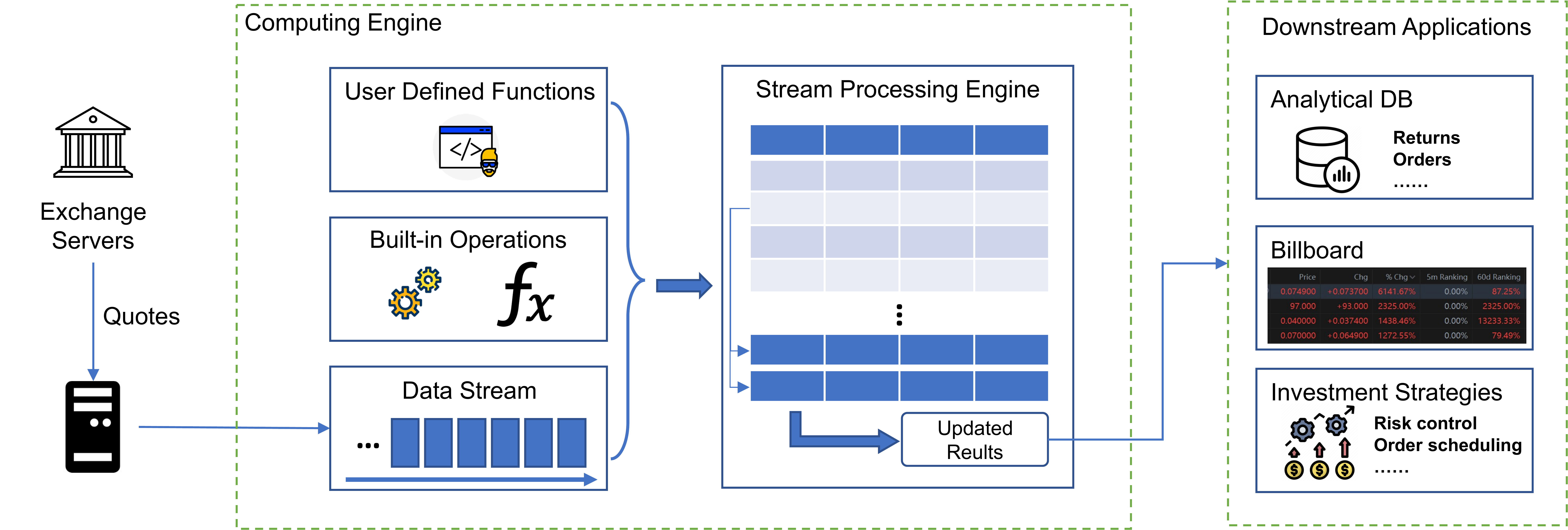}
    \caption{Stream processing workflow}
    \label{fig:stream_processing}
\end{figure}
}

\newcommand{\figureAssetMgmtEcosys}{
\begin{figure}[!t]
    \centering
    \includegraphics[width=\linewidth]{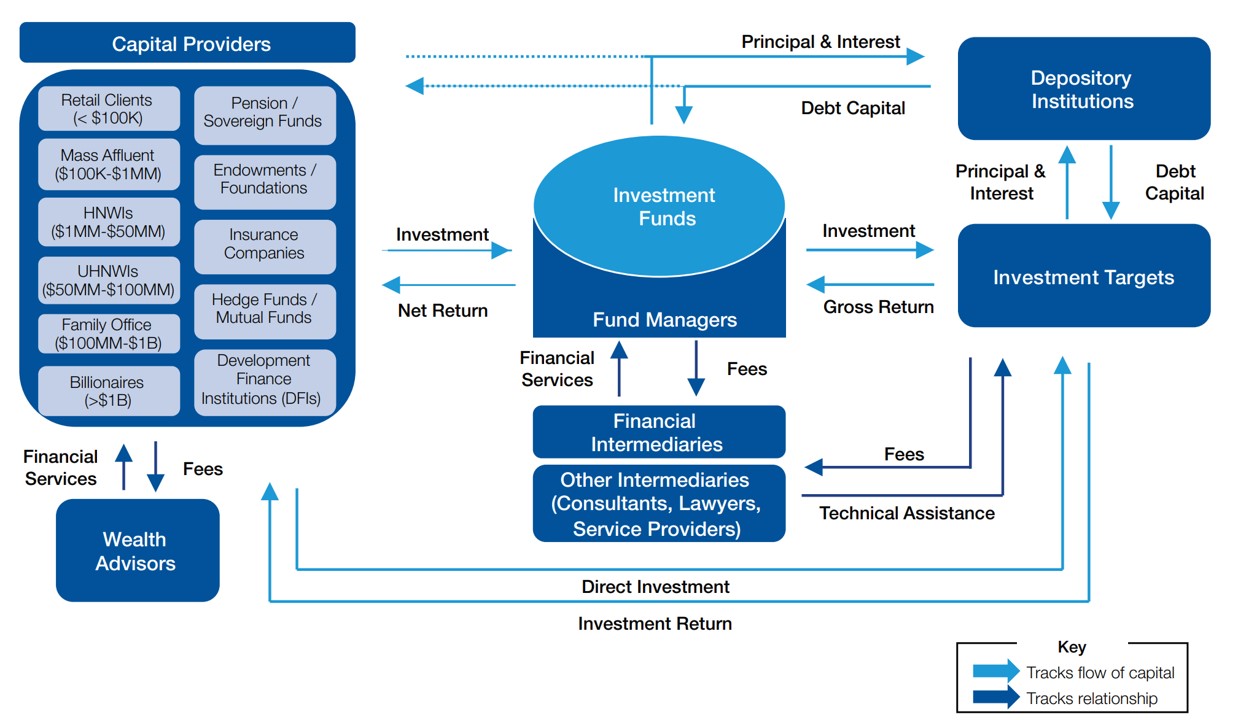}
    \caption{The asset management ecosystem \cite{wef_margins_nodate}}
    \label{fig:asset_mgmt_ecosys}
\end{figure}
}

\newcommand{\figureFinKG}{
\begin{figure*}[!ht]
    \centering
    \includegraphics[width=\textwidth]{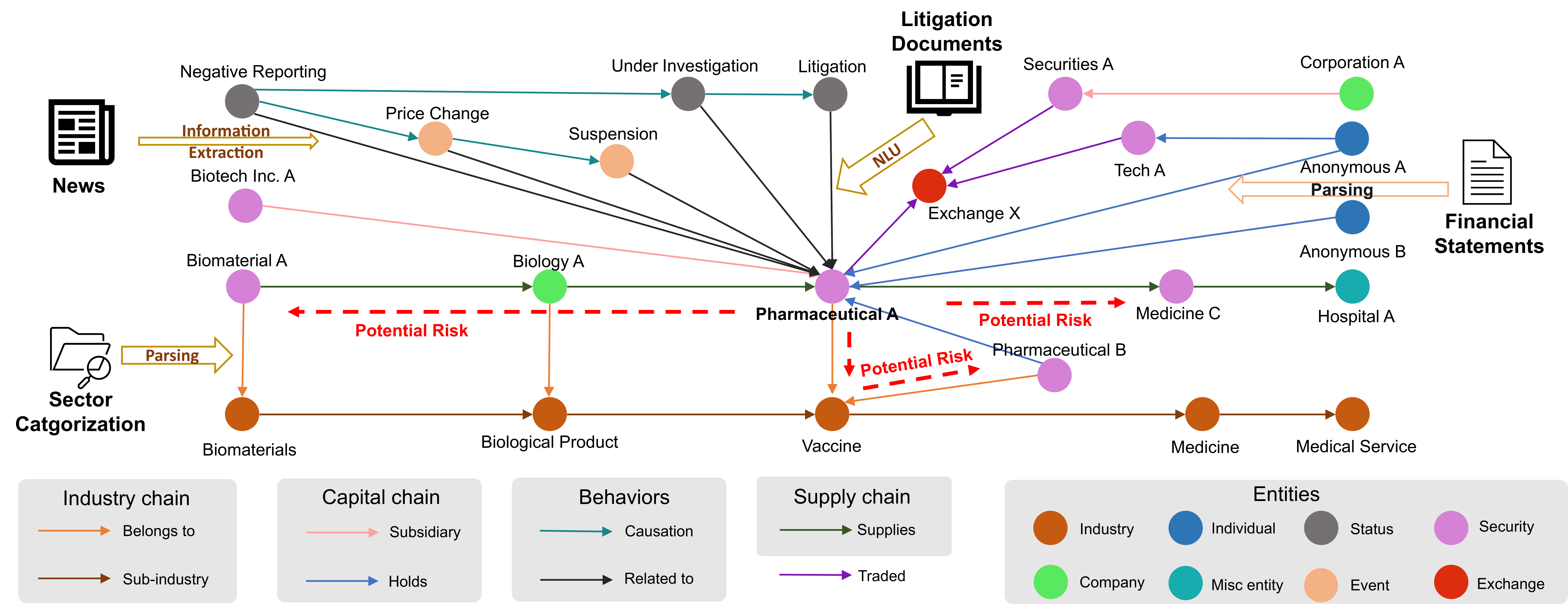}
    \caption{An example of financial knowledge graph that contains behavioral information. All the financial entities and events are fictitious and only for illustration purposes.}
    \label{fig:FinKG}
\end{figure*}
}

\newcommand{\figureTradingSystem}{
\begin{figure}[!ht]
    \centering
    \includegraphics[width=\linewidth]{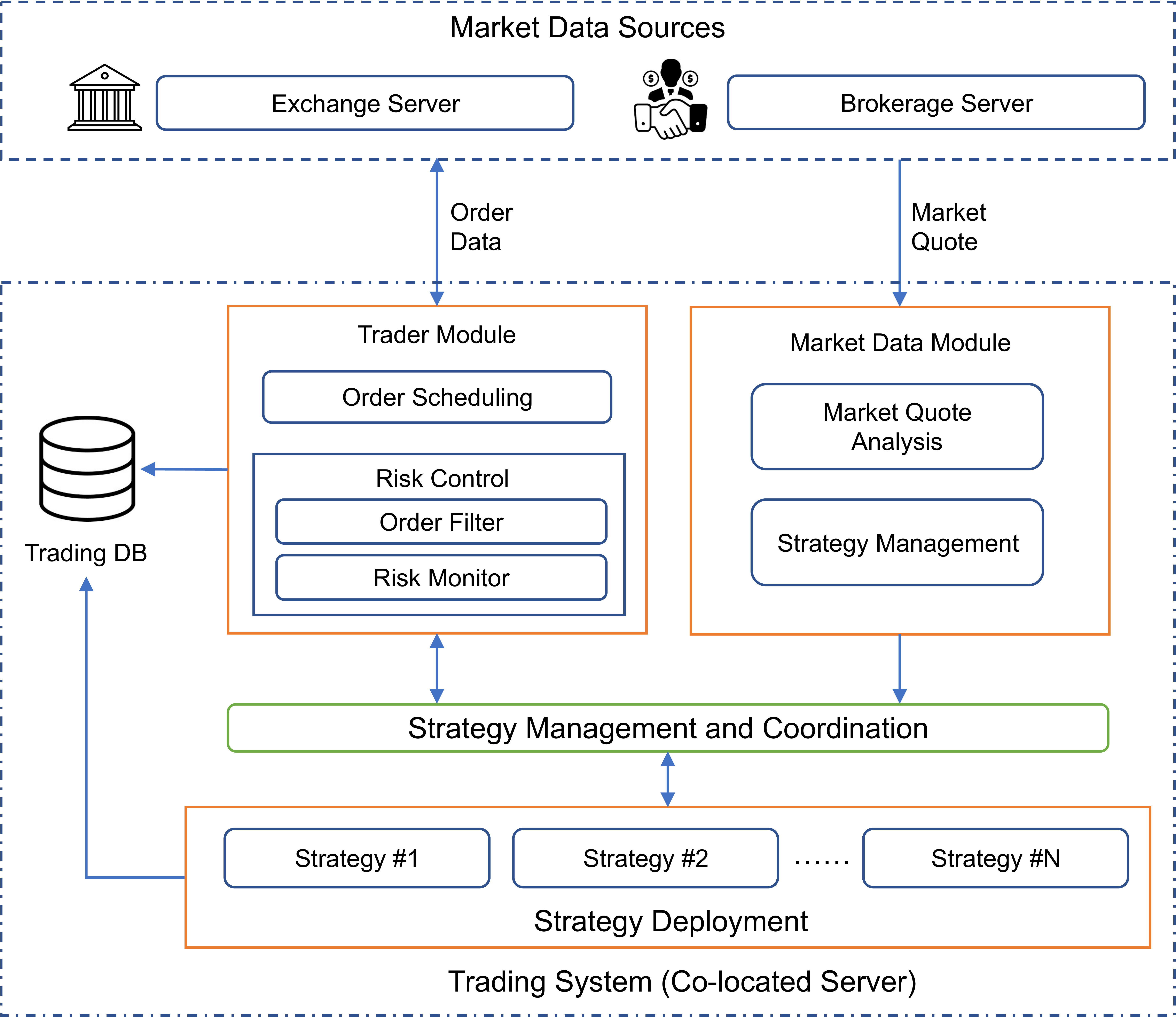}
    \caption{Architecture of the trade execution system}
    \label{fig:arch_trading_sys}
\end{figure}
}

\newcommand{\figureMLExplainabilityPlot}{
\begin{figure*}
    \centering
    \includegraphics[width=\textwidth]{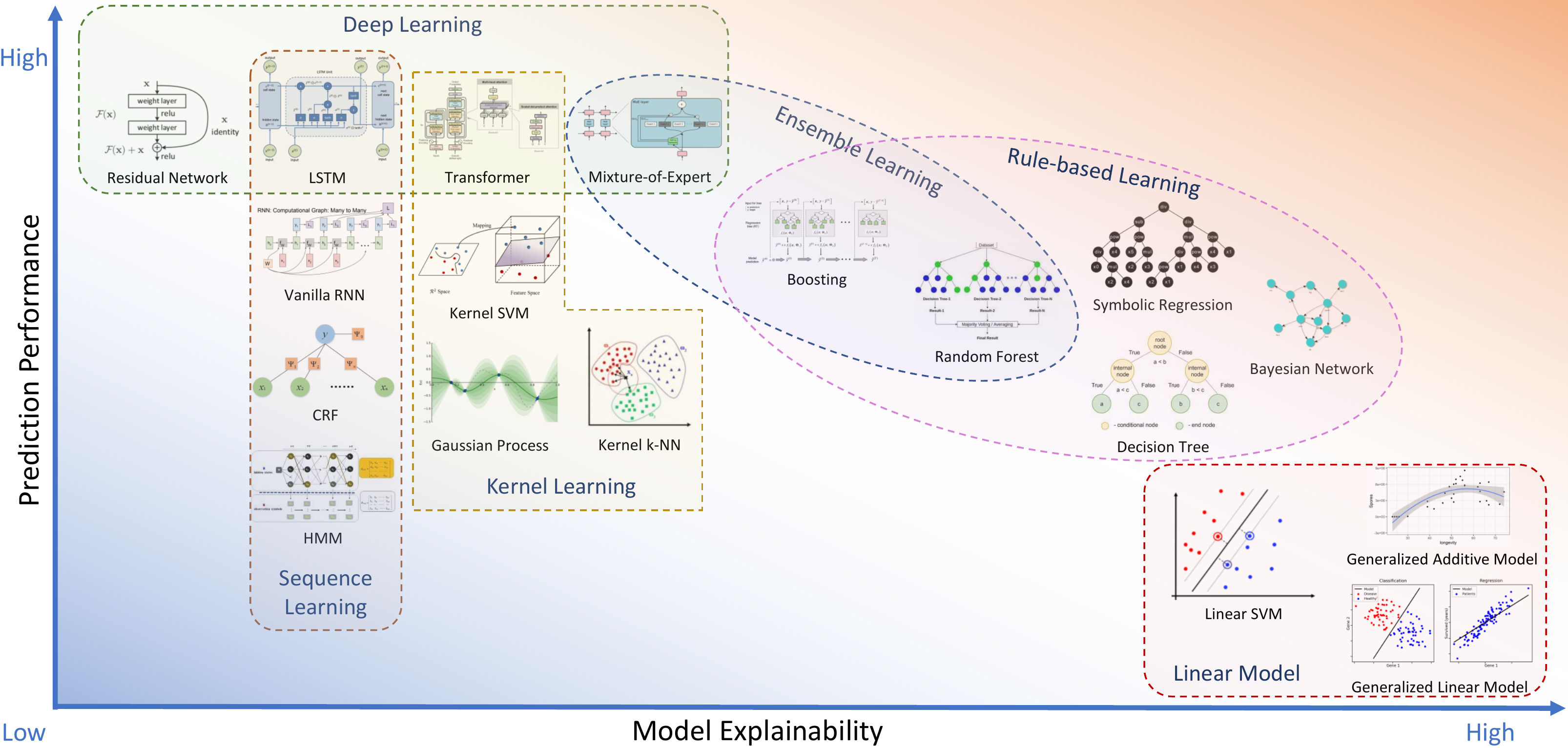}
    \caption{Comparison of popular machine learning algorithms according to prediction performance and model explainability. Part of this figure is cited from \cite{towfighi_pysrurgs_2019, meshalkin_robust_2020, wang_ss-xgboost_2020}.}
    \label{fig:explainable_ml_comparison}
\end{figure*}
}

\newcommand{\figureNeuralFactor}{
\begin{figure}[!t]
    \centering
    \includegraphics[width=\linewidth]{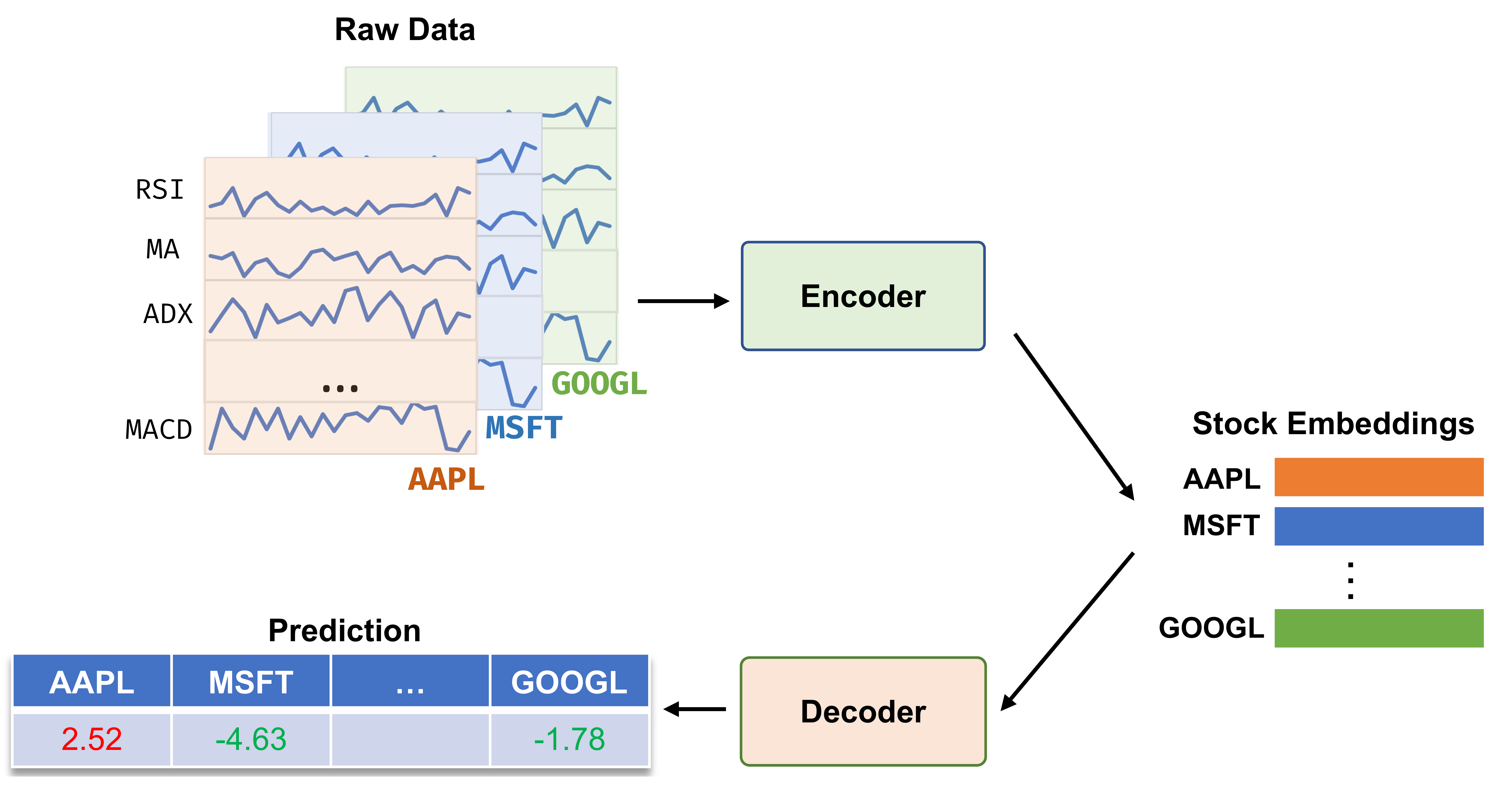}
    \caption{Illustration of the encoder-decoder architecture used in stock prediction. Both embeddings and predictions can be used as factors.}
    \label{fig:neural_factor}
\end{figure}
}

\newcommand{\figureNeuralSymbolicRegression}{
\begin{figure}[t]
    \centering
    \begin{subfigure}{\linewidth}
        \centering
        \includegraphics[width=\linewidth]{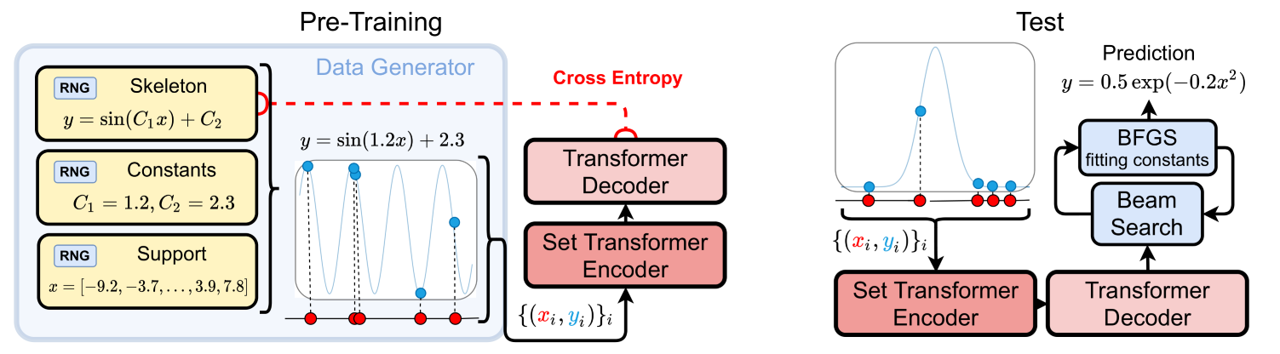}
        \caption{Neural symbolic regression in a sequence generation manner \cite{biggio_neural_2021}.}
        \label{fig:neural_symbolic_regression} 
    \end{subfigure}
    \begin{subfigure}{0.8\linewidth}
        \centering
        \includegraphics[width=\linewidth]{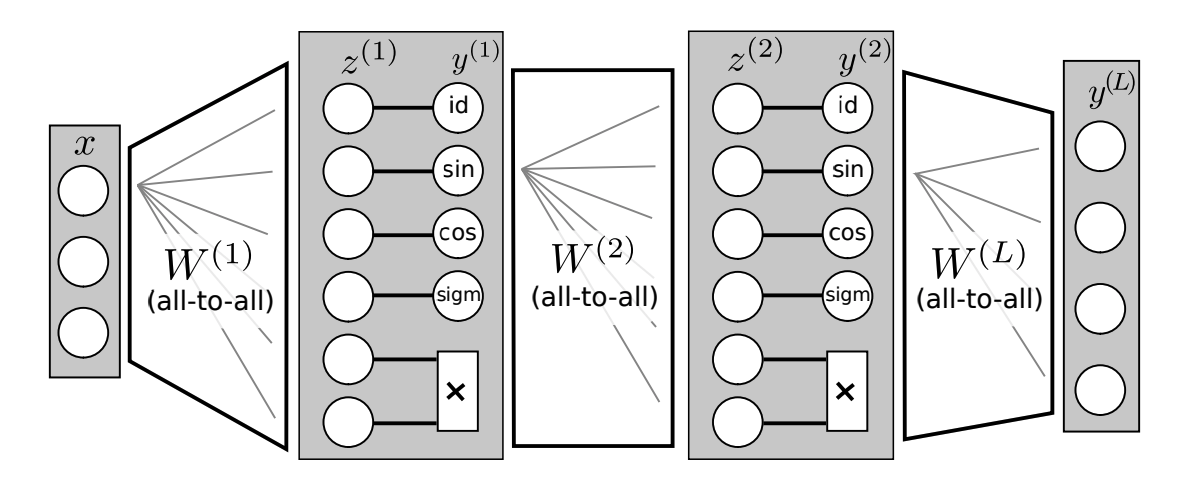}
        \caption{Neural symbolic regression that directly uses the neural network as symbolic expressions \cite{martius_extrapolation_2017}.}
        \label{fig:neural_symbolic_regression_2} 
    \end{subfigure}
    \caption{Illustrations of neural symbolic regression algorithms.}
\end{figure}
}

\newcommand{\figureGeneticProgramming}{
\begin{figure}[!t]
    \centering
    \includegraphics[width=\linewidth]{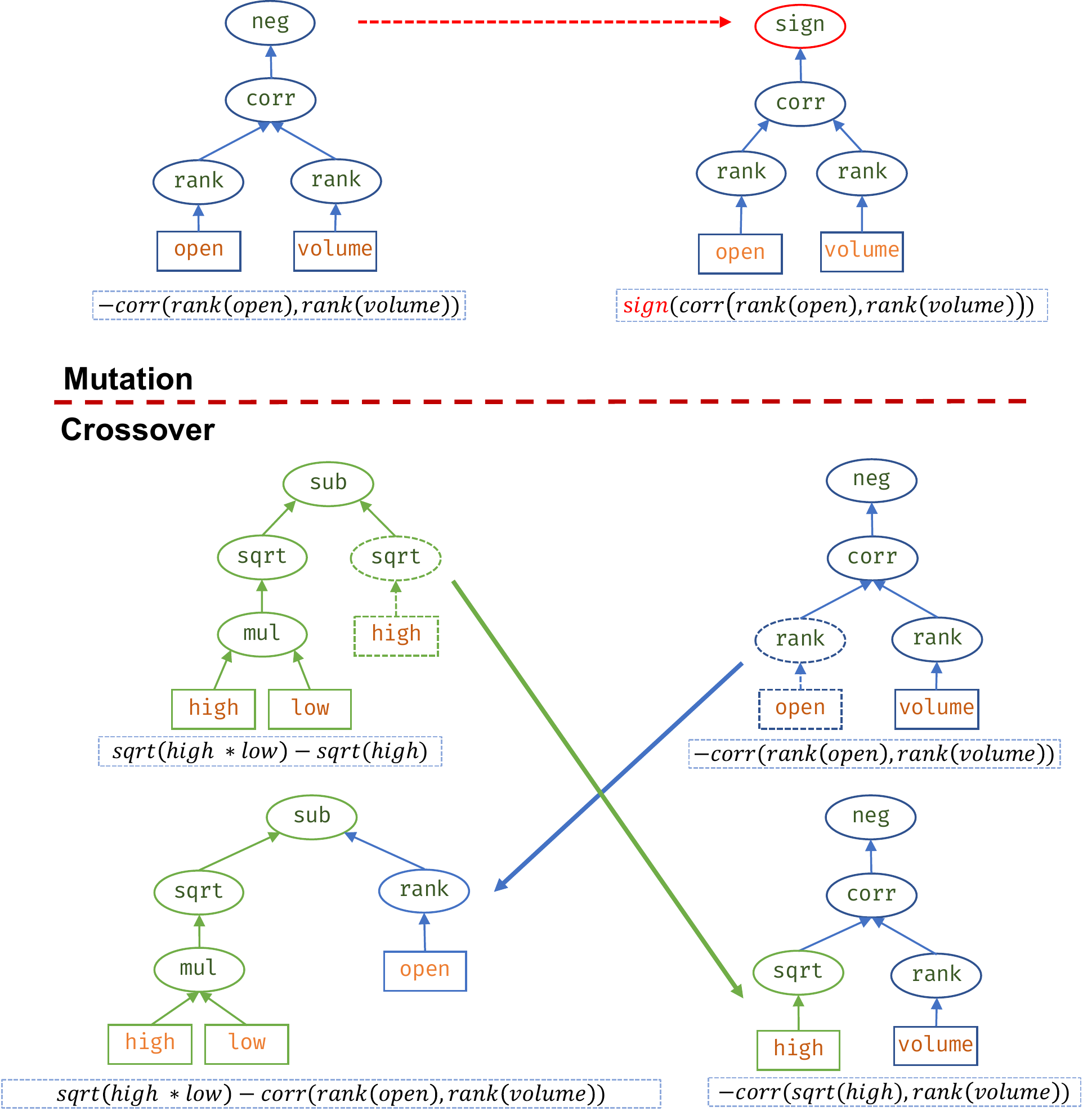}
    \caption{Illustrations of two evolutionary mechanism used in genetic programming.}
    \label{fig:genetic_programming}
\end{figure}
}

\newcommand{\figureTimeExplainability}{
\begin{figure*}
    \centering
    \begin{subfigure}{0.45\textwidth}
        \centering
        \includegraphics[width=\linewidth]{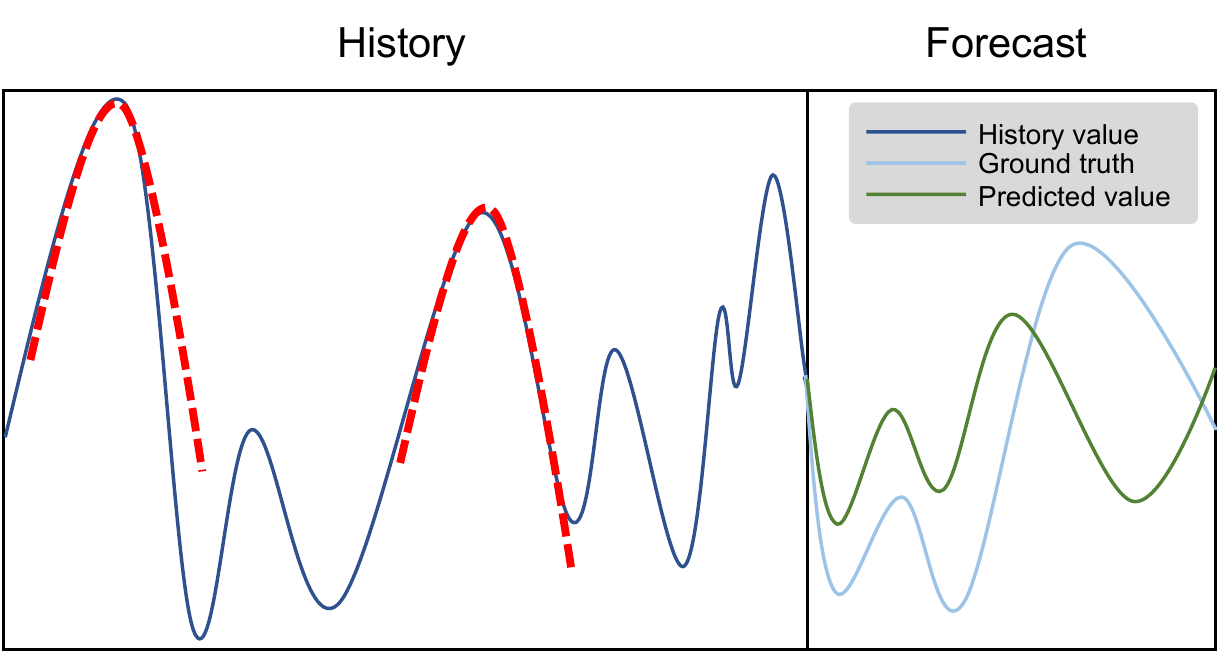}
        \caption{ABC}
        \label{fig:time_explain_single}
    \end{subfigure}
    \begin{subfigure}{0.45\textwidth}
        \centering
        \includegraphics[width=\linewidth]{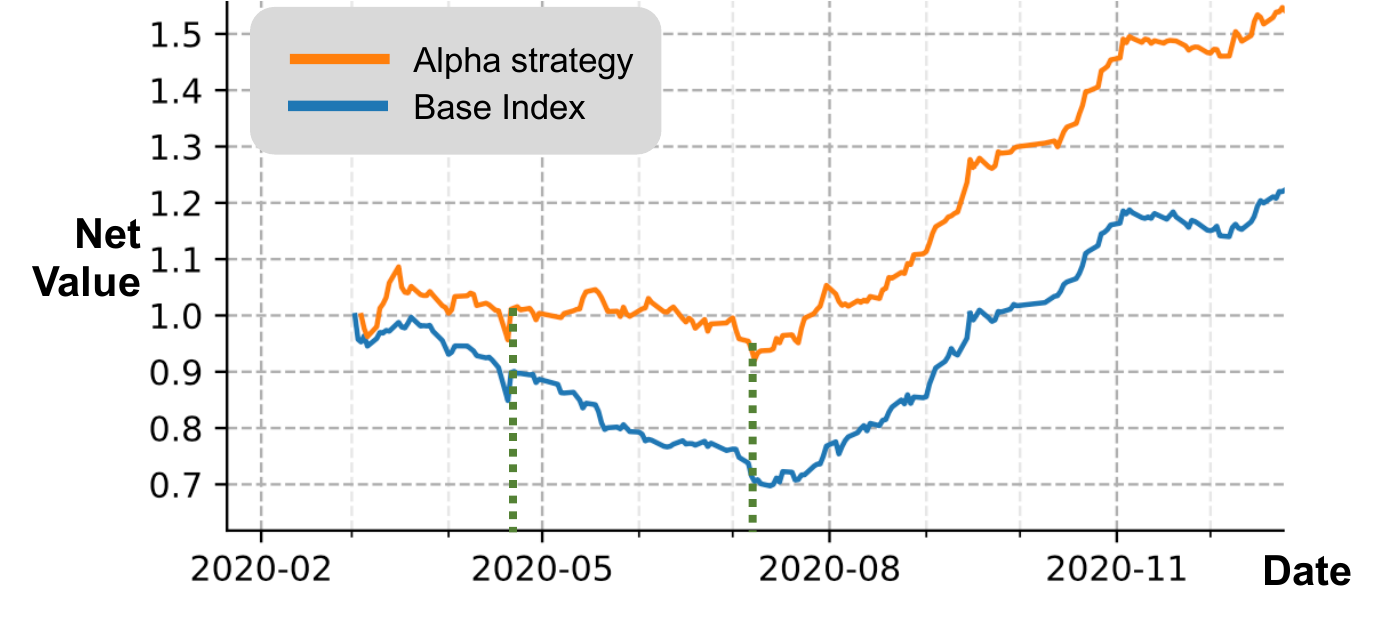}
        \caption{DEF}
        \label{fig:alpha_explain_time}
    \end{subfigure}
    \caption{Time explainability}
    \label{fig:time_explainability}
\end{figure*}
}

\newcommand{\figureAssetMgmtWorldwide}{
\begin{figure*}[ht!]
    \centering
    \begin{subfigure}{0.35\textwidth}
        \centering
        \includegraphics[width=\linewidth]{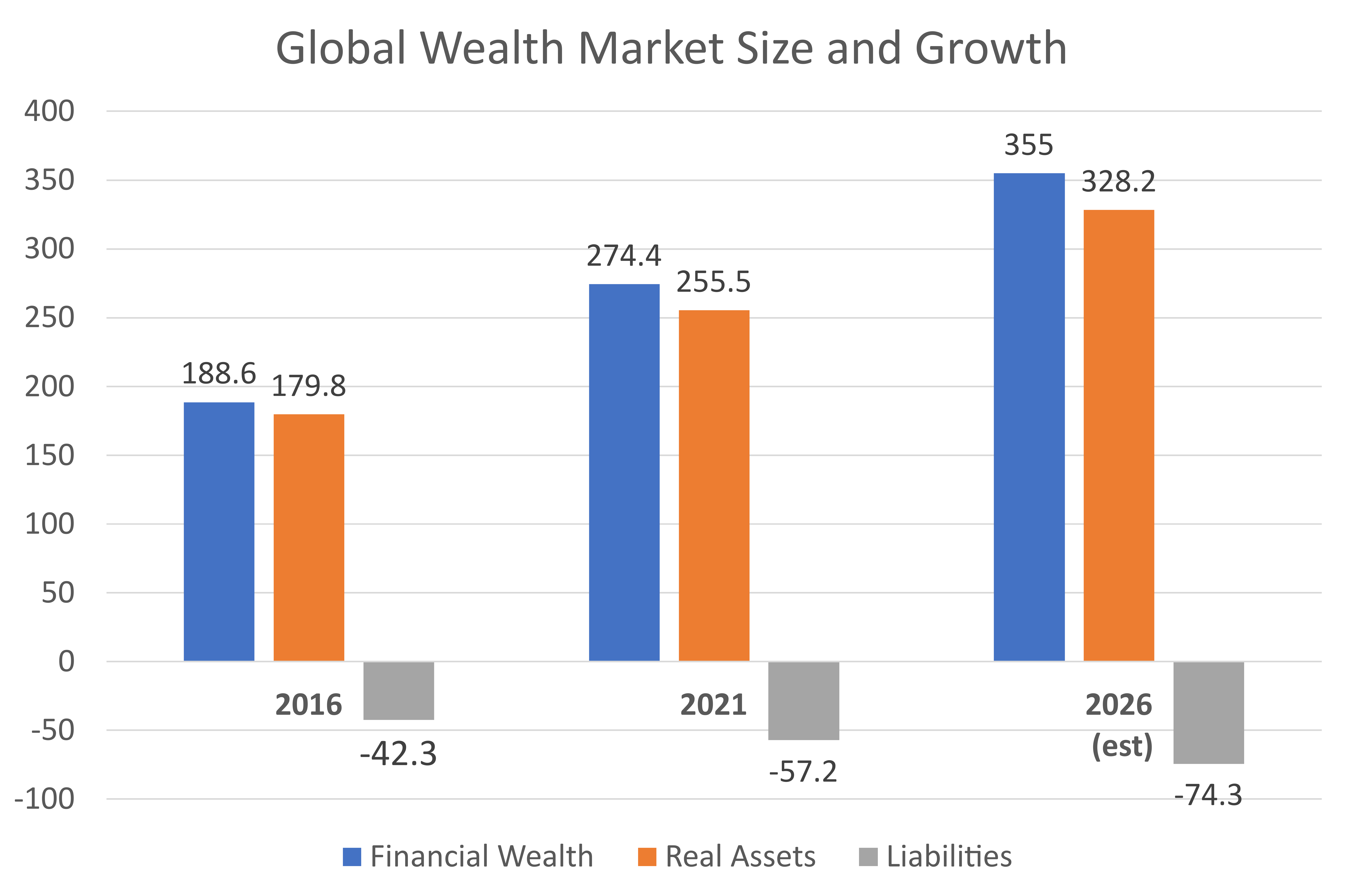}
        \caption{Global market sizes of wealth management.}
    \end{subfigure}
    \begin{subfigure}{0.64\textwidth}
        \centering
        \includegraphics[width=\linewidth]{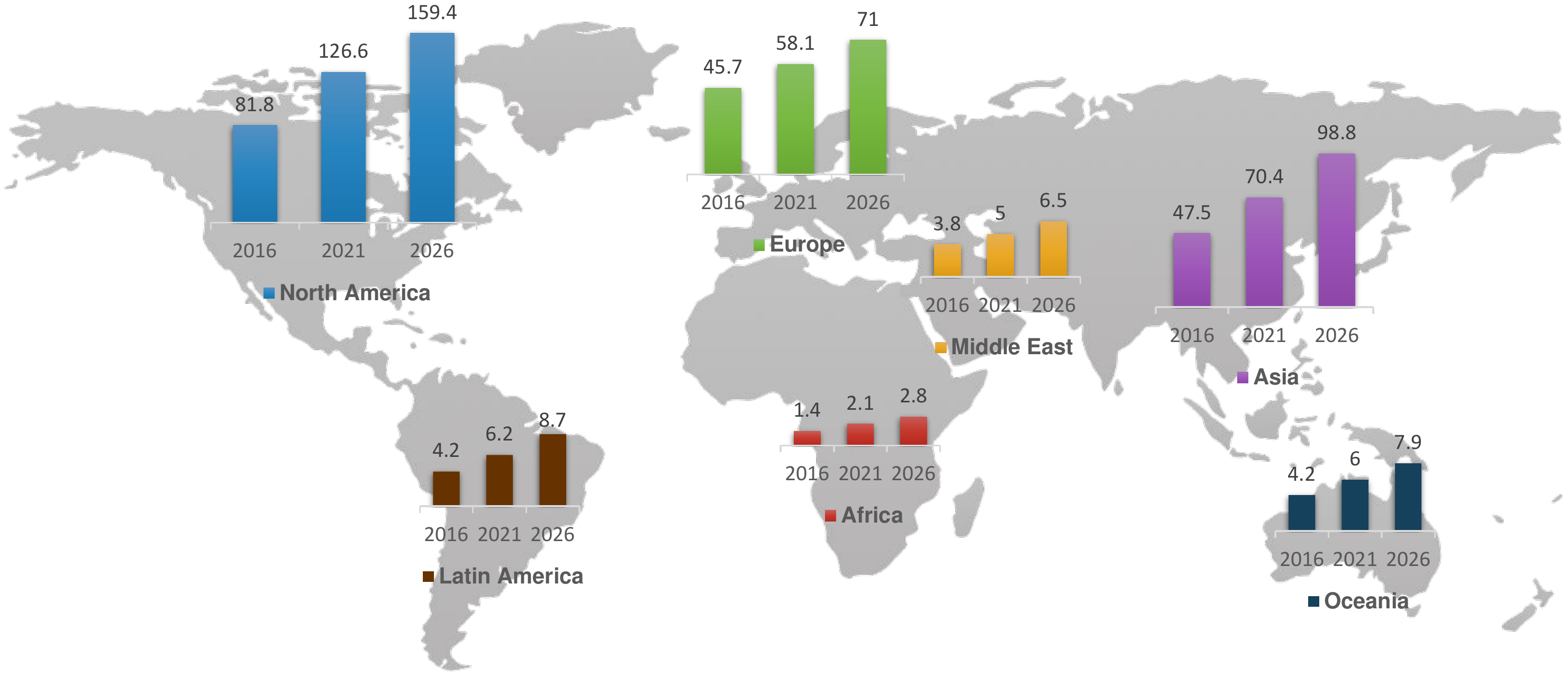}
        \caption{Regional market sizes of wealth management.}
    \end{subfigure}
    \caption{Global and regional market sizes of wealth management industry (unit: trillion USD). Panel (a) illustrates the volume of financial wealth, real assets and liabilities in 2016, 2021 and 2026 (estimated) in the world. Panel (b) shows the distribution of financial wealth in seven regional markets around the world in 2016, 2021 and 20226 (estimated). Data come from the report of BCG \cite{zakrzewski_global_2022}.}
    \label{fig:asset_mgmt_worldwide}
\end{figure*}
}

\newcommand{\figureXAIStock}{
\begin{figure*}[!ht]
    \centering
    \begin{subfigure}{0.25\textwidth}
       \centering
       \includegraphics[width=\linewidth]{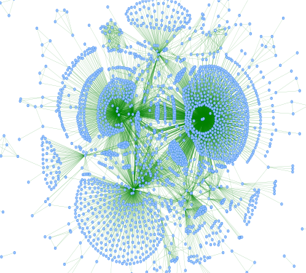}
       \caption{Stock similarity graph computed via return correlation.}
       \label{fig:xai_stock_similarity}
    \end{subfigure}
    \hfill
    \begin{subfigure}{0.37\textwidth}
       \centering
       \includegraphics[width=\linewidth]{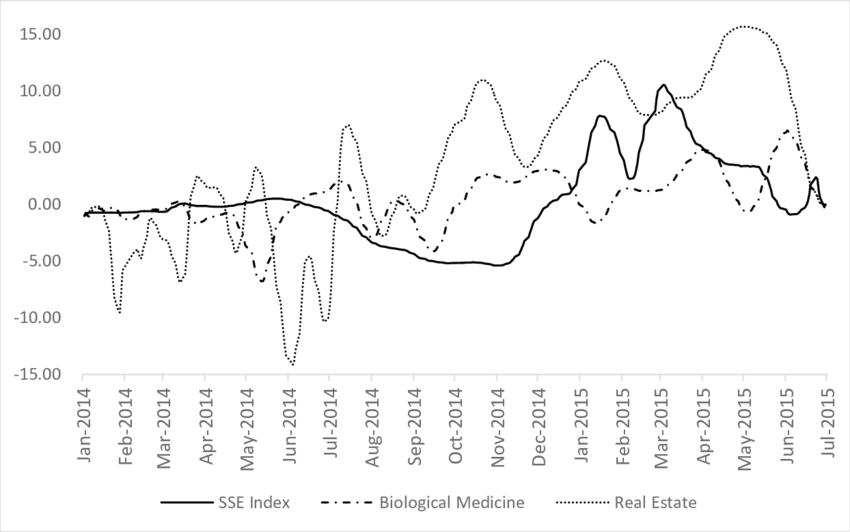}
       \caption{Lead-lag effect in Chinese stock market. Figure from \cite{guo_can_2017}.}
       \label{fig:xai_stock_lead-lag}
    \end{subfigure}
    \hfill
    \begin{subfigure}{0.28\textwidth}
       \centering
       \includegraphics[width=\linewidth]{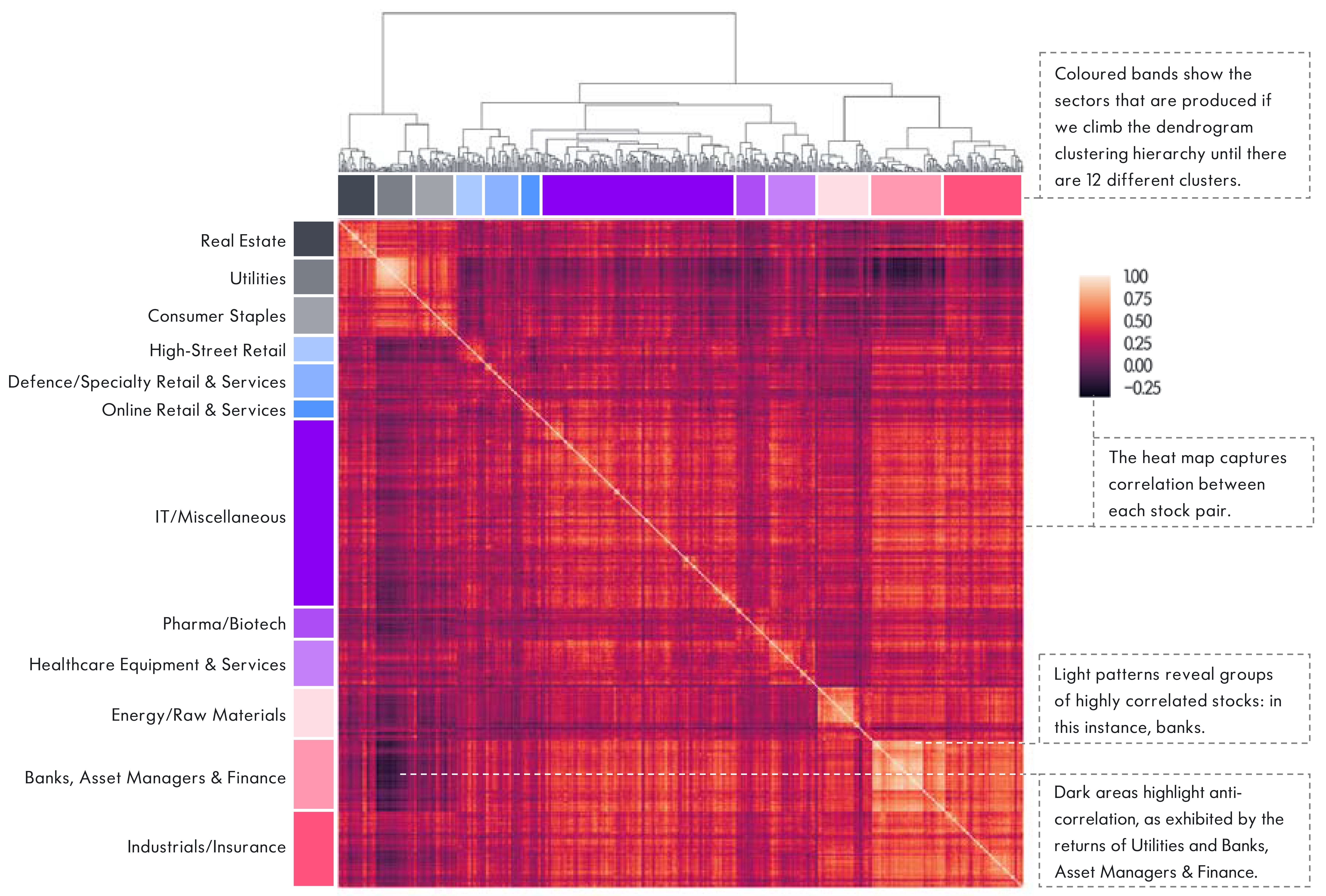}
       \caption{Stock correlation matrix. Potential clustering can be visualized from this matrix. Figure from \cite{winton_systematic_2018}.}
       \label{fig:xai_stock_latent-concept}
    \end{subfigure}
    \caption{XAI in stock.}
    \label{fig:xai_stock}
\end{figure*}
}

\newcommand{\figureXAITime}{
\begin{figure*}[!ht]
    \centering
    \begin{subfigure}{0.3\textwidth}
       \centering
       \includegraphics[width=\linewidth]{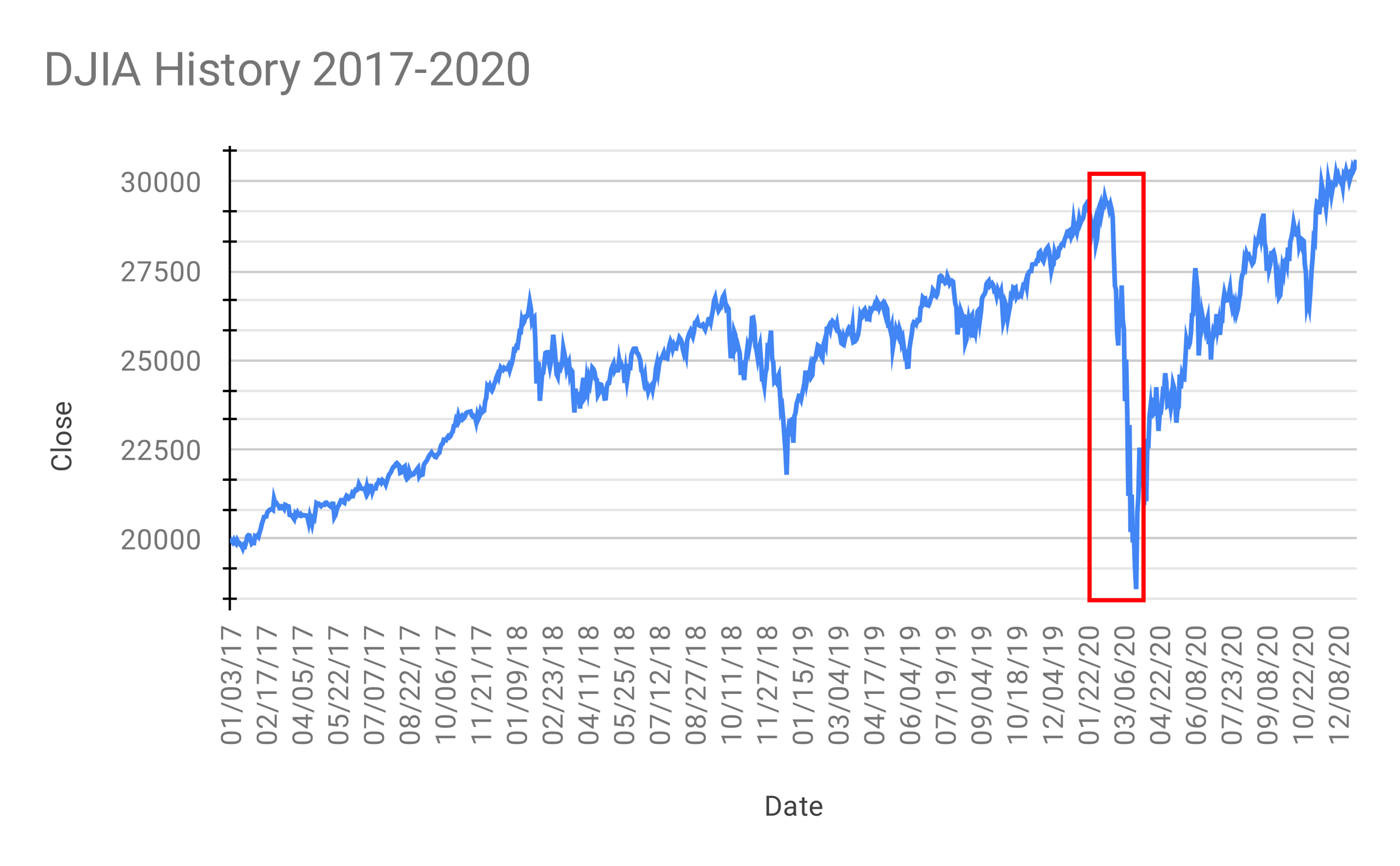}
       \caption{Extreme market condition in 2020 of Dow Jones Industrial Average. Figure from \cite{wikipedia_2020_2022}.}
       \label{fig:xai_time_extreme-market}
    \end{subfigure}
    \hfill
    \begin{subfigure}{0.33\textwidth}
       \centering
       \includegraphics[width=\linewidth]{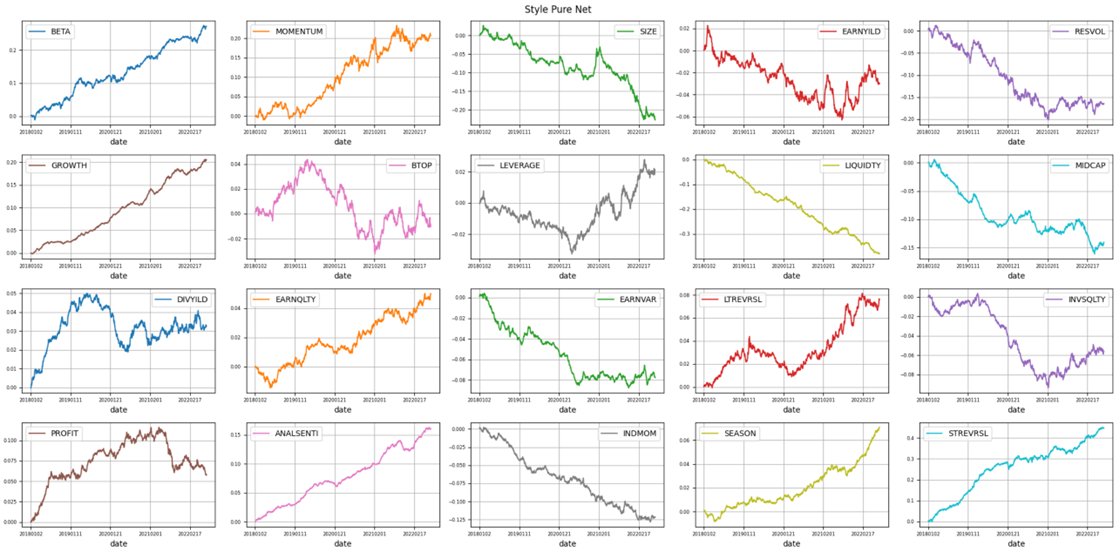}
       \caption{Return curves of BARRA risk factors.}
       \label{fig:xai_time_style}
    \end{subfigure}
    \hfill
    \begin{subfigure}{0.28\textwidth}
       \centering
       \includegraphics[width=\linewidth]{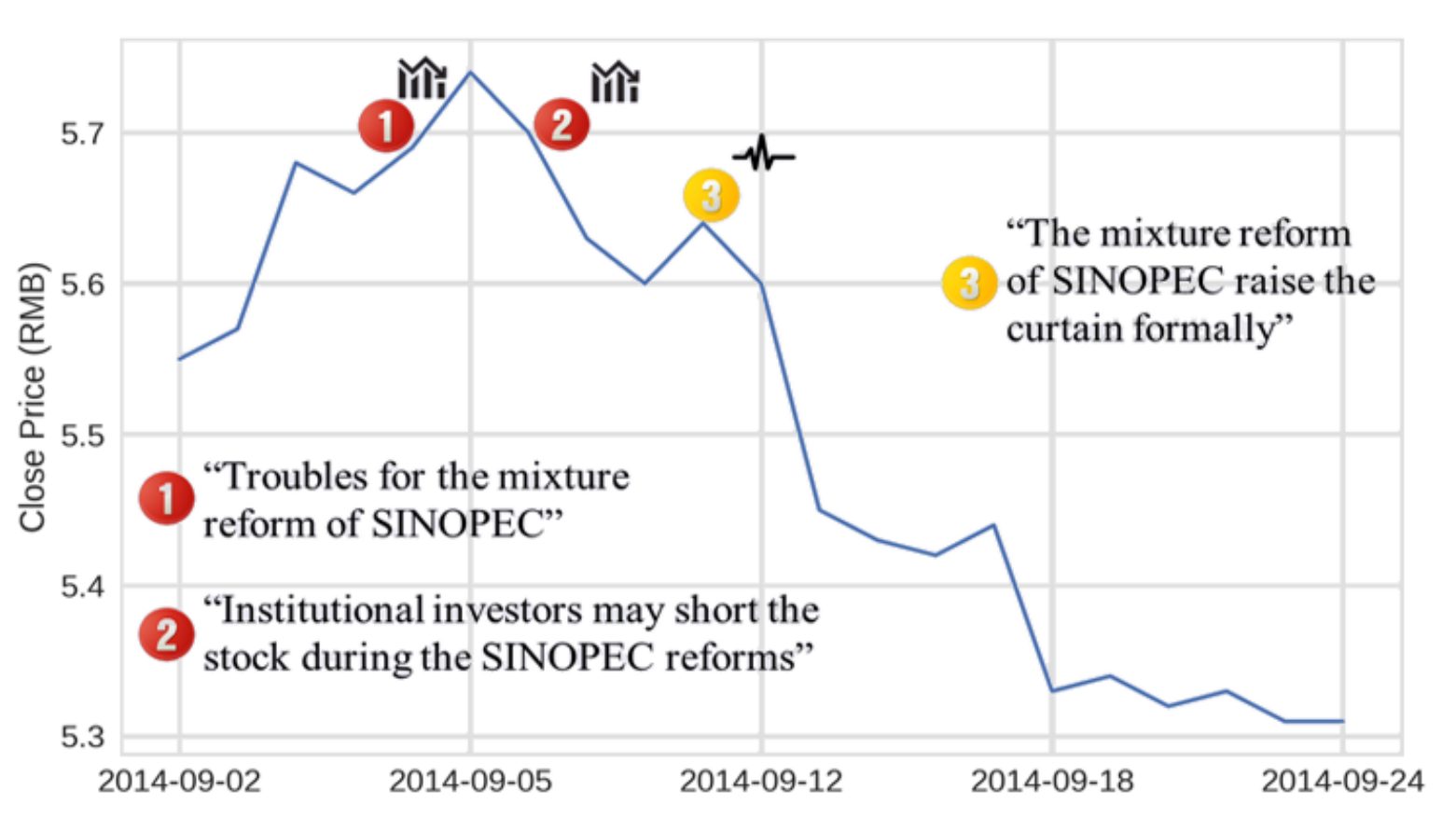}
       \caption{Influence of breaking news across time. Figure from \cite{hu_listening_2019}.}
       \label{fig:xai_time_event}
    \end{subfigure}
    \caption{XAI in time.}
    \label{fig:xai_time}
\end{figure*}
}

\newcommand{\figureXAIFactor}{
\begin{figure*}[!ht]
    \centering
    \begin{subfigure}{0.24\textwidth}
       \centering
       \includegraphics[width=\linewidth]{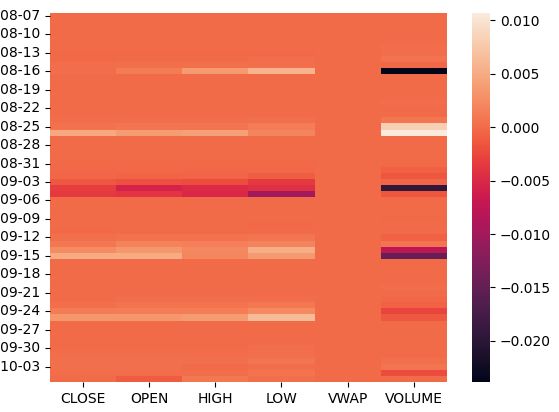}
       \caption{Feature importance for basic volume-price factors}
       \label{fig:xai_factor_single}
    \end{subfigure}
    \hfill
    \begin{subfigure}{0.28\textwidth}
       \centering
       \includegraphics[width=\linewidth]{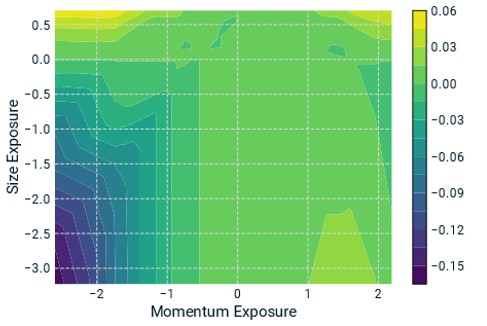}
       \caption{Interaction between size and momentum factors. Figure is cited from \cite{bonne_machine_2021}.}
       \label{fig:xai_factor_interaction}
    \end{subfigure}
    \hfill
    \begin{subfigure}{0.33\textwidth}
       \centering
       \includegraphics[width=\linewidth]{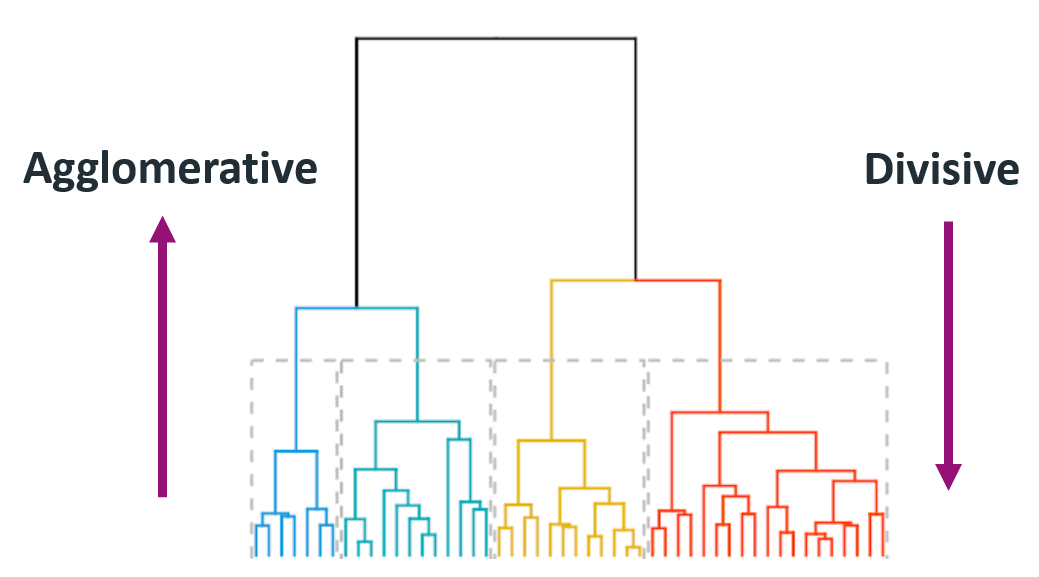}
       \caption{Hierarchical clustering. Figure is cited from \cite{sharma_hierarchical_2021}.}
       \label{fig:xai_factor_evolution}
    \end{subfigure}
    \caption{XAI in factors.}
    \label{fig:xai_factor}
\end{figure*}
}

\newcommand{\figureLatentConcept}{
\begin{figure}
    \centering
    \includegraphics[width=\linewidth]{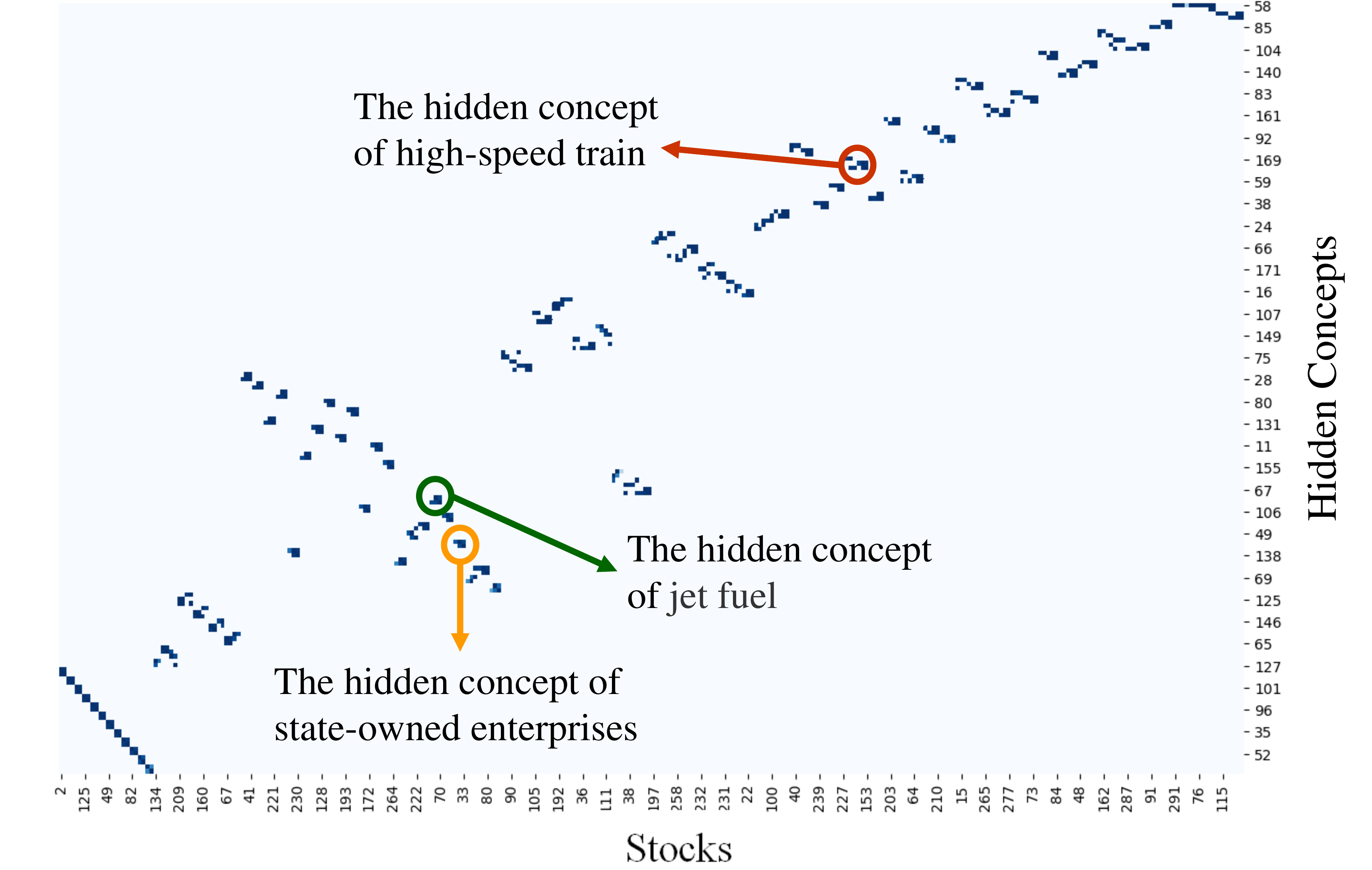}
    \caption{Latent concepts \cite{xu_hist_2022}.}
    \label{fig:latent_concepts}
\end{figure}
}

\newcommand{\figureSectorRotation}{
\begin{figure}
    \centering
    \includegraphics[width=\linewidth]{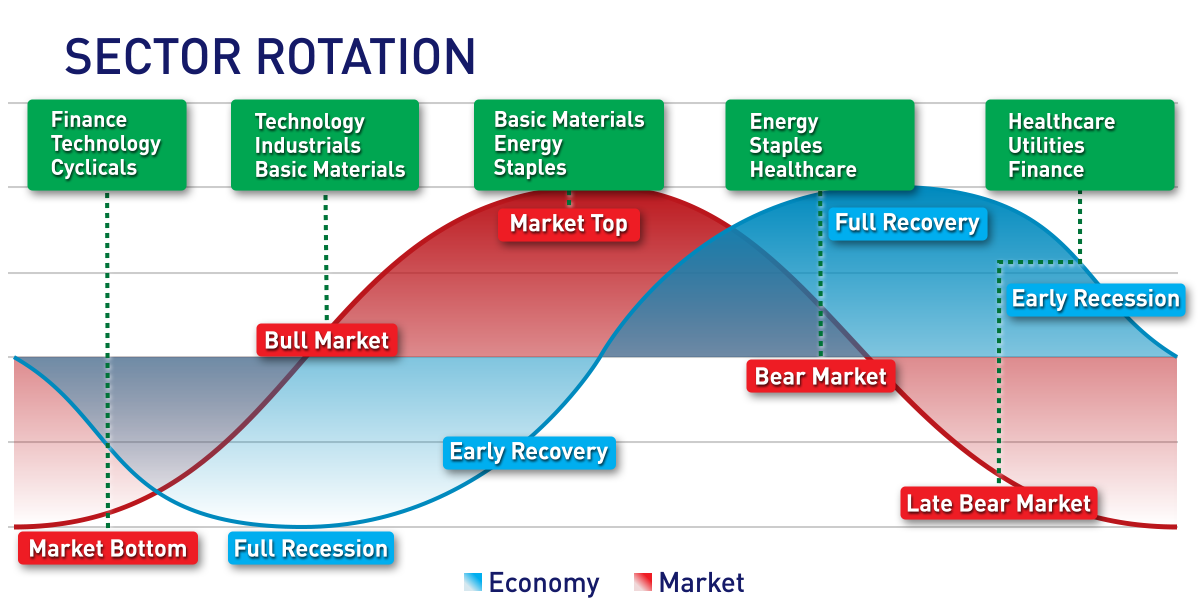}
    \caption{Sector rotation}
    \label{fig:sector_rotation}
\end{figure}
}

\newcommand{\figureExtremeMarket}{
\begin{figure}[!ht]
    \centering
    \includegraphics[width=\linewidth]{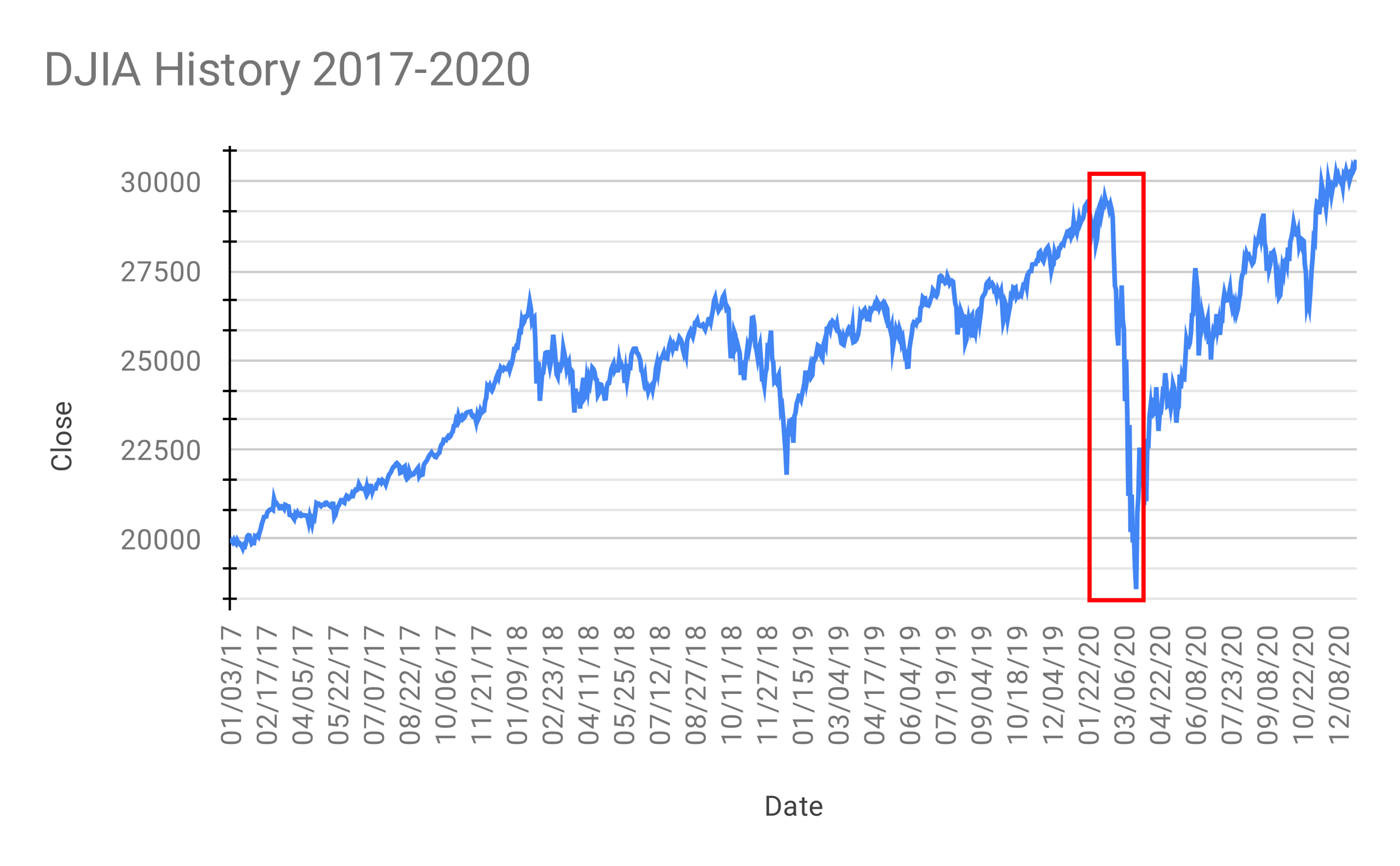}
    \caption{An example of extreme market behavior: The 2020 US stock market crash.}
    \label{fig:fin_crisis}
\end{figure}
}

\newcommand{\figureModelCostCompare}{
\begin{figure}[h]
    \centering
    \includegraphics[width=\linewidth]{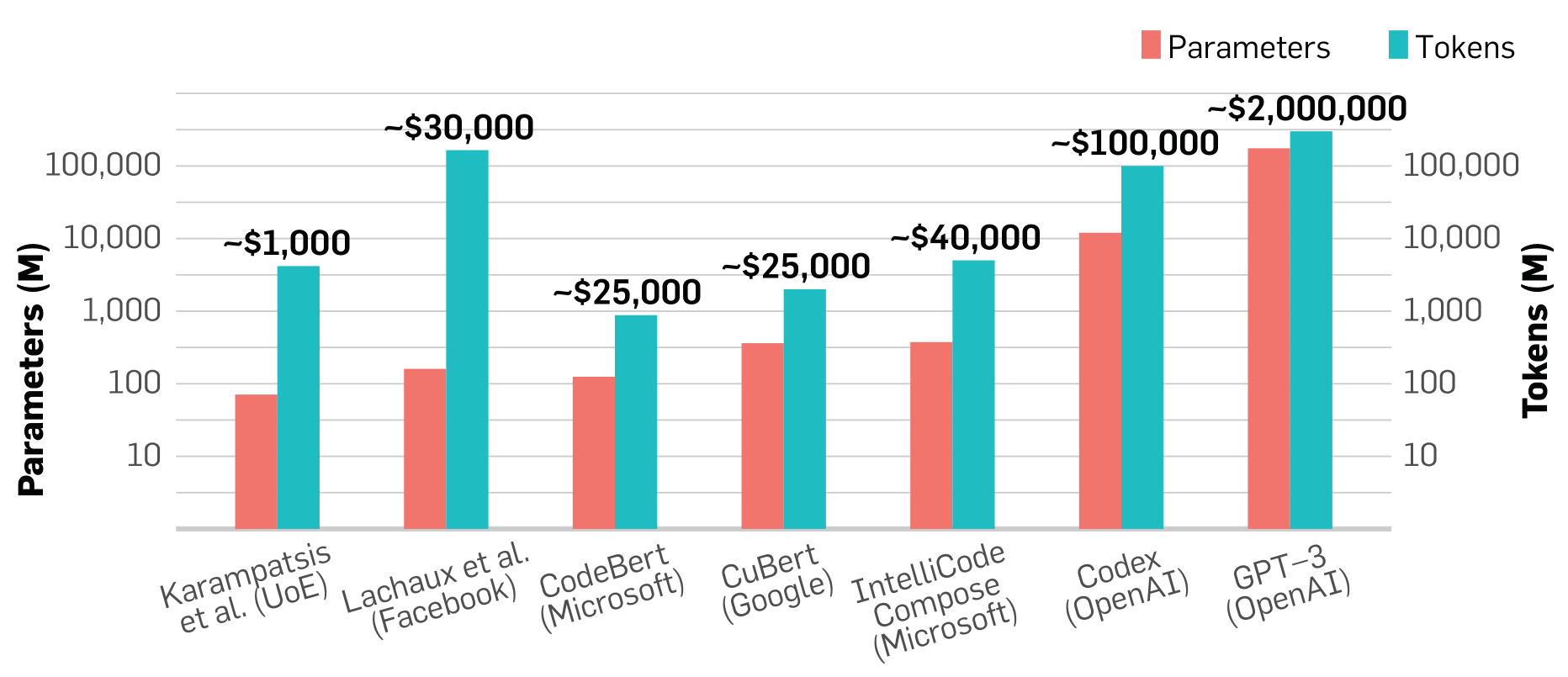}
    \caption{Comparison of large-scale deep learning pretraining models for code generation from model parameter size, data (token) size and estimated training cost. Figure is cited from \cite{hellendoorn_growing_2021}.}
    \label{fig:model_cost_compare}
\end{figure}
}

\newcommand{\figureModelSizeGrowth}{
\begin{figure}[h]
    \centering
    \includegraphics[width=0.9\linewidth]{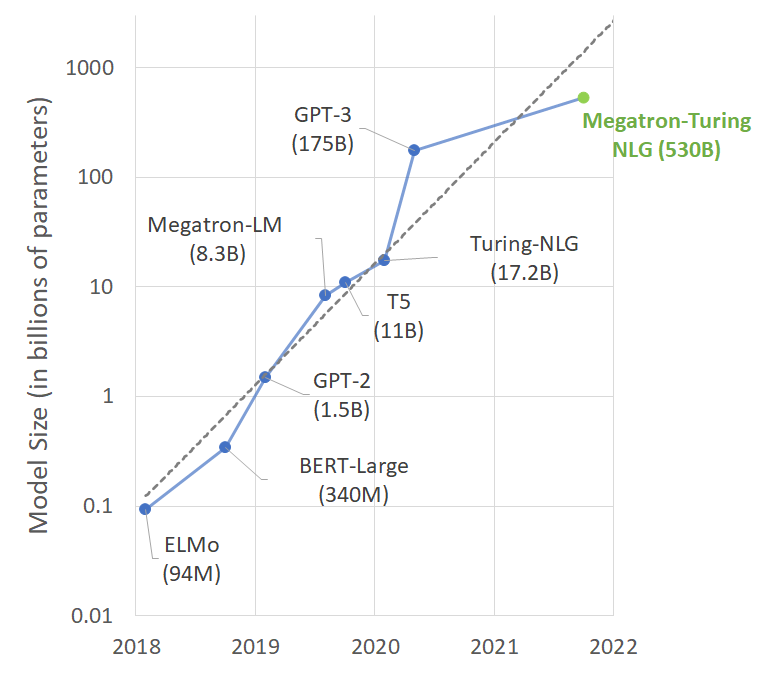}
    \caption{Growth of model sizes (measured by number of parameters) from 2018 to 2022. Figure is cited from \cite{smith_using_2022}.}
    \label{fig:model_size_growth}
\end{figure}
}

\newcommand{\figureXAIDimensions}{
\begin{figure}[!t]
    \centering
    \includegraphics[width=\linewidth]{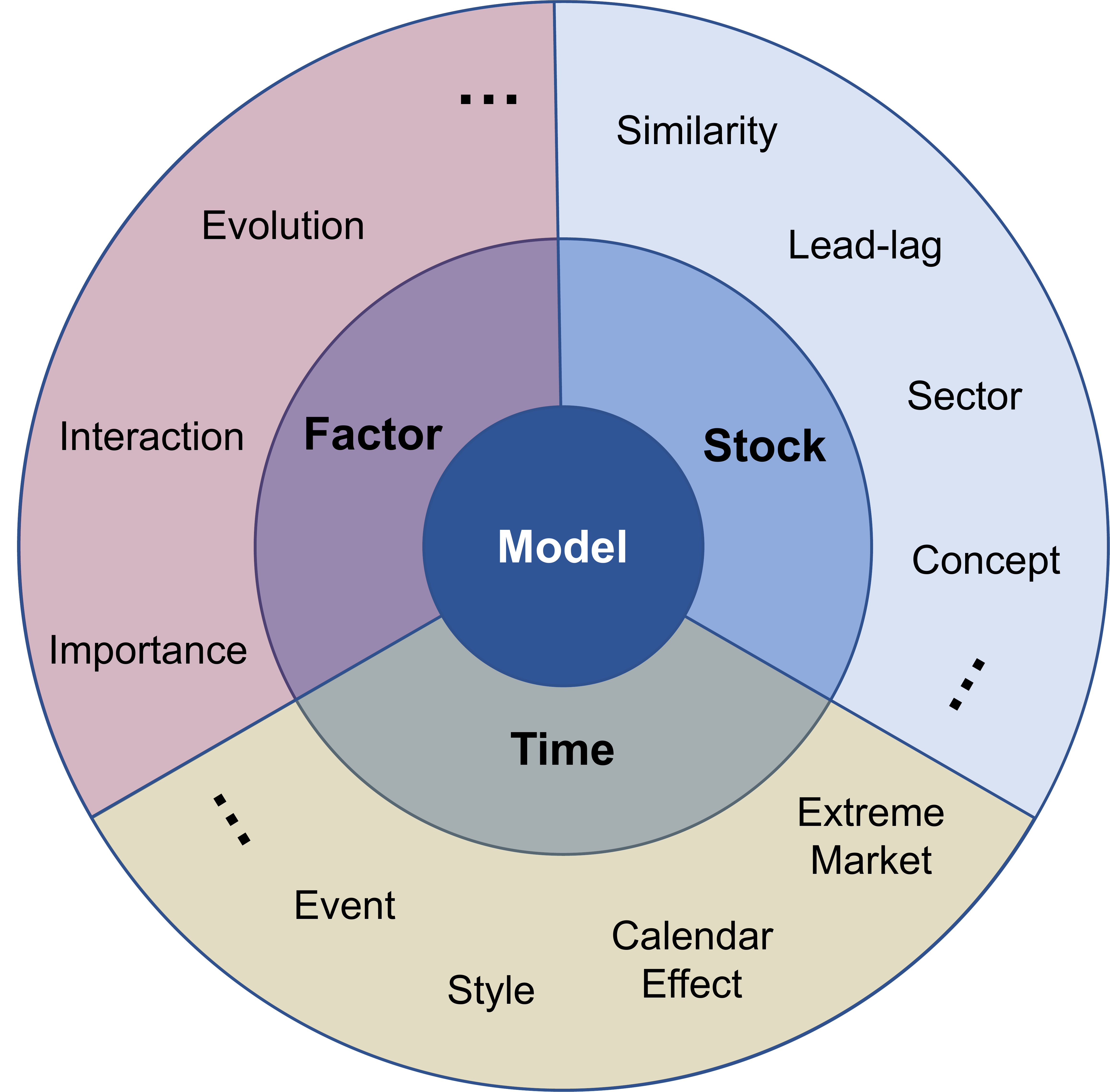}
    \caption{XAI Dimensions}
    \label{fig:xai_dim}
\end{figure}
}

\newcommand{\figureKBHistory}{
\begin{figure*}[!ht]
    \centering
    \includegraphics[width=\textwidth]{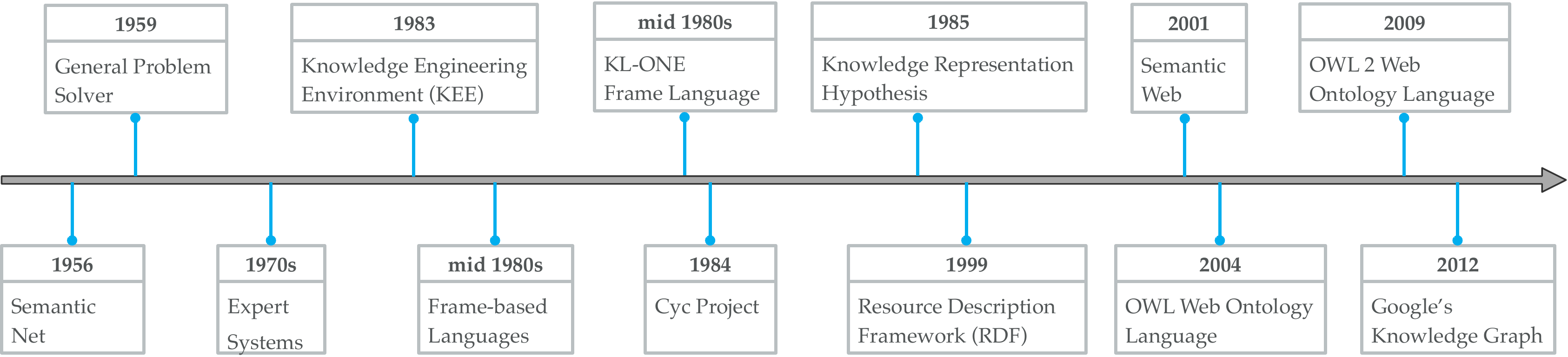}
    \caption{Knowledge base history. Figure cited from \cite{ji_survey_2022}.}
    \label{fig:kb_history}
\end{figure*}
}

\newcommand{\figTenChallenge}{
\begin{figure*}[!ht]
    \centering
    \includegraphics[width=\textwidth]{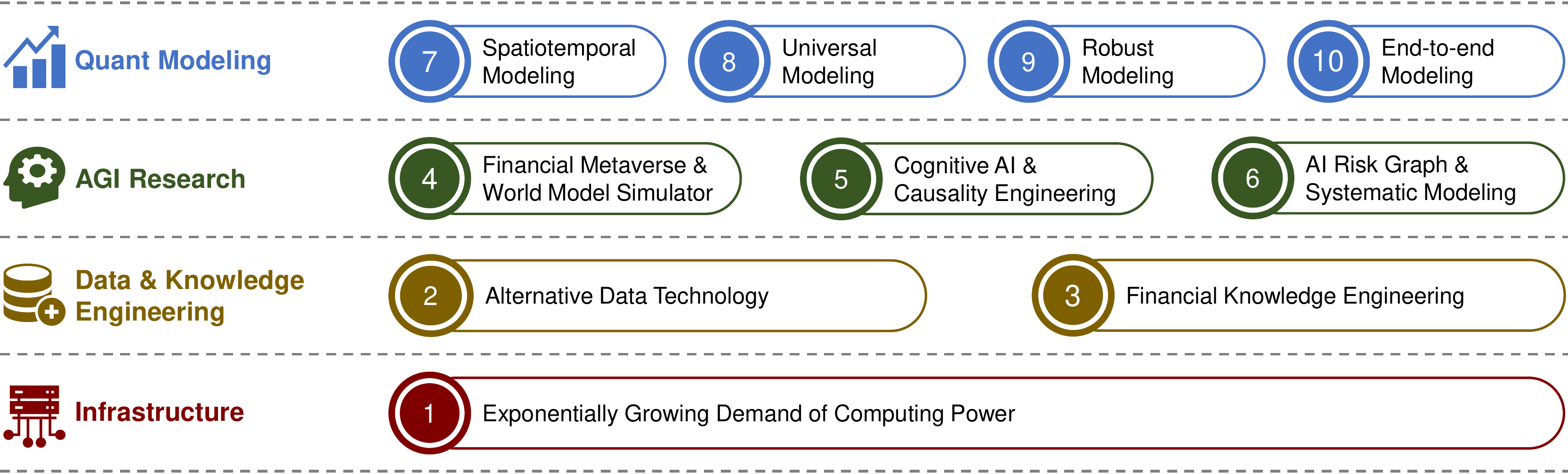}
    \caption{The 10 challenges in quantitative investment technology.}
    \label{fig:TenChallenge}
\end{figure*}
}

\newcommand{\figureKBTechniques}{
\begin{figure*}[!ht]
    \centering
    \begin{subfigure}{0.2\textwidth}
        \centering
        \includegraphics[width=\linewidth]{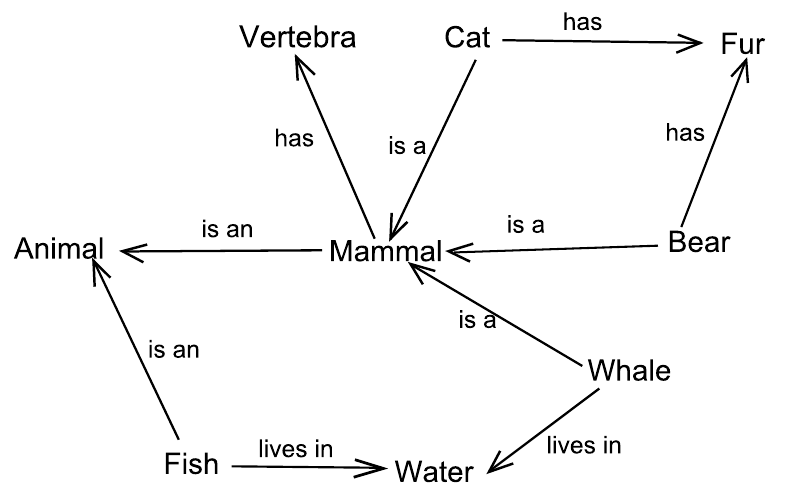}
        \caption{Semantic network}
        \label{fig:kb_semantic_network}
    \end{subfigure}
    \caption{Knowledge base techniques}
    \label{fig:kb_techniques}
\end{figure*}
}

\newcommand{\figureKGReasoningSIGIR}{
\begin{figure*}[!h]
    \centering
    \includegraphics[width=\textwidth]{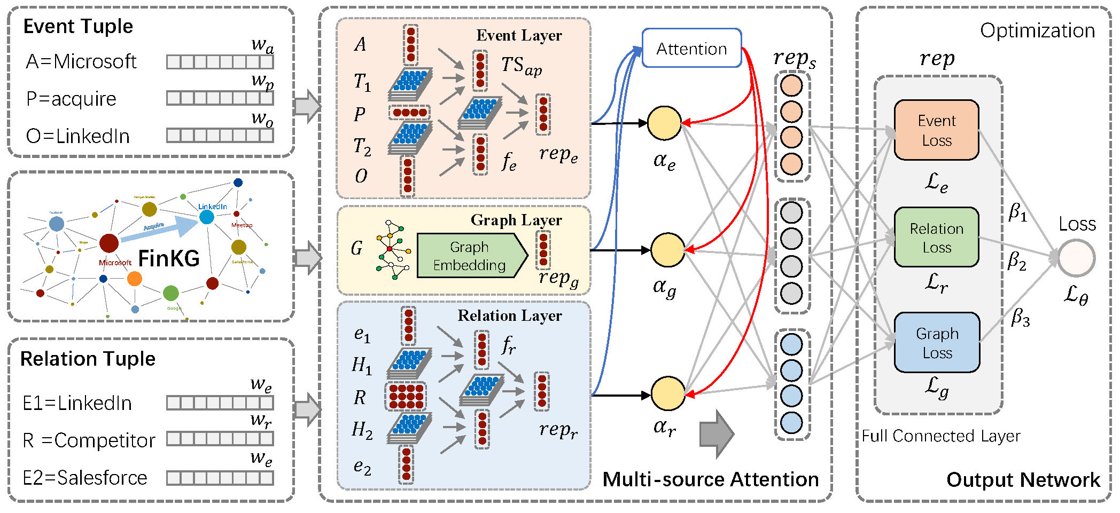}
    \caption{Knowledge graph reasoning for stock prediction. Figure is cited from \cite{cheng_knowledge_2020}.}
    \label{fig:kg_reasoning_sigir}
\end{figure*}
}

\newcommand{\figureWorldModel}{
\begin{figure}[!h]
    \centering
    \includegraphics[width=\linewidth]{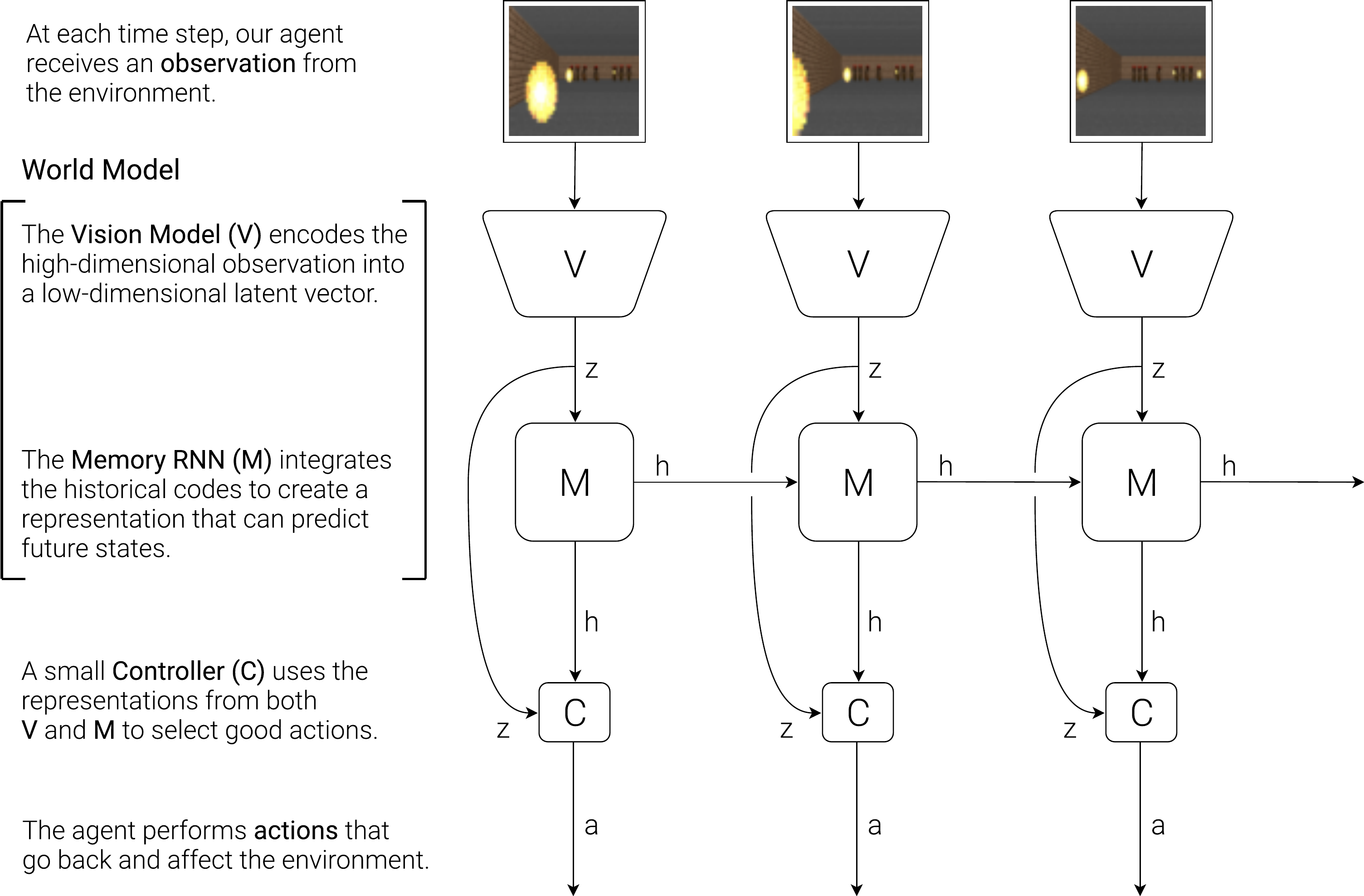}
    \caption{An example of world model architecture. Figure is cited from \cite{ha_worldmodel_2018}.}
    \label{fig:world_model}
\end{figure}
}

\newcommand{\figAutoMLSimple}{
\begin{figure}[!h]
    \centering
    \includegraphics[width=\linewidth]{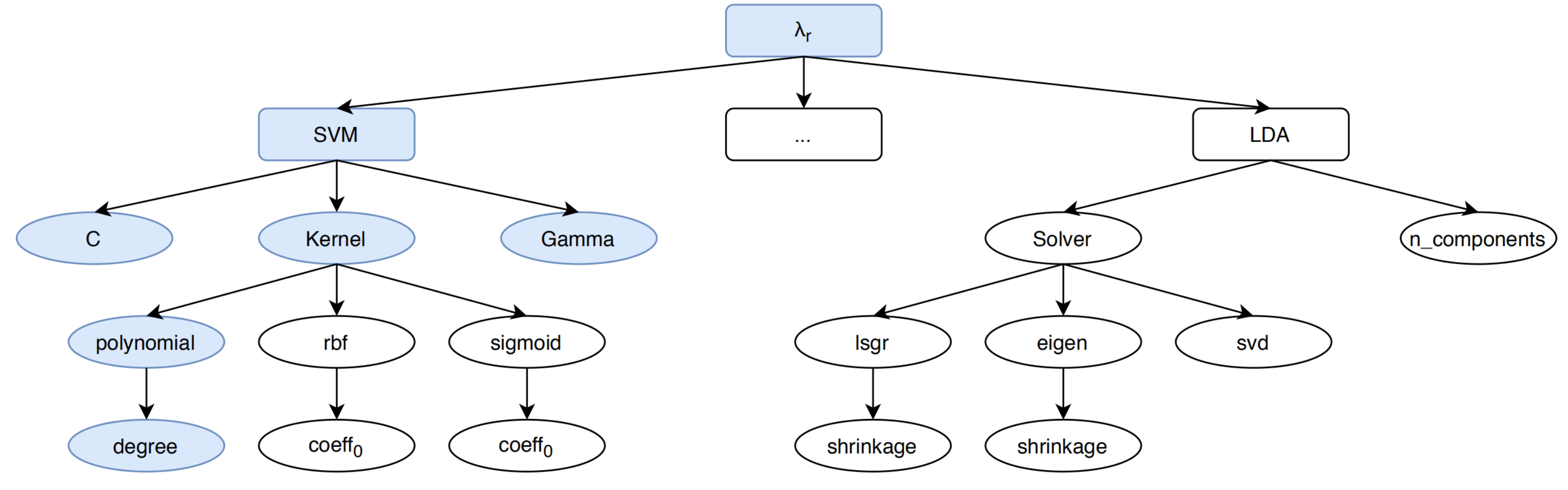}
    \caption{Illustration of early-stage automated machine learning based on brute-force search of algorithms and hyperparameters. Figure is cited from \cite{zoller_benchmark_2021}.}
    \label{figAutoMLSimple}
\end{figure}
}

\newcommand{\figureGlobalExplanationTaxonomy}{
\begin{figure*}[!htbp]
    \centering
    \includegraphics[width=\textwidth]{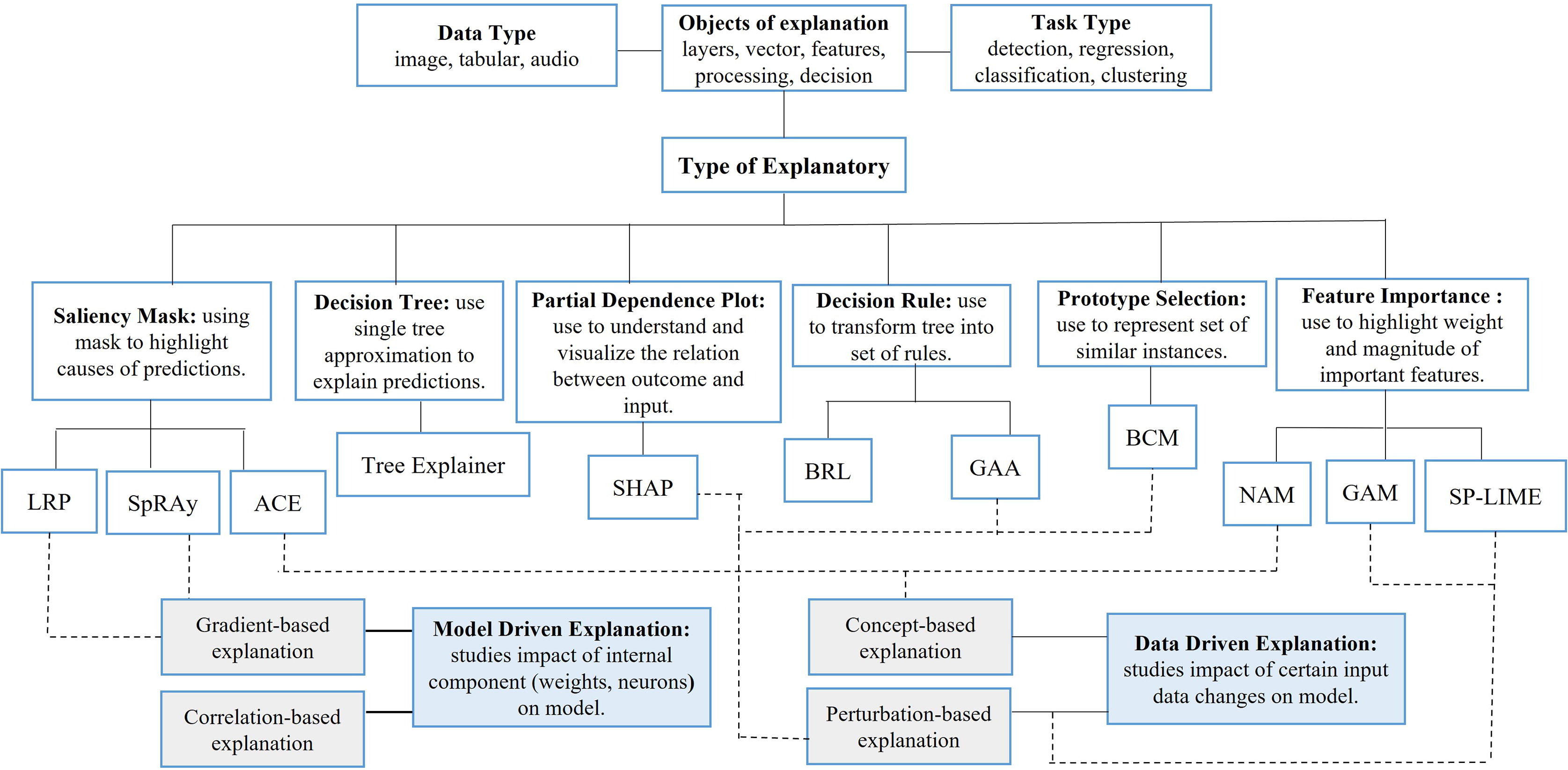}
    \caption{Taxonomy of global explanation methods. Figure cited from \cite{saleem_explaining_2022}.}
    \label{fig:GlobalExplanationTaxonomy}
\end{figure*}
}

\newcommand{\figureLocalToGlobalExplanation}{
\begin{figure*}[!ht]
    \centering
    \includegraphics[width=\textwidth]{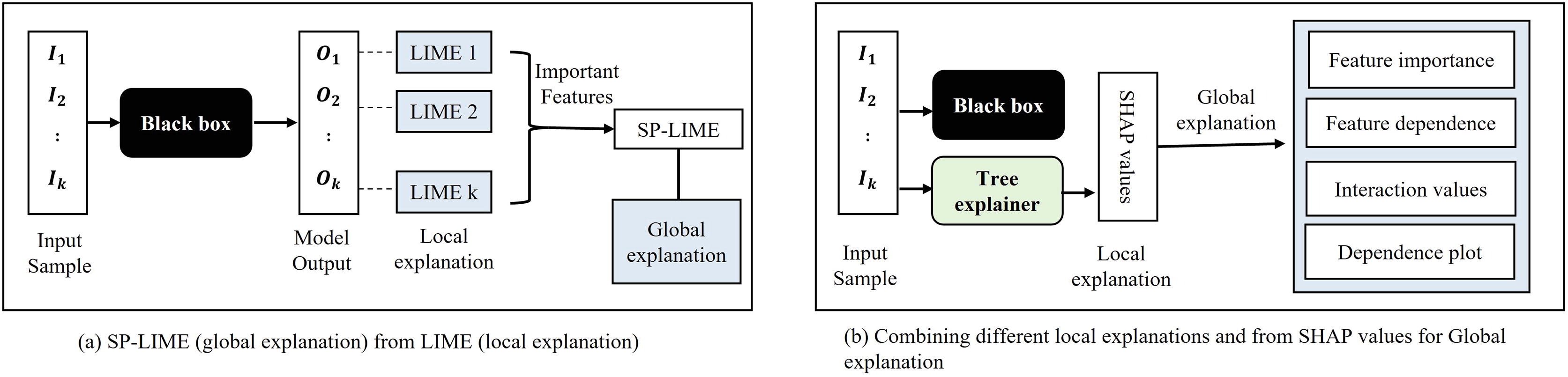}
    \caption{Illustration of how local explanations can be aggregated to form global explanation. Figure cited from \cite{saleem_explaining_2022}.}
    \label{fig:LocalToGlobalExplanation}
\end{figure*}
}

\newcommand{\figureFinKGWWW}{
\begin{figure*}[!ht]
    \centering
    \includegraphics[width=\textwidth]{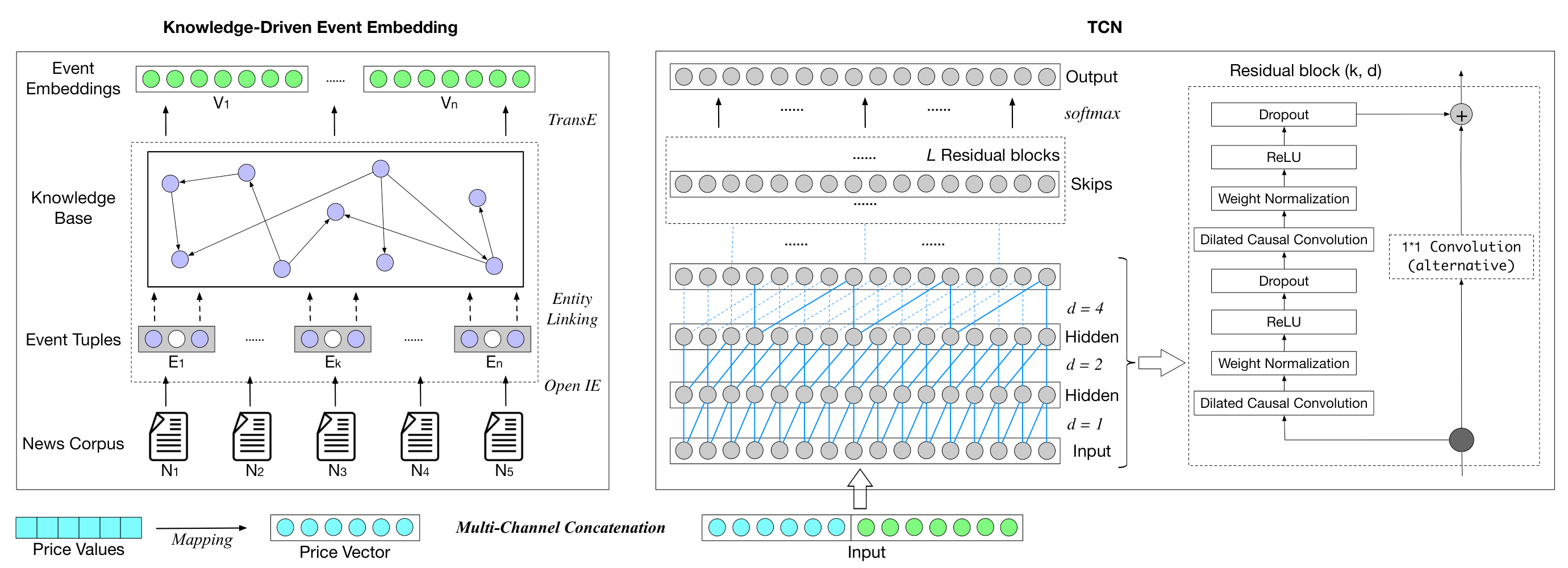}
    \caption{Knowledge graph event embedding combined with other factors in stock prediction. Figure cited from \cite{deng_knowledge-driven_2019}.}
    \label{fig:FinKGWWW}
\end{figure*}
}
\newcommand{\tableQPquant}{
\begin{table}[ht]
    \footnotesize
    \centering
    \caption{Comparisons of P-quant and Q-quant \cite{meucci_p_2011}.}
    \resizebox{\linewidth}{!}{
        \begin{tabular}{l|ll}
            \toprule
            & \textbf{Q-quant} & \textbf{P-quant} \\
            \midrule
            Goal  & Extrapolate the present & Model the future \\
            Scenario & Derivatives pricing & Portfolio management  \\
            Measure &  Risk-neural measure & Probability measure \\
            Modeling & Continuous stochastic process & Discrete time series \\
            Example & Black-Scholes model & Multifactor model \\
            Algorithm & Ito calculus, PDEs & Statistics, Machine Learning \\
            Challenge & Calibration & Estimation/Prediction \\
            Business & Sell-side & Buy-side \\
            \bottomrule
        \end{tabular}
    }
    \label{tab:pq-quant}
\end{table}
}

\newcommand{\tableSurveyComparison}{
\begin{table}[ht]
    \centering
    \begin{tabular}{cccc}
        \toprule
        \textbf{Work} & \textbf{Review} & \textbf{Perspective} & \\
         & 
    \end{tabular}
    \caption{Caption}
    \label{tab:my_label}
\end{table}
}

\newcommand{\tableQuantWorkflow}{
\begin{table*}[ht]
  \centering
  \caption{The input, output, goal, and typical methods of each stage in the quantitative investment workflow. \label{comparison_stages}}
  \resizebox{\textwidth}{!}{
    \begin{tabular}{|P{0.2\textwidth}|P{0.15\textwidth}|P{0.2\textwidth}|L{0.3\textwidth}|L{0.25\textwidth}|P{0.2\textwidth}|}
    \toprule
    \centering
    \textbf{Stage} & \textbf{Inputs} & \textbf{Outputs} & \multicolumn{1}{c|}{\textbf{Goals}} & \multicolumn{1}{c|}{\textbf{Typical Methods}} & \textbf{References} \\
    \midrule
    Pre-processing & Raw data & Data & Clean and transform the original data. & Batch processing & \cite{noauthor_pandas_2022} \\
    \midrule
    Factor mining & Data & Signal factors & Extract predictive factors.  & Human labor, heuristic search, machine learning & \cite{kakushadze_101_2016, cui_alphaevolve_2021, fang_alpha_2020} \\
    \midrule
    Asset selection & Signal factors & Predicted asset attributes & Make use of the factors to predict asset attributes such as return and volatility. & Machine learning, statistical modelling & \cite{zhang_stock_2017, lin_learning_2021} \\
    \midrule
    Portfolio optimization & Predicted asset attributes & Position assignments & Find the optimal asset allocation given the capital budget and return/risk expectations. & Numerical optimization, reinforcement learning & \cite{ye_reinforcement-learning_2020, xu_relation-aware_2020} \\
    \midrule
    Order execution & Position assignments & Orders & Convert the asset allocation weights to market orders. & Reinforcement learning & \cite{nevmyvaka_reinforcement_2006} \\
    \midrule
    Risk analysis & Market returns & Strategy adjustments & Measure the loss and volatility of the strategy based on the feedback and make adjustments to the strategy. & Machine learning, probabilistic modelling, statistical analysis & \cite{lin_deep_2021} \\
    \bottomrule
    \end{tabular}%
  }
  \label{tab:workflow_goals}%
\end{table*}%
}

\newcommand{\tableQuantTasks}{
\begin{table*}[ht]
  \centering
  \caption{Three prototypical tasks of quantitative investment.}
  \resizebox{\textwidth}{!}{
    \begin{tabular}{|P{0.2\textwidth}|L{0.4\textwidth}|L{0.4\textwidth}|P{0.2\textwidth}|}
    \toprule
    \centering
    \textbf{Tasks} & \multicolumn{1}{c|}{\textbf{Goals}} & \multicolumn{1}{c|}{\textbf{Profit}} & {\textbf{References}} \\ \midrule
    Trend trading & Predict whether the price of an asset/portfolio is going up or down. & The difference between buy and sell prices. & \cite{wang_alphastock_2019} \\
    \midrule
    Statistical arbitrage & Find correlated assets where there are price gaps. & Profit when the price gap is closed. & \cite{guijarro-ordonez_deep_2021, gatev_pairs_2006, rad_profitability_2016} \\
    \midrule
    Market making & Capture the micro price fluctuations in the market and provide enough liquidity. & The price difference between the sell and buy orders. & \cite{abernethy_adaptive_2013, zhong_data-driven_2020, gasperov_market_2021, haider_predictive_2022, spooner_robust_2020} \\
    \bottomrule
    \end{tabular}%
  }
  \label{tab:actual_tasks}%
\end{table*}%
}

\newcommand{\tableFundamentalData}{
\begin{table}[htbp]
  \centering
  \caption{An example of Microsoft's Fundamental data}
    \begin{tabular}{llll}
    \toprule
          & \textit{(Q3)FY2021} & \textit{(Q2)FY2021} & \textit{(Q1)FY2021} \\
    \midrule
    \textbf{Total Revenue} & 49.36B & 51.73B & 45.32B \\
    \textbf{Cost of revenue} & 15.62B & 16.96B & 13.65B \\
    \textbf{Gross profit} & 33.75B & 34.77B & 31.67B \\
    \textbf{Operating profit} & 20.36B & 22.25B & 20.24B \\
    \textbf{Net income} & 16.73B & 18.77B & 20.51B \\
    \bottomrule
    \end{tabular}%
  \label{tab:fundamental}%
\end{table}%
}

\newcommand{\tableDataCollection}{
\begin{table*}[!htbp]
  \centering
  \caption{Categorization of financial data. \qi{Removed the usage column. Reformat this table.} \qi{Add a comma for the sentences. Also add commas for other tables.}}
    \resizebox{\textwidth}{!}{
    \begin{tabular}{|L{0.15\textwidth}|L{0.15\textwidth}|L{0.2\textwidth}|L{0.4\textwidth}|L{0.3\textwidth}|L{0.15\textwidth}|}
    \toprule
    \multicolumn{3}{|c|}{\textbf{Modality}} & \multicolumn{1}{c|}{\textbf{Features}} & \multicolumn{1}{c|}{\textbf{Example}}  &  \multicolumn{1}{c|}{\textbf{References}} \\
    \midrule
    \multirow{3}{*}{Numerical} & \multirow{2}{*}{Quote data} & Regular interval & Quotes that are generated at regular time intervals & 1-minute candlestick chart &  \cite{sezer_financial_2020, wiese_quant_2020} \\ 
    \cmidrule{3-6}
    &  & Irregular interval & Quotes that are generated at irregular time intervals & Tick-level\footnote{\url{https://bit.ly/3Ob8Gjc}} order book data &  \cite{briola_deep_2020} \\
    \cmidrule{2-6}          
    & \multicolumn{2}{l|}{Fundamental data} & Fundamental data such as revenues and profits. & Financial statements &  \cite{tadoori_introduction_2020} \\
    \midrule
    \multirow{2}{*}{\parbox{0.08\textwidth}{Relational}} & \multicolumn{2}{l|}{Pairwise edges} & The relation between a pair of entities & Knowledge graph &   \cite{feng_temporal_2019, kim_hats_2019, xu_hist_2022, xu_rest_2021} \\
    \cmidrule{2-6}    
    & \multicolumn{2}{l|}{Hyperedges} & The relation involving a set of entities & Sector categorization &   \cite{sawhney_spatiotemporal_2020} \\
    \midrule
    \multirow{3}{*}{Alternative} & \multicolumn{2}{l|}{Text} & Information expressed in natural language. & Social media posts &  \cite{hu_listening_2019, li_modeling_2020, xu_stock_2018, sawhney_fast_2021} \\
    \cmidrule{2-6}
    & \multicolumn{2}{l|}{Images} & Images that are related to the traded asset & Satellite images & \cite{lutz_how_2018, partnoy_stock_2019} \\
    \cmidrule{2-6}          
    & \multicolumn{2}{l|}{Other modalities} & Anything & WiFi traffics, cell phone signals & \cite{jha_implementing_2018} \\
    \bottomrule
    \end{tabular}%
    }
  \label{tab:data_collection}%
\end{table*}%
}

\newcommand{\tableEvalMetrics}{
\begin{table*}[!t]
  \centering
  \caption{A summary of evaluation metrics. \qi{reformat this table}}
  \resizebox*{\textwidth}{!}{
    \begin{tabular}{|L{0.2\textwidth}|L{0.2\textwidth}|L{0.35\textwidth}|L{0.35\textwidth}|L{0.15\textwidth}|}
      \toprule
      \multicolumn{2}{|c|}{\textbf{Metrics}} & \multicolumn{1}{c|}{\textbf{Examples}} & \multicolumn{1}{c|}{\textbf{Motivations}}  & \multicolumn{1}{c|}{\textbf{References}} \\
      \midrule
      \multirow{2}{*}{\parbox{0.15\textwidth}{Machine learning metrics}} & Prediction-based metrics & Root mean squared error (RMSE), area under the ROC curve (AUC), F1 Score & Evaluating the model with prediction metrics. & \cite{li_modeling_2020, wang_clvsa_2019} \\ 
      \cmidrule{2-5}          
      & Information retrieval-based metrics & Precision and recall, mean reciprocal rank (MRR), normalized discounted cumulative gain (NDCG) & Evaluating the model with retrieval metrics. & \cite{feng_temporal_2019, wang_alphastock_2019} \\
      \midrule
      \multirow{2}{*}{\parbox{0.15\textwidth}{Investment-specific metrics}} & Correlation metrics & Information coefficient, information ratio & Calculating the correlation with the ground truth. & \cite{xu_hist_2022} \\
    \cmidrule{2-5}
    & Portfolio metrics & Annualized return, maximum drawdown, Sharpe ratio & Evaluating model profitability. & \cite{kim_hats_2019, wang_alphastock_2019} \\
      \bottomrule
      \end{tabular}%
  }
  \label{tab:eval_metrics}%
\end{table*}%
}

\newcommand{\tableLibraryComparison}{
\begin{table*}[!htbp]
  \centering
  \caption{Library Summary}
    \begin{tabular}{|c|p{8.835em}|c|c|c|c|}
    \toprule
    \multicolumn{2}{|p{15.25em}|}{Library} & \multicolumn{1}{p{7.665em}|}{vn.py} & \multicolumn{1}{p{5em}|}{qlib} & \multicolumn{1}{p{5em}|}{QuantLib} & \multicolumn{1}{p{9.5em}|}{FinRL} \\
    \midrule
    \multicolumn{2}{|p{15.25em}|}{Developer} & \multicolumn{1}{p{7.665em}|}{VeighNa} & \multicolumn{1}{p{5em}|}{Microsoft} & \multicolumn{1}{p{5em}|}{QuantLib Community} & \multicolumn{1}{p{9.5em}|}{AI4Finance Foundation} \\
    \midrule
    \multicolumn{2}{|p{15.25em}|}{License} & \multicolumn{1}{p{7.665em}|}{MIT} & \multicolumn{1}{p{5em}|}{CLA} & \multicolumn{1}{p{5em}|}{BSD} & \multicolumn{1}{p{9.5em}|}{MIT} \\
    \midrule
    \multicolumn{1}{|c|}{\multirow{4}[8]{*}{Data}} & Data Source &       & \multicolumn{1}{p{5em}|}{Yahoo! Finance} &       & \multicolumn{1}{p{9.5em}|}{Yahoo! Finance, WRDS} \\
\cmidrule{2-6}          & Data Type &       & \multicolumn{1}{p{5em}|}{Stocks} & \multicolumn{1}{p{5em}|}{Stocks, Forex, Futures, Options} &  \\
\cmidrule{2-6}          & Supported Data Modality &       & \multicolumn{1}{p{5em}|}{Quote Data, Relational Data} &       &  \\
\cmidrule{2-6}          & Factor Engine &       & \multicolumn{1}{p{5em}|}{abc} &       &  \\
    \midrule
    \multicolumn{1}{|c|}{\multirow{3}[6]{*}{Modelling}} & Supported Algorithm Types &       &       &       &  \\
\cmidrule{2-6}          & Backend Engine &       &       &       &  \\
\cmidrule{2-6}          & GPU Acceleration &       &       &       &  \\
    \midrule
    \multicolumn{1}{|c|}{\multirow{3}[6]{*}{Evaluation}} & Backtest Engine &       &       &       &  \\
\cmidrule{2-6}          & Analysis Engine &       &       &       &  \\
\cmidrule{2-6}          & Visualization &       &       &       &  \\
    \midrule
    \multicolumn{1}{|c|}{\multirow{3}[6]{*}{Deployment}} & Database Engine Support &       &       &       &  \\
\cmidrule{2-6}          & Trading API &       &       &       &  \\
\cmidrule{2-6}          & Exchange Interface &       &       &       &  \\
    \bottomrule
    \end{tabular}%
  \label{tab:lib_summary}%
\end{table*}%
}

\newcommand{\tableAutoModelingHorizontal}{
\begin{table}[!ht]
    \centering
    \caption{Algorithms for different search targets in automated modeling.}
    \resizebox{\linewidth}{!}{%
    \begin{tabular}{|l|p{0.2\linewidth}|p{0.3\linewidth}|p{0.3\linewidth}|p{0.3\linewidth}|p{0.3\linewidth}|}
        \toprule
        \diagbox[]{Target}{Algorithm} & Grid/Random Search & Evolutionary Algorithm & Reinforcement Learning & Bayesian Optimization & Gradient-based Methods \\
        \midrule
        Architecture & $\boldcheckmark$ & $\boldcheckmark$ & $\boldcheckmark$ & & $\boldcheckmark$ \\
        Hyperparameter & $\boldcheckmark$ & & & $\boldcheckmark$ &  \\
        Objective & $\boldcheckmark$ & & & & \\
        \bottomrule
    \end{tabular}    
    }
    \label{tab:auto_modeling}
\end{table}
}

\newcommand{\tableAutoModeling}{
\begin{table}[!htbp]
    \centering
    \caption{Search algorithms and their applicable search targets.}
    \resizebox{\linewidth}{!}{%
    \begin{tabular}{|l|c|c|c|c|}
        \toprule
        \diagbox[]{Algorithm}{Target} & NAS & HPO & TOS & References \\
        \midrule
        Grid/Random Search & & $\boldcheckmark$ & $\boldcheckmark$ & \cite{bergstra_random_2012, li_random_2020} \\
        Evolutionary Algorithm & $\boldcheckmark$ & $\boldcheckmark$ & $\boldcheckmark$ & \cite{real_large-scale_2017, xie_genetic_2017, suganuma_genetic_2018, real_regularized_2019, tani_evolutionary_2021} \\
        Reinforcement Learning & $\boldcheckmark$ & $\boldcheckmark$ &  & \cite{zoph2016neural, zoph_learning_2018, zhong_blockqnn_2021, jomaa_hyp-rl_2019} \\
        Bayesian Optimization & $\boldcheckmark$ & $\boldcheckmark$ &  & \cite{falkner_bohb_2018,  klein_fast_2017, white_bananas_2021, kandasamy_neural_2018} \\
        Gradient-based Method & $\boldcheckmark$ & $\boldcheckmark$ &  & \cite{liu_darts_2018, cai_proxylessnas_2019, bengio_gradient-based_2000, maclaurin_gradient-based_2015, franceschi_forward_2017} \\
        \bottomrule
    \end{tabular}    
    }
    \label{tab:auto_modeling}
\end{table}
}

\section{Introduction}
\label{sec:introduction}
Quantitative investment is an important part of wealth management (a.k.a. asset management) industry. This section contains introductory knowledge about quant, including market situation, classification and principles of strategy development, historical landmarks, and concepts of Quant1.0--Quant4.0.

\subsection{Wealth Management and Quant}
\label{sec:introduction_market}
\figureAssetMgmtWorldwide
The wealth management industry is one of the largest sectors of the world's economy. According to a global wealth report from Boston Consulting Group (BCG)~\cite{zakrzewski_global_2022} and the illustration in Figure~\ref{fig:asset_mgmt_worldwide}, the volume of global financial wealth has grown from 188.6 trillion USD in 2016 to 274.4 trillion USD in 2021, almost three times as the global nominal GDP in 2021. Moreover, the company predicts this number will increase to 355 trillion USD in 2026. It is not surprising that North America, Asia, and Europe are the three biggest regional markets of wealth management in the world, with approximately 46\%, 26\%, and 21\% of the global market size in 2021, respectively. We also see the stable and sustainable growth of the wealth management market, both globally and regionally. Figure \ref{fig:asset_mgmt_ecosys} shows an ecosystem of the wealth management industry, where investment funds as well as fund managers (a.k.a. investment managers) play core roles. They raise money from various capital providers, such as endowment foundations, fund of funds (FOF), family offices, billionaires, insurance companies, pension/sovereign funds and retail clients, and invest this money into financial markets to bet return and profit for their customers. Many types of investment instruments are liked by fund managers, such as stocks, exchange-traded funds (ETFs), bonds, futures, options, and foreign exchange \cite{fabozzi_handbook_2002}. Some investment funds even borrow money from depository institutions such as banks or peer-to-peer lending companies for investment and profit from the difference between investment return and loan interest. With the rapid development of digital economy, big data, and artificial intelligence, more and more new technologies are applied in the wealth management industry, leading to a branch of financial technology/engineering, called ``investment engineering'' \cite{blakey_introduction_2005}. Consequently, the pipeline of investment research, trading execution, and risk management is becoming a systematic, automated, and intelligent process, and this philosophy has been practiced in the recent evolution of quant. 
\figureAssetMgmtEcosys

As an important family of players in financial markets and the wealth management industry, contemporary quant applies rigorous mathematical and statistical modeling techniques, machine learning techniques, and algorithmic trading techniques to discover asset pricing abnormalities in financial markets and make money from the following arbitrage or investment opportunities. Compared with traditional fundamental and technical investment, quantitative investment has a number of advantages. Firstly, the performance of quant strategies can be examined and evaluated beforehand using back-test experiments based on historical data before the beginning of real trading. Secondly, quant trading has speed superiority in bidding orders with the best price. Thirdly, it eliminates the negative effect of human emotion in decision-making. Finally, quant research has significant advantages in data analysis with much deeper, broader and diversified coverage of information about financial markets and sectors. In the past 30 years, information infrastructure and computer technology are widely applied by financial exchange markets around the world. Nowadays, massive financial data are generated and millions of orders are executed every second, leading to the rapid growth of the quant industry. Taking the U.S. stock market as an example, over 60\% of overall trading volumes comes from the orders placed by computer trading algorithms rather than human traders \cite{cheng_just_2017}.


\subsection{Quant Strategies}
\label{sec:introduction_strategies}
A quant strategy is a systematic function or trading methodology used for trading securities in financial markets based on predefined rules or trained models for making trading decisions. Strategies are usually the core intelligent property of a quantitative fund. 
\subsubsection{Components of Quant Strategy}
A standard quant strategy contains a series of components, such as investment instrument, trading frequency, trading mode, strategy type and data type, and we introduce them one by one (see Figure~\ref{fig:strategy_classification}). 
\figureStrategyClassification
\begin{itemize}[leftmargin=*]
    \item \underline{Investment instrument} specifies which financial instruments are put in the universe by the strategy. Popular candidate instruments include stocks, ETFs, bonds, foreign exchanges, convertible bonds, and cryptocurrencies, as well as more complicated financial derivatives such as futures, options, swaps, and forwards \cite{hull_options_2006}. An investment strategy could trade either a single type of instrument (e.g., a strategy for trading ETFs) or multiple types of instruments (e.g., an alpha hedging strategy that longs stocks and shorts index futures to eliminate market risks). 
    \item \underline{Trading frequency} specifies how to hold your asset in portfolio and how frequently to trade. Usually, high-frequency trading holds a position in several minutes or seconds, while low-frequency trading may hold an asset over several months or years. Comparing high-frequency trading and low-frequency trading, the dramatic discrepancy of holding periods result in very different consideration in strategy design. For example, asset capacity limitations and trading costs are big issues for high-frequency trading, while how to control the risk of drawdown \cite{guo_quantitative_2016} is what we should carefully think about for low-frequency trading.  
    \item \underline{Model type} characterizes how to formally model the trading problem. Examples include cross-sectional trading, time-series trading, and event-driven trading \cite{guo_quantitative_2016}. Cross-sectional trading is used commonly in stock selection, where all stocks in a universe are ranked according to their scores of expected future returns predicted by a model, and portfolio managers could long stocks with the highest scores and short those with the lowest scores. Time series trading is relatively simple, where long/short trading operates only on a single instrument such as a certain stock or a certain future contract. Event-driven trading differs from time-series trading because the time intervals between events are not evenly distributed over time, while investment decisions and trading executions are triggered by the occurrence of events. 
    \item \underline{Trade type} is a series of thinking templates for us to design a strategy quickly. Examples include momentum trading \cite{chan_momentum_1996}, mean-reversion trading \cite{poterba_mean_1988}, arbitrage trading \cite{pole_statistical_2007}, hedging \cite{koziol_hedging_1990}, market making \cite{baird_option_1993}, etc. By leveraging these strategy types, traders can explore profit chances from different aspects of financial markets. Specifically, momentum trading assumes the price trend is sustainable in the following time window and it follows this trend direction to trade. Mean-reversion trading, on the contrary, bets the price trend will move towards the opposite direction in recent future and buy opposite positions. Hedging is the purchase of one asset with the intention of reducing the risk of loss from another asset. Arbitrage is simultaneously longing and shorting the same asset in different markets or a pair of highly correlated assets in order to profit from the convergence of price discrepancy. Market making is a liquidity-providing trade that quotes both a buy and a sell price in a tradable asset held in inventory, hoping to make a profit on the bid–ask spread. 
    \item \underline{Data type} means what type of data is used in a strategy. Typical data types include quote data, limit order books \cite{gould_limit_2013} b, news data, financial statements, analysts' reports, and alternative data such as sentimental data, location data, satellite images, etc. A strategy researcher must consider what kind of data he has and what kind of data he needs in a strategy development process. For example, limit order book streams are usually used in building high-frequency trading strategies, while news data are used more commonly in event-driven strategies.  
\end{itemize}

\subsubsection{Examples of Popular Strategies}
Figure \ref{fig:strategy_classification} also list a number of popular strategies as examples. For example, stock hedging strategy based on multifactor model \cite{fama_cross-section_1992} is very popular in many main markets around the world. This strategy hedges market risk by longing the most favorable stocks and shorting the other end (in some markets prohibiting shorting, short the corresponding index future or index option instead). If we trade stocks with multifactor models in a long-only way without shorting and constrain the risk exposure between selected portfolios and certain stock indices, it is an enhanced indexing strategy, which is almost the most popular quantitative strategy in China's stock market if measured by assets under management (AUM). 

\subsection{Fundamental Principles of Asset Management}
\label{sec:introduction_principle}
Similar to the situation that learning law of energy conservation could help avoid the trap of perpetual motion machine, it is beneficial to learn some fundamental principles of asset management so as to get rid of some common traps in strategy development.  
\subsubsection{Fundamental Law of Active Management}
The first principle is the \textit{fundamental law of active management} developed by Richard Grinold and Ronald Kahn \cite{grinold_fundamental_1989}. This principle states that the performance of an active investment manager (or equivalently quant model) depends on the quality of investment skills and, consequently, the frequency of investment opportunities. This law can be expressed mathematically as follows:
\begin{equation}
    IR = IC \times \sqrt{Breadth}
\end{equation}
where $IC$ is the information coefficient (correlation between the predicted return and true return in a future time window) evaluating investment quality, $Breadth$ means the number of independent investment decisions in a year, and $IR$ is the ratio of portfolio returns above the returns of a benchmark to the volatility of returns, measuring the performance of asset management. Mathematically, the fundamental law of active management can be regarded as an application of the central limit theorem in mathematical statistics \cite{noauthor_central_2008}. When applying this law in practice, we have to notice that $IC$ and $Breadth$ are usually not independent. For example, given a strategy, we may increase its $Breadth$ by relaxing the threshold of trading signals, but in this way, $IC$ may decrease because more false-positive noise is introduced to our decisions. Therefore, a good strategy should find an optimal trade-off between these two coupled variables. Figure \ref{fig:law_of_active_mgmt} illustrates the distribution of various popular strategies on $IC$ and $Breadth$, and their corresponding $IR$ performance.  
\figurePrincipleActiveManagement

\subsubsection{Impossible Trinity of Investment}
The second principle is the impossible trinity of asset management. Specifically, any investment strategy can not meet the following three conditions simultaneously, i.e., high return, low risk (or equivalently high stability), and high capacity. Figure \ref{fig:impossible_trinity} illustrates the impossible trinity using a radar chart with three variables return, stability and capacity. For example, high-frequency market making and calendar arbitrage strategy could reach high return and stability (low portfolio volatility), but the capacity of its AUM is usually small, typically hard to exceed several billions of USD even in global trading. On the contrary, stock fundamental strategy has high capacity up to trillions of USD, but its return and stability are not as good as those of high-frequency trading.


\subsection{History of Quantitative Investment}
\label{sec:introduction_history}
The origin of quant can trace back to over a century ago when French mathematician Louis Bachelier published his Ph.D. thesis ``The Theory of Speculation'' in 1900 \cite{bachelier_theorie_1900} and he exhibited how to use probability law and mathematical tools to study the movement of stock prices. As a pioneer exploring the application of advanced mathematics in financial markets, Bachelier's work inspired academic research of quantitative finance despite the lack of industry application due to data scarcity at his age. Quantitative investment was first practiced by American mathematics professor Edward Thorp, who used probability theory and statistical analysis to win blackjack games, and his research was subsequently used to seek systematic and consistent returns in stock markets \cite{thorp_beat_1967}. In this subsection, we introduce the history and landmarks in the development of quantitative finance through two routes: research landmarks in academia and evolution of quant in industry practice.  

\subsubsection{Q-Quant and P-Quant}
People in academia and investment industry classify quantitative finance into two branches, which are usually referred to as ``Q-quant'' and ``P-quant''. These two branches are named after their differentiation in modeling based on risk-neural measure and probability measure, respectively. Generally speaking, Q-quant studies the problem of derivative pricing and \textit{extrapolate the present}, using a model-driven research framework where data is usually used to adjust the parameters of models. On the other hand, P-quant studies quantitative risk and portfolio management to \textit{model the future}, using a data-driven research framework where different models are built to improve the fitting of historical data. Usually, Q-quant research is conducted in sell-side institutes such as investment banks and security companies, while P-quant is popular in buy-side institutes such as mutual funds and hedge funds. Table~\ref{tab:pq-quant} compares the characteristics of these two types of quant. 
\tableQPquant

\subsubsection{Landmarks in Q-Quant}
\figureAcademicTimeline
In 1965, Paul Samuelson, American economist and the winner of 1970 Nobel Memorial Prize in Economic Sciences, introduced stochastic process and stochastic calculus tools in analyzing financial markets and modeling the stochastic movement of stock prices~\cite{samuelson_proof_1965}, and in 1965, he published a paper studying the lifetime portfolio selection problem using a stochastic programming method~\cite{samuelson_life_1969}. In the same year, another American economist Robert Merton published his work about lifetime portfolio selection as well. Different from Samuelson's work using discrete-time stochastic process, Merton's work modeled the random uncertainty of portfolio using continuous-time stochastic calculus~\cite{merton_lifetime_1969}. Almost in the same year, economists Fischer Black and Myron Scholes demonstrated that the expected return and risk of assets under management could be removed by dynamically revising a portfolio, and thus inventing the risk-neutral strategy for derivative investment~\cite{taleb_dynamic_1997}. They applied the theory to real market trading and published it in 1973. The risk-neutral formula was later named in honor of them and called \textit{Black-Scholes Model}~\cite{black_pricing_1973}, a partial differential equation (PDE) tool for pricing a financial market containing derivative investment instruments. Specifically, the Black–Scholes model establishes a partial differential equation governing the price evolution of a European option call or European option put, as follows:
\begin{equation}\label{equ:BSM}
    \frac{\partial V}{\partial t} + \frac{1}{2} \sigma^2 S^2 \frac{\partial^2 V}{\partial S^2} + r S \frac{\partial V}{\partial S} - rV = 0
\end{equation}
where $V$ is the price of the option as a function of stock price $S$ and time $t$, $r$ is the risk-free interest rate, and $\sigma$ is the volatility of the stock. This PDE has a closed-form solution called Black-Scholes Formula. Since Robert Merton was the first to publish a paper expanding the mathematical understanding of the options pricing model, he was usually credited with the contribution of this theory as well. Merton and Scholes received the 1997 Nobel Memorial Prize in Economic Sciences for their discovery of the risk-neutral dynamic revision. The original Black-Scholes model was extended later for deterministically variable rates and volatilities, and was extended to characterize the price of European options on instruments paying dividends, as well as American options and binary options. 

As a pioneering work in risk-neutral theory, Black-Scholes model has many limitations, one of which is the assumption that the underlying volatility is constant over the life of the derivative, and is unaffected by the changes in the price level of the underlying security. This assumption usually contradicts the phenomenon of the smile and skew shapes of implied volatility surfaces. A possible solution is to relax the constant volatility assumption. By characterizing the volatility of the underlying price using stochastic process, it is possible to model derivatives more accurately in practice, and this idea leads to a series of works about stochastic volatility, such as the Heston model~\cite{heston_closeform_1993} and the SABR model~\cite{hagan_managing_2002}. As a commonly used stochastic volatility model, the Heston model assumes the variation of the volatility process varies as a square root of the variance itself, and it exhibits a reversion trend towards the long-term mean of variance. Another popular stochastic volatility model is the SABR model, commonly used in interest rate derivative markets. This model uses stochastic differential equations to describe a single forward (such as a LIBOR forward rate, a forward swap rate, or a forward stock price) as well as its volatility, and has the ability to reproduce the effect of volatility smile. In recent years, deep learning and reinforcement learning techniques are applied to integrate with risk neural Q-quant modeling and a concept \textit{learning to trade} was introduced by Hans Buehler \cite{buehler_deep_2019}, who proposed the deep hedging model, a framework for hedging a portfolio of derivatives in the presence of market frictions such as transaction costs, market impact, liquidity constraints or risk limits and for modeling the volatility stochastic process using deep reinforcement learning and market simulation.  It does not use the Greeks anymore and naturally captures co-movements of relevant market parameters.

In addition to derivative pricing models, market efficiency theory and risk modeling theory are also very important in Q-quant, both in academia and industry. In 1980s, Harrison and Pliska established the fundamental theorem of asset pricing \cite{harrison_martingales_1981}, which provides a series of necessary and sufficient conditions for an efficient market to be arbitrage free as well as complete. In 2000, David X. Li introduced the statistical model Gaussian copula \cite{li_default_2000} to evaluate the value-at-risk (VaR) of derivative pricing and portfolio optimization, especially the collateralized debt obligations (CDO). Gaussian copula quickly became a tool for financial institutions to correlate associations between multiple financial securities since it is relatively simple in modeling even for those assets too complex to price previously, such as mortgages.

\figureEfficientFrontier

\subsubsection{Landmarks in P-Quant}
Q-quant plays an extremely important role in quantitative finance. In this article, however, we stand on a buy-side point of view and focus on asset prediction and portfolio optimization problems, and thus all discussions about quantitative investment in the following content assume a P-quant statement unless otherwise specified. The origin of P-quant started from the establishment of modern portfolio theory introduced by Harry Markowitz. The theory was initialized in his Ph.D. thesis ``Portfolio Selection'' and later published in Journal of Finance in 1952 \cite{markowitz_portfolio_1952}, and an extension published in his book \emph{Portfolio Selection: Efficient Diversification of Investments} \cite{markowitz_portfolio_1959} in 1959. According to the old adage “Don’t put all your eggs in one basket”, Markowitz came up with the concept of efficient frontier of asset investment in financial market and formalized it mathematically as a quadratic optimization problem by maximizing the expected return of the portfolio given its risk (usually measured by the variance of the assets in a portfolio) at a certain level. Figure \ref{fig:markowitz_theory} illustrates the application of Markowitz's theory for allocating the best positions for assets in portfolio and illustrates the concept of efficient frontier. 

Based on the modern portfolio theory, the Capital Asset Pricing Model (CAPM) was later introduced by Jack Treynor (1961, 1962) \cite{treynor_market_1961}, William F. Sharpe (1964) \cite{sharpe_capital_1964}, John Lintner (1965) \cite{lintner_valuation_1965} and Jan Mossin (1966) \cite{mossin_equilibrium_1966} independently. CAPM aims to describe the relationship between systematic risk from the market and expected return for assets. 
\begin{equation}
    E(R_p) - R_f = \alpha + \beta \cdot (E(R_m) - R_f) 
\end{equation}
where $E(R_p)$ is the expect return of portfolio, $R_f$ is the risk-free return, $E(R_m)$ is the expected return of market. Specifically, CAPM decomposes asset return and risk into two separate parts, alpha and beta. Alpha measures the performance of a portfolio compared to a benchmark index (e.g., S\&P500 index), while beta measures the variance of the portfolio in relation to a benchmark index, characterizing the risk from market volatility. One of the main contributors to CAPM, William Sharpe, shared the 1990 Nobel Prize with Harry Markowitz. A following important step in quantitative finance is the establishment of arbitrage pricing theory (APT) by MIT economist Stephen Ross in 1976 \cite{ross_arbitrage_1976}. APT improved its predecessor CAPM by further introducing the multifactor model framework to build the relationship between asset price and various macroeconomic risk variables. Under the multifactor model framework, Nobel Prize-winning economist Eugene Fama proposed the famous Fama–French Three-Factor Model with his colleague Kenneth French at the University of Chicago in 1992 \cite{fama_cross-section_1992}. 
\begin{equation}
E(R_p) - R_f= \beta_0 + \beta_{1} \cdot (E(R_m) - R_f) + \beta_2 \cdot SMB + \beta_3 \cdot HML
\end{equation}
This model establishes the relationship between the expected portfolio return (up to subtracting a risk-free return) $E(r_p) - r_f$ with respect to three systematic risk factors: expected market return $E(R_m) - R_f$, size $SMB$ (the spread between small capitalization stocks and large capitalization stocks), book-to-market values $HML$ (the spread between high book-to-market companies and low book-to-market companies). The three-factor model was then extended to Fama and French Five Factor Model in 2015 \cite{famafrench_fivefactor_2015}, by adding two more factors: profitability (return spread of the most profitable firms minus the least profitable) and investment aggressiveness (the return spread of firms that invest conservatively minus aggressively). 

Parallel with the progress of multifactor models, a number of significant research about time-series analysis appears in 1980s. In 1980, Nobel Prize winner Christopher Sims introduced the Vector Autoregression (VAR) model into economics and finance. As an extension of single sequence autoregressive (AR) model and autoregressive-moving-average (ARMA) model commonly used in time-series analysis, VAR characterizes the autoregressive properties over time across multiple times series and it assumes constant variance of error terms in the regression formula. In 1982, Robert Engle introduced the Autoregressive Conditional Heteroskedasticity (ARCH) model and extend it to Generalized Autoregressive Conditional Heteroskedasticity (GARCH) model to characterize the pattern of financial volatility in the market by specifying stochastic variance in the model. In 1987, he introduced co-integration method with Clive Granger (inventor of Granger Causality for modeling lead-lag patterns among multiple time series) for testing the significance of mean-reversing patterns in financial time series, and co-integration test has been widely used in discovering promising asset pairs for statistical arbitrage strategies. Both Engle and Granger received the 2003 Nobel Prize for their contribution to time series analysis which has been widely applied in quantitative finance for market forecasting and investment research. In 2018, three pioneers in deep learning techniques, Yoshua Bengio, Geoffrey Hinton and Yann LeCun, are granted the Turing Award. Nowadays, deep learning has been widely used by academic researchers in finance and quant researchers in financial institutions to build complex nonlinear models in order to learn the relationship between financial signals and expected returns and to predict asset prices, and its powerful ability in fitting big data significantly improves the performance of market prediction and portfolio management.

Although accurately predicting the future trend of asset price is a very important task in P-Quant, how to explain the effect of model prediction and interpret how a model is really working seems more important for quant researchers since ``know how'' is more crucial than ``know what'' in risk management for portfolio managers. Causal effect analysis \cite{eichler_causal_2012} and factor importance analysis are two core tasks in quant model interpretation. Clive Granger invented the Granger Causality Test in 1969 \cite{granger_investigating_1969} for determining whether one time series is useful in forecasting another. The original Granger causality test does not account for latent confounding effects and does not capture instantaneous and non-linear causal relationships, though several extensions have been proposed to address these issues. Although there is an argument about whether Granger causality test can evaluate ``real'' causality in terms of statistics, this method has been widely applied in quant research such as searching and evaluating pairs of stocks with significant lead-lag effect and trading with corresponding strategies. In 1994, Guido Imbens and Joshua Angrist introduced the local average treatment effect (LATE) model to characterize the statistical causal effect in economics, finance and social sciences, and they shared the 2021 Nobel Prize in economics. Another important contributor in causal inference is the Turing Award winner Judea Pearl, who invented the causal diagram (Bayesian network) and it can be used to mine the causal effect among factors and returns in multifactor model. On the other hand, in the area of factor importance analysis, Shapley Value has become an important criterion for measuring the contribution of single feature in a complex nonlinear machine learning model. In fact, it is interesting that this criterion was invented originally to measure the contribution of individual player/agent in a cooperative gaming process when it was first proposed by Lloyd Shapely, a Nobel Prize winner and a pioneer in game theory research.

\subsubsection{Development of Quant in Industry}
\figureQuantHistory
The blooming era of quantitative investment funds started from 1990s, along with the emergence of the Internet and the development of electronic trading in exchanges. Here we briefly introduce the evolution of quant operating models and classify them into three generations, denoted as Quant 1.0--3.0, and summarize their characteristics in Figure \ref{fig:quant_history}. 
\begin{itemize}[leftmargin=*]
\item \underline{Quant 1.0} appeared in the early age of quantitative investment but it is still the most popular quant operating model in contemporary market. The features of Quant 1.0 includes: 1) Small but elite team, typically led by an experienced portfolio manager and composed of a few genius researchers and traders with strong mathematics, physics or computer science background; 2) applying or even inventing mathematical and statistical tools to analyze financial market analysis and discover mispriced assets for trading; 3) trading signals and trading strategies are usually simple, understandable and interpretable to reduce the risk of in-sample over-fitting in modeling. This operating model has high efficiency in quant trading but low robustness in management. Especially, the success of a Quant 1.0 team relies too much on particular genius researchers or traders, and such a team may decline or even bankrupt rapidly with the departure of genius. In addition, such a small ``strategy workshop'' limits the research efficiency on complex investment strategies such as quantitative stock alpha strategy which depends on diversified financial data types, extremely large data volume, and complex modeling techniques such as super large deep learning model.  
\item \underline{Quant 2.0} changes quant operating model from small \emph{genius' workshop} to an industrialized and standardized \emph{alpha factory}. In this model, hundreds or even thousands of investment researchers work on the same pipeline to mine effective alpha factors \cite{tulchinsky_introduction_2019} out of the plethora of financial data, using standardized evaluation criteria, standardized back-test processes and standardized parameter configurations. These alpha mining researchers are rewarded by submitting qualified alpha factors which usually have high back-test returns, high Sharpe ratio, reasonable turnover rate and low correlation with existing factors in the alpha database. Traditionally, each alpha factor is a mathematical expression characterizing some pattern or profile of stocks, or some relationship between stocks, although more and more complicated machine learning factors are mined as well. Typical alpha factors include momentum factors, mean-reversion factors, event-driven factors, volume-price dispersion factors, growth factors, etc. Many alpha factors submitted by alpha researchers are combined into statistical models or machine learning models by portfolio managers to find the optimal asset positions after appropriate risk neutralization, expecting to obtain a stable and promising excess return in the market. However, large-scale team work results in huge costs for human resources, and the situation gets more and more serious with the team growing larger and larger. Specifically, we could expect the number of discovered effective alphas follows an approximately linear trend with the team size (actually in practice, discovering new effective alphas is more and more difficult when the size of accumulated factors is already large), but the portfolio return grows significantly lower than the expand of alpha volume and team size, and this results in the profit margin getting smaller and smaller. This phenomenon is caused by a number of reasons such as the limitation of strategy market capacity, the growing difficulty in discovering new effective alphas, and even the limitation of human intelligence in searching all possibilities in strategy space. 
\item \underline{Quant 3.0} emerges with the rapid development of deep learning techniques which have exhibited success in many domain areas such as computer vision and natural language processing. Different from Quant 2.0 which puts more research efforts and human labor into mining sophisticated alpha factors, Quant 3.0 pays more attention to deep learning modeling. With relatively simpler factors, deep learning still has the potential to learn a prediction model performing as well as a Quant 2.0 model, by leveraging its powerful end-to-end learning ability and its flexible model fitting ability. In Quant 3.0, the cost of human labor of alpha mining is at least partially replaced by the cost of computing power, especially for the expensive GPU servers. But generally speaking, it is a more efficient way for quant research in the long run. 
\end{itemize}

\subsection{Quant 4.0: Why and What}
\label{sec:introduction_quant4.0}
\subsubsection{Limitations of Quant 3.0}
Although Quant 3.0 has demonstrated its success in some strategy scenarios such as high-frequency stock and future trading, it has three primary limitations. 
\begin{tightenumerate}
    \item Traditionally, building a ``good'' deep neural network is time-consuming and labor-intensive, because of the heavy work in network architecture design and model hyperparameter tuning, as well as the tedious work in model deployment and maintenance in trading ends.
    \item It is a challenge to read understandable messages from a model encoded by deep learning black box, making it very unfriendly to investors and researchers who care much about the mechanism of financial markets and expect to know the source of profit and loss.
    \item The good performance of deep learning relies heavily on extremely large volumes of data, and thus only high-frequency trading (or at least medium cross-sectional alpha trading with large breadth) belongs to the strategy pool that deep learning favorites. This phenomenon prevents deep learning techniques from application in low-frequency investment scenarios such as value investing, fundamental CTA and global macro.
\end{tightenumerate}
New research and new techniques are needed to address these limitations, and this leads to our proposal for Quant 4.0 in this article. 
\figureQuantFourKeys

\subsubsection{What is Quant 4.0?}
We believe the limitations of Quant 3.0 are very likely to be solved or at least partially solved in the future with the quick development of the artificial intelligence (AI) technology frontier. Quant 4.0, the next-generation quant technology, is practicing the philosophy of ``end-to-end going all-in on AI'' and ``AI creates AI'' by incorporating the state-of-the-art automated AI, explainable AI and knowledge-driven AI and plotting a new picture for quant industry. 
\begin{itemize}[leftmargin=*]
    \item \underline{Automated AI} aims to build end-to-end automation for quant research and trading, in order to significantly reduce the cost of labor and time for quant research including data preprocessing, feature engineering, model construction and model deployment, and to dramatically improve R\&D's efficiency and sustainability. In particular, we introduce state-of-the-art AutoML \cite{he_automl_2021} techniques to automate every module in the whole strategy development pipeline. In this way, we propose to change traditional hand-craft modeling to an automated modeling workflow in an ``algorithm produces algorithm, model builds model'' manner, and eventually move towards a technical philosophy of ``AI creates AI''. Besides AI automation, another important task is to make AI more transparent, which is essentially important for investment risk management. 
    \item \underline{Explainable AI}, usually abbreviated as XAI in machine learning area, attempts to open the black box encapsulating deep learning models. Pure black-box modeling is unsafe for quant research because people can not calibrate the risk accurately. It is difficult to know, for example, where returns come from and whether they rely on certain market styles, and what the reason for a specific drawdown is, under black-box modeling. More and more new techniques in the field of XAI could be applied in quant to enhance the transparency of machine learning modeling, and thus we recommend quant researchers to pay more attention to XAI. We have to notice that improving model explainability has costs. Figure~\ref{fig:VersatilityAccuracyExplainability} shows an impossible trinity of versatility, accuracy and explainability, and tells us that we have to sacrifice at least one apex in the triangle to obtain the benefit from the other two. For example, physical law $E=mc^2$ establishes an explainable and accurate relationship among energy, mass and speed of light, but this formula can be only applied in specific domains of physics and sacrifices versatility. Imagining that we provide more prior knowledge or domain experience in a model, it is equivalent to reducing the versatility to protect the performance of accuracy and explainability simultaneously. 
    \figureVersatilityAccuracyExplainability
    \item \underline{Knowledge-driven AI} differs from the data-driven AI which heavily depends on large volumes of data samples and thus is appropriate for investment strategies with large breadth such as high-frequency trading or stock cross-sectional trading. It is an important complement to data-driven AI techniques such as deep learning (illustrated in Figure~\ref{fig:posterior} using Bayes' theorem). In this paper, we introduce knowledge graph which represents knowledge with a network structure composed of entities and relations, and stores knowledge with semantic triples. A knowledge graph of financial behaviors and events could be analyzed and inferred for investment decisions using symbolic reasoning and neural reasoning techniques. This implies potential applications to those investment scenarios with low trading frequency but intensive fundamental information in collection and analysis, including value investing and global macro investment. 
    \figurePosterior
\end{itemize}

\figureQuantWorkflow
\section{Automated AI for Quant 4.0}
\label{sec:autoai}
Automated AI for Quant 4.0 covers the automation of the full quant pipeline. In this section, we will first give an overview of the pipeline and then introduce how to upgrade it to an automated AI pipeline.

\subsection{Automating Quant Research Pipeline}
\label{sec:autoai_pipeline}
\subsubsection{Traditional Quant Pipeline}
Over decades of development, quant research has formed a standard workflow as shown in Figure \ref{fig:quant_workflow} (blue part). This workflow consists of a number of modules, including data preprocessing, factor mining, modeling, portfolio optimization, order execution, and risk analysis. 
\begin{tightitemize}
    \item \underline{Data preprocessing} is usually the first step in quant research. Original raw data may have many issues.
    Firstly, financial data usually have missing records, more or less. For example, in technical analysis, you may not receive price data at some time points due to packet loss during communication, or you may miss the price data on some trading days because of stock suspension. Similarly, in fundamental analysis, you may miss part of financial statement data since they are not reported on time. Although conventional statistical data imputation methods could be used to estimate and fill in missing records, we must avoid using future information in the imputation process.
    Secondly, financial data contain extreme values and outliers which may come from misrecording, data storage issues, data transfer issues, or extreme markets, and these outliers may lead to risky biases in investment decisions. Outliers could be eliminated by data winsorization methods \cite{adams_identifying_2019} which limit extreme values in a certain percentile range, but we have to notice that some outliers are actually strong signals for quant trading rather than noise, and must differentiate the two during data preprocessing.
    Thirdly, many financial data, such as news event data, have low data coverage and irregular updating frequency. We must align these types of data with high coverage and regular frequency such as quotes data for the convenience of downstream factor mining and modeling tasks.
    Fourthly, different data features have quite different scales in value range and thus some ``large'' features may dominate ``small'' features in modeling. Therefore, data standardization methods are used to normalize the range of features. We have to take care of the way to standardize the data in order to reduce information loss.
    \item \underline{Factor mining} is a task of feature engineering \cite{zheng_feature_2018}, which uses financial and economic domain knowledge to design, search, or extract factors (features for downstream modeling) from raw data. Usually, a larger factor value indicates a more significant trading signal. The motivation of factor mining is to find those signals from raw data for market prediction and improve the quality of downstream modeling tasks.
    Traditionally, financial factors could be represented as either algebraic formulas or rule-based expressions. Let's take a simple stock alpha factor as an example.
    \begin{equation}
        factor = -\hbox{ts\_corr}(\hbox{rank}(close), \hbox{rank}(volume), 50)
    \end{equation}
    where the $\hbox{ts\_corr}()$ function computes the correlation of daily close price and volume along time using the data from the previous 50 trading days, representing how similar the trend of $close$ time series and $volume$ time series are.
    The $\hbox{rank}()$ function maps the values in a cross-section to their orders and normalizes them to the range of $[-1,+1]$ evenly according to their descending order, in order to remove the effect of extreme values.
    This factor prefers to select those stocks when their price and volume move in opposite directions, and the idea behind it is based on the assumption that a price trend can not sustain without the support of volume growth.
    Traditionally, factor mining is a labor-intensive job. Most quant researchers can only discover a limited number of ``good'' factors in a year. Different financial institutions have different definitions or criteria for a ``good'' factor, but most of them consider a few common aspects, such as return, Sharpe ratio, maximum drawdown, turnover rate, and correlation with other factors \cite{tulchinsky_introduction_2019}, and moreover, some institutions require the factors must be meaningful, understandable and explainable in economics.
    \item \underline{Modeling} is a task to build statistical or machine learning models using factors and to predict market trends, asset price movements, best trading times, or most/least valuable assets. Usually, prediction models are evaluated through back-test experiments which simulate the prediction and trading process using historical data. Choices of models must consider a number of factors, such as prediction accuracy, model explainability, model robustness, and computational complexity, and find the best tradeoff according to the ultimate goal.
    In particular, we must notice that most statistical or machine learning models are not specifically developed for financial time series, and we have to adjust the application of these models in quant modeling.
    Firstly, financial time series prediction must avoid using future information, and thus we prefer forward-validation \cite{schnaubelt_comparison_2019} (splitting the time series into training, validation, and test blocks over time) rather than cross-validation in model hyperparameter optimization.
    Secondly, financial time series are usually significantly nonstationary, far from the \textit{independent and identically distributed} (i.i.d.) assumption required by many machine learning models. Therefore, data transformation is needed to make the data distribution closer to i.i.d. and if possible, look more like a normal distribution.
    Thirdly, market style moves over time and it results in the shift of financial time-series distribution. Therefore, periodic model retraining is necessary for keeping the model adapted to market style variation.  
    \item \underline{Portfolio optimization} aims to find the optimal asset allocation to expect high return and low risk simultaneously. While prediction models tell us what or when to buy/sell, portfolio optimization specifies how much to buy/sell. A typical portfolio optimizer attempts to solve a constrained convex quadratic programming problem which is extended from Markowitz's efficient frontier theory. 
    \begin{eqnarray*}
    \max_{w_t} & &  w_t^T r_t \\
    \hbox{subject to} & &   w_t^T \Sigma w_t \le C_1 \\
    & & | w_t - w_{t-1} | \le C_2 \\
    & & 0 \le w_{i,t} \le C_3 \le 1, \hbox{for}~i=1,2,\ldots,n
    \end{eqnarray*}
    where $r_t = (r_{1,t}, r_{2,t}, \ldots, r_{n,t})^T$ is the returns of $n$ assets (e.g., stocks) at time $t$, and $w_t = (w_{1,t}, w_{2,t}, \ldots, w_{n,t})^T$ is the corresponding position weights (percentages of capital allocation). $C_1, C_2, C_3$ are positive constraint bounds. $\Sigma$ is the volatility matrix of the $n$ asset returns at time $t$. The target function tries to maximize the portfolio return and control the upper bound of risk and turnover rate (to reduce transaction cost). The key in this optimization problem is how to estimate the volatility matrix $\Sigma$ whose estimation is usually unstable if historical data is not long enough, and in this case dimension reduction tricks such as regularization and factorization can be helpful to improve estimation robustness. 
    \item \underline{Order execution} is a task that buys or sells orders with optimal prices and minimal market impact. Usually buying (or selling) a big order at one time will push the price of the target asset in a harmful direction (market impact by this big order), and therefore increase the trading cost. A widely used solution is order splitting, which divides a big order into a number of small orders to reduce market impact. Algorithmic trading provides a series of mathematical tools for order splitting, from the simplest time-weighted average price (TWAP) and volume-weighted average price (VWAP) to the complicated reinforcement learning methods \cite{nevmyvaka_reinforcement_2006} in which optimal order flow is modeled as a (partially observable) Markov decision process. 
    \item \underline{Risk analysis} is an indispensable task for quant research and quant trading. We must discover and understand every possible risk exposure in order to better control unnecessary and harmful risks in quant research and trading \cite{coleman_practical_2011}. In the monitor module, risks are measured in real-time and these messages and analysis are sent back to help quant researcher improve their strategies. The most popular risk model in stock trading is the BARRA model \cite{barra_handbook} which decomposes portfolio volatility into the exposures of a number of predefined risk factors, including style factors (size, growth, liquidity, etc.) and industry factors. However, the BARRA model could explain only about 30\% of total volatility, leaving the risk hidden in the remaining 70\% part still unknown.  
\end{tightitemize}
\subsubsection{Automated AI Quant Pipeline}
The automated pipeline of Quant 4.0 is shown in Figure \ref{fig:quant_workflow} (orange part), where modules in the pipeline are automated by applying state-of-the-art AI technology. In the following part of this section, we will concentrate on three core modules in the automated pipeline.
\begin{tightenumerate}
    \item Automated factor mining (\S\ref{sec:autoai_feature}) applies automated feature engineering techniques to search and evaluate significant financial factors generated from meta factors. We will introduce popular search algorithms and demonstrate how to design the algorithmic workflow. 
    \item Automated modeling (\S\ref{sec:autoai_model}) applies AutoML techniques to discover optimal deep learning models, automatically selecting the most appropriate models and the optimal model structures, and tuning the best hyperparameters; 
    \item One-click deployment (\S\ref{sec:autoai_deployment}) builds an automated workflow to deploy trained large models on trading servers with limited computing power. It executes model compression, task scheduling, and model parallelization automatically, saving a lot of labor and time for tedious ``dirty'' work. 
\end{tightenumerate}

\subsection{Automating Factor Mining}
\label{sec:autoai_feature}
Feature engineering for quant refers to the process of extracting financial factors from original data, on which effective pattern recognition is difficult due to their intrinsic noisiness \cite{black_noise_1986, ait-sahalia_high_2009, mancini_measuring_2013}. Traditionally, financial factors with significant ``alpha'' are explored and developed by quant researchers manually, they rely on professional domain expertise and comprehensive knowledge of financial markets. Although some financial institutions started using random search or generic programming algorithms, these techniques are mainly used as small-scale auxiliary tools to help improve the productivity of quant researchers. In Quant 4.0, We propose to automate the factor mining process by formulating feature engineering as a search problem and utilize corresponding algorithms to generate factors with satisfactory backtest performance at scale. In particular, according to their expression form, we classify factors as 1) symbolic \cite{kakushadze_101_2016} factors which are symbolic equations or symbolic rules, and 2) machine learning factors which are expressed by neural networks, and we will elaborate on the details in the following part of \S\ref{sec:autoai_feature}.

\subsubsection{Symbolic Factors}
\label{sec:autoai_feature_symbolic}
\figureFactorMiningPipeline
Symbolic factor mining can be regarded as a special case of symbolic regression \cite{cava_contemporary_2021}. Traditional symbolic regression algorithms usually generate a large number of symbolic expressions from given operands and operators and select the symbolic expressions that maximize the predefined objective function.
Figure \ref{fig:factor_mining} shows a framework for automated symbolic factor mining, which consists of four core parts: operand space, operator space, search algorithm, and evaluation criterion.
\begin{itemize}[leftmargin=*]
    \item \underline{Operand Space} defines which meta factors could be used for factor mining. Meta factors are fundamental components for factor construction. Typical meta factors include basic price and volume information, sector categorizations, basic features extracted from limit/order books, common technical indices, basic statistics from financial analysts, important signals from financial reports, announcements and other research reports from public companies, sentiment signals from investor emotions \cite{rashid_investor_2019, abdul_karim_market_2022}, etc. 
    \item \underline{Operator space} defines which operators could be used in the factor mining process. For example, in cross-sectional stock selection, the operators could be classified as main operators for constructing symbolic factors and post-processing operators for standardizing the factors for different trading environments. Main operators could be classified further as element-wise operators such as $\sqrt()$ and $\log()$, time-series operators such as $ts\_rank()$ and $ts\_mean()$ which compute the rank order and mean along each stock respectively, cross-sectional operators such as $rank()$ and $quantile()$ which compute the rank and quantile along the cross-section at a specific trading time, and group operators such as $group\_rank()$ which compute rank order in each group (e.g., industry or sector) respectively. Post-processing operators are used to ``fine-tune'' the generated factors. Typical post-processing operators are standardization operators such as winsorization for outlier clipping \cite{adams_identifying_2019} and normalization for unifying data range, neutralization operators for risk balancing, grouping operators for restricting the universe of stock selection, and decay operators for controlling turnover rate so as to reduce transaction cost.
    \item \underline{Search algorithms} aim to search and find effective or qualified factors as efficiently as possible. 
    A simple way to generate new factors is the Monte Carlo (MC) algorithm which randomly picks the elements in the operand and operator spaces and generates a symbolic expression tree recursively. Unfortunately, the search time may grow exponentially with the length and complexity of the generated formula, and push us to consider more efficient alternatives. The first option is Markov-chain Monte Carlo (MCMC) algorithm \cite{andrieu_introduction_2003}, which generates factors in sampling with importance way from a posterior distribution \cite{jin_bayesian_2020}, and thus it is more efficient than MC. 
    The second option is genetic programming \cite{chen_empirical_2021}, which is a special evolutionary algorithm for sampling and optimizing tree-type data. The third option is about gradient-based methods such as neural networks, which approximate the discrete symbolic formulas with continuous nonlinear functions and search along the gradient direction, significantly more efficient than random search. 
    \item \underline{Evaluation criteria} measure the quality of factors found by search algorithms. The performance of the generated factors is evaluated using backtest experiments. Typical evaluation criteria include information coefficient (IC), information ratio based on information coefficient (ICIR), as well as annualized return, maximum drawdown, Sharpe ratio, and turnover rate. In addition, it is very important to keep information diverse among factors by filtering out redundant factors which highly correlated with other factors.
\end{itemize}
Due to their importance in factor mining, we introduce about two types of search algorithms in detail. 

\figureGeneticProgramming
\figureNeuralSymbolicRegression

\begin{tightitemize}
\item \underline{Genetic programming (GP)} \cite{ahvanooey_survey_2019} (Figure \ref{fig:genetic_programming}), an extension of genetic algorithm \cite{katoch_review_2021}, is a metaheuristic algorithm for searching tree-structured symbolic factor expressions. In an algorithmic loop, GP starts from a number of initial factors, and it uses an evolutionary mechanism to produce the next generation of factors, aiming to improve factor performance measured by the fitness function. There are two types of evolutionary mechanism in GP: \textit{mutation} which replace a node randomly with another operand of an operator, and \textit{crossover} where two trees swap their subtrees randomly. In each iteration, all factors are evaluated using IC or alternatives and only the best-performing factors are kept. This process is repeated until convergence. 

\item \underline{Neural symbolic regression} utilizes gradient information to accelerate the search process. Neural networks are used to learn a continuous and nonlinear function to approximate those discrete symbolic expressions and use this function to generate new formulas. We introduce two works about neural symbolic regression. The first paper \cite{biggio_neural_2021} (Figure \ref{fig:neural_symbolic_regression}) builds a transformer generative model from a number of existing symbolic expressions. In the training stage, a special transformer model (called set transformer \cite{lee_set_2019}) encodes the formulas in the training set into embedded vectors, and they are delivered to a transformer decoder to update symbolic expressions in an autoregressive way using beam search, and this process is repeated until convergence. The generative model is trained by minimizing the cross-entropy loss between the generated expression and the original one. In the test stage, the trained model is used to generate new symbolic expressions. The second paper \cite{martius_extrapolation_2017} (Figure \ref{fig:neural_symbolic_regression_2}), designed a new neural network specifically for expressing symbolic formulas. In particular, the activation functions in this network are replaced with symbolic operators such as $sin(\cdot)$ and $\sqrt(\cdot)$. This special neural network has the flexibility to generate almost all formula expressions need to use in factor mining. 
    
\end{tightitemize}

\subsubsection{Machine Learning Factors}
\label{sec:autoai_feature_neural}
\figureNeuralFactor
Symbolic factors have their advantages in simplicity and understandability, and are thus widely used in practice. However, their representation ability is limited by the richness of operands and operators. Machine learning factors, on the other hand, have more flexibility in representation to fit more complicated nonlinear relationships \cite{DBLP:journals/nn/HornikSW89}, and thus they have the chance to perform better in market prediction. 
In particular, mining machine learning factors \cite{thakkar_comprehensive_2021, hu_survey_2021, jiang_applications_2021, sezer_financial_2020} is a process to fit neural networks, where gradients provide the optimal direction for fast search of solutions. As shown in Figure \ref{fig:neural_factor}, most deep neural networks for stock prediction follows the encoder-decoder architecture \cite{sutskever_sequence_2014}, where the encoder maps meta factors to a latent vector representation and the decoder transforms this embedding to some outcome such as future return \cite{wang_alphastock_2019}. In fact, not only the final outcome, but also the embedding itself could be used as a (high-dimensional) machine learning factor \cite{wang_deeptrader_2021}, and further applied to various downstream tasks. 

Machine learning factors have some limitations as well. Firstly, they are usually hard to interpret and understand because of the black-box nature of machine learning. Secondly, gradient search used by neural networks may be stuck at some local optima and result in model instability problems. Finally, neural networks may suffer more serious overfitting due to their flexibility, and this situation gets worse in quant because data are extremely noisy.


\figureNAS
\subsection{Automated Modeling}
\label{sec:autoai_model}
The automation of statistical machine learning such as SVM, decision tree and boosting has been extensively researched. A simple and direct automation method is the brute-force enumeration of all possible configurations, each including the choice of machine learning algorithms and the corresponding hyperparameters (Figure \ref{figAutoMLSimple}). 
\figAutoMLSimple

In this article, we focus on the state-of-the-art deep learning automation problem, which is more complex due to the end-to-end property and network architecture issue in modeling. The configuration of a deep learning model consists of three parts: architecture, hyperparameters, and objectives, and they jointly determine the final performance of the models. Traditionally, these configurations are tuned manually. In Quant 4.0, they are searched and optimized using various AutoML algorithms. A standard AutoML system needs to answer the following three questions: what to search (i.e., search space \S\ref{sec:autoai_model_space}), how to search (i.e., search algorithm \S\ref{sec:autoai_model_strategy}), and why to search (i.e., performance evaluation \S\ref{sec:autoai_model_estimation}).

\subsubsection{Search Space}
\label{sec:autoai_model_space}
Search space is designed for the three configuration settings that need to be optimized in an automatic way. 
\begin{tightitemize}
\item \underline{Architecture} configures a network structure. For example, the architecture of a multi-layer perceptron is specified by the number of hidden layers and the number of neurons at each layer. The architecture of a convolution neural network needs to consider more configurations such as the number of convolution kernels as well as their strides and receptive fields. The architecture of large-scale models such as Transformer is composed of a number of predefined blocks (e.g. self-attention blocks, residual blocks) linked together. As discussed above, architecture is complex and may have a hierarchical structure at different scales. Accordingly, the search space can be defined at various granularities, ranging from low-level operators such as convolutions and attentions to high-level modules such as LSTM cells. Early search algorithms run on the finest granularity and optimize the low-level structure of the neural network \cite{mendoza_towards_2016, baker_accelerating_2018}. Such a search process is flexible in network structure but inefficient in incorporating prior knowledge and abstraction. One solution is to assume a hierarchical structure in network architecture. Specifically, at a high level, the network is designed to be a graph of cells (a.k.a. blocks/motifs \cite{he_automl_2021, elsken_neural_2019}), each of which is a subnetwork. Many cells share the same internal structure at a low level in order to reduce the computational cost. Cell-based search algorithms \cite{zoph_learning_2018, zhong_blockqnn_2021} need to find both high-level structures between cells as well as low-level structures within the cells.  

\item \underline{Hyperparameter} controls the overall training process. For example, learning rates determines the step size moving towards a minimum of a loss function. A smaller learning rate is more accurate in solution but slower in convergence. The batch size determines the number of samples involved in a batch for gradient estimation, which also has an influence on training efficiency and stability. The search space for hyperparameters is simpler than that for architecture since most hyperparameters are continuous (e.g., learning rate) or approximately continuous values (e.g., batch size).   

\item \underline{Objective} specifies the loss functions and labels used for training models. The loss function is the key component of machine learning models since it provides a goal towards which a model should be trained. Besides classic loss functions such as mean square loss and cross-entropy loss, new loss functions specifically designed for quant tasks can be also selected. Labels define the ``ground-truth'' target the model aims to fit. For example, either price raise/fall or future returns in different holding time windows can be considered in the search space.   
\end{tightitemize}

\subsubsection{Search Algorithm}
\label{sec:autoai_model_strategy}
Given the search space, we could use search algorithms to find the best model configuration.
Table \ref{tab:auto_modeling} lists various types of search algorithms and their corresponding tasks: network architecture search (NAS) \cite{elsken_efficient_2019}, hyperparameter optimization (HPO) \cite{feurer_hyperparameter_2019} and training objective selection (TOS). 
\tableAutoModeling

\begin{itemize}[leftmargin=*]
    \item \underline{Grid search} is a brute-force algorithm that searches on a grid of configurations and evaluates all of them. It is a good choice when the search space is small due to ease of implementation and parallelization \cite{bergstra_random_2012}. However, most NAS and HPO problems in deep learning have extremely large search spaces and grid search can not scale well for them. Moreover, grid search is used more popular in HPO and TOS than in NAS whose search space is difficult for grid layout except enumerating all possibilities.
    \item \underline{Random search} generates a number of candidate configurations using some stochastic sampling mechanisms, such as Monte Carlo or MCMC. It is very straightforward to implement and parallelize (mainly for independent sampling mechanisms such as Monte Carlo sampling or importance sampling). Random search is very flexible and can be used for NAS, HPO, and TOS. Although random search is usually faster than grid search \cite{bergstra_random_2012}, it is still difficult to handle high-dimensional search space as the number of potential configurations grows exponentially with the number of hyperparameters.

    
    \item \underline{Evolutionary algorithm} is an extension of random search. It utilizes evolution mechanisms to improve model configurations iteratively. It encodes the architecture of network networks as a population and performs the evolution steps on them to improve the model iteratively. Specifically, the models are first encoded according to their underlying computation graph. Then, a set of pre-defined evolutionary operators are applied to the encoded models, including architectural modifications such as inserting or deleting several operations and adding skipping connections, as well as hyperparameter-related operations such as learning rate adjustment. At each iteration, the best-performing models are selected via tournaments and combined via mutation and crossover operations to form the next generation. Evolutionary algorithms inherently support weight inheriting among generations, which helps accelerate the convergence in training and increase the searching efficiency.
    
    \item \underline{Reinforcement learning} models the architecture search problem as a Markov decision process. In each step, an RNN controller chooses an action to sample a new architecture and the corresponding deep neural network model is trained. Then the performance of the model evaluated on the validation set is used as a reward which is forwarded to compute the policy gradient and update the RNN controller. This loop is iterated until convergence. The reinforcement learning framework is very universal for most optimization problems and it could be used for NAS, HPO, and TOS.
    
    \item \underline{Bayesian optimization} explores the search space more efficiently by leveraging surrogate models to approximate the objective function that couldn't be expressed explicitly. Specifically, for a black-box objective function for HPO, Bayesian optimization initializes a prior distribution using a surrogate function such as Gaussian process or tree-structured Parzen estimator. Then it samples new data points from the prior distribution (with importance) and calculates their values using the underlying objective function. Given these new samples and prior, the posterior function can be calculated and is used as an updated surrogate function to replace the original prior function. This process is repeated until the optimal solution is found. Traditionally, Bayesian optimization is used for continuous search tasks such as HPO, but recent works have extended it to NAS tasks as well.

    
    \item \underline{Gradient-based methods} is very efficient when the gradient of the objective function exists. However, for NAS, the search space is discrete and couldn't define a gradient directly. One solution is to ``soften'' the architecture and define an over-parameterized ``super-architecture'' which covers all possible candidates and is differentiable. A typical gradient-based NAS method is DARTS \cite{liu_darts_2018} which constructs an over-parameterized network where all types of candidate operations are present on the computation graph. The resulting value is the weighted sum of the results of all the operations, where the weights are the softmax values of a parameterized probability vector. Both model parameters and architectural parameters are trained via a bi-level optimization problem. In the inference process, the architecture and hyperparameters with the highest probability are selected. DARTS is substantially faster than random research and reinforcement learning in NAS and HPO tasks. 
\end{itemize}

\subsubsection{Accelerating Evaluation}
\label{sec:autoai_model_estimation}
The computational cost of automated model search comes from two parts: search algorithm and model evaluation, and the latter is usually the bottleneck of the computation because it is very time-consuming to train a deep neural network until convergence under a given configuration. Several methods are introduced in previous research to address this issue. Firstly, the training process of neural networks can be early-stopped before convergence to reduce the computational time for evaluation \cite{zoph_learning_2018}. Secondly, the model can select fewer samples to accelerate the training process \cite{klein_fast_2017}. Thirdly, warm-start model training can be used to leverage the information from existing selected models \cite{liu_progressive_2018} or inherit the information from an over-parameterized ``parent'' model \cite{wei_network_2016, liu_darts_2018, cai_proxylessnas_2019} to accelerate the search loop. 



\subsection{Automated One-click Deployment}
\label{sec:autoai_deployment}
Model deployment is the task of transferring the developed model from offline research to online trading. 
It is not only simply transferring code and data, but also synchronizing data and factor dependency, adapting trading server and system, debugging model inference, testing computing latency, etc. In the following part, we focus on one important problem in model deployment: how to accelerate deep learning inference for high-frequency trading and algorithmic trading scenarios. We propose an automated one-click deployment solution utilizing techniques such as model compilation \cite{li_deep_2021} and model compression \cite{cheng_model_2018, choudhary_comprehensive_2020} to realize inference acceleration \cite{li_deep_2021, cheng_model_2018, choudhary_comprehensive_2020}. The former makes the inference faster without changing the model itself, and the latter seeks smaller and lighter alternative models to save inference time.

\subsubsection{Acceleration by Model Compilation}
\figureInferenceOpt
At the development stage, deep learning models' functionality is the top priority for the underlying framework which implements the computations. Hence, at this stage, the framework strictly maps all the operations to the computation graph. However, such direct mapping introduces large room for optimization at the deployment stage where the computations are fixed. Therefore, the model's computation can be greatly simplified and adapted to hardware features without hurting its original semantics. Such optimization is one of the major topics in deep learning compilers \cite{chen_tvm_2018, lattner_mlir_2021, cyphers_intel_2018}, which can be categorized as \textit{front-end optimizations} and \textit{back-end optimizations}, which work on high-level and low-level intermediate representations (IRs) for deep learning models respectively. Following the summary in \cite{li_deep_2021}, we will briefly introduce relevant optimization techniques.

Front-end optimization, as illustrated in Figure \ref{fig:front_end_opt}, focuses on simplifying the structure of the computation graphs.
For example, algebraic simplification techniques such as constant folding \cite{aho_compilers_1986} and strength reduction \cite{cooper_operator_2001} convert expensive operations into cheaper ones via transformation or merging.
Common subexpression extraction (CSE) \cite{kildall_unified_1973} techniques identify repeated nodes in the computation graph and merge them into one node to avoid duplicate computations.

Back-end optimization, as illustrated in Figure \ref{fig:back_end_opt}, is performed with an emphasis on the features of hardware architectures, such as locality and memory latency.
For example, pre-fetching techniques \cite{vanderwiel_data_2000} load data from main memory to GPU before they are needed, and speed up fetch operations.
Loop-based optimizations \cite{sarkar_general_1992} reorder, fuse and unroll operations inside loops to enhance locality among neighboring instructions.
Memory latency hiding \cite{volkov_understanding_2016} techniques aim to increase instruction throughput so as to mitigate the problem of high latency in accessing memory.
Parallelization techniques such as loop splitting \cite{allen_optimizing_2002}, automatic vectorization \cite{wolfe_high_1996} and loop skewing \cite{karp_organization_1967, lamport_parallel_1974}, can also be applied to maximize the parallelism provided by modern processors.

\subsubsection{Acceleration by Model Compression} 
Model compression aims to reduce model size for inference acceleration while minimizing drops in performance. In this way, the compressed model can be regarded as an approximation of the original model.
Generally speaking, model compression can be performed at both micro- and macro-level, where the former focuses on the precision of individual model parameters and the latter focuses on simplifying the overall model structure. 

At the micro-level, pruning and quantization techniques can be applied to reduce both the number of parameters and the bit size of the individual parameters. Model pruning \cite{han_learning_2015}, as shown in Figure \ref{fig:model_pruning}, removes unimportant connections and neurons in neural networks that have little influence on the activation of the neural. The identification of candidate parameters is usually based on their weights, where those with smaller weights are considered for pruning. Model quantization \cite{han_deep_2016}, as shown in Figure \ref{fig:model_quantize}, converts the parameters from floating-point numbers to low-bit representations. Specifically, a codebook is constructed to store the approximated values of the original parameters according to the distribution of all parameter values. The parameters are then quantized according to the codebook and thus the bit size for the parameters is reduced. Due to the inevitable performance drop induced by precision reduction, the compressed model is usually re-trained to approach its original performance.

At the macro-level, the model can be significantly compressed into a smaller model with simpler architectures via knowledge distillation \cite{hinton_distilling_2015} and low-rank factorization \cite{zhang_efficient_2015, yu_compressing_2017}.
Knowledge distillation (Figure \ref{fig:knowledge_distill}) compresses a model by transferring useful knowledge from the original large model (called the teacher model) to a small and simple model (called the student model) with minimal knowledge loss. 
Low-rank factorization techniques (Figure \ref{fig:low_rank}) assume the sparsity of model parameters and then split the parameter matrix of the original neural networks into products of low-rank matrices \cite{zhang_efficient_2015}, and thus reduce the model complexity.
\figurePruneQuantize 

\section{Explainable AI for Quant 4.0}
\label{sec:xai}
XAI \cite{belle_principles_2021, murdoch_definitions_2019}, as an attractive research direction for decades, is critical to the trustworthiness and robustness of AI models.
In the case of quant, improvement in the explainability of AI can make the decision process more transparent and easy to analyze, providing useful insights to researchers and investors and discovering potential risk exposures. 
In this section, we will discuss how to leverage XAI in Quant 4.0. \S\ref{sec:xai_overview} introduces common XAI techniques and \S\ref{sec:xai_scenario} connects these techniques to real quant scenarios. 

\subsection{Overview of Explainable AI}
\label{sec:xai_overview}
XAI is an emerging interdisciplinary research area covering machine/deep/reinforcement learning, statistics, game theory and visualization. Here we focus on two types of XAI: model-intrinsic explanation \cite{rudin_stop_2019} and model-agnostic explanation \cite{ribeiro_model-agnostic_2016}.

\subsubsection{Model-intrinsic Explanation in XAI}
\label{sec:xai_overview_model-intrinsic}
Risk control and management is the top priority of financial industry. When AI models are deployed in real-world applications, their decision process is usually required to be transparent by regulatory authorities for the safety of transactions. Moreover, model-intrinsic explainability is the requirement of many large financial institutions such as banks and insurance companies.

A machine learning model is intrinsically explainable if its internal structure or mechanisms can be easily explained. Some machine learning algorithms such as linear models and decision trees are inherently explainable, as many other algorithms such as deep neural networks and kernel learning methods (SVM, Gaussian process, etc.) are black boxes with poor explainabilities. Figure \ref{fig:explainable_ml_comparison} illustrates many popular machine learning methods arranging along their general performance and explainability. We can see The increase in model-intrinsic explainability usually leads to a decrease in the model's prediction performance, and therefore the selection of machine learning algorithms is essentially a trade-off between explainability and performance. 
We briefly introduce a few typical machine learning methods in terms of explainability and predictive performances and discuss their applicable scenarios.
\figureMLExplainabilityPlot
\begin{tightitemize}
    \item \underline{Linear Models}, such as linear regression, logistic regression, linear discriminant analysis, linear SVM and addition model, is a family of methods where features or transformation of a group of features are in an additive form and thus the performance of final prediction can be easily deposed to the effects from individual features or feature groups. Therefore, linear models are intrinsically understandable and explainable. For example, linear regression explicitly encodes the importance of each feature in their corresponding regression coefficients (assuming every feature is normalized to eliminate the effect from scales and units). Although linear models are easy to explain, they are suffering the poor prediction performance since they couldn't encode complicated nonlinear relationships between prediction outputs and features. 
    \item \underline{Rule-based Learning} is another type of easy-to-explain methods. Different from a linear model which fits a linear decision boundary, a rule-based learning method fits a stepwise decision boundary characterized by decision rules combining a number of logical expressions. Examples of rule-based learning include decision tree \cite{quinlan_induction_1986} and symbolic regression, as well as ensemble models such as random forest \cite{breiman_random_2001} and boosting \cite{friedman_greedy_2001, ke_lightgbm_2017, chen_xgboost_2016}. Rule-based learning models are intrinsically explainable decision rules that are close to the logical thinking process of human beings. However, to better fit the training data and improve prediction performance, the decision rule is usually complicated, and it reduces the explainability and increases the risk of overfitting. 
    \item \underline{Ensemble Learning} combines multiple machine learning algorithms to achieve better decision performance than single models. Typical examples of ensemble learning include random forest and boosted trees that combine multiple tree models and make predictions based on the aggregation of individual decisions. Although there is controversy, in this article, we classify mixtures of experts (MoE) \cite{masoudnia_mixture_2014} as an ensemble method as well. MoE combines multiple expert networks in parallel in a layer and decides which expert (or experts) participates in the decision of a specific data point via a gating mechanism.
    Compared with other machine learning methods, ensemble learning provides high-level explainability by demonstrating the relative importance of single models or experts.
    \item \underline{Kernel Learning}, also known as kernel method or kernel machine, is a family of nonparametric learning methods which make predictions by computing the similarity between samples. The similarity is characterized by a kernel function, which is a special inner product defined in a high-dimensional Hilbert space where original data samples are mapped \cite{hofmann_kernel_2008}. For example, kernel SVM \cite{cortes_support-vector_1995} transforms original positive/negative samples into another space where they can be easily separated using a linear decision boundary. 
    In principle, kernel functions can be of arbitrary forms that satisfy the Mercer's condition \cite{vapnik_statistical_1998}, and they determine the nonlinear relationships between inputs and outputs. 
    Moreover, the idea of kernel functions is extended to the self-attention mechanism \cite{tsai_transformer_2019} used in neural networks such as Transformer \cite{vaswani_attention_2017}.
    Traditionally, we think the kernel trick improves the performance of models but weakens their explainability. From another point of view, however, the definition of kernel itself encodes the prior insight of users and could help understand the model. 
    \item \underline{Sequence Learning} refers to a family of machine learning methods that work with sequential data such as time series or sentences. They are widely used in speech recognition, language understanding, DNA sequence analysis, and stock price prediction. Sequence learning methods characterize the underlying structure hidden in sequential data and discover implicit patterns. For example, hidden Markov model (HMM) \cite{rabiner_introduction_1986} assumes that the underlying structure is a homogeneous Markov chain determined by a transition matrix (or transition kernel for continuous state space) and assumes the observed sequence is randomly generated from this chain through emission probabilities. The transition probabilities and emission probabilities are estimated during model training. Although HMM is generally a black-box model, its transition probability matrix provides some insight into the auto-regressive structure in prediction. 
    Conditional random field (CRF) \cite{lafferty_conditional_2001} extends the first-order Markov assumption of HMM and characterizes longer-range time dependency using graphical model for probability modeling, and this extra flexibility usually brings better prediction performance for CRF.
    Recurrent neural networks (RNN) such as LSTM \cite{hochreiter1997long} and GRU \cite{cho_learning_2014} exhibit better performance in sequence prediction, but it is harder to explain their internal mechanism. 
    \item \underline{Deep Learning} usually has superior prediction performance \cite{goodfellow_deep_2016} but its shortage in explainability is clear. Some special operators in deep neural networks such as convolution and attention provide partial and local explanations about their mechanisms. For example, the self-attention layer in a Transformer \cite{vaswani_attention_2017} characterizes the relative importance of each position in a sequence with respect to other positions.
\end{tightitemize}
The model-intrinsic explainability for machine learning is always a contradiction with its prediction capability. However, before the appearance of a brand-new machine learning model satisfying both high prediction accuracy and high explainability, we could rebuild and improve current machine learning methods. We could either start from an explainable model such as a linear or rule-based model and improve its prediction performance by incorporating more local nonlinear structures with better predictive power. For example, starting from a decision tree model, we could replace the decision rule in each leaf node with a neural network \cite{kontschieder_deep_2015}, thus improving the model's flexibility. As another example, starting from a deep neural network, we could also improve its partial explainability by incorporating a special layer (e.g. self-attention layer \cite{song_autoint_2019}) identifying which important feature interacts more frequently.

\figureGlobalExplanationTaxonomy
\figureXAIMethods
\subsubsection{Model-agnostic Explanation in XAI}
\label{sec:xai_overview_model-agnostic}
To address the contradiction between performance and explainability, a suboptimal solution is to weaken the requirement and shift from model-intrinsic explanation to model-agnostic explanation.  
Based on the scope of explanations, model-agnostic XAI can be categorized as global methods (explanation applied to all samples) \cite{saleem_explaining_2022} and local methods (explanation applied to part of samples) \cite{molnar_interpretable_2022}.

Global methods explain the characteristics of features with respect to all samples in the dataset. These characteristics include the importance of features, the importance of feature sets, the interactive effect of features, and other high-order effects of features.  
There are various types of methods for estimating global feature importance. 
\begin{tightitemize}
\item \underline{Feature marginalization methods} estimate the importance of a specific feature or a specific feature set by marginalizing all other features in the model. Specifically, the importance of the first feature is calculated by integrating out all other features, and similarly, we calculate the second feature, the third feature, etc.  
For example, partial dependence plot (PDP) \cite{friedman_greedy_2001} computes the marginalized model function w.r.t. the features of interest, and it visualizes the corresponding feature importance. Accumulated local effect (ALE) plot \cite{apley_visualizing_2020} provides an unbiased estimate of marginal effects using conditional distributions that consider the correlation between features.    
\item \underline{Feature leave-one-out methods} evaluate the importance of concerned features by comparing the difference in the model performance before and after leaving these features out of data. 
Specifically, the feature of a model can be left out by shuffling its values in samples \cite{breiman_random_2001, fisher_all_2019}. Leave-one-out methods can be extended to evaluate the interactive effects of features as well. 
For example, based on the partial dependence function, H-statistic \cite{friedman_predictive_2008} is proposed to test for the interaction between features. Other alternative techniques for evaluating and visualizing feature interactions have also been proposed in \cite{hooker_discovering_2004, greenwell_simple_2018}. Besides, we can also perform functional decomposition \cite{hooker_discovering_2004} on the original model to explore interactions between all possible sets of features. 
\item \underline{Feature surrogate methods} interpret the model by learning a globally explainable surrogate model \cite{schwartzenberg_fidelity_2020} that approximates the original model. The surrogate model is trained under the supervision of the original model using a dataset where the inputs remain the same while the outputs are produced by the original model. 
\end{tightitemize}

Figure \ref{fig:GlobalExplanationTaxonomy} summarizes popular global explanation methods, some of which have been introduced above.
Moreover, these methods can be categorized as data-driven and model-driven, where the former methods treat the models as black boxes and query the model for explanations, and the latter methods treat models as white boxes and provide explanations using internal information such as gradients.

Local methods explain the feature importance at the sample level, i.e., how important a feature is for specific samples. 
Similar to the partial dependence plot, we can draw the individual conditional expectation (ICE) plot \cite{goldstein_peeking_2015} for each sample that illustrates the effect of the concerned features when the values of other features are fixed at specific values. 
To interpret a black-box machine learning model at a specific sample, we can learn a surrogate model in the vicinity of a data sample to explain the original model locally.
Typical examples include LIME \cite{ribeiro_why_2016} and Anchors \cite{ribeiro_anchors_2018}.
In LIME (Figure \ref{fig:xai_methods}), LASSO regression \cite{tibshirani_regression_1996} is used as the surrogate model to fit the samples randomly generated by perturbing the original samples around the specified one.
In this way, the selected samples by this local LASSO model contribute most to the fitness of the specified sample.
Analogous to LIME, anchors are explanations on individual samples in the form of IF-THEN rules \cite{ribeiro_anchors_2018} that involve features. These anchors are generated by adding features to the rules iteratively using beam search. At each iteration, the candidate with the highest estimated precision is kept and used as the seed for the next iteration.
Furthermore, we can also use feature importance to provide local explanations.
For example, SHAP \cite{lundberg_unified_2017} proposes a unified framework for computing the importance of each feature in a sample using Shapley values \cite{shapley_17_1953}.
Since the exact computation of SHAP values is computationally expensive, SHAP also proposes several approximation methods to accelerate the estimation.
Gradient information can also be applied \cite{selvaraju_grad-cam_2017} in differentiable models such as deep neural networks to illustrate the importance of input features to the model's prediction.
As shown in Figure \ref{fig:LocalToGlobalExplanation}, local explanations such as LIME and SHAP can also serve as global explanations using aggregations.
Such global explanations can be attained by aggregating explanations for all samples to form explanations at the dataset level.
For example, by computing the average importance of each feature across the whole dataset, we can identify the important features that make great contributions to model prediction for most samples in the dataset.
\figureLocalToGlobalExplanation

\subsection{Explainable AI for Quant}
\label{sec:xai_scenario}
\figureXAILoops
In this part, we take stock alpha investment as an example.
The input of deep learning models has three directions: stock, time, and factors (inner circle in Figure \ref{fig:XAILoops}).
Based on these directions, various tasks of interest can be formed to provide practical insights (outer circle in Figure \ref{fig:XAILoops}).
These tasks can be further instantiated as specific examples to provide explanations for realistic problems in investment. 

\figureXAIStock
\subsubsection{Explanation on Stock}
\label{sec:xai_scenario_stock}
Explanations can be provided for individual stocks to illustrate their sensitivity to different factors at different times and their relationships with each other.
Some tasks for explanations on stock are provided as follows:
\paragraph{Stock similarity}
Stocks are ubiquitously correlated with each other from many aspects (as shown in Figure \ref{fig:xai_stock_similarity}), and correlated stocks are expected to share common properties.
In this sense, utilizing the relationships between financial instruments can bring advantages to our analysis and prediction over traditional methods that treat stocks individually.
We can also better understand what the model has learned by analyzing the similarities between stock embeddings.
However, the challenge lies in determining an appropriate similarity metric, which is expected to have enough flexibility and effectiveness.
This problem is related to metric learning \cite{kulis_metric_2013, kaya_deep_2019} and graph structure learning \cite{zhu_deep_2021}.
A good similarity metric between stock embeddings is needed.
Based on this metric, a graph structure can be constructed by computing an adjusted adjacency matrix based on pairwise similarity.

\paragraph{Lead-lag effect}
In a lead-lag effect \cite{hou_industry_2003}, the trend of a stock is followed by some other stocks with a lag in time.
Following lead-lag effects, investors can observe the trends (i.e. price going up/down) of the leading stocks and take corresponding positions on the lagging stocks before the same trends duplicate.
In this way, investors can profit by precisely identifying lead-lag effects on the market \cite{li_detecting_2022, fan_does_2022}.
However, it is not a trivial job to identify lead-lag effects, since duplicated trends appear frequently in financial markets but only few of them are actually caused by lead-lag effects.
Strict identification of lead-lag effects needs to be conducted via causal inference that requires counterfactual explanations \cite{bottou_counterfactual_2013}: What will the trends of the lagging stocks be if the lead stock didn't go in this way?
Nevertheless, counterfactual reasoning is usually infeasible in real-world financial markets.

\paragraph{Sector trends}
Sectors are stock categorizations defined according to certain criteria such as industry and market capitalization.
Stocks in the same sector share certain common properties, and the trend of individual stocks can be influenced by their sectors.
Therefore, it is important to identify sectors' contributions to individual stocks.
To do this, we can treat stocks' membership to different sectors as categorical features and compute the importance of these factors via feature importance algorithms.
Besides, investors can also gain insight into the sensitivity of sectors to different types of features by evaluating feature interactions between sector memberships and other ordinary features.


\figureXAITime

\subsubsection{Explanation on Time}
\label{sec:xai_scenario_time}
Explanation can be computed on individual time points to illustrate the situation of stocks and factors at that cross-section, and explanations across multiple cross-sections can be further combined to provide insights for market features in a time interval.

\paragraph{Extreme market}
In stock markets, there are extreme conditions where nearly all stocks on the market experience severe price drops (Figure \ref{fig:xai_time_extreme-market}).
In extreme markets, it is hard for quant strategies to obtain excess return since the prices of all stocks drop together, and there is little room for arbitrage.
Therefore, in extreme markets, it is important to identify the less affected stocks and trade them to earn excess returns.
To achieve this, we can decompose stock returns from two aspects: those contributed by market trends and those contributed by stock-specific features
The decomposition can be computed by categorizing factors into market factors and stock-specific factors.
Then, the importance of these two types of factors can be computed via feature importance algorithms.
And we need to select the stocks where the importance of stock-specific factors outweighs that of market factors.

\paragraph{Calendar effect}
Calendar effects \cite{sakalauskas_research_2009} refer to market anomalies that are related to calendars, such as days in a week, months in a year, and event-related periods such as the U.S. presidential cycle.
Calendar effects are caused by market participants' anticipations toward future trends and have a great influence on market trends.
Thus, it is important in quant to identify calendar effects and make use of them to adjust investment strategies.
Such identification can also be achieved via feature importance algorithms.
By computing the importance of calendar factors, such as categorical features representing weekdays and days in the month, we can see whether model predictions heavily rely on these features. Stronger importance on calendar factors usually indicates potential calendar effects.

\paragraph{Style transition}
Style factors are used in multiple factor models (as introduced in \S\ref{sec:introduction_history}) such as BARRA \cite{barra_handbook} to describe the intrinsic features of stocks such as size, volatility, growth, etc.
In such models, the returns of stocks are contributed by their exposures to these style factors, and the return contribution per unit exposure, also called factor return, differs across style factors.
Moreover, the return of each factor also changes over time (Figure \ref{fig:xai_time_style}) because of the transitions in the market's preference for different styles.
If such transitions can be accurately recognized, investors can adjust their strategies accordingly to focus on stocks with large exposures to the dominating style factors.
To detect style transitions, we can regard style exposures as factors and compute their contribution to stock returns using feature importance algorithms.
We can then observe the distribution of factor contributions across time and detect shifts in this distribution as signals for style transitions.

\paragraph{Event influence}
Breaking events (Figure \ref{fig:xai_time_event}) usually have a great influence on stock markets.
Investors need to have a good understanding of the influences of breaking events to reduce the negative impacts or profit from the events.
Usually, an event is associated with two pieces of information: the timestamp of its occurrence and the specific contents, which are usually represented in natural languages.
Event contents can be encoded as specific factors using NLP techniques \cite{hu_listening_2019}, and the effect of an event can be computed as the importance of the content factors concerning the market trends after its timestamp.
Besides, we can also compute causal explanations to show the causal effect of the event.

\figureXAIFactor
\subsubsection{Explanation on Factors}
\label{sec:xai_scenario_factor}
Explanations can be computed on each factor to illustrate the sensitivity of different stocks to it at different times.
The explanations can be combined to show the interactive effects among factors for specific stocks.

\paragraph{Factor type}
Factors can be categorized from various aspects.
For example, in terms of data sources, factors can be categorized as volume-price factors, sentiment factors, fundamental factors, etc.
In terms of financial features, factors can be categorized as momentum factors, mean-reversion factors, lead-lag factors, etc.
In terms of the time scales, factors can be categorized as tick-level factors, minute-level factors, day-level factors, etc.
The contribution of different types of factors to portfolio returns can be computed by feature importance algorithms, and it provides investors with a better understanding of the investment strategy generated by AI, 
For example, in Figure \ref{fig:xai_factor_single}, the contribution to the model prediction of each factor at different positions in a time window is illustrated as a heatmap.

\paragraph{Factor interaction}
Deep learning model is good at capturing the complicated associations between factors, and some weak factors can be combined to form strong factors.
Such interactions reflect intriguing patterns among factors and provide new insights into finding new factors. 
Factor interactions can be revealed using feature crossing techniques.
For example, in Figure \ref{fig:xai_factor_interaction}, the landscape of model prediction with respect to the value of two factors (exposure to size and momentum style factors) is illustrated in a contour map.
And it can be seen from the map that model prediction decreases as both factor values drop.

\paragraph{Factor hierarchy}
We can depict the semantic similarity among factors in a hierarchical way.
Leveraging relevant techniques such as hierarchical clustering \cite{mullner_modern_2011}, a factor evolution graph demonstrates factor relations by arranging factors with higher similarities in lower-order neighborhoods (lower-level subtrees in the example of Figure \ref{fig:xai_factor_evolution}). 

\section{Knowledge-driven AI for Quant 4.0}
\label{sec:knowledge}
\figureFinKG
As discussed in \S\ref{sec:introduction_quant4.0}, knowledge-driven AI is an important complementary technology to data-driven AI, especially in low-frequency investment scenarios such as value investing and global macro investment.
In this section we attempt to answer two technical questions: 1) how to build a practical knowledge-driven AI system, and 2) how to apply knowledge-driven AI to quant research.

The first question is about the components of a typical knowledge-driven AI system \cite{wikipedia_knowledge-based_2022}. Generally speaking, such a system consists of two parts: a knowledge base which stores all knowledge we need in analysis, reasoning and decision, and a knowledge reasoning engine which analyzes and makes decisions based on the knowledge \cite{hayes-roth_building_1983}.
These two parts correspond to the problems of knowledge representation and knowledge reasoning \cite{cheng_knowledge_2020}\cite{shinavier_panel_2019}, respectively, both are hot research directions in artificial intelligence.
In practice, knowledge graph techniques \cite{singhal_introducing_2012} are widely used to build a knowledge base due to their simplicity and scalability. The second question considers how knowledge graph techniques can be applied to quant from the perspective of knowledge representation and reasoning. We introduce how financial knowledge is extracted from various sources and incorporated into a financial knowledge graph. We will also introduce some potential application scenarios where knowledge graph reasoning brings extra benefits for investment.

Figure \ref{fig:FinKG} illustrates what a financial behavior knowledge graph looks like with an example extracted from a large knowledge graph. In this example, various entities including public companies, securities, sectors, individuals, events, as well as relationships such as supply chain, capital chain, and behavioral relations are characterized in the financial behavior knowledge graph. Information constituting the knowledge graph is collected from various data such as news reports, litigation documents, financial statements, research reports, and sectors, and extracted using NLP techniques such as natural language understanding and information extraction. Knowledge reasoning techniques can be applied to this graph for further analysis and decision. For example, negative new events or lawsuit outcomes may substantially affect the stock price of the company under study (\textit{Pharmaceutical A}). Moreover, when the potential risk of this company is discovered from the knowledge graph, it may propagate to related entities such as important shareholders of this company and those companies on the same supply chain. 

In this section, we introduce knowledge representation and reasoning techniques in \S\ref{sec:knowledge_repr} and \S\ref{sec:knowledge_reason}, respectively, and introduce the application in quant in \S\ref{sec:knowledge_app}.

\subsection{Knowledge Representation}
\label{sec:knowledge_repr}
The goal of knowledge representation is to encode human knowledge into machine-readable forms. It is the foundation of knowledge-driven AI and it determines the modeling method for downstream knowledge reasoning tasks.

\subsubsection{Knowledge Base Techniques}
\label{sec:knowledge_repr_history}
\figureKBHistory
In the beginning, we briefly review the history of knowledge base techniques (Figure \ref{fig:kb_history}) \cite{wikipedia_knowledge_2022}. Early knowledge base concepts started from the semantic network which originates from ancient philosophers centuries ago \cite{sowa_semantic_1992} and it was first implemented by computer scientists in 1956 \cite{lehmann_semantic_1992}.
A semantic network is a special graph where the nodes represent concepts of interest and the edges represent semantic relationships between these concepts.
These semantic graphs can be formalized equivalently as a set of semantic triples \cite{noauthor_semantic_2022} (a base of the popular series of technical standards such as RDF \cite{noauthor_resource_nodate}), and it is moreover applied widely in the later knowledge graph techniques.
Construction of semantic networks uses many NLP techniques such as semantic parsing \cite{kamath_survey_2019}, which use entity recognition and relation extraction to parse text data and build the semantic network.

Later on, logic-based knowledge representation techniques such as propositional logic \cite{noauthor_propositional_nodate}, descriptive logic \cite{mann_description_2003} and first-order logic \cite{wikipedia_first-order_2022} were introduced, and they quickly became the mainstream techniques for decades.
As an early logic-based knowledge system, \textit{general problem solver} was proposed in 1959 \cite{newell_report_1959}. It formalizes problems of interest as Horn clauses \cite{horn_sentences_1951} and applies logic-based methods to solve them. 
The application of general problem solver is limited because most real-world problems are too complex, and this shortcoming leads to the development of expert systems \cite{jackson_introduction_1999} which restrict the knowledge to domain-specific tasks only. 
In addition to expert systems, frame-based languages \cite{minsky_framework_1975} is an alternative technique for logic-based knowledge representation. Frame-based languages characterize real-world objects with concepts such as classes and inheritance relationships, similar to the abstraction in object-oriented programming. 
Successful examples of frame-based expert systems include the Knowledge Engineering Environment \cite{martin_knowledge_1988}, KL-ONE framing language \cite{woods_kl-one_1992} and the Cyc system \cite{lenat_building_1989}.

Traditional knowledge base techniques faced new challenges in the big data era, where data volume and data exchange exploded. The requirement of new protocols results in the emergence of various knowledge standards such as resource description framework (RDF) \cite{noauthor_rdf_nodate} and web ontology language (OWL) \cite{noauthor_owl_nodate}, and these new techniques are summarized and redefined as a new concept Semantic Web \cite{berners-lee_semantic_2001}. 
Semantic web techniques are further extended to satisfy large-scale applications in practice and it leads to the development of knowledge graph techniques which is the mainstream methods for building knowledge-based systems in data-driven scenarios.



\subsubsection{Knowledge Graph Techniques}
\label{sec:knowledge_repr_kg}
We introduce the structure of a knowledge graph and how to build it in practice. 

\paragraph{Structure of Knowledge Graph} 
A knowledge graph typically consists of two parts \cite{noauthor_what_nodate}: ontology and instances. 

\begin{tightitemize}
    \item \underline{Ontology} is the schema of a knowledge graph, which specifies the types and semantic meanings of the entities and relationships \cite{kendall_ontology_2019}.
    In practice, an ontology is usually encoded with formal languages such as OWL and RDF.
    For example, in the knowledge graph shown in Figure \ref{fig:FinKG}, the ontology specifies the eight types of financial entities (industry, individual, status, etc.) and the various types of relationships between them (supplies, shareholding, subsidiary, etc.).
    In addition to those regular elements, ontology can further define other attributes for entities and relationships such as timestamps, hyperlinks, etc.

    \item \underline{Instances} are the main body of a knowledge graph.
    They can be expressed as semantic triples, which consist of subjects, predicates, and objects.
    For example, in Figure \ref{fig:FinKG}, the semantic triple (\textit{Pharmaceutical A}, \textit{supplies}, \textit{Medicine C}) indicates the fact that \textit{Pharmaceutical A} is a supplier of \textit{Medicine C}.
    In a knowledge graph, the subjects and objects are entities and the predicates are relations.
    In practice, semantic triples are usually implemented in RDF and stored in graph databases.
\end{tightitemize}

\paragraph{Knowledge Acquisition}
In practice, knowledge graphs are usually very large, with millions or even billions of entities.
Therefore, knowledge acquisition techniques are required to automatically construct the knowledge graph.
As summarized in \cite{ji_survey_2022}, knowledge acquisition can be performed via knowledge extraction and knowledge graph completion.
\ItemizeGlobal{
    \item \underline{Knowledge Graph Completion} aims to fill up the missing links in a knowledge graph based on existing information, and this problem has been extensively studied.
    For example, \cite{galarraga_amie_2013} proposes a rule learning model that extracts symbolic rules from knowledge bases with large-scale data. The rules can be further used to infer missing facts between entities.
    In \cite{neelakantan_compositional_2015}, an RNN model is proposed to compose paths in the knowledge graph into embedding vectors. The composed embedding is then used to infer the missing relationship between the starting and ending entities of the path.
    \item \underline{Building from scratch} extracts structural information from raw data.
    Most works on knowledge graph construction focus on the task of information extraction from text data \cite{yangarber_automatic_2000, welty_towards_2006, muhammad_open_2020}.
    This task consists of two major steps: entity recognition \cite{yadav_survey_2018} and relationship extraction \cite{pawar_relation_2017}).
    Entity recognition involves semantic role labeling \cite{jurafsky_semantic_2021} which identifies entity roles (e.g., subject, object, etc.) in semantic triples, and entity disambiguation \cite{shen_entity_2015} which aligns text tokens to entity names in the knowledge graph.
    Relationship extraction can be formalized as a prediction problem where the inputs are entity pairs of interest and the output is the relationships between entities.
    Various machine learning methods such as convolutional neural networks \cite{zeng_relation_2014}, attention mechanism \cite{shen_attention-based_2016}, and graph neural network \cite{zhang_graph_2018} can be applied to this problem.
    In addition, some works \cite{miwa_modeling_2014, miwa_end--end_2016, katiyar_going_2017}    unify the entity recognition task and relationship extraction task in one end-to-end learning framework, where the inputs are token sequences (e.g., sentences) and the outputs are semantic triples.  
    
}

\subsection{Knowledge Reasoning}
\label{sec:knowledge_reason}
Knowledge reasoning refers to the process of analysis, inferences, proofs and decisions (e.g., inferring new facts, generating new conclusions, extracting new rules, etc.) based on existing knowledge and data. 
Reasoning can be conducted in various ways, including symbolic logic methods, neural methods, and neuro-symbolic methods.

\subsubsection{Symbolic Reasoning}
\label{sec:knowledge_reason_symbolic}
Symbolic reasoning can be conducted in either a deterministic way or a probabilistic way. Deterministic reasoning is performed by applying inference rules on given facts recursively until reaching the desired conclusion.
Meanwhile, probabilistic methods are also applied in practice for more flexible reasoning.
Probabilistic symbolic reasoning methods model the distribution of the existence of fact triples given the knowledge on the graph.
Therefore, logic rules are represented in a softer way, which allows more flexibility in the reasoning process.
Various methods have been proposed for probabilistic reasoning.
For example, probabilistic logic programming \cite{ng_probabilistic_1992} models a logic program as a computation graph similar to Bayesian networks.
Markov logic programming \cite{richardson_markov_2006} involves a first-order knowledge base where each rule is assigned as scalar weight. In this way, the joint distribution of all the observed and hidden facts is modeled as the normalized exponential weighted sum of the grounds of each rule.
There are also other probabilistic reasoning techniques such as stochastic logic programming \cite{cussens_parameter_2001} and TensorLog \cite{cohen_tensorlog_2016}.

Symbolic reasoning has the advantage of transparency and explainability compared with neural methods.
Specifically, it models knowledge in an explicit way by expressing facts and rules with logic chains or formulas.
However, symbolic reasoning depends on curated knowledge bases that require enormous human effort in construction and maintenance.
Moreover, it also suffers from high computation costs since most symbolic reasoning algorithms have to search in a high-dimensional search space.
This fact usually limits their application in big data.

\subsubsection{Neural Reasoning}
\label{sec:knowledge_reason_neural}
Neural reasoning methods learn decision rules with deep neural networks and can represent non-linear associations.
The information in entities and relations is embedded into neural networks.
Neural methods utilize gradient descent search in the training process, and achieve better efficiency, especially in big data scenarios.
Besides, neural methods also achieve better reasoning performance due to stronger expressiveness.



Neural methods train a deep learning model using knowledge graph structure described by semantic triples as well as the attributes of entities and relationships. During the inference process, the trained model predicts the fact of the input triple of interest.  
The following question is how to encode the input including both structures and attributes. 
Different embedding spaces can be used to represent entities and relationships. Many works represent entities as embedding vectors in Euclidean space but they differ in the modeling of relationships.
\cite{bordes_translating_2013} represents relationships as points in Euclidean space that translate subject embedding towards object embedding.
\cite{lin_learning_2015} encodes relationships as projection matrices that map entity embeddings to low-dimensional space.
\cite{socher_reasoning_2013} defines relationships as 3-D tensors that represent the bilinear similarities between entities from multiple dimensions. 
In addition to Euclidean space, other types of embedding space can also be used, including complex vector space \cite{trouillon_complex_2016, sun_rotate_2019}, Gaussian distribution \cite{he_learning_2015, xiao_transg_2016}, and manifolds \cite{xiao_one_2016}.
The possibility of semantic triples can be computed directly from the representations of the corresponding entities and relationships.
For example, in \cite{bordes_translating_2013}, the possibility is computed as the distance between object embedding and subject embedding through relationship embedding.
The computation of possibility can be extended to capture more complex interactions between entities and relationships using neural networks.
For example, \cite{dettmers_convolutional_2018} concatenates subject embedding, object embedding and relationship embedding and feeds them to a convolutional neural network.
Moreover, neural networks can also leverage the structural information of the knowledge graph.
For example, in \cite{guo_learning_2019}, a recurrent neural network encodes paths on the graph (similar to logic chains) involving multiple semantic triples.
In \cite{schlichtkrull_modeling_2018}, a graph neural network leverages the structural information of the knowledge graph and computes graph convolution to improve reasoning.

\subsubsection{Neurosymbolic Reasoning}
\label{sec:knowledge_reason_neurosymbolic}
Symbolic reasoning and neural reasoning have different advantages, which can be combined in a neurosymbolic reasoning framework.
Neurosymbolic reasoning can be conducted by either injecting logic structures into the embedding framework, or vice versa \cite{zhang_knowledge_2022}.

For the former idea, \cite{lin_modeling_2015} incorporates conjunction rules into the computation of relation embedding in multi-hop paths. Following the framework of TransE \cite{bordes_translating_2013}, the relationships of semantic triples on a reasoning path are composited as one single embedding vector, which is used to translate subject embedding towards object embedding. 
\cite{ding_improving_2018} put constraints on the representation learning process to enhance the prediction confidence of the conclusions in entailments.
Besides, other works also propose various ways to inject logical structures \cite{guo_jointly_2016, guo_knowledge_2018} or ontological schemas \cite{wang_tagat_2021, zhang_knowledge_2018, chang_typed_2014} into the knowledge graph embedding framework.

For the latter idea, \cite{qu_probabilistic_2019} proposes to infer the missing facts using neural networks and then reason over the queries with Markov logic networks. The whole model is optimized via the expectation-maximization algorithm \cite{dempster_maximum_1977}, where the E-step corresponds to inferring hidden facts, and the M-step corresponds to maximizing the likelihood of the given facts.
\cite{wei_large-scale_2015} uses knowledge graph embedding techniques to help shrink the size of candidate sets for fact inference in large-scale knowledge bases. Then, inference via ground network sampling is performed on a Markov logic network to compute the final results.
Moreover, there are also other works utilizing knowledge graph embedding for logic reasoning \cite{rocktaschel_end--end_2017} and rule learning \cite{wang_learning_2016, yang_differentiable_2017}.

\figureKGReasoningSIGIR
\subsection{Application in Quant}
\label{sec:knowledge_app}
In this part, we will exhibit how knowledge-driven AI is applied in Quant 4.0 from two aspects: construction of financial behavioral knowledge graph and knowledge graph reasoning for quant.

\subsubsection{Building a Financial Knowledge Graph}
\label{sec:knowledge_app_build}
\paragraph{Ontology Design}
The ontology of a financial behavioral knowledge graph should cover information from the following aspects: 1) the fundamental information of financial entities, 2) financial events happening between financial entities, and 3) the causal relationships between entities and events.
Correspondingly, categories for entities include but are not limited to:
\begin{itemize}[leftmargin=*]
    \item \underline{Financial entities}, including stocks, bonds, banks, public companies, important individuals, commodities, etc. 
    \item \underline{Concepts}, reflecting the fundamental information about financial entities, such as sectors, industries, exchanges, regions and countries, currencies, etc.
    \item \underline{Events}, which are economic behaviors such as administrative punishment, illegal actions, litigation states, shareholding changes, personnel changes, etc. 
\end{itemize}

Similarly, categories for relationships include but are not limited to:
\begin{itemize}[leftmargin=*]
    \item \underline{Relationship between entities}, such as subsidiaries, belongs, shareholding, etc. These relationships are thus associated with timestamps indicating their beginning and ending times.
    For example, in Figure \ref{fig:FinKG}, the industrial chain and capital chain relationships describe the sector categorization and capital relationships between entities.
    \item \underline{Relationship between events}, such as co-occurrence, lead-lag. 
    \item \underline{Relationship between events and entities} between events. For example, in Figure \ref{fig:FinKG} the negative reporting leads to the price change and further investigation on ``Pharmaceutical A'', leading to a suspension in trading and litigation. Causal relations are usually inferred from existing knowledge and serve as important auxiliary information in downstream reasoning tasks.
    For example, in Figure \ref{fig:FinKG}, a ``related to'' relation connects the event of ``negative reporting'' and the corresponding financial entity ``Pharmaceutical A'' to indicate a negative reporting happening to the publicly traded company ``Pharmaceutical A''. These relations are usually associated with timestamps indicating the specific time point that the events happen.
\end{itemize}
\figureFinKGWWW
\paragraph{Knowledge Acquisition}
Knowledge that constitutes the financial behavioral knowledge graph can be acquired from various sources, and the most challenging part is to extract useful structural knowledge from unstructured data (text data as a representative example).
Natural language expression has high flexibility and substantial personality, and thus it is a challenge for machines to accurately understand the information in documents and extract the most useful knowledge to build a knowledge graph. 
There is a lot of contradictory information in news and documents, leading to the difficulty in fact extraction. Therefore, probabilistic models and machine learning models play important roles in knowledge extraction with confidence evaluation. 
Moreover, the information is usually incomplete and extrapolation is needed to infer the knowledge that we are actually interested in. Graph completion is thus required to infer the missing knowledge from the given facts.
Fourthly, different data sources share inconsistent update frequencies, which brings new challenges for an appropriate representation of the knowledge graph.
Both snapshot-based representations and growing flat knowledge graphs can capture temporal information.
A growing flat graph is friendly for temporal updates, but it is hard to perform temporal analysis directly on it.
On the contrary, snapshot-based representations are naturally suitable for temporal analysis, while also bringing extra costs in storage and management.


\subsubsection{Knowledge Reasoning for Quant}
\label{sec:knowledge_app_reason}
Given a knowledge graph, we can get meaningful representations of knowledge by reasoning on the graph.
For quant, the knowledge representations can be incorporated into existing factors as external information and fed to deep learning models for better predictions.
Figure \ref{fig:kg_reasoning_sigir} demonstrates a typical pipeline of knowledge reasoning in quant.
Specifically, events and relations between stocks are represented as semantic triples, where the entities and relationships are embedded into vectors.
Then, neural reasoning is performed on these semantic triples to compute the embedding of events, relations and the whole knowledge graph.
After training on historical data, the knowledge representations are used in investment strategies to generate trading decisions.

There are also other works studying the application of knowledge reasoning for quant.
\cite{ding_knowledge-driven_2016} proposes incorporating relational and categorical knowledge for better event embeddings. Given a semantic triple representing an event, external information about the entities in the semantic triple is retrieved from a knowledge graph and involved in the computation of event embedding.
\cite{deng_knowledge-driven_2019} extracts events from news texts and uses entity linking techniques \cite{sil_re-ranking_2013} to align the extracted information with the knowledge graph. Then event embeddings are generated using TransE. The embeddings are then combined with volume-price data in a temporal convolutional network \cite{bai_empirical_2018} for stock prediction, as shown in Figure \ref{fig:FinKGWWW}.
\cite{feng_temporal_2019} leverages fundamental information such as sector categorizations and supply chains to build a knowledge graph and uses temporal graph convolution to compute the embedding of each stock. The embeddings are then used to predict the stock returns, and the whole model is trained by maximizing the stock ranking loss.
\cite{long_integrated_2020} uses node2vec \cite{grover_node2vec_2016} to generate stock embeddings based on a knowledge graph and uses these embeddings to compute the similarity between stocks. In this way, the top-K nearest neighbors are computed for each stock, and the factors from neighbors are used to enhance the original factors.
Other works \cite{ang_learning_2021, xu_rest_2021} also uses knowledge graph to generate better stock embeddings or perform event-driven investment.
\section{Building Quant 4.0: Engineering \& Architecture}
\label{sec:system}

Sections \ref{sec:autoai}, \ref{sec:xai} and \ref{sec:knowledge} introduce the three components of Quant 4.0 for the algorithmic perspective. In this section, we  ``retrospect'' Quant 4.0 from a system point of view and study how to put all these components together in one system. Figure \ref{fig:system_arch} illustrates the architecture of a proposed Quant 4.0 system framework, including the offline system for quant research and the online system for quant trading.

\figureSystemArch

\subsection{System for Offline Research}
\label{sec:system_offline}
Quant 4.0 offline quant research system aims to improve the efficiency of quant research. It contains several layers (hardware layer, raw data layer, meta factor layer, factor layer, and model layer) and modules (high-performance computing clusters, data system, cache system, data preprocessing, automated factor mining, knowledge-based system, large-scale data analytics, AutoML, and risk simulation). 


\subsubsection{Hardware Platform Architecture}
\label{sec:system_offline_hardware}
The underlying hardware platform for offline research is a high-performance computing cluster, consisting of many computing nodes that are mounted with shared storage \cite{weil_ceph_2006} and interconnected by high-bandwidth network \cite{infiniband_white_paper}. The combination of multiple nodes aggregates the computing power distributed in various single nodes to support large-scale quant computing tasks.
However, communication bottleneck usually serves as one of the constraints for scaling up computing power \cite{zhang_is_2020}. To address this problem, the network topology of the cluster adopts a hierarchical structure, where neighboring computing nodes are connected via lower-level switches to increase the overall throughput of the cluster.

\subsubsection{Design of Data System}
\label{sec:system_offline_raw-data}
The data layer for Quant 4.0 system is to collect a tremendous amount of financial data and provide data management and query service for applications in upper layers. Since financial data are heterogeneous and multimodal, different types of database systems are involved in this layer to manage different types of data. Examples include SQL database \cite{codd_relational_1970}, time-series database \cite{namiot_time_2015, jensen_time_2017}, NoSQL database \cite{gessert_nosql_2017} and graph database \cite{kumar_kaliyar_graph_2015}.

\begin{tightitemize}
    \item \underline{SQL database} stores and manages data following the relational model proposed in \cite{codd_relational_1970}, where data are stored in tables that represent relations among attributes. Most traditional financial data such as quote data and financial statements can be represented in this type and are suitable for SQL databases.     
    \item \underline{Graph database} \cite{angles_survey_2008} is designed to store and manage graph-structured data composed of nodes and edges. Such graph structures widely exist in financial data since financial entities are connected with each other through various relationships such as money flow and supply chain relations. Such links may indicate some latent patterns that are shared in neighborhoods. Graph database can be used to store and manage economic graphs, financial knowledge graphs and financial behavior graphs. 
    \item \underline{NoSQL database} \cite{gessert_nosql_2017} is used for storing non-tabular data such as key-value pair, document, and wide-column. It is appropriate to manage financial text data and image data, such as financial report text, news text, social media text, satellite and drone images etc. 
    \item \underline{Time series database} \cite{jensen_time_2017} is a special database type designed for quickly accessing, computing and managing time series data. Such kind of database engines is optimized to accelerate time series data processing, e.g., computing stock stream data collected in real-time, and computing time-series factors for high-frequency trading etc.  
\end{tightitemize}
Moreover, the large volume of financial data requires a highly efficient distributed storage system to accelerate data access. To further improve the read/write speed for high-frequency tick data (limit order book etc), in-memory database \cite{tan_-memory_2015} could be used as data cache to store the most frequently read data. It saves the data transfer time between hard disk and memory. 

In addition to layer-specific components, the data layer, meta factor layer, and factor layer share a large-scale data processing platform which provides a complete solution and a convenient model for various data-driven tasks.
The key components of this platform include a distributed file system that provides convenient and reliable data access on distributed storage systems, and a distributed computing engine that provides simple yet efficient programming interfaces for parallel computing on a large number of computing nodes.
Specifically, distributed file systems such as HDFS \cite{shvachko_hadoop_2010} adopts the architecture consisting of name nodes and data nodes, where data replication is performed on multiple data nodes for fault tolerance.
On the other hand, distributed computing engines such as MapReduce \cite{dean_mapreduce_2008} and its open-source implement including Hadoop \cite{apache_hadoop} and Spark \cite{zaharia_resilient_2012, zaharia_spark_2010} provide programming models for parallel computing tasks.
By abstracting parallel computing tasks into a set of primitive operations (e.g. map and reduce in MapReduce and transformation in Spark), the programming tool is ease to use by developers and is generic enough to handle many common parallel programming tasks.


\subsubsection{Factor Mining System}
\label{sec:system_offline_factor}
Raw data have different formats, but factor mining requires unified input format. Therefore, corresponding to the workflow in Figure \ref{fig:factor_mining}, the meta factor layer is involved to preprocess raw data with various modalities into meta factors with unified formats and appropriate values. 

The factor layer builds automated factor mining engine and automated factor mining pipeline. The factor mining algorithms have been introduced in \S\ref{sec:autoai_feature_symbolic}. Here we introduce how to implement factor mining at scale from a system point of view. In particular, we concern how to improve the system efficiency to discover more ``good'' factors per unit time. 
\begin{tightitemize}
    \item Factor mining system needs a parallelization architecture to improve computational efficiency. 
    \item Syntactic validity of factors should be checked in real time during the factor generation process to reduce the CPU time wasted by invalid factors. 
    \item Diversity of factors should be controlled in real time during factor generation process in order to reduce the CPU time consumption for redundant factors.  
\end{tightitemize}
The whole system is backed by some key techniques in distributed execution and computing acceleration.
Distributed execution tools such as message queues and distributed caches enable asynchronous parallel execution on multiple nodes, thus guaranteeing system scalability.
Computing acceleration techniques such as vector and streaming SIMD instructions on CPU and massive parallel computing on general-purpose graphics processing unit (GPGPU) \cite{hennessy_computer_2011} significantly improve the computation efficiency for data frame operations and thus increasing the productivity of the whole system.

Meta factor layer, factor layer, and model layer are connected to the factor base, which is an integrated platform for the storage, computation, dependency management, backtest, tracking, and analysis of all factors (the output of models are also factors). Factors generated in these layers are committed to the factor base in various forms. Specifically, meta factors pre-processed from raw data are directly committed to the factor base in data frames, with appropriate descriptions about their data sources and pre-processing methods. Factors generated via automated factor mining are committed to the factor based in the form of symbolic expressions where the operands are other factors in the factor base. The model outputs generated in the model layer can also be regarded as speical machine learning factors, which can be committed to the factor base together with specifications of input factors, model architectures, hyperparameters and training descriptions, etc.

\subsubsection{Knowledge-based System}
\label{sec:system_offline_knowledge}
In parallel with the factor mining system, a knowledge-based system is also involved in the overall architecture to provide knowledge-driven AI practice.
As discussed in \S\ref{sec:knowledge}, the knowledge-based system consists of two modules: a knowledge base for knowledge representation and an inference engine for knowledge reasoning. Specifically, a distributed graph computing platform is used as the knowledge base to store large financial behavior graph and the inference engine is built upon for downstream financial knowledge reasoning and decision making.

Compared with traditional small knowledge bases, financial behavior knowledge graph in Quant 4.0 has the potential to grow up to a scale with billions of nodes and edges, and thus it requires a flexible and scalable architecture \cite{lu_large-scale_2014, xiao_tux_2017} to store and manage large-scale dynamic graph data. Therefore, we need a distributed graph computing and management platform, which partitions the whole knowledge graph into a number of subgraphs distributed in different nodes of a computing cluster. Due to the sparse nature of knowledge graph, this partition is specifically arranged to maximize data locality, where graph nodes located in the same partition are more densely connected than those located in different partitions \cite{yang_edges_2021}.
For the inference engine, both knowledge graph embedding and rule-based symbolic reasoning algorithms can be implemented on this platform. 

\subsubsection{Modeling System}
\label{sec:system_offline_model}
The model layer is in charge of automatic generation of machine learning models and the corresponding risk evaluation and backtest simulation procedures before they are deployed into real-world environments. Therefore, this layer involves two major components: an AutoML module and a pre-trade risk analysis and simulation module.

The AutoML module implements automated model generation algorithms upon large-scale distributed deep learning systems.
The bottom layer of its technology stack consists of parallel computing platforms such as CUDA \cite{noauthor_cuda_nodate}, and communication backends such as message passing interface (MPI) \cite{noauthor_mpi_1993} that provide standard interfaces for communications between computing nodes in a distributed system.
In the second layer of the technology stack, deep learning frameworks such as PyTorch \cite{paszke_pytorch_2019} provide basic interfaces for training and inference of deep neural networks via hardware-accelerated linear algebra operations and automatic differentiation engines.
In addition, computing orchestration systems such as pathways \cite{barham_pathways_2022} combine low-level communication primitives with deep learning framework functionalities to implement higher-level parallelisms such as model parallelism and pipeline parallelism, which is further wrapped as standard interfaces for upper-level programs.
The top layer of the technology stack consists of implementations of model generation algorithms such as neural architecture search and hyperparameter optimization (NAS+HPO) \cite{jin_auto-keras_2019}, mixture of experts (MoE) \cite{fedus_switch_2021} and large-scale pretraining \cite{shoeybi_megatron-lm_2020, rajbhandari_zero_2020}.

The risk and simulation module identifies and analyzes the potential risk exposures of the models before they are deployed to real-world trading environments.
This module implements explainable AI techniques applied to analyzing factor, model, and causality to reveal nonlinear risk exposures that are more complex than ordinary risk exposures explained by the BARRA model.
In addition, market simulator \cite{byrd_abides_2020} is used to test the performance of trading strategies with higher precision than backtest, whose results may be biased by historical data. Specifically, the simulation environment system \cite{byrd_abides_2020} can be built using multi-agent reinforcement learning \cite{amrouni_abides-gym_2021, karpe_multi-agent_2020} where agents imitate the behavior of various market participants in real world \cite{coletta_towards_2021}.

\subsection{System for Online Trading}
\label{sec:system_online}
The online trading system focuses on deploying investment strategies for real trading and executing post-trade analysis, and its major goal is to achieve low trading latency and high execution efficiency.
The trading system consists of three modules: deployment, trading and analysis, and we introduce their functions in the following part.

\subsubsection{Model Deployment}
\label{sec:system_online_deployment}
The deployment layer aims to implement the philosophy of technology ``one-click deployment''. It involves a computation scheduling module, an automated deployment module, and an optional hardware acceleration module. The computation scheduling module arranges a reasonable and efficient computation order for factors based on their intrinsic data dependencies which form a directed acyclic graph (DAG) over factors. In the computation scheduling system, each factor starts its computation if and only if all its preceding factors on the DAG have finished their computation. 
Computation scheduling is a tedious work and it should be automated by system due to the following reasons. 
\begin{tightitemize}
    \item We must synchronize offline factor dependencies with online factor dependencies and keep their consistency in real time. 
    \item The number of factors may grow up quickly. Imagine what happens when you accumulate millions of factors? It is a nightmare for any quant researcher to deploy so many factors by hand.
    \item Adding, deleting and updating factors is the daily job of factor maintenance, which relies on correctly managing factor dependencies. 
\end{tightitemize}
From the algorithmic point of view, the problem of computation scheduling can be regarded as topological sorting on the dependency DAG.
In practice, the scheduler is designed to coordinate system components and schedule their executions according to the computation order on DAG.
In common implementations \cite{noauthor_apache_2022} of such scheduling systems, asynchronous scheduling is adopted to improve the overall execution efficiency. In this way, the pending steps can start their execution immediately after the previous steps finish.

The automated deployment module aims to deploy deep learning models trained from offline research to online trading. It implements the algorithmic techniques discussed in \ref{sec:autoai_deployment}. 
In addition to deep learning compilers and model compression engines, this module also involves optimized hardware kernel libraries, which provide implementations of common data processing and modeling functions that are highly optimized based on hardware features.
Popular examples of such libraries include cuDNN \cite{chetlur_cudnn_2014} and MKL \cite{wang_intel_2014}.
Functions in kernel libraries typically include basic linear algebra subprograms (BLAS) \cite{van_de_geijn_blas_2011} that are standard interfaces for scientific computing, and some optimized operators for deep learning such as $3 \time 3$ convolution.
The implementations are specifically optimized based on domain-specific architectures \cite{jouppi_domain-specific_2018} on devices (e.g. Tensor Core \cite{markidis_nvidia_2018}) to maximize the potential of hardware in use. 



The hardware acceleration module aims to improve the computation efficiency of data processing and model inference using special hardware technology such as field-programmable gate array (FPGA) acceleration. Strategy components such as network protocol stack or machine learning models implemented on FPGA with customized logic can bypass redundant logics that are inevitable on generic hardware, thus achieving lower latency and winning an advantage for traders over other market participants.  However, to deploy strategies on specific hardware, it usually takes a huge amount of development effort to complete strategy migration with satisfactory speed. Therefore, high-level synthesis techniques \cite{rupnow_study_2011, coussy_introduction_2009, cong_fpga_2022} are developed to address this problem. They directly generates register transfer languages for high-level representations in which strategies are originally implemented.

\subsubsection{Trading Execution}
\label{sec:system_online_execution}
The execution layer converts trading decisions to actual orders that are executed in exchanges, and its goal is to reduce the trading latency as much as possible in order to capture the fleeting trading chances in the market. The latency can be decomposed into two parts: transmission latency is the delay of signal communication between trading servers and market servers (exchange server or brokerage server), and computation latency is the delay between quotes receiving and order sending. To reduce transmission latency, the trading system is usually deployed on servers that are colocated with market servers (such as racks and cabinets provided by brokers). To reduce computation latency, a trading system must be optimized in full strategy pipeline from data collection to order execution through various software and hardware acceleration techniques.

\subsubsection{Trading Analysis}
\label{sec:system_online_analysis}
The analysis layer monitors the execution of investment strategies and performs analysis for further adjustments. It involves a post-trade risk control module responsible for ordinary performance monitor and a post-hoc XAI module for explaining strategy behavior from the perspective of AI. The post-trade risk control module performs return and risk attribution, aiming to analyze strategy's performance and revealing intrinsic risk structure hidden by machine learning blackboxes. Furthermore, the post-hoc XAI module provides in-depth analysis and explanation of investment strategies. 
It provides thorough risk analysis by analyzing all strategy components in terms of importance and sensitivity, and visalizes risk results.

\subsection*{\textbf{Remarks:}}
Risk control is one of the core tasks of quantitative investment and is also the primary consideration in system design. In our proposed engineering framework, the idea of risk control runs throughout the whole architecture. Specifically, system-level risk control is reflected in hardware and raw data layers, where the reliability and stability of the underlying hardware and raw data are the top priority. Data-level risk control is reflected in meta factor and factor layers, where quality control and management of factors and knowledge are emphasized. Pre-trade risk control is reflected in the model layer, where model robustness and explainability are required. Real-time risk control is reflected in deployment and execution layers, where the system's trading behavior is confined within constrained areas to avoid unexpected situations. Post-trade risk control is reflected in the analysis layer, where detailed analysis is conducted to provide comprehensive and reasonable insights into strategy performance.

\section{Discussion on 10 Challenges in Quant Technology}\label{sec:open_problems}
We have summarized 10 challenges in the development of next-generation quant technology. These challenges range from computing and data infrastructure, investment modeling, risk modeling, market simulation and cognitive AI technology, and they provide new research directions for researchers interested in AI technology and quant. We believe some problems might be solved in the next couple of years with the rapid development of AI technology, while others may remain challenging for a long time. 
\figTenChallenge

\subsection{Exponentially Growing Demand of Computing Power}\label{subsec_ComputingPower}
\figureModelSizeGrowth
With the rapid development of GPU technology and parallel computing technology, AI models have been scaling up year by year. The number of parameters in deep learning models has grown exponentially from 94 million (ELMo in 2018 \cite{matthew-deep-2018}) to over 500 billion (Megatron in 2022 \cite{smith_using_2022}) (Figure \ref{fig:model_size_growth}). The rapid growth in model size not only improves models' performances, but also leads to a paradigm shift which is deeply affecting the technical roadmap of AI research. Accordingly, as an important domain going all-in on AI, Quant 4.0 heavily relies on ultra large-scale computing power and related engineering technology. 
\subsubsection{Quant 4.0 and Supercomputers}
Investment Research of Quant 4.0 requires super-computing infrastructure as a fundamental support for large-scale factor mining, large-scale modeling, back-test and evaluation. One important idea behind Quant 4.0 is to ``replace human power with computing power'' in quant research. We have to notice that Quant 4.0 has much more demand on computing power than what we could usually image ever before, because of the following reasons:
\begin{itemize}[leftmargin=*]
    \item \underline{Large-scale automated factor mining} requires large distributed computing clusters with either homogeneous CPU parallelization or heterogeneous CPU-GPU parallelization. Even on a relatively simple homogeneously structured computing cluster, a high-frequency alpha factor mining task needs millions of CPU cores in order to achieve enough efficiency, performance and diversity in intensive factor search and evaluation. In particular, as more and more factors are discovered and accumulated, it will take longer and longer computing time (or equivalently computing power) to find a new qualified factor which not only performs well but also weakly correlated with existing factors discovered before. The same computation power problem happens similarly for heterogeneous clusters.  
    \item \underline{Large-scale deep learning} is necessary in Quant 4.0 research as some billion-parameter-level or even trillion-parameter-level deep learning models with large data volume do exhibit superior performance in many AI scenarios \cite{bommasani_opportunities_2021}. More and more large models are applied to quant tasks such as market prediction, portfolio position computing, risk forecasting and real trading by hedge funds and other institutional investors. However, the training process of large deep learning model requires extremely high computing power. For example, given a scenario of cross-sectional alpha model on 4000+ stocks and hundreds of daily factors of 10 years' history, a Transformer-type deep neural network with about ten billion parameters needs about 1-5 days to finish a single training/validation cycle on a cluster with 100 Nvidia A-100 GPUs, without counting the additional computational cost due to rolling training, model ensemble and autoML.
    \item \underline{Rolling training}, i.e., iteratively retraining the same model with data sampled from a time window shifting towards the future, is a characteristic of quant modeling since the patterns of financial market and the patterns of investment instruments vary over time. In fact, financial time series are strongly non-stationary, and thus data features exhibit different distributions in different market styles and different time periods, and the performance of a model usually decays over time without retraining on recently incremented data. This phenomenon means a one-time model training on data like what people do in many other AI scenarios such as natural language process and image recognition doesn't work well in quant research. Therefore, rolling training is widely adopted in quant research for back-test simulation, paper trading and real trading. Imaging an experiment rolling training once a week, a ten years' back-test process retrain the model $10 \times 50 = 500$ times, significantly increasing the computational cost. 
    \item \underline{Model ensemble} \cite{zhou_ensemble_2012} is a common way to improve performance and robustness of a quant strategy and reduce the risk of portfolio asset loss. Moreover, ensemble of diversified or uncorrelated models could usually improve strategy performance and asset return. However, the computational cost increases linearly with the number of models combined in a strategy, and computer power becomes an important foundation for strategy stability and risk management.  
    \item \underline{NAS and HPO} are another two modules in quant pipeline heavily consuming computing power. Both of them are formalized as search-and-optimization problems in the network architecture space and the hyper-parameter space, respectively, and the computational cost is nonnegligible even if more and more fast algorithms are developed to improve the search efficiency. 
    \item \underline{Feature selection} (or feature importance computation) plays an important role in improving model performance and enhancing model explainability. Unfortunately, feature selection algorithms for deep learning are usually computationally intensive as well due to complicated feature interactive effects in the model, and they require sufficient computing power to identify significant factors within affordable time. 
\end{itemize}

\subsubsection{Solving Computing Power Dilemma}
Learning large model is very expensive. Figure \ref{fig:model_cost_compare} compares a number of deep learning models on code generation tasks in three aspects: parameter size, sample (token) size and estimated cost. We can see that GPT-3 model has about 170 billion parameters and costs about 2 million USD to complete the computation, which is obviously too expensive to be afforded for most financial institutions. Given the expensive and limited computing power, what are the solutions to this dilemma? We attempt to provide a few suggestions and research directions.

\figureModelCostCompare

\begin{itemize}[leftmargin=*]
    \item \underline{Research of faster algorithms} is the most direct and most significant way to reduce computational cost. However, it is a difficult problem to reduce the computational complexities of deep neural network algorithms, NAS/HPO algorithms and feature selection algorithms without losing their performance and it needs the joint effort of researchers from multiple relevant areas including but not limited to applied mathematics, theoretical computer science and machine learning. 
    \item \underline{Online learning \cite{hoi_online_2021} and incremental learning \cite{de_lange_continual_2022}} are machine learning techniques where data are acquired sequentially along time and the model is updated once new data are received without retraining the model on all data, as opposed to batch learning techniques such as traditional deep learning which train a model using the entire training data set at once. The computation time can be saved by model updating rather than model retraining.  
    \item \underline{Pretraining-finetuning paradigm} could help save overall computation time as well. Specifically, in the pretraining stage, we build a general but computationally intensive deep learning model and in the finetuning stage, we adapt it to various domains, scenarios and tasks in trading.
    After the computationally expensive pretraining process has been finished, the following finetuning tasks cost significantly less computation power and can be repeatedly performed.
    \item \underline{Model diversification} reduces model similarity and reduces information redundancy in model representation, thus decreasing the number of required models in the ensemble. New algorithms are required to improve the diversity of a new model compared with existing ones in model bases. In particular, we need new techniques to enforce the model diversity before or during the training process directly. 
    \item \underline{Utilizing informative low-frequency data} is a way to enhance computational efficiency without too much effort on engineering techniques or algorithmic tricks. Many types of low-frequency data such as various alternative data, fundamental data and event data are informative for investment decisions, complementing data with relatively high frequency including quote data and limit order book data.
    Due to the sparsity and small size nature of these low-frequency data, and moreover, easy to compute for modeling.  
    \item \underline{Sharing computing power and large model} is a potential business/operating model to avoid overlapping investment in computing and modeling infrastructure and to dilute the cost across the whole quant investment industry. Although cloud computing service is an option for computing power sharing, it is not designed specifically for quant investment. In particular, almost all architectures and software systems of cloud computing clusters are not optimized towards the computational demand of quant. Therefore, this industry needs its own professional quant computing sharing mechanism, platform and service. A similar sharing mechanism could be applied to large models as well. As a business model, financial institutes such as hedge funds and mutual funds could buy licenses for the usage of supercomputers and pretrained large models, just like buying financial data. 
\end{itemize}



\subsection{Alternative Data Technology}
In principle, any data about economic activities have the potential to be used for investment and might be considered by quant researchers. Alternative data, a concept opposite to conventional financial data such as financial statements, stock quotes and limit order books, provides a much broader space for quant research. Although the concept of alternative data has appeared for a decade and is getting more and more popular in quant industry, it is still in the early stages of industry application because of a few reasons. Firstly, many alternative data sets such as news and research reports have very narrow coverage and low breadth since a news event usually only relates to a limited number of stocks, and you couldn't expect a public company has news or significant events every day. Secondly, some alternative data such as satellite images are too expensive for small funds, and relevant satellite image detection and recognition techniques are not well-established due to limited labeled data and expensive labeling costs. Thirdly, processing and cleaning raw alternative data is a time-consuming ``dirty work'', and some informative alternative data can not be supplied legally and sustainably. In this subsection, we briefly introduce some examples of alternative data and the corresponding processing techniques, and discuss the difficulties in data acquisition and data aggregation. 

\subsubsection{Examples of Alternative Data}    
We list a few types of alternative data as examples and simply discuss the technical difficulties and opportunities for extracting informative signals from them. To avoid repetition, those alternative data already introduced in \S\ref{sec:knowledge}, such as supply chain data, are omitted here. 
\begin{tightitemize}
    \item \underline{Text data from news and social media} are one of the most popular alternative data used by investment institutions. The key is to extract traders' sentiment signals correlated with future market trends. Natural language processing techniques such as opinion mining, sentiment modeling and trend tracking could be used to build useful alpha signals or factors. 
    \item \underline{Satellite image data} have been used by hedge funds and other institutional investors for many purposes. For example, investors could use satellite images to estimate how busy the parking lots of retailers are, providing investment signals for longing or shorting the retailers' stocks. Currently most satellite images are read and analyzed by human, but we believe computer vision techniques such as image recognition and object detection have huge application potential in the future when main technical issues (e.g., cheap labeled data) are solved. 
    \item \underline{Merchandise sales data from e-commerce} track sale records of consumer products companies from their online channels and estimate their actual income before an upcoming financial statement is reported. This kind of data is usually fragmented and so correct data aggregation is important. In addition, we have to follow relevant data protection laws to make sure the data legitimacy.  
    \item \underline{Logistic and inventory data} track transportation and logistic activities of a company to estimate their sales performance which might be helpful for predicting the corresponding stocks. 
    \item \underline{Credit card or e-payment transaction data} is another angle to learn the operation and sales performance of a company or an industry, providing a possible fine-grained analysis through the information of micro-economic activities. Data legitimacy and data sensitivity are the main concerns before using them.  
    \item \underline{Geolocation data} collect foot-traffic information by GPS or cellphone locators. For example, an investor could use foot-track data for pair-trading arbitrage of Adidas and Nike by analyzing their geolocation activities in different retail shops and predicting spreads between the stock prices of these two companies.
\end{tightitemize}

\subsubsection{Problems in Data Acquisition}
Traditionally, hedge funds and other institutional investors acquire data from third-party data providers or data vendors. With the rapid accumulation of internet data, more and more hedge funds started data collection from websites using web crawlers. However, collecting data by web crawling is suffering unprecedented restrictions because more and more websites are attempting to protect their data using protocol securities and anti-crawler techniques. In addition, many informative data sets such as transaction data are distributed in various banks, e-payment institutes and credit card companies, and few of them allow users to take data out of their own servers considering intellectual property protection, data security and privacy protection. Here we provide a few possible ideas to collect and utilize more protected data legally. 
\begin{tightitemize}
    \item \underline{Certificate techniques for data asset ownership} are the foundation for data owners who are willing to share their data and obtain legal income from data users. Decentralized data management techniques such as blockchain techniques provide a potential solution for this mechanism design problem. 
    \item \underline{Data encryption techniques} \cite{acar_survey_2018}, including data storage encryption, data transfer encryption and data computation encryption, provide technical solutions for protecting the interest of data owners and data users (e.g., investors), and potentially encourage the willing of data exchange and data sharing. 
    \item \underline{Federated learning} \cite{zhang_survey_2021, li_survey_2021} is a machine learning technique for modeling data sources distributed across multiple decentralized servers each only holding a part of the whole data samples and it realizes information fusion without exchanging those local data. Federal learning techniques have the potential to help investors improve their quant models without spending too much time dealing with data security and privacy issues with massive data owners. 
\end{tightitemize}

\subsubsection{Problems in Data Aggregation}
Data aggregation is important for quant to find alpha signals across different types of data. However, data structures of alternative data are diverse and heterogeneous, resulting in difficulties in data aggregation, especially when integrating with traditional financial data. 
\begin{tightitemize}
    \item \underline{Heterogeneity in frequency} is a characteristic of many alternative data types such as news data and geolocation data, since they are collected irregularly.
    Therefore, it's important to align time-series signals occurring at unaligned time points across different instruments and different data types.
    A series of problems naturally arise. For example, how to build prediction models on these irregular and sparse data samples? How to better estimate and impute missing information in alternative data?
    All these data processing problems need to be further explored.
    For example, it is worthwhile to consider whether data embedding techniques learning ``good'' representations in lower-dimensional latent spaces help fuse heterogeneous data. 
    \item \underline{Differentiating signals from noise} is a difficult problem for the extraction of alpha signals from alternative data and should be carefully examined during data processing. Due to limited sample sizes and irregular sample timestamps, identifying true signals out of false positive patterns is more complicated than ever before. Therefore, it is demanding to develop new robustness techniques and significance test techniques to help evaluate the truth of signals and the significance of patterns.  
\end{tightitemize}

\subsection{Financial Knowledge Engineering}
As we have introduced in \S\ref{sec:knowledge}, knowledge-driven AI will play an important role in future quantitative finance. Researching new knowledge representation methods, building complete and reliable knowledge bases and developing new knowledge reasoning and decision algorithms are crucial problems in knowledge engineering for financial investment. 
\subsubsection{Difficulties in Knowledge Engineering}
Financial knowledge engineering is an engineering pipeline and research area for constructing effective AI knowledge systems covering knowledge acquisition, knowledge representation, knowledge management, knowledge reasoning, risk analysis and investment decision-making. Given the limitations of popular knowledge techniques, it is necessary for researchers to explore more efficient and more sophisticated methods to better represent and leverage different types of knowledge such as declarative knowledge \cite{burgin_theory_2017}, structured knowledge \cite{jonassen_structural_1993}, procedural knowledge \cite{pavese_knowledge_2022} etc. In particular, exploring and researching particular knowledge representation techniques for quant and other financial applications are valuable for contributing to the development of investment industry and for finding a broader investment field for quant. Before the wide application of knowledge engineering in quant, a number of technical difficulties need to be solved in the future. 
\begin{tightitemize}
    \item \underline{More sophisticated knowledge representation methods} are extremely demanding for building an effective financial knowledge engineering in the future. In particular, we need better data structure and model structure to encode knowledge relevant to all types of economic and financial theories and practical activities. 
    \item \underline{More advanced knowledge management system} is the requirement of financial knowledge engineering. It requires more research on how to build an automatic and self-updating pipeline of knowledge acquisition, knowledge update, knowledge aggregation and knowledge correction, and on how to implement a reliable and scalable knowledge management system.
    \item \underline{Next-generation knowledge reasoning algorithms} should be more explainable, providing more reliable analysis and prediction results. We encourage machine learning researchers to pay more attention to this area, which is part of next-generation AI core techniques. 
\end{tightitemize}

\subsubsection{Knowledge Engineering vs Large Model}
Large-scale pretraining models have been widely used in many AI fields especially in natural language processing and computer vision and in multi-modal pattern recognition tasks. For example, large model GPT-3 \cite{brown_language_2020} has about 175 billion parameters, and it could be imagined as a knowledge ``crystal ball'' for natural language generation and natural language reasoning. In this respect, large pretraining models play a similar role as knowledge engineering, and both of them could provide decision support and content generation function for downstream tasks. So whether large model is a good substitute for knowledge engineering? In our opinion, it is not true. Knowledge engineering has its particular advantage in explicit knowledge representation and knowledge logical reasoning, making the decision process transparent and understandable. Moreover, as the memory module in knowledge engineering, knowledge base has its advantage in flexibility, storing and managing all types of popular data structures including entities, facts and even rules. As a matter of fact, these two technology roadmaps are complementary to each other and have the potential to be integrated into a unified framework to improve the final performance of investment decisions. 

\subsection{Financial Metaverse \& World Model Simulator}
In quant research, it is very important to understand the underlying logic and micro-structure of financial markets. For example, we would like to know how the market will react to specific news about a financial statement fraud event, or how much a big order affects the asset price in the market. Unfortunately, empirical studies using historical data usually result in biased conclusions because they do not provide experimental access to all relevant information.
In particular, even if certain extreme market events have never happened in history, it doesn't mean they will not happen in the future.
What will happen if our trading strategy meets these assumed extreme events? Back-test experiments based on historical data can't answer this question, but we expect that financial metaverse can do it. Financial metaverse aims to build a simulated financial market parallel to real-world financial markets and use it as an experimental environment to simulate situations other than what have happened in real markets. 
\subsubsection{Financial Metaverse Market Simulator}
Daniel Freidman, UCSC Economics Professor, expressed his view that simulation of markets provides a powerful tool to analyze not only individual participant behavior, but also overall market reactions that emerge from the interaction of individual agents \cite{byrd_abides_2020}. Financial metaverse should be able to support various kinds of research experiments (about traders' behavior and market phenomenon) that are difficult to complete using historical data or trading experiments in real markets. Such a simulator could be used in a number of different scenarios. We list a number of examples as follows.
\begin{tightitemize}
    \item Accurately estimating the market impact of a big order at certain market conditions and certain time point
    \item Analysis of trader behaviors reacting to a particular market event
    \item Understand the impact of a certain type of traders in markets
    \item High-precision simulation of transaction cost
    \item Running treatment-control random experiments to test causal effect to answer ``what if'' questions against particular historical dates
\end{tightitemize}
Besides the above applications, financial metaverse can also be applied to accurately justify causal effects of economic factors. This motivation is achieved by designing and conducting randomized experiments in this simulated market. Moreover, financial metaverse can also provide a fundamental experimental platform for causality engineering discussed in \S\ref{cognitiveai_causalityengineering}. Although financial metaverse is of great value for quant research and financial market research, there are many technical difficulties we have to face. 
\begin{tightitemize}
    \item There is no way to collect fine-grained data identifiable to individual traders, with which market simulation will become much easier.
    \item The data we have are lack information about the motivation and intent of every trader. 
    \item A simulator can not enumerate and involve all possible factors affecting a market.
    \item Current computing power can not support high-precision simulation with a large number of trading agents in financial metaverse.
\end{tightitemize}

\subsubsection{World Model for Simulation}
A technical problem of financial metaverse is the complexity of the simulation environment from which agents in the same reinforcement learning system are hard to learn, and it makes traditional reinforcement learning algorithms hard to learn millions of weights of a large model. Usually a financial metaverse requires at least thousands of agents, each of which is modeled by a large neural network and plays a specific group of traders with some typical style, to participate in simulated market trading and it aims to recurrent the real-market by a reinforcement learning simulation as precise as possible. This makes regular computing power incapable of completing such accurate simulations. World model \cite{ha_worldmodel_2018} provides a potential solution for this difficulty in financial metaverse. It is inspired by the function of human brain which develops a mental model of the world by learning an abstract representation of complex information flushing into the brain.
In a world model, the dimension of outputs from the simulator (environment) is reduced using unsupervised learning such as variational autoencoder (VAE).
We can use this abstraction to train small network controllers which let the training algorithm search on a small space for credit assignment task, and thus the computation can be accelerated. Figure \ref{fig:world_model} illustrates the structure of a world model for computer vision problems. 
\figureWorldModel
The cognition abstraction idea in world model could be used in financial metaverse to solve the simulation computation problem for financial markets. 

\subsection{Cognitive AI \& Causality Engineering}\label{cognitiveai_causalityengineering}
Cognitive AI has been regarded as the future direction of artificial general intelligence (AGI) \cite{alattas_overview_2021} by many domain experts. In this subsection, we discuss its advantages and technical challenges in investment application, and we propose a new concept \textit{causality engineering} as a potential solution for causal machine learning in AGI. 
\subsubsection{Cognitive AI for Investment}
In 2011, Israeli-American psychologist and economist Daniel Kahneman published his book \textit{Thinking, Fast and Slow}~\cite{kahneman2011thinking}, introducing two modes of thought for the first time. Specifically, system 1 thinking, driven by instinct and experiences, is a near-instantaneous process and happens automatically, intuitively, and with little effort. On the other hand, system 2 thinking, driven by comprehensive logical reasoning, is slower, more conscious, more deliberative, and requires more effort. Contemporary mainstream AI technology such as deep learning and reinforcement learning is running on a fast track approaching system 1 thinking, and we refer to it as \textit{perceptive AI}. On the contrary, cognitive AI techniques aim to simulate system 2 thinking of human, providing more sophisticated, more logical and more understandable AI solutions. Many pioneering researchers have proposed their opinions and/or solutions for cognitive AI. For example, Gary Marcus proposes ``a hybrid, knowledge-driven, cognitive-model-based approach'' towards robust artificial intelligence \cite{marcus_next_2020}. Yoshua Bengio identifies system 2 deep learning as being able to ``understand, reason and generalize beyond training distributions'' \cite{bengio_msr-talk_2022} and proposes corresponding techniques such as causal machine learning and generative flow networks \cite{bengio_gflownet_2022} to achieve this goal.
\FigureSystemOneTwo
Figure~\ref{FigureSystem12} illustrates the concepts of system 1 and system 2, and explains their appropriate scenarios and tasks for investment decisions. Different from perceptive AI mainly applied in high-frequency and high-breadth trading tasks, cognitive AI gives quant opportunities to touch those high-capacity but low-frequency investment strategies, including value investing which buys/sells securities that appear underpriced/overpriced through multi-dimensional fundamental analysis and usually holds the position over months or even over years, as well as global macro investment which selects assets and establishes portfolio based on the interpretation and prediction of large-scale events related to national economies (interest rate trends, CPIs growth, GDPs growth, unemployment rates, policies, etc.), and international relations (inter-governmental relations, international trade, cross-border payments, etc.) to decide long/short positions in various equity, fixed income, currency, commodities, and futures markets.

\subsubsection{Causality Engineering}
Causal inference \cite{yao_survey_2021} and causal machine learning \cite{pearl_book_2018, scholkopf_toward_2021} have been regarded as one potential technical route towards AGI \cite{bengio_msr-talk_2022}. For investment tasks, understanding the true causal relationships among numerous factors is extremely challenging, due to the difficulty in enumerating all possible confounding variables that affect the future trend in one single experiment and give us a misleading conclusion. Therefore, we propose a potential solution through causality engineering, a proposed concept which aims to build and maintain a large-scale causal diagram database storing and managing all known/inferred causal effect relationships and potential confounding variables \cite{vanderweele_definition_2013} and causal variables. Different from regular alpha factor bases, a causal diagram database collects as many economic factors and computes their intrinsic conditional probability of causality between each other. Moreover, causality engineering will develop a series of algorithmic tools for different purposes, including testing the statistical significance of causal pairs, testing the effect of potential confounders, and searching significant causal pairs or causal clauses. Leveraging causality engineering, we expect that most main confounding factors disturbing the correct justification of true causal effect factors could be discovered in experiments and could be eliminated correctly using appropriate statistical methods.

\subsection{AI Risk Graph \& Systematic Modeling}
\label{subsec:AIRiskGraph}
The rapid growth of various types of financial big data brings us opportunities to model and analyze financial risk systematically, ranging from macroeconomy to micro-market. We propose the concept of \textit{AI risk graph}, a special financial knowledge graph for recording, computing, analyzing and forecasting financial risks at different hierarchical dimensions, including country, district, industry, sector, etc. 
\subsubsection{Risk Graph for Systematic Modeling}\label{subsubsec:RiskGraph}
An AI risk graph should be able to represent the causal dependency of different types of risk among public companies, private companies, banks, important individuals and many other economic entities, and represent risk transfer between various entities.
Such a graph could be used to systematically model various types of risks at different levels by computing the conditional probability of risk for a specified object (public company, sector, industry, etc) at certain conditions (time, market environment, debt leverage, etc).
Various statistical graphical models and machine learning algorithms could be applied to estimate risk conditional probability. Moreover, AI risk graph could help decompose observed risk value in a more scientific and more intuitive way in order to improve the interpretability of risk analysis and risk management. 
\subsubsection{Complex Risk Measure for Investment}
Classic risk management techniques such as BARRA \cite{barra_handbook} risk factor analysis model are used in measuring the overall risk associated with securities relative to the market risks. Specifically, the model decompose the overall risk into a number of exposures from different risk factors with a linearly additive interpretation. However, the limitation of linear risk modeling is obvious, in particular when the prediction model is built with highly nonlinear machine learning sample fitting. How to characterize, measure, and evaluate nonlinear risk is an important research topic in quant. In particular in nonlinear modeling scenarios. 
\begin{tightitemize}
    \item How to define a reasonable and practically useful nonlinear risk measure?
    \item How to make sure the nonlinear risk by definition exists and is identifiable?
    \item How to tell the difference between risk and noise in an extreme market?
    \item What proportion of overall risk could be explained by linear and nonlinear risk?
\end{tightitemize}

\subsection{Spatiotemporal Modeling}\label{subsec:SpatiotemporalModeling}
The data structure for stock modeling is typically a tensor with three orthogonal axes: time, stock, and factor. Traditionally, stock strategies are developed either along the time axis (called time-series modeling or temporal modeling) or along the stock axis (called cross-sectional modeling or spatial modeling). These two modeling types have significant differences in strategy development.
Specifically, cross-sectional modeling only compares relative strengths of investment signals within the same cross-section at some time point, and the signal strengths of the same stock at different time points are usually not comparable. Cross-sectional modeling has its advantage in neutralizing market risk automatically by longing top stocks and shorting bottom stocks ranked by cross-sectional trading signals. On the other hand, time-series modeling treats each stock individually, and the trading signals of different stocks at the same time point are not comparable.
\subsubsection{Unifying Cross-section \& Time-series}
A technically difficult but practically feasible idea is to merge cross-sectional modeling and time-series modeling in a unified framework, in order to absorb the advantages from both sides. For this aim, we attempt to provide a few tips on how to build a unified model and what potential difficulties may exist.
\begin{tightitemize}
    \item A unified model needs to update stock cross-sections in a very high frequency (in seconds, for example) in order to match the prediction paces of time-series modeling. 
    \item A unified model should be very selective in longing and shorting stock positions at each trading point in order to reduce the turnover rate of strategy so as reduce transaction costs. 
    \item The model should provide a hyperparameter for users to tune the balance between neutralization (alpha-oriented strategy) and absolute return (beta-oriented strategy), according to the design of the portfolio style and the demand of customers. 
\end{tightitemize}
\subsubsection{Spatiotemporal Graph for Quant}
Although a unified model aims to combine the advantage of cross-sectional modeling and time-series modeling, the relationships (or interactions) among stocks are hard to be incorporated in it. Spatiotemporal graph modeling could help complement this part of the information. In particular, a spatiotemporal graph \cite{atluri_spatio-temporal_2018, wang_deep_2020} can be embedded as a vector of latent factors using graph convolution or graph attention representation learning techniques, the latent vector could be concatenated with the factor vector used in unified modeling to improve the prediction performance.   

\subsection{Universal Modeling}\label{subsec:UniverssalModeling}
As data volume and computing power is growing rapidly, large-scale pretrained model has become one of the mainstream AI paradigms in practice. Successful examples such as BERT \cite{devlin_bert_2019}, GPT-3 \cite{brown_language_2020}, CLIP \cite{radford_learning_2021}, Codex \cite{chen_evaluating_2021} and DALL-E \cite{ramesh_zero-shot_2021} have demonstrated the effectiveness of large-scale pretrained models on a number of application domains such as machine translation, multi-modal understanding, creative content generation and so on. Many successful pretraining models have exhibited their universal power to help improve many downstream tasks with different tasks and data sources. We expect this phenomenon could be reproduced in quant applications. 

\subsubsection{Pretraining-Funetuning Paradigm}
The pretraining-finetuning paradigm in AI has demonstrated its success in natural language processing \cite{brown_language_2020} and computer vision \cite{dosovitskiy_image_2022,wang_image_2022}. By pretraining a large model on as much data in a self-supervised manner, one can obtain a model extracting the common information across various tasks and use it to achieve superior performances on a series of downstream tasks, compared to task-specific training. 
\figurePretrain
We think the pretraining-finetuning paradigm may be transferred to quant scenarios due to a number of reasons.
\begin{tightitemize}
    \item Many finance prediction tasks have sufficiently large volumes of data with diversified types and this helps train a large-scale neural network pretraining model during the pretraining stage.
    \item Many different quant prediction tasks may share the same pattern and information in modeling, especially for high-frequency trading where trading decisions mainly depends on market trends and traders' behaviors in the market. A general pretraining model could learn common patterns across different markets and even across different types of instruments, and these common patterns may help improve the downstream prediction tasks. 
    \item Pretraining-finetuning improves the automation of the modeling process and saves computational cost since we only need to maintain and update a pretrained main model which is a large neural network and adapt it to solve different downstream tasks by finetuninng the main model. 
\end{tightitemize}
\subsubsection{Challenge in Quant Pretraining}
The pretraining-finetuning paradigm has to face a number of difficulties if it is applied to quant. 
\begin{tightitemize}
    \item Investment data have extremely low signal-to-noise ratio, and make the pretraining process difficult to converge to a satisfying solution.  
    \item The design of the quant pretraining process must be careful to avoid using future information. Therefore, the masks in self-supervised pretraining or labels in supervised pretraining can be allocated only on the right-hand side like what GPT-3 does. 
    \item Pretraining should retain both flexibility and versatility. Specifically, we should figure out how to define appropriate labels or how to set appropriate masking structures. 
\end{tightitemize}

\subsection{Robust Modeling}
Data noise is the biggest issue in quant modeling. It leads to three problems that negatively affect the robustness and correctness of our model. 
\begin{tightenumerate}
    \item The signal-to-noise ratio in financial data is extremely low. For example, empirically, the information coefficient of an effective stock cross-sectional alpha model for daily trading lies around the level of 0.1, indicating a signal-to-noise ratio at 10\% level, and thus the low signal-to-noise ratio makes it more difficult for machine learning algorithms to tell true patterns out of false positive noise when the sample size is not sufficiently large. 
    \item It is difficult to correctly and robustly model the distribution of data with heavy noise. For example, the performance of many machine learning models is sensitive to model configurations as well as model hyperparameters when they are trained on very noisy financial data. This makes models very sensitive to sample outliers, time-series change points and the option of parameters, and thus increases the risk of overfitting.
    \item The noisy financial data do not follow an independent and identical distribution (i.i.d.) condition which is considered as the underlying assumption of most machine learning algorithms. Since market style is constantly changing as a consequence of an enormous number of factors, it is hard to find a model that persistently generates effective trading signals forever. Such noisy nature of financial data hinders the effectiveness of transferring existing algorithms directly, requiring both theoretical and practical innovations in the study of AI technology.
\end{tightenumerate}
To address these problems, we suggest researchers considering the following directions in the future.
\begin{tightitemize}
    \item \underline{Causal effect modeling} \cite{guo_survey_2020} can be applied to explore the causal effect between factors and financial decisions. By removing the interference from confounding variables, we have the chance to approximate the causal effect by estimating an underlying stable relationship between input factors and output decisions, and reduce the uncertainty of the model. Moreover, causal effect learning techniques could be applied to machine learning modeling in order to improve their out-of-distribution generalization. 
    \item \underline{Continual learning techniques} \cite{de_lange_continual_2022} aim to cumulatively acquire new information from data without forgetting the knowledge obtained from previous tasks. They could be used to improve out-of-distribution generalization \cite{shen_towards_2021} and thus increase model robustness through iteratively retraining with accumulated data over time and frequently updating the prediction model.
    \item \underline{Model ensemble technique} is useful in many scenarios to improve prediction stability by combining multiple single models, especially when single models have sufficiently differentiable outputs. Classic ensemble methods include bootstrap aggregating (bagging) \cite{breiman_bagging_1996}, boosting \cite{schapire_strength_1990, breiman_arcing_1998}, stacking \cite{wolpert_stacked_1992, breiman_stacked_1996}, Bayesian model averaging \cite{hoeting_bayesian_1999}, etc. 
    \item \underline{Model diversification for ensemble} expands the diversity of single models prepared to be combined, and the robustness of the ensemble model could benefit from diversification among single models. Various techniques could be applied to increase model diversity and discrepancy, such as model randomization (bootstrap, permutation, dropout \cite{srivastava_dropout_2014}, etc.), model neutralization, model decorrelation, mixture-of-experts \cite{masoudnia_mixture_2014}, etc. 
    \item \underline{Incorporating diversified data} from different sources and different patterns helps improve model robustness as well. For example, models trained from fundamental data extracted from financial statements usually complement well with models trained from stock prices and volumes, and a combination of the two types of models may help improve stability. 
\end{tightitemize}

\subsection{End-to-end Modeling}\label{subsec:End2EndModeling}
As we have introduced in \S\ref{sec:autoai_pipeline}, the traditional quant research pipeline consists of a number of steps (as shown in the blue blocks in Figure~\ref{fig:EndtoEndModeling}), and each step has its own optimization direction. For example, the factor mining module searches ``good'' factors aiming to find effective alpha signals with significant single-factor IC or back-test profit-and-loss ratio. The modeling module trains machine learning models aiming to minimize some sort of loss functions that measure the difference between labels and prediction outcomes. The portfolio optimization module allocates assets with optimal positions aiming to maximize some sort of value-at-risk (VaR) target. And the trading execution module computes the optimal order size in real-time aiming to minimize the market impact and reduce transaction costs.
We can see the optimization goals of these steps are somewhat different from each other.
For example, a ``good'' factor with high IC doesn't necessarily contribute positively to the final model output in a complicated nonlinear relationship. Therefore, it is natural to consider if there exists an end-to-end model taking meta factors as input and trading orders as output, and if its performance has an advantage compared with traditional separate modeling. 
\FigureEndtoEndModeling

\subsubsection{End-to-end Consistent Optimization}
It is a difficult task to build a consistently optimal model due to technical reasons. Firstly, end-to-end modeling is usually hard to be formulated as a typical supervised learning problem because there is no clear label definition for the whole pipeline. In particular, the steps of portfolio optimization and algorithmic trading are dynamic optimization problems relevant to Markov time dependency which are difficult to be incorporated into a supervised learning framework. Secondly, trading decisions in the order execution step are as frequent as a few seconds, while meta factors are usually updated every couple of minutes or days, and therefore it is difficult to construct ``$(x,y)$ pairs'' of samples for machine learning models. Thirdly, the noisy financial data make it difficult to train machine learning models difficult, and in particular, it is easy to be stuck at some local optimum during the training process, making the model sensitive to disturbance, noise and outliers. Finally, the computational cost of an end-to-end model is extremely high, which redirects to the challenge of computing power discussed in \S\ref{subsec_ComputingPower}. 

The opportunity for solutions to consistent modeling relies on the development of new machine learning paradigms satisfying the following requirements.
\begin{tightenumerate}
    \item The machine learning model should have a hierarchical structure supporting data and decisions at various time granularities including millisecond-level, second-level, minute-level, day-level, week-level, or even month-level. 
    \item The computational complexity should be controlled to finish the computation in an affordable time. 
    \item The optimization direction of the model should depend on predefined labels if available, and should depend on the effect of trading executions measured by typical evaluation criterion for quant. 
\end{tightenumerate}
Based upon the above analysis, we would recommend researchers interested in the problem to think from an existing baseline model under the reinforcement learning framework \cite{wang_alphastock_2019, wang_deeptrader_2021}, and pay more attention to fusing data, factors, decisions and executions in multiple time granularities. 

\subsubsection{Learning Unstructured Data}
Another problem in end-to-end learning is how to model unstructured data automatically. Many raw financial data are unstructured in format, says, they are impossible or inappropriate to be formatted as matrices, tensors, or data frames. Examples include order volume distributions on various bid/ask prices from limit order book data, interactions or causal relationships from economic behavior and event data, and investor emotion from news text data. How to effectively train a deep learning quant model end-to-end with a mixture of structured and unstructured financial data is still an open problem to us. New representation learning techniques (especially new information extraction and embedding methods) for unstructured data are needed for future investment research.

\section{Conclusion and Perspective}
\label{sec:conclusion}
In this article, we proposed the concept of Quant 4.0 which describes what the next-generation quant looks like. We think AI technology will become the core of quant research in the future. In particular, we claim that Automated AI, explainable AI, and knowledge-driven AI are three key components in Quant 4.0, and we think the development of these research areas will not only drive the evolution of quant research but also promote the progress of next-generation AI technology. We emphasize the importance of engineering in Quant 4.0. All three key components can not be implemented at scale without excellent system architecture and powerful computation infrastructure. Furthermore, we summarize 10 main challenges in quant technology, including one challenge about infrastructure, two challenges about financial data, three challenges about AGI technology and quant application, and four challenges about AI quant modeling. 

We must emphasize that Quant 4.0 is a dynamic concept and it will evolve and improve itself with the emergence of more and more new technology in the future. We encourage all researchers from related research areas to pay more attention to this interdisciplinary field which may become one of the demands driving the development of next-generation AI technology. To be honest, there is still a long way to go before achieving an ideal Quant 4.0 level since a lot of technical challenges in both artificial intelligence and quant engineering need to be solved. However, the rapid growth of AI technology in the past decade always moves beyond our expectations and brings us exciting progress, brand-new ideas, and more importantly, more confidence to explore this direction. Finally, we hope this perspective article could provide some insights to both academia and industry and encourage more researchers to study and contribute to this interdisciplinary field.  

\addcontentsline{toc}{section}{Acknowledgement}
\section*{Acknowledgement}
This work would not have been possible without the support of International Digital Economy Academy (IDEA). We would like to thank Prof. Qi Liu at Hong Kong University who provides helpful suggestions for the automated AI part of this article, thank Mr. Hang Yuan at IDEA Research for advice on the engineering part and thank Dr. Zhouchi Lin at IDEA Research for advice on the introduction part.  
\vskip 0.2in
\addcontentsline{toc}{section}{References}
\bibliographystyle{elsarticle-num}
\bibliography{references}
\clearpage
\addcontentsline{toc}{section}{Author Biographies}
\section*{\Large Author Biographies}

\authorBib{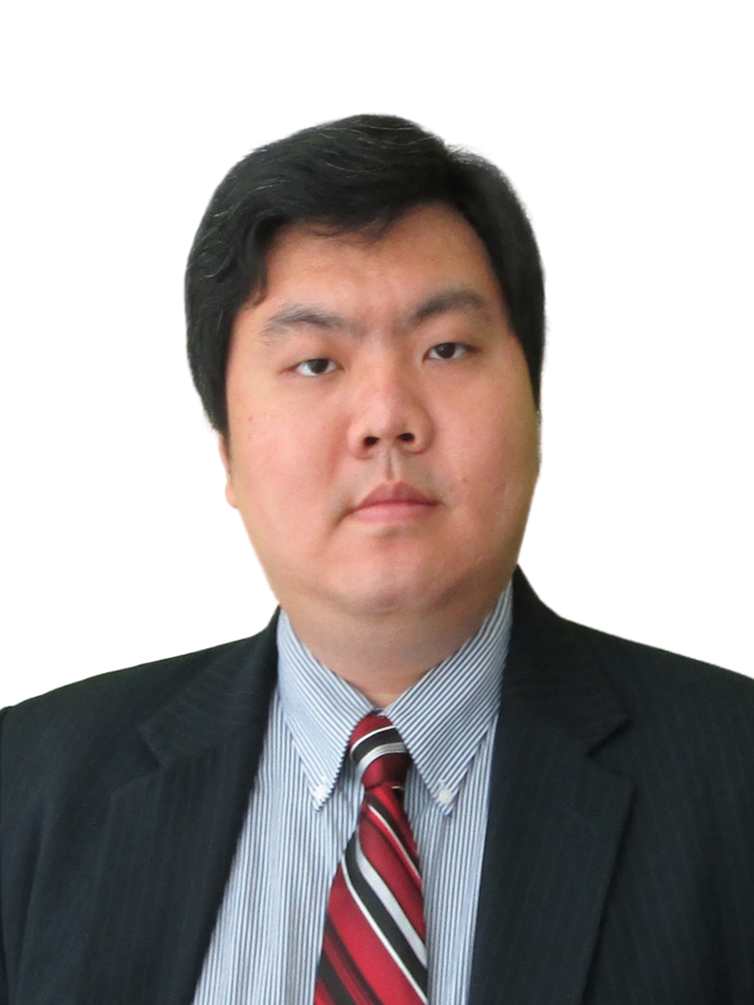}{Jian Guo}{is currently the Executive President and a Chief Scientist at International Digital Economy Academy (IDEA). As a founding member of IDEA, he also serves as the head of IDEA Research Center of AI Finance \& Deep Learning and a Professor of Practice at the Hong Kong University of Science and Technology (Guangzhou). Dr. Guo received his B.S. in mathematics from Tsinghua University, and received his Ph.D. in statistics from University of Michigan in 2011. He started his professorship (tenure-track) at Harvard University since 2011. He published a number of research papers in deep/reinforcement/statistical learning, including theory and application. Dr. Guo is one of the pioneering AI finance researchers, and is an entrepreneur in quantitative investment industry. }

\authorBib{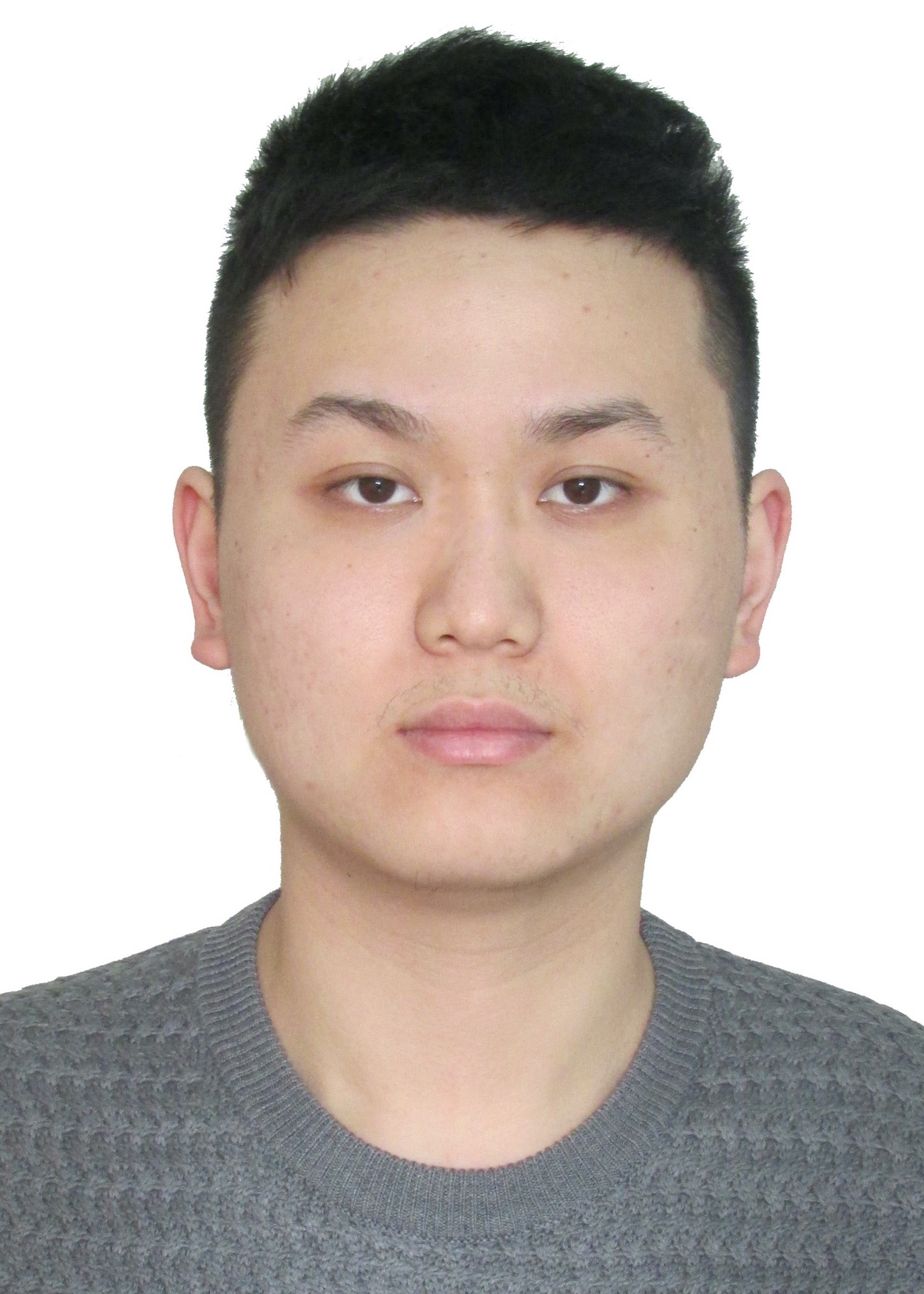}{Saizhuo Wang}{is currently a Ph.D. candidate in Department of Computer Science and Engineering at the Hong Kong University of Science and Technology, under the supervision of Prof. Harry Heung-Yeung Shum and Prof. Lionel Ming-Shuan Ni and working with Prof. Jian Guo. He received his bachelor of engineering degree in computer science from Chu Kochen Honors College at Zhejiang University. His research interest mainly in the interdisciplinary field of artificial intelligence and financial technology.}

\authorBib{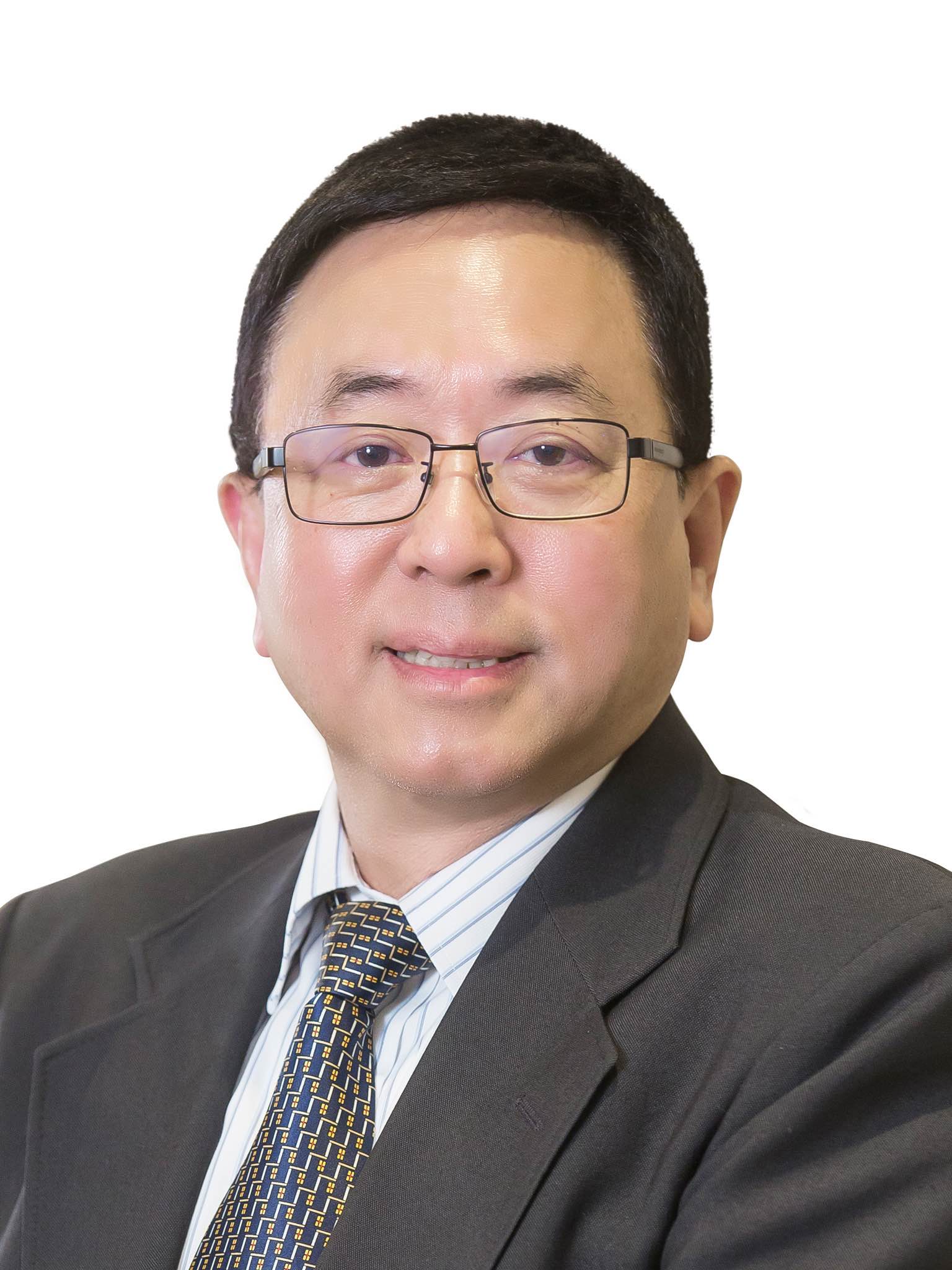}{Lionel M. Ni}{is currently the Founding President of the Hong Kong University of Science and Technology (Guangzhou) and Chair Professor in the university's Data Science and Analytics Thrust, as well as Chair Professor of Computer Science and Engineering at the Hong Kong University of Science and Technology. He is a Life Fellow of IEEE, and a Fellow of the Hong Kong Academy of Engineering Science. Prof. Ni’s research includes high-performance computing, mobile computing, wireless networking, big data, and intelligent computing. He has published three books and 350+ refereed journal and conference articles. He has chaired 30+ professional conferences and has received eight awards for authoring outstanding papers. Prof. Ni received his Ph.D. in Electrical Engineering from Purdue University in 1980.}

\authorBib{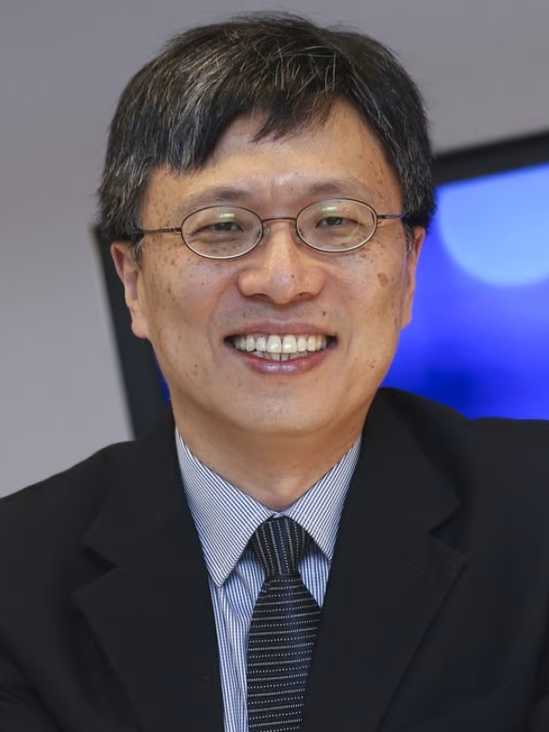}{Heung-Yeung Shum}{is the Founding Chairman of International Digital Economy Academy (IDEA), and a Professor-at-Large at the Institute for Advanced Study, Hong Kong University of Science and Technology. He is a Foreign Member of National Academy of Engineering of the US, International Fellow of Royal Academy of Engineering of the UK, ACM Fellow and IEEE Fellow. Until March 2020, he was the Executive Vice President of Microsoft Corporation, responsible for AI and Research. Dr. Shum received his Ph.D. in Robotics from School of Computer Science at Carnegie Mellon University.}

\end{document}